\documentclass{report}
\usepackage[final,nonatbib]{nips_2018}
\usepackage[utf8]{inputenc}
\usepackage{paralist}
\usepackage{hyperref}
\usepackage[dvipsnames]{xcolor}
\usepackage{url}
\usepackage{graphicx}
\graphicspath{{fig/}}

\usepackage{type1cm}        
\usepackage{makeidx}         
\usepackage{graphicx}        
\usepackage{multicol}        
\usepackage[bottom]{footmisc}
\usepackage{hyperref}
\usepackage{microtype}
\usepackage[linesnumbered,ruled,vlined]{algorithm2e}
\usepackage{bbm}
\usepackage[dvipsnames]{xcolor}
\usepackage{enumitem}
\usepackage{caption}
\usepackage{subcaption}

\usepackage{tabularx}
\usepackage{colortbl}

\definecolor{midnightgreen}{rgb}{0.0, 0.29, 0.33}

\usepackage{newtxtext}       %
\usepackage{newtxmath}       
\usepackage[square]{natbib}
\usepackage{multirow}

\title{Neural Approaches to \\ Conversational Information Retrieval}

\author{
{\bf Jianfeng Gao}\\
Microsoft\\
{\tt jfgao@microsoft.com}
\And
{\bf Chenyan Xiong}\\
Microsoft\\
{\tt cxiong@microsoft.com}
\AND 
{\bf Paul Bennett}\\
Microsoft\\
{\tt pauben@microsoft.com}
\And 
{\bf Nick Craswell}\\
Microsoft\\
{\tt nickcr@microsoft.com}
}

\usepackage{mwe}
\usepackage{amsmath}
\usepackage{amssymb}
\usepackage{wasysym}
\usepackage{bbm}

\begin{document}

\maketitle

\begin{abstract}
A conversational information retrieval (CIR) system is an information retrieval (IR) system with a conversational interface which allows users to interact with the system to seek information via multi-turn conversations of natural language, in spoken or written form. Recent progress in deep learning has brought tremendous improvements in natural language processing (NLP) and conversational AI, leading to a plethora of commercial conversational services that allow naturally spoken and typed interaction, increasing the need for more human-centric interactions in IR. As a result, we have witnessed a resurgent interest in developing modern CIR systems in both research communities and industry. 

This book surveys recent advances in CIR, focusing on neural approaches that have been developed in the last few years. 
This book is based on the authors' tutorial at SIGIR'2020 \citep{gao2020recent}, with IR and NLP communities as the primary target audience. However, audiences with other background, such as machine learning and human-computer interaction, will also find it an accessible introduction to CIR. 
We hope that this book will prove a valuable resource for students, researchers, and software developers.

This manuscript is a working draft. Comments are welcome.
\end{abstract}

\tableofcontents
\newpage

\chapter{Introduction}
\label{chp:introduction}

A conversational information retrieval (CIR) system is an information retrieval (IR) system with a conversational interface which allows users to interact with the system to seek information via multi-turn conversations of natural language (in spoken or written form). CIR provides a more natural user interface for information seeking than traditional, single-turn, search engines, and is particularly useful for search on modern devices with small or no screen.
 
CIR is a long-standing topic, which we can trace back to the 1960s. However, the research in CIR remained in its infancy until the 2010s due to the lack of large amounts of conversational data, sufficient natural language processing (NLP) technologies, strong commercial needs etc. Even today, popular commercial search engines, such as Google and Bing, provide only limited dialog capabilities. 
 
Recent progress in machine learning (e.g., deep learning) has brought tremendous improvements in NLP and conversational AI, leading to a plethora of commercial conversational services that allow naturally spoken and typed interaction, increasing the need for more human-centric interactions in IR. As a result, we have witnessed a resurgent interest in developing modern CIR systems in both research communities and industry. 

This book surveys recent advances in CIR, focusing mainly on neural approaches that have been developed in the last 10 years. This book is based on the authors' tutorial presented at SIGIR 2020~\citep{gao2020recent}, with IR and NLP communities as the primary target audience. However, audiences with other background, such as machine learning and human-computer interaction, will also find it an accessible introduction to CIR. We hope that this book will prove a valuable resource for students, researchers, and software developers.
 
 
\section{Book Organization}
\label{sec:book-organization}
 
In the rest of Chapter~\ref{chp:introduction}, we motivate the research of CIR by reviewing the studies on how people search, showing that information seeking can be cast in a framing of human-machine conversations. We then describe the properties of CIR, which led to the definition of a CIR system~\citep{radlinski2017theoretical} and a reference architecture (Figure~\ref{fig:cir-system-arch}) which we will describe in detail in the rest of the book.
To provide the background for the discussions, we also review recent advances in conversational AI~\citep{gao2019neural}, and brief the early works in CIR~\citep{croftCIRTutorial2019}.

In Chapter~\ref{chp:eval} we provide a detailed discussion of techniques for evaluating a CIR system -- a goal-oriented conversational AI system with a human in the loop. We present two approaches. System-oriented evaluation captures the user’s requests and preferences in a fixed dataset. User-oriented evaluation studies the interactions of a real user with the search system. Then, we describe two emerging forms of evaluation: CIR user simulation and responsible CIR. 
 
Chapters \ref{chp:CIR} to \ref{chp:proconv} describe the algorithms and methods for developing main CIR modules (or sub-systems), as shown in the reference architecture of Figure~\ref{fig:cir-system-arch}.
In Chapter \ref{chp:CIR} we discuss conversational document search, which can be viewed as a sub-system of the CIR system shown in Figure~\ref{fig:cir-system-arch}. We start with an introduction to the task and public benchmarks, review Transformation-based pre-trained language models which are the building blocks of many CIR modules, then describe the main components of the conversational search system, including contextual query understanding, sparse and dense document retrieval, and neural document ranking. 

In Chapter \ref{chp:qmds} we discuss algorithms and methods for query-focused multi-document summarization, which aim at producing a concise summary of a set of documents returned by the document search module in response to a user query. This is one of the key components of the result generation module of a CIR system.

In Chapter \ref{chp:c-mrc} we describe various neural models for conversational machine comprehension (CMC), which generate a direct answer to a user query based on retrieved query-relevant documents. Equipped with CMC, a CIR system can be used as an open-domain conversational question answering system. 

In Chapter \ref{chp:c-kbqa} we discuss neural approaches to conversational question answering over knowledge bases (C-KBQA), which is fundamental to the knowledge base search module of a CIR system. We introduce the C-KBQA task, describe the forms of knowledge bases and the open benchmarks, and discuss in detail a modular C-KBQA system that is based on semantic parsing.  
We then present a unitary (non-modular) system that is based on a Transformer-based language model which unifies the C-KBQA modules.

In Chapter \ref{chp:proconv} we discuss various techniques and models that aim to equip a CIR system with the capability of proactively leading a human-machine conversation by asking a user to clarify her search intent, suggesting the user what to query next, or recommending a new topic for the user to explore.

In Chapter \ref{chp:case-study} we review a variety of commercial systems for CIR and related tasks. We first present an overview of research platforms and toolkits which enable scientists and practitioners to build conversational experiences. Then we review historical highlights and recent trends in a range of application areas.
 
Chapter \ref{chp:conclusion} concludes the book with a brief discussion of research trends and areas for future work.

\section{How People Search}
\label{sec:how-people-search}


This section reviews search tasks and theoretical models of IR. A good survey of early works is presented by Marti Hearst in Chapter 2 of \cite{baeza2011modern}. A discussion on recent works is reported in~\cite{collins2017search}.  

\subsection{Information Seeking Tasks}
\label{subsec:information-seeking-tasks}

People search for various reasons, ranging from looking up disputed news or a weather report to completing complex tasks such as hotel booking or travel planning.

\citet{marchionini2006exploratory} group information seeking tasks into two categories, \emph{information lookup} and \emph{exploratory search}. Lookup tasks are akin to factoid retrieval or question answering tasks which modern Web search engines and standard database management systems are largely engineered and optimized to fulfill.

Exploratory tasks include complex search tasks, such as learning and investigating searches. Compared to lookup tasks, exploratory searches require a more intensive human-machine interaction over a longer-term iterative \emph{sensemaking} process \citep{pirolli2005sensemaking} of formulating a conceptual representation from a large collection of information. 
For example, learning searches involve users reading multiple information items and synthesizing content to form new understanding. In investigating searches, such as travel planning and academic research, users often take multiple iterations over long periods of time to collect and access search results before forming their personal perspectives of the topics of interest.  

Despite that more than a quarter of Web search are complex \citep{collins2017search}, modern Web search engines are not optimized for complex search tasks. Since in these tasks human users heavily interact with information content, the search engine needs to be designed as an intelligent task-oriented conversational system of facilitating the communication between users and content to achieve various search tasks. CIR systems we discussed in this book are mainly developed for complex searches. 

\subsection{Information Seeking Models}
\label{subsec:information-seeking-models}

Many theoretical models of how people search have been developed to help improve the design of IR systems. A classic example is the cognitive model of IR proposed by \citet{sutcliffe1998towards}, where the information seeking process is formulated as a cycle consisting of four main activities: 
\begin{enumerate}
    \item problem identification,
    \item articulation of information needs,
    \item query (re-)formulation, and
    \item results evaluation.
\end{enumerate}

Early models mainly focus on information lookup tasks which often do not require multiple query-response turns. These models assume that the user's information need is static and the information seeking process is one of successively refining a query until enough relevant documents have been retrieved. 

More recent models emphasize the dynamic nature of the search process, as observed in exploratory searches, where users learn as they search and adjust their information needs as they read and evaluate search results. A typical example is the \emph{berry-picking} model \citep{bates1989design}, which presents a dynamic search process, where a berry-picker (searcher) may issue a quick, under-specified query in the hope of getting into approximately the right part of the information space or simply to \emph{test the water}, and then reformulate her query to be more specific to get closer to the information of interest.

Some information seeking models focus on modeling the search strategy or policy that controls the search process. For example, \citet{bates1979information} suggests that searchers' behaviors can be formulated as a hierarchical decision making process which is characterized by search strategies (high-level policies) which in turn are made up of sequences of search tactics (low-level policies), and that searchers often monitor the search process, evaluate the cost and benefit of each step, and adjust the policies to optimize the return. Bates's model bears a strong resemblance to the cognitive model that motivates the development of the classic modular architecture of task-oriented dialog systems illustrated in Figure~\ref{fig:modular-task-bot-arch}, which will be discussed next.

The most relevant to CIR discussed in this book is the theoretical framework for conversational search, proposed by \citet{radlinski2017theoretical}.  Summarizing all the requirements of CIR, they propose five properties to measure the extent to which an IR system is conversational.
These properties are:
\begin{enumerate}
    \item \text{User revealment}: The system helps the user express or discover her information need and long-term preferences.
    \item \text{System revealment}: The system reveals to the user its capabilities and corpus, building the user’s expectations of what it can and cannot do.
    \item \text{Mixed initiative}: The system and user both can take initiative as appropriate.
    \item \text{Memory}: The user can reference past statements, which implicitly also remain true unless contradicted.
    \item \text{Set retrieval}: The system can reason about the utility of sets of complementary items.
\end{enumerate}

Then, taking together these properties, they define a CIR system as a task-oriented dialog system ``for retrieving information that permits a mixed-initiative back and forth between a user and agent, where the agent’s actions are chosen in response to a model of current user needs within the current conversation, using both short- and long-term knowledge of the user.''


\section{CIR as Task-Oriented Dialog}
A CIR process can be viewed as a task-oriented dialog, with information seeking as its task. This section describes the mathematical model and the classical modular architecture of task-oriented dialog systems, reviews how popular search engines support human-system interactions in Web search through the lens of a task-oriented dialog system, and summarizes the research topics being actively studies to make IR systems more conversational.   

The classical modular approach to building task-oriented dialog systems (or task bots) is motivated by the theories of human cognition. Cognition is formulated as an iterative decision making process \citep{marcus2020next}: organisms (e.g., humans) take in information from the environment, build internal cognitive models based on their perception of that information, which includes information about the entities in the external world, their properties and relationships, and then make decisions with respect to these cognitive models which lead to human actions that change the environment. Cognitive scientists generally agree that the degree to which an organism prospers in the world depends on how good those internal cognitive models are \citep{gallistel1990organization,gallistel2011memory}. 

\begin{figure}[t] 
\centering 
\includegraphics[width=0.99\linewidth]{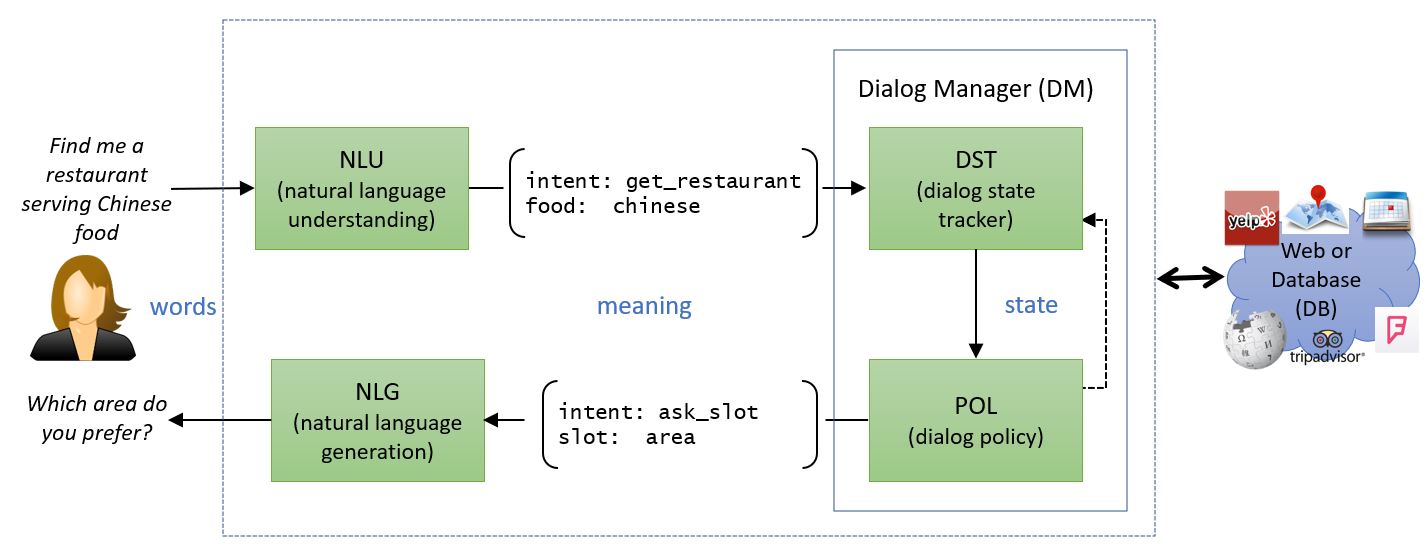}
\vspace{-2mm}
\caption{A modular architecture for multi-turn task-oriented dialog systems. It consists of the following modules: NLU (natural language understanding), DM (dialog manager), and NLG (natural language generation). DM contains two sub-modules, DST (dialog state tracker) and POL (dialog policy). The dialog system, indicated by the dashed rectangle, has access to an external database or Web collection. Adapted from \citet{gao2019neural}.} 
\label{fig:modular-task-bot-arch} 
\vspace{0mm}
\end{figure}

Similarly, the classical modular architecture of task bots, as shown in Figure~\ref{fig:modular-task-bot-arch}, views multi-turn conversations between a system and a human user as an iterative decision making process, where the system is (the agent of) the organism and the user the environment. The system consists of a pipeline of modules that play different roles in decision making. At each iteration, a natural language understanding (NLU) module identifies the user intent and extracts associated information such as entities and their values from user input. A dialog state tracker (DST) infers the dialog belief state (the internal cognitive model of the dialog system). The belief state is often used to query a task-specific database (DB) to obtain the DB state, such as the number of entities that match the user goal. The dialog state and DB state are then passed to a dialog policy (POL) to select the next system action. A natural language generation (NLG) module converts the action to a natural language response. Like cognitive scientists, dialog researchers also believe that the quality of a task bot depends to a large degree upon the performance of dialog state tracking (or its internal cognitive model), which had been the focus of task-oriented dialog research for many years
\citep[e.g.,][]{gao2019neural,young2013pomdp}.

\begin{figure}[t] 
\centering 
\includegraphics[width=0.99\linewidth]{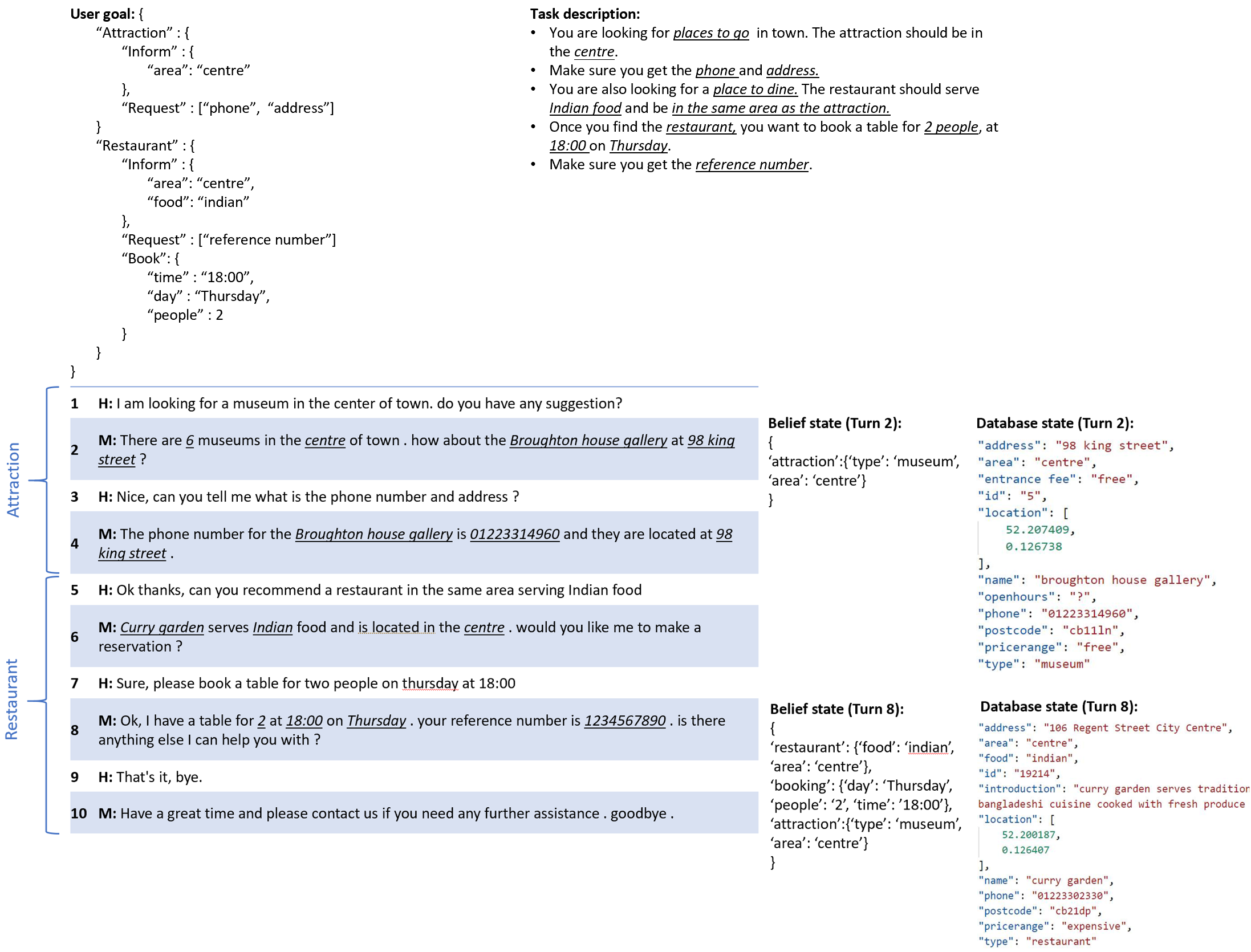}
\vspace{-2mm}
\caption{An example of a task-oriented dialog. (Top) A user goal and a task description. (Bottom) Multiple turns of user-system utterances, and the dialog belief states and database states at Turns 2 and 8. Adapted from \citet{gao2020robust}.} 
\label{fig:multiwoz-example} 
\vspace{0mm}
\end{figure}

Figure~\ref{fig:multiwoz-example} is a dialog of completing a multi-domain task produced by a user and a dialog system \citep{gao2020robust,peng2020soloist}. The user starts the conversation by asking for a recommendation of a museum in the center of town. The system identifies the user request, and provides a recommendation based on the search result from an \text{attraction} DB. Then, the user wants to book a table in a restaurant in the same area. 
We can see that through the conversation, the system develops belief states, which can be viewed as the system’s understanding of what the user needs and what is available in the DB. Based on belief state, the system picks the next action, either asking for clarification or providing the user with information being requested.  
This example also presents some challenges in conversational search. For example, the agent needs to understand that the ``same area'' refers to ``centre of town'', (i.e., the so-called co-reference resolution problem), then identifies a proper entity from the \text{restaurant-booking} DB to make the reservation.

The example shows that in a task-oriented dialog the user and the system play different roles.
The user knows (approximately) what she needs, but not what is available (in the DB).
The system, on the other hand, knows what is available, but not the user's information needs.
Dialog is a two-way process in which the user and system get to know each other to make a deal.

\begin{figure}[t] 
\centering 
\includegraphics[width=0.99\linewidth]{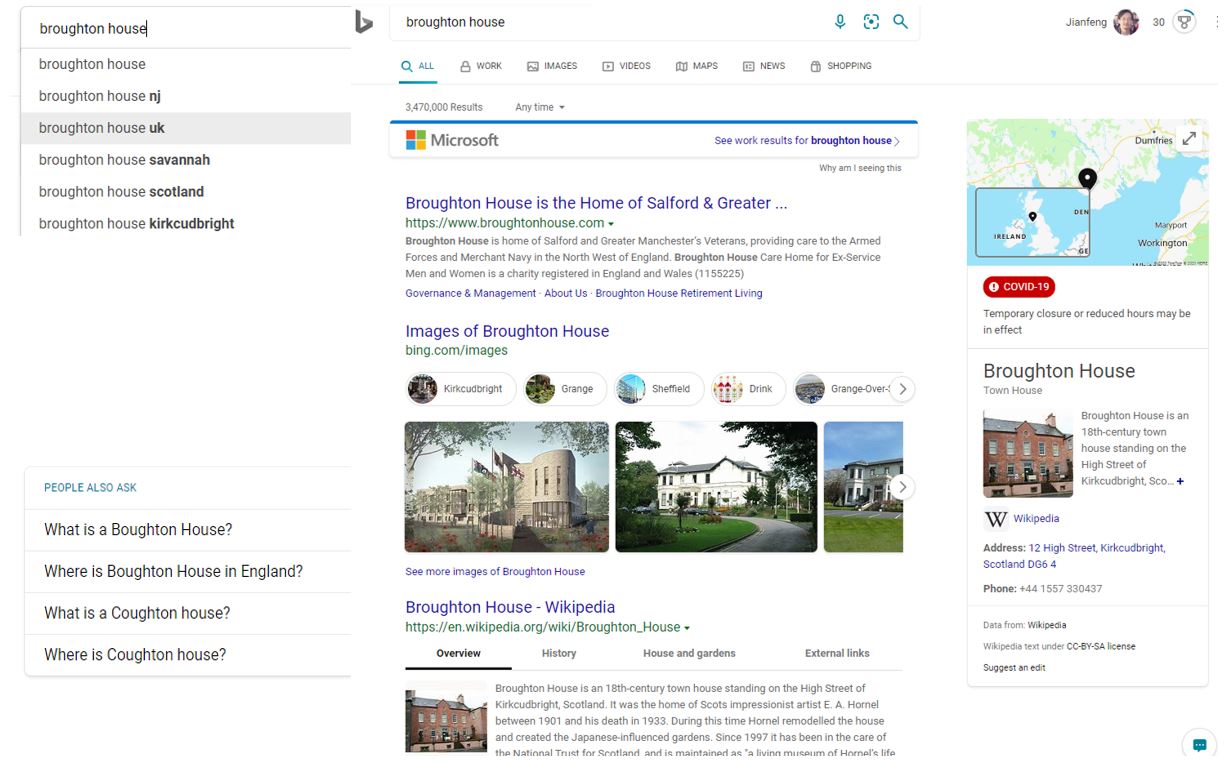}
\vspace{-2mm}
\caption{An example of Bing Web search interface. The system makes search more effective by auto-completing an input query (Top-Left), organizing search results in the SERP (Right), and suggesting related queries that people also ask (Bottom-Left).} 
\label{fig:bing-serp-example} 
\vspace{0mm}
\end{figure}

Now, consider the user-system interactions in Web search. 
The problem setting resembles that of task-oriented dialogs. 
The user knows (approximately) what she needs, but not what is available (on the Web). The system, on the other hand, knows what is available, but not the user's search intent.
Unlike the dialog system demonstrated in Figure~\ref{fig:multiwoz-example}, most popular search engines mainly treat Web search as a one-way information retrieval process. 
In the process, the user plays a proactive role to iteratively issue a query, inspect search results, and reformulate the query;
while the system, taking the Bing Search engine as an example as illustrated in Figure~\ref{fig:bing-serp-example}, plays a passive role to make search more effective by 
auto-completing a query, 
organizing search results in Search Engine Results Pages (SERP) and 
suggesting related queries that people also ask.

It is generally agreed that that effective information seeking requires multi-turn user-system interactions where both parties can take initiative as appropriate, and that users' search experiences can be significantly improved, especially on devices with small or no screen, if the system can play a more active role.  
Therefore, there have been many studies that explicitly model and support the interaction by
tracking belief state (user intent),
asking clarification questions,
providing recommendations, 
understanding natural language input, and generating natural language output, 
and so on.
In this book, we will discuss methods and technologies, with a focus on neural approaches developed in the last ten years, which can be incorporated into IR systems to make search experiences more conversational, effortless, and enjoyable. 
We start our discussion with a reference architecture of CIR systems in the next section.


\section{CIR System Architecture}
\label{sec:ch1-cir-system-architecture}

The development of CIR systems is more challenging than building typical task bots because information seeking is an open-domain task while most task bots, whose architecture is shown in Figure~\ref{fig:modular-task-bot-arch}, are designed to perform domain-specific tasks.

The dialog in Figure~\ref{fig:multiwoz-example} is domain-specific, consisting of two domains, attraction-booking and restaurant-book. 
For each domain, a set of slots are defined by domain experts. 
In the restaurant-booking domain, for example, slots like \texttt{restaurant-name}, \texttt{location}, \texttt{food-type}, \texttt{number-people}, \texttt{phone-number}, \texttt{date}, \texttt{time}, etc. are necessary.
Such a domain-specific dialog can be viewed as a process of \emph{slot-filling}, where a user specifies the values for some slots to constrain the search, such as \texttt{location} and \texttt{food-type}, and the system tries to look for entities (restaurants) in its DB which meet the constraints and fills the slots whose values are asked by the user such as \texttt{restaurant-name}.  
Since the slots are pre-defined, the possible actions that the task bot can take at each dialog turn can also be pre-defined. 
For example, the system response in Turn 6 in the dialog of Figure~\ref{fig:multiwoz-example}
\[
\text{``Curry Garden serves Indian food.''}
\]
is generated from the action template defined in the form of dialog act \citep{austin1975things} as:
\[
\texttt{inform(restaurant-name = ``...'', food-type = ``...'')}.
\]

\begin{figure}[t] 
\centering 
\includegraphics[width=0.99\linewidth]{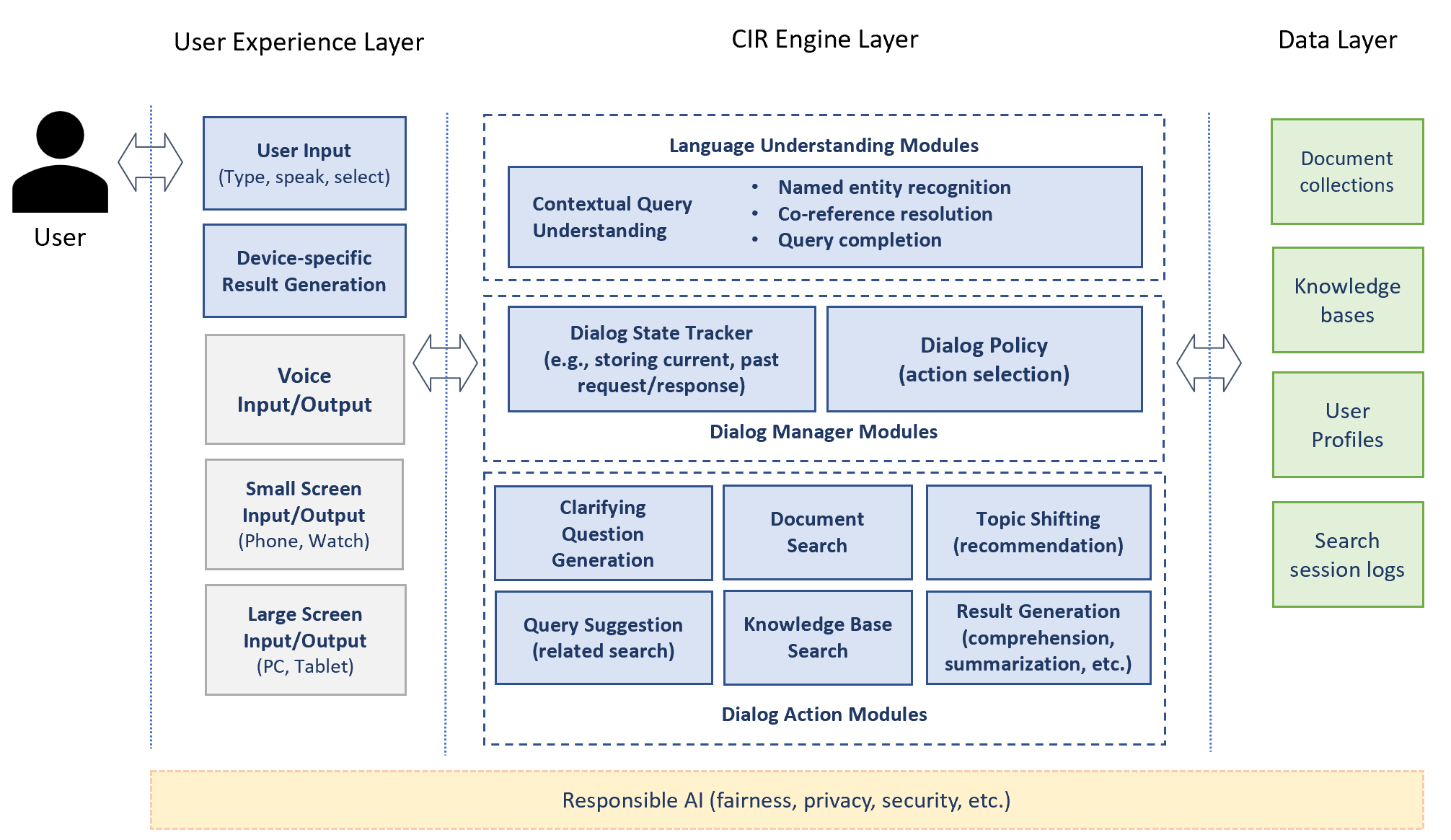}
\vspace{0mm}
\caption{A reference architecture of CIR systems.} 
\label{fig:cir-system-arch} 
\vspace{0mm}
\end{figure}

A CIR system, however, deals with open-domain dialogs with a much larger (or infinite) action space since users might search any information by issuing free-form queries. As a result, it is impossible for system designers to pre-define a set of actions that the system can take. Instead, we group actions into several action classes based on high-level user intents, such as asking clarifying questions, document search, shifting topics, and develop an action module for each class to generate responses.  

Figure~\ref{fig:cir-system-arch} shows a reference architecture of CIR systems that we will describe in detail in this book. The architecture is not only a significant extension of the task-oriented dialog system architecture in Figure~\ref{fig:modular-task-bot-arch} to deal with open-domain information seeking tasks, but also an extension to popular Web search engines in that it explicitly models multi-turn user-system conversations (e.g., via dialog manager modules). 
It consists of three layers: the CIR engine layer, the user experience layer, and the data layer.

\subsubsection*{CIR Engine Layer}

\begin{figure}[t] 
\centering 
\includegraphics[width=0.95\linewidth]{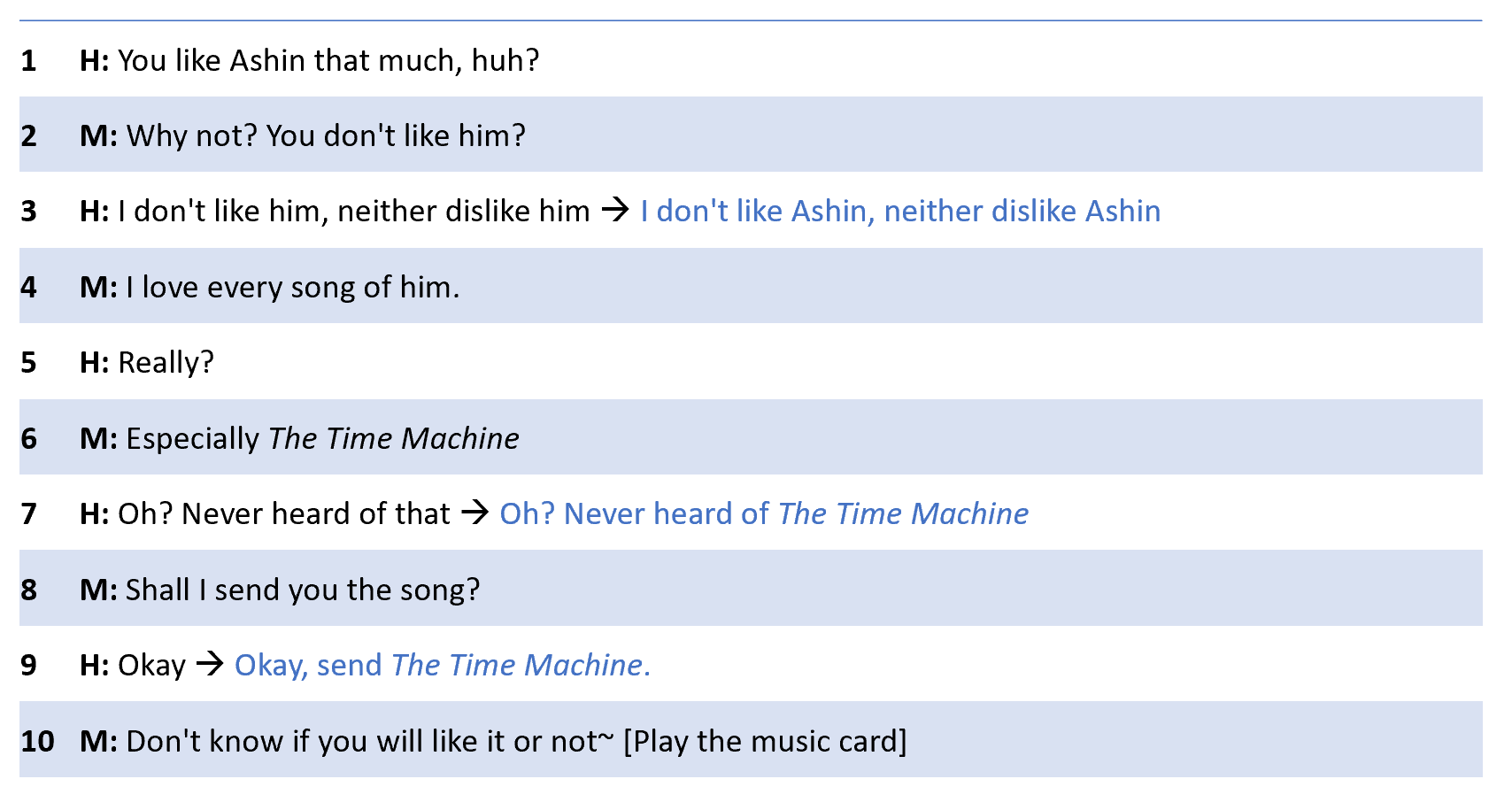}
\vspace{0mm}
\caption{An example conversational search session, where the contextual query understanding modules rewrite user queries into context-independent queries as indicated by the arrows. Adapted from \citet{zhou2020design}.} 
\label{fig:context-query-understanding-example} 
\vspace{0mm}
\end{figure}

This layer consists of all the major modules of the system. We group them into three categories. 
The first category is a set of language understanding modules for contextual query understanding.
Unlike the NLU module of task bots that performs slot filling based on a pre-defined set of slots, contextual query understanding conceptually acts as a query rewriter. 
In conversational search, ellipsis phenomena are frequently encountered. The rewriter uses contextual information of dialog history (within the same search session) to rewrite the user input query at each dialog turn to a \emph{de-contextualized} query which can be used to retrieve relevant documents via calling search APIs\footnote{The search API of most commercial search engines (e.g., Bing) only take a single query, not a dialog session, as input.} or retrieve answers from a knowledge base.
These modules need to identify common types of name entities (e.g., person names, places, locations etc.), replace pronouns with their corresponding entity mentions in a query (co-reference resolution), and complete the query.
As shown in Figure~\ref{fig:context-query-understanding-example}, user queries are rewritten to include context by, for example, replacing ``him'' in Turn 3 with the detected entity name ``Ashin'', ``that'' with ``The Time Machine'' in Turn 7, and adding “send The Time Machine” in Turn 9. 

The second category is a set of dialog manager (DM) modules. Similar to DM in task bots, it consists of a dialog state tracker, which keeps track of the current dialog state (e.g., by storing and encoding search interactions, including current and past user requests and system responses, in working memory), and a dialog policy, which selects action modules based on the dialog state. 
Since the dialog policy only picks action classes by activating the corresponding action modules (e.g., topic shifting) but allows the selected action module to generate primary actions (e.g., what the next topic to shift), the CIR system uses a hierarchical policy: (1) a high-level policy to select action modules, and (2) a set of low-level policies, one for each action module, to generate system responses.
In addition, we might include in DM a personalization module to tailor retrieval results to an individual's interests by incorporating information about the individual (e.g., from her user profile) beyond the specific query provided. Although personalization recently becomes a controversial topic due to the concerns about privacy and effectiveness
\footnote{It was reported that Google steps back on personalization because they found it did not really help improve search results and had a risk of cause users lose trust in Google due to data privacy. See e.g., https://www.cnbc.com/2018/09/17/google-tests-changes-to-its-search-algorithm-how-search-works.html},
the topic needs to be revisited in the context of CIR because a dialog agent is more describable to be personal \citep{zhou2020design,dinan2019second,roller2020open}. One challenge is around Responsible AI, which needs to be considered in developing all the CIR modules and data files, as illustrated in Figure~\ref{fig:cir-system-arch} (Bottom) to defend against harms and mitigate potential toxicity and bias. 

The third category is a set of action modules. Document search (to be discussed in Section~\ref{chp:CIR}) and knowledge base search (Chapter~\ref{chp:c-kbqa}) are the two most important modules. 
They are almost always activated for every user input during conversational search because the dialog state needs to be updated based on the retrieval results produced by these two modules. Other modules are only activated by the dialog policy for particular requests raised with respect to the dialog state. 
The Clarifying Question Generation module (Section~\ref{sec:clari_q}) is activated when DM believes it is necessary to help a user clarify her search intent. 
Query Suggestion (Section~\ref{sec:suggest_q}) is activated to improve the productivity of information seeking.  
Topic Shifting (Section~\ref{sec:shift_topic}) is activated for the system to lead the conversation by recommending a new topic for the user to explore next. 
Search Result Generation is only activated when the CIR system is ready to share the results with the user. This module consists of several sub-modules that generate results of different forms, including a machine reading comprehension module (Chapter~\ref{chp:c-mrc}) that generates a direct answer to a user question, a summarization module (Chapter~\ref{chp:qmds}) that generates a concise summary of the retrieved documents.


\subsubsection*{User Experience Layer}

This layer connects the CIR engine to various search platforms providing different user experiences, such as voice assistant, 
mobile search on the devices with small screens, and desktop search on the devices with big screens.
This layer also provides a direct search interface with a user, taking user input in different modalities (e.g., type, speak, select) and presenting search results in an appropriate form according to the user's device (e.g., a concise answer or a summary of search results for voice search and mobile search, or a SERP for desktop search as shown in Figure~\ref{fig:bing-serp-example}).

\subsubsection*{Data Layer}

This layer consists of collections of indexed documents and structured knowledge bases in forms of knowledge graphs, relational databases etc. from which the CIR system retrieves answers to user queries. It might also contains user profiles of registered users for personalized search. Most search engines also log search sessions for the purpose of evaluating and revising the system.

\subsection{An Example: Macaw}

\begin{figure}[t] 
\centering 
\includegraphics[width=0.90\linewidth]{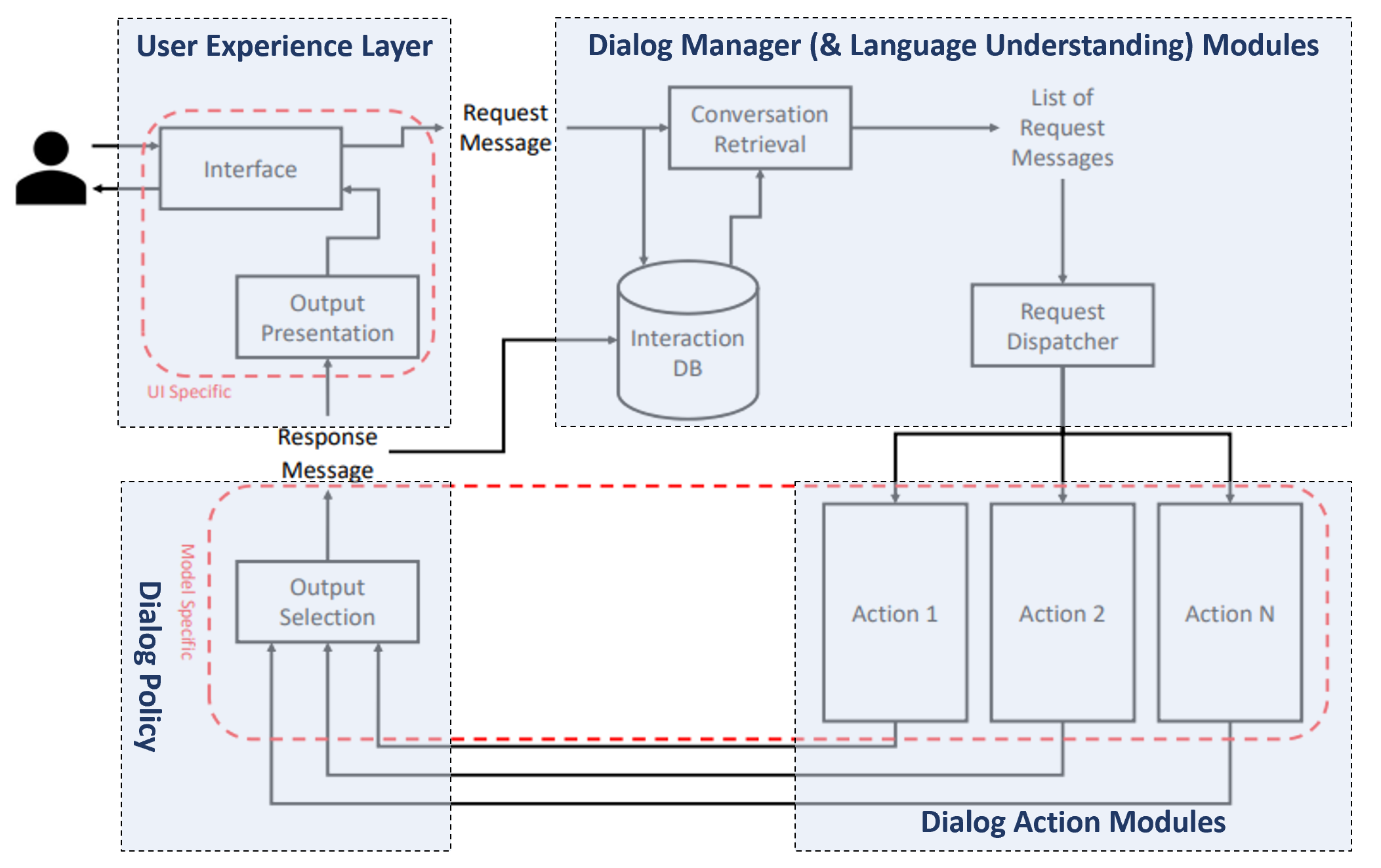}
\vspace{0mm}
\caption{The high-level architecture of Macaw, an open-source framework with a
modular architecture for CIR research \citep{sigir-2020:zamani-craswell-macaw}, viewed as an implementation of the reference architecture of CIR systems of Figure~\ref{fig:cir-system-arch}.} 
\label{fig:macaw-arch} 
\vspace{0mm}
\end{figure}

Macaw \citep{sigir-2020:zamani-craswell-macaw} is an open-source framework with a modular architecture for CIR research.  Macaw supports multi-turn, multi-modal, and mixed-initiative interactions, and enables research for tasks such as document retrieval, question answering, recommendation, and structured data exploration.  

The modular design of Macaw, as shown in Figure~\ref{fig:macaw-arch}, can be viewed as an implementation of the reference architecture of Figure~\ref{fig:cir-system-arch}.
For example, the user experience layer is implemented in Macaw as UI specific modules. The dialog action modules correspond to Macaw's Action modules. Dialog state is stored in the interaction DB in Macaw. Dialog policy is implemented by the Request Dispatcher and Output Selection modules in Macaw. For every user interaction, Macaw runs all actions in parallel, and then chooses one or combine multiple of these outputs to generate the response to the user. Therefore, the action selection function of dialog policy is implemented by the Output Selection module in Macaw. 

The modular design of Macaw also makes it easy to study new CIR methods. For example, we might add new retrieval modules or a new user interface to Macaw, and evaluate them in a batch mode or perform user studies in an interactive mode.

\section{Remarks on Building an Intelligent CIR System}

\citet{minsky1988society} portrays intelligence as a \emph{combination} of tiny pieces that are themselves not intelligent.  
The architecture of Figure~\ref{fig:cir-system-arch} presents a CIR system as an assemble of many modules. Each by itself is simply a function of mapping input to output, learned from pre-collected training data, and is not intelligent.

Although there has been a plethora of research on improving individual modules e.g., using modern deep learning approaches, as we will describe in the rest of the book, how to make these modules work together as an intelligent whole remains largely unexplored. For example, the dialog manager is the central controller of the dialog system. It is expected to be intelligent enough to guide a multi-turn search session on the fly, picking proper actions at each turn, such that the combined is optimal for achieving user's specific goals. In this setting, it is insufficient to optimize each module individually without the knowledge of user's goal.  However, most research on DM is carried out in the context of domain-specific task bots (as we will discuss in Section~\ref{sec:dialog-policy}), not for open-domain conversational tasks such as CIR.

The challenge is that we are building a CIR system for a fast-evolving world where user interests and world knowledge (e.g., Web documents) change constantly. Thus, the intelligence of the system are measured by how fast it can adapt to its environment. An intelligence CIR system has to explore very efficiently to adapt using very few experiences, since exploration often leads to sub-optimal user experience that incurs real-world cost. 

While humans are extremely good at learning new things quickly from seeing a couple of examples, AI systems often require a large amount of training data to adapt. Many scientists believe that the ability of making good credit assignment is among the most important that contributes to human's extremely efficient learning capability. As pointed out by \citet{minsky2007emotion}, to gain more from each experience (e.g., a logged search session), it is not wise to remember too many details, but only those aspects that were relevant to the (user's search) goals. Learning could be far more efficient if we assign the credit for the success or failure not only to the final action that led to the result, but to earlier choices the DM made that selected the winning strategy.

\section{Early Works on CIR}
This section briefly reviews early works on CIR before deep learning became popular, and draws the connection between early approaches to CIR and modern neural approaches which will be described in this book.
Most of these early works are based on traditional, keyword-based IR systems. But the concepts and design principles that were explored are instrumental in building new generations of CIR systems. Our discussion follows \citet{croftCIRTutorial2019}.

\subsection{System-Oriented and User-Oriented IR Research}

\begin{figure}[t] 
\centering 
\includegraphics[width=0.80\linewidth]{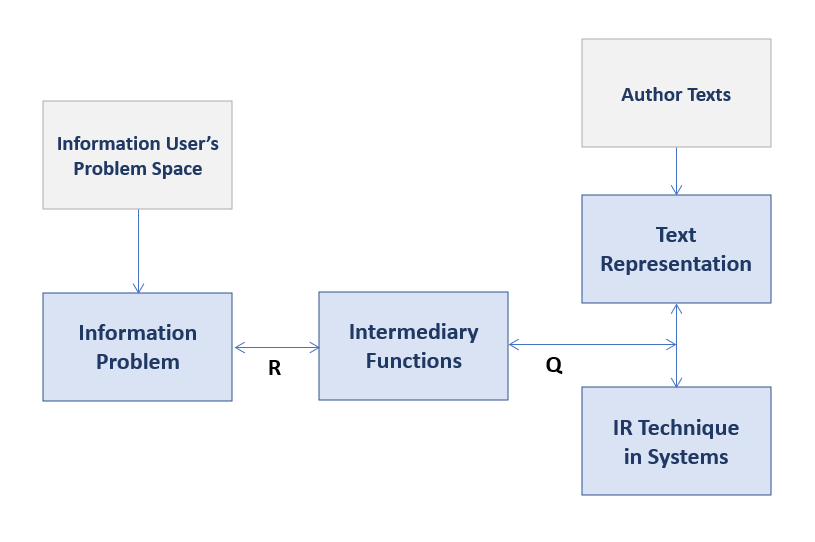}
\vspace{0mm}
\caption{IR research areas \citep{ingwersen1989modern}.} 
\label{fig:two-ir-research-areas} 
\vspace{0mm}
\end{figure}

\citet{ingwersen2006turn} present a review of IR research from the 1960s to the 1990s, grouping IR research into two complementary areas, system-oriented IR research and user-oriented IR research, as illustrated in Figure~\ref{fig:two-ir-research-areas}.

The system-oriented IR research, shown on the right side of Figure~\ref{fig:two-ir-research-areas}, focuses on authors' text, their representation, queries, and IR techniques that determines the behavior of the IR system.
The user-oriented IR research, shown on the middle and right-side components in Figure~\ref{fig:two-ir-research-areas}, focuses on user's problem spaces, information problems, request, interaction with intermediaries, interface design and query (re)formulation. 

Figure~\ref{fig:two-ir-research-areas} also demonstrates the cognitive view of IR with interaction as the central process \citep{ingwersen2006turn}: Interaction takes place between an (machine) intermediary and a searcher having a desire for information, whereby request formulation and reformulation (R) occurs. The query denotes reformation or transformation (Q) of the request for information, in the form required by the IR technique. At search time, the IR technique, user request and query (re)formulation interact.
As pointed out by \citet{ingwersen1992information}, ``the primary task of IR is to bring into harmony the cognitive structures of authors, system designers, and indexers with those of the intermediary (human or computer) and the user, in order to cope with the actually information need.'' 

The CIR methods discussed in this book aim to achieve the primary IR task by unifying system-oriented and user-oriented research using neural approaches. 

\subsection{System Architecture}

CIR has always been one of the ultimate goals of IR research since the development of the first generation IR systems, such as the SMART document retrieval system, in the 1960s.  

Before the development of any full-fledged CIR systems, there were studies of exploring the potential benefit of CIR.  
For example, \citet{lesk1969interactive} present a systematic evaluation of the effectiveness of various types of interactive search methods used in conjunction with SMART, including both pre-search (query reformulation) and post-search (relevance feedback) methods, and semi- or fully-automatic query reformulation methods. The authors argue that it is important in an interactive search environment to take into account the amount of effort required from the user to obtain satisfactory search results. Thus, the standard performance of fully-automatic retrieval operations is compared against the improvements obtainable through interactive (conversational) procedures at additional cost in user effort and computer time. 

The design of early CIR systems is inspired by the studies of human information search strategy. 
\citet{bates1979information} presents 29 search tactics or actions made by human searchers to further a search, in four categories: 
\begin{enumerate}
    \item Monitoring: Tactics to keep the search on track and efficient 
    \item File structure: Techniques for threading one's way through the file structure of information facility to desired file, source, or information within the source.
    \item Search formulation: Tactics to aid in the process of designing or redesigning the search formulation.
    \item Term: Tactics to aid in the selection and revision of specific terms within the search formulation.
\end{enumerate}

These tactics are fundamental to the design the CIR system action modules, as in the reference architecture of Figure~\ref{fig:cir-system-arch}.

\citet{brooks1986research} present a modular architecture based on a study of real-life human-user - human intermediary information interaction. The architecture consists of a list of functions that the intermediary (human or computer) needs to perform properly given the CIR context in order to provide appropriate responses to achieve an IR task. As shown in Table~\ref{tab:functions-of-retrieval-interface}, many of these functions resemble the modules shown in Figure~\ref{fig:cir-system-arch}.

\begin{table}[th]
    \caption{The functions of an intelligent interface for retrieval \citep{brooks1986research}.}
    \label{tab:functions-of-retrieval-interface}
\begin{tabular}{ll}
\hline
\textbf{Name of function}                & \textbf{Description}  \\
\hline
\begin{tabular}[c]{@{}l@{}} 1. Problem State \\ ~~~~~(PS) \end{tabular}
& \begin{tabular}[c]{@{}l@{}} Determine position of user in problem treatment process, \\
e.g. formulating problem, problem well-specified etc. \end{tabular} \\
\begin{tabular}[c]{@{}l@{}} 2. Problem Mode \\ ~~~~~(PM) \end{tabular}
& \begin{tabular}[c]{@{}l@{}} Determine appropriate mechanism capability,\\ 
e.g. document retrieval \end{tabular} \\
\begin{tabular}[c]{@{}l@{}} 3. User Model \\ ~~~~~(UM) \end{tabular}
& \begin{tabular}[c]{@{}l@{}} Generate description of user type, goals, beliefs, \\ 
knowledge, etc., e.g. graduate student, thesis etc. \end{tabular} \\
\begin{tabular}[c]{@{}l@{}} 4. Problem Description \\ ~~~~~(PD) \end{tabular}
& \begin{tabular}[c]{@{}l@{}} Generate description of problem type, topic, structure,\\ 
environment, etc. \end{tabular} \\
\begin{tabular}[c]{@{}l@{}} 5. Dialog Mode \\ ~~~~~(DM) \end{tabular}
& \begin{tabular}[c]{@{}l@{}} Determine appropriate dialog type and level for situation,\\ 
e.g., menu, natural language. \end{tabular} \\
\begin{tabular}[c]{@{}l@{}} 6. Retrieval Strategy \\ ~~~~~(RS) \end{tabular}
& \begin{tabular}[c]{@{}l@{}} Choose and apply appropriate strategies to knowledge\\ 
resource. \end{tabular} \\
\begin{tabular}[c]{@{}l@{}} 7. Response Generator \\ ~~~~~(RG) \end{tabular}
& \begin{tabular}[c]{@{}l@{}} Determine propositional structure of resource to the user,\\ 
appropriate to the situation. \end{tabular} \\
\begin{tabular}[c]{@{}l@{}} 8. Explanation \\ ~~~~~(EX) \end{tabular}
& \begin{tabular}[c]{@{}l@{}} Describe mechanism operation, restrictions etc. to user\\ 
as appropriate. \end{tabular} \\
\begin{tabular}[c]{@{}l@{}} 9. Input Analyst \\ ~~~~~(IA) \end{tabular}
& \begin{tabular}[c]{@{}l@{}} Convert input from user into structures usable by\\ 
functional experts. \end{tabular} \\
\begin{tabular}[c]{@{}l@{}} 10.Output Generator \\ ~~~~~(OG) \end{tabular}
& \begin{tabular}[c]{@{}l@{}} Convert propositional response to the form appropriate\\ 
to user, situation, dialog mode. \end{tabular} \\
\hline
\end{tabular}
\end{table}

\begin{figure}[t] 
\centering 
\includegraphics[width=0.80\linewidth]{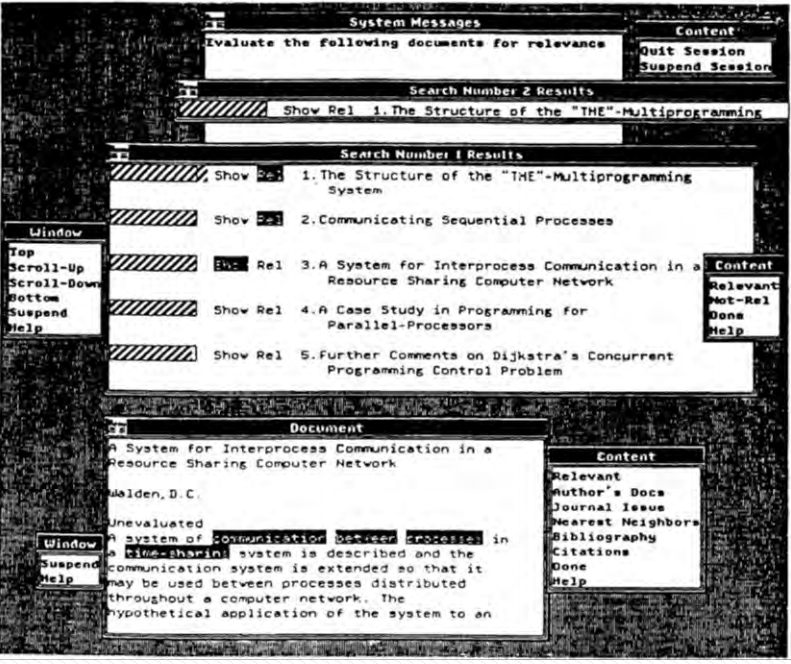}
\vspace{0mm}
\caption{A sample search session of I3R \cite{Croft1987I3RAN}.} 
\label{fig:I3R} 
\vspace{0mm}
\end{figure}

Inspired by these studies, \citet{Croft1987I3RAN} have developed the I3R system where search sessions are structured based on interactions with a human intermediary. The authors argue that the most effective method of improving the retrieval performance of an IR system is to acquire a detailed specification of the user's information need. Thus, I3R provides a number of facilities and search strategies, serving for different stages of a search session, where users can influence the system actions by stating goals they wish to achieve, by evaluating system output, and by choosing particular facilities directly, etc., as illustrated in Figure~\ref{fig:I3R}.

Similarly, \citet{oddy1977information} presents a prototype CIR system where the machine takes a more active role to form a specification of user's information needs through human-machine dialogs, without obligating the user to issue a precise query in the first place. The work demonstrates the potential of a (future) CIR system to support complex search activities, such as exploratory search, which require intensive human-machine interactions when forming and evolving search intents.

\citet{Schneiderman1992DesigningTU} shows that user search behaviors vary greatly due to experience among users, and thus the system is expected to interact with users accordingly to meet users' information needs. ``First-time users need an overview to understand the range of services...plus buttons to select actions. Intermittent users need an orderly structure, familiar landmarks, reversibility, and safety during exploration. Frequent users demand shortcuts or macros to speed repeated tasks and extensive services to satisfy their varied needs.''

Thus, \citet{shneiderman1997clarifying} propose a four-phase framework for user-interface design. The phases are (1) formulation (what happens before the user starts a search, including query formulation); (2) action (starting the search); (3) review of search results (what the user sees resulting from the search); and (4) refinement (what happens after review of results and before the user goes back to formulation with the same information need).

In what follows, we review early works on efficient search result representation, search refinement via user feedback and exploratory search, since they are critical to CIR and have attracted lots of research since the 1990s. 

\subsection{Search Result Presentation}

A good presentation of search results could significantly reduce the cognitive overhead of user judgments about the relevance of the retrieved documents. There have been lots of efforts to offer complementary views to the flat-ranked list returned by conventional search engines, including information visualization, query-biased summarization, search results clustering, and so on. Many of these technologies have been used by modern search engines (such as Bing and Google) to generate SERPs. 

\begin{figure}[t] 
\centering 
\includegraphics[width=0.80\linewidth]{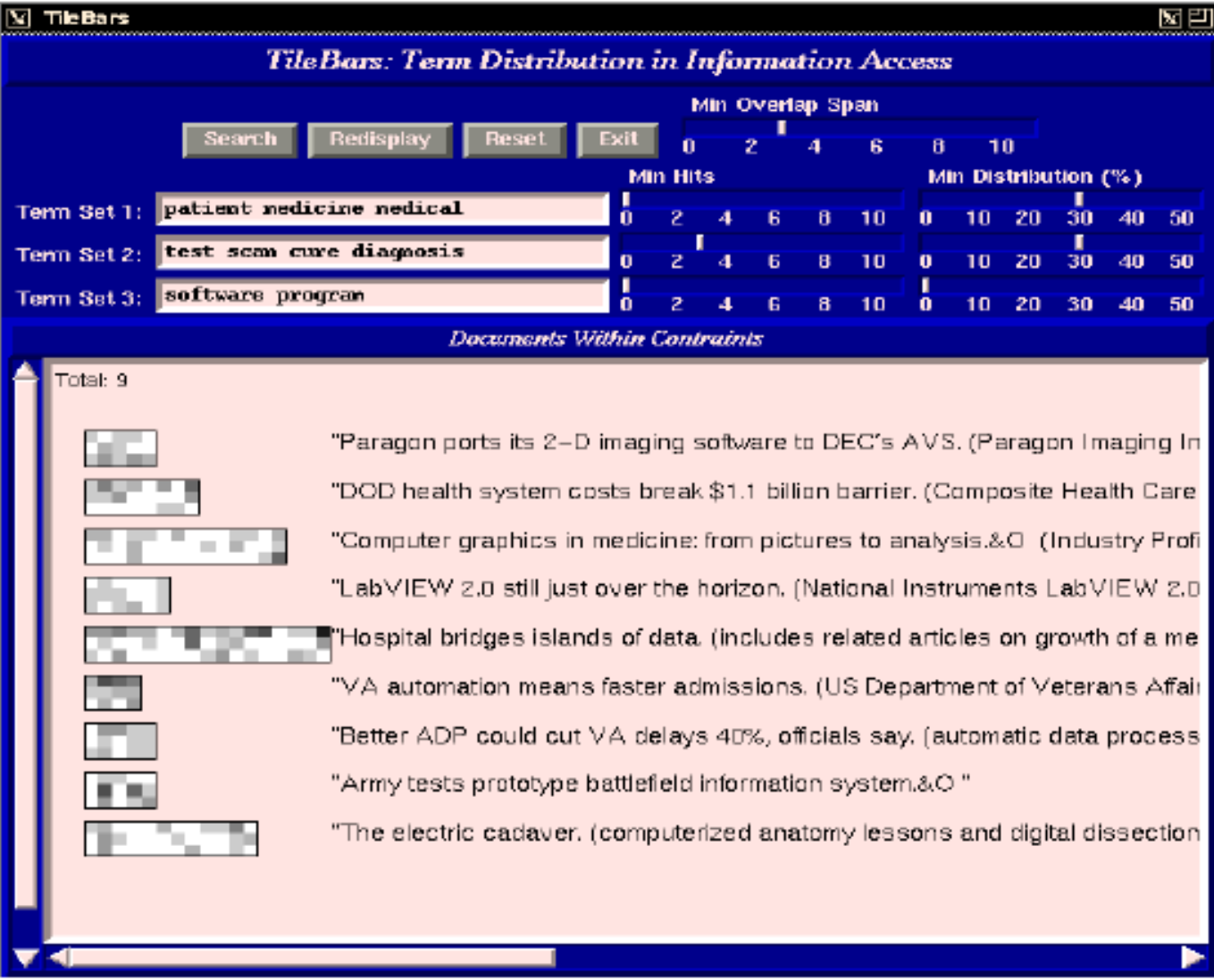}
\vspace{0mm}
\caption{The Tilebar search visualization \cite{Hearst1995TileBarsVO}. Retrieved documents are shown as rectangles, text segments as squares, the darkness of a square indicates the frequency of query terms in the segment, and the patterns in a column can be quickly scanned and deciphered, helping users in making relevance judgments about the documents.} 
\label{fig:tilebar} 
\vspace{0mm}
\end{figure}

Most search results are text documents. But reading text is a cognitive burden since this has to be done linearly. Visual representations of text enable users to scan textual information more quickly and effectively, and thus have been widely explored in presenting retrieved documents. One of the best known is TileBars \citep{Hearst1995TileBarsVO}, which allows users to simultaneously view the relative length of the retrieved documents, the relative frequency of the query terms, and their distributional properties with respect to the document and each other. As shown in Figure~\ref{fig:tilebar}, retrieved documents are shown as rectangles, text segments as squares, the darkness of a square indicates the frequency of query terms in the segment, and the patterns in a column can be quickly scanned and deciphered, helping users in making relevance judgments about the documents. 
Variations of TileBars is proposed by \citet{Heimonen2005VisualizingQO,Hoeber2006ACU}. 
A good survey of visualization in search interfaces is \citet{hearst2009search}.

In addition to Visualization, another widely used approach is query-biased summarization. By producing a summary focused on the topic guided by a user's query, this approach aims to minimize the user’s need to refer to the full document text, while at the same time to provide enough information to support her retrieval decisions. An early investigation of the approach is \citet{tombros1998advantages}. The authors show that compared to full texts and static pre-defined summaries (e.g., the title and first few lines of a document), displaying query-biased summaries of retrieved documents significantly improves the accuracy and speed of user relevance judgments. 

When a user query is exploratory or ambiguous, the search result of Web search engines is often a mix of documents retrieved based on different subtopics or interpretations of the query, implying that the user has to go through the long list to identify the items of her interest. Search results clustering is a family of methods of grouping the results returned by a search engine into a hierarchy of labeled clusters to help users efficiently navigate search results \citep{croft1978organizing, Cutting1992ScatterGatherAC, allen1993interface, croft1996evaluation, wang2006exploring, carpineto2009survey}. Compared to regular document clustering, search results  clustering in the CIR setting presents unique challenges \citep{wang2006exploring}, including:

\begin{itemize}
    \item Speed: As the clusters are generated on-the-fly, it is crucial that clustering does not introduce apparent delay to the search. Thus, many methods use query-biased summaries (or snippets) rather than the full documents for clustering. 
    \item Coherent clusters: Topically similar documents should be grouped together. Overlapping is allowed due to documents having multiple topics. 
    \item Ease-of-browsing: Concise and accurate cluster descriptions are provided.
\end{itemize}

\begin{figure}[t] 
\centering 
\includegraphics[width=0.90\linewidth]{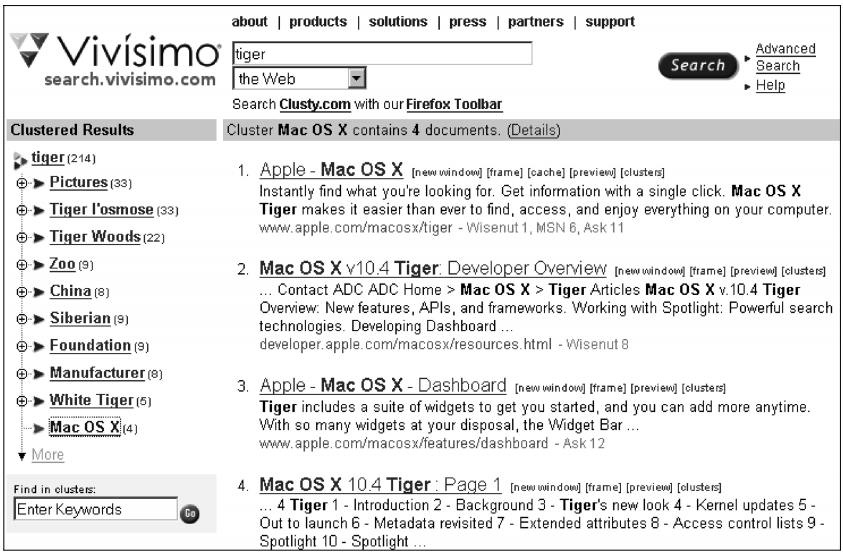}
\vspace{0mm}
\caption{Clustered search results for the query ``tiger''. A screenshot from the commercial engine Vivisimo. Figure credit: \citet{carpineto2009survey}.} 
\label{fig:vivisimo} 
\vspace{0mm}
\end{figure}

The IR systems that perform clustering of Web search results, also known as clustering engines, have been commercialized since the 1990s \citep{carpineto2009survey}. The first commercial clustering engine is Northern Light, launched in the end of the 1990s.  It is based on a pre-defined set of categories, to which the search results were assigned. 
An early major breakthrough is made by Vivisimo, whose clusters and cluster descriptions (labels) are dynamically generated from the search results, as shown in Figure~\ref{fig:vivisimo}. Vivisimo won the ``best meta-search engine award'' by SearchEngineWatch.com from 2001 to 2003. However, up to today, major search engines, such as Google and Bing, have not fully embraced clustering engines, except for some vertical search results, due to the latency and quality issues. 

\subsection{Relevance Feedback Interactions}

An important research topic of CIR is how to make use of user's feedbacks. Consider the following iterative relevance feedback process \citep{manning2008introduction}:

\begin{enumerate}
    \item The user issues a query.
    \item The system returns an initial set of retrieval documents.
    \item The user gives feedback on the relevance of documents.
    \item The system revises the retrieval results based on user feedback.
    \item Go back Step 3 until the user satisfies with the results.
\end{enumerate}

In Step 3, the retrieval results can be revised via query refinement or improving the ranking model directly using user feedback as training data. 

The query refinement approach is motivated by the idea that while it may be difficult for a user to formulate a good query when she does not know the collection well (e.g., in exploratory search), it is easy to judge the relevance of returned documents. The iterative query reformulation process can be effective in tracking a user's evolving information need \citep{manning2008introduction,baeza2011modern}.
The Rocchio algorithm \citep{rocchio1971relevance} is a classic query refinement method where the feedback data is used to re-weight query terms in a vector space. 

In probabilistic ranking models \citep{robertson1976relevance,robertson1977probability,robertson2009probabilistic}, user feedback data can be used as additional training data to revise the models' estimate of the likelihood of a document being relevant via either probabilistic term re-weighting or probabilistic term expansion. 

The click data has been used to improve the ranking models of modern search engines. For example, Deep Structured Similarity Models (DSSMs) \citep{huang2013learning,shen2014latent,gao2014modeling} are neural network ranking models that are trained and revised using large amounts of click data. Using contrastive learning, these models are learned to rank higher the documents with user clicks given a query than the documents without user click. 

\subsection{Exploratory Search}

As discussed in Section~\ref{sec:how-people-search}, exploratory search is the most important information seeking task that motivates the development of CIR systems.
An exploratory search process consists of learning and investigating \citep{marchionini2006exploratory}, requiring intensive interactions between a user and the machine. In exploratory search, a user usually submits a tentative query to navigate proximal to relevant documents in the (Web) collection, then explores retrieved results to better understand what is available it, selectively seeking and passively obtaining cues about her next actions \citep{white2009exploratory}.

Exploratory search systems can be viewed as an instance of CIR systems. Unlike traditional IR systems that return a list of documents or an answer in response to a query, exploratory search systems need to instigate cognitive change through learning and improved understanding \citep{white2009exploratory}, supporting aspects of sense-making \citep{pirolli2005sensemaking} and berry-picking \citep{bates1989design}. 

\citet{white2009exploratory} give a very readable lecture of exploratory search, reviewing the research up to 2009 and proposing eight features that guide the design of exploratory search systems. These features, as listed below, are fundamental to building CIR systems.

\begin{enumerate}
    \item Support querying and rapid query refinement: Systems must help users formulate queries and adjust queries and views on search results in real time. 
	\item Offer facets and metadata-based result filtering: Systems must allow users to explore and filter results through the selection of facets and document metadata. 
	\item Leverage search context: Systems must leverage available information about their user, their situation, and the current exploratory search task. 
	\item Offer visualizations to support insight and decision making: Systems must present customizable visual representations of the collection being explored to support hypothesis generation and trend spotting. 
	\item Support learning and understanding: Systems must help users acquire both knowledge and skills by presenting information in ways amenable to learning given the user’s current knowledge and skill level
	\item Facilitate collaboration: Systems must facilitate synchronous and asynchronous collaboration between users in support of task division and knowledge sharing. 
	\item Offer histories, workspaces, and progress updates: Systems must allow users to backtrack quickly, store and manipulate useful information fragments, and provide updates of progress toward an information goal. 
	\item Support task management: Systems must allow users to store, retrieve, and share search tasks in support of multi-session and multi-user exploratory search scenarios.
\end{enumerate}


\chapter{Evaluating Conversational Information Retrieval}
\label{chp:eval}

A key challenge in CIR is evaluation, to compare different CIR designs and approaches, and carefully consider the different forms of evaluation to draw conclusions on what works best. In each turn of a conversation, the system updates dialog state and performs action selection, choosing what to present from a variety of action responses (Figure~\ref{fig:cir-system-arch}). Some actions involve retrieval of data types such as phrases, passages, documents, knowledge base entities, and multimedia. There are also non-retrieval actions, such as initiating chit chat, recommending follow-up questions that the user might ask, and asking the user a clarifying question.



The user’s satisfaction with this iterative process is our primary concern. How do we know whether the conversational system is working well? Is it selecting the right actions? Is each action producing optimal outputs? Does the overall approach give the user a positive, useful, and friction-free CIR experience?



\begin{table}[ht]
    \caption{Forms of observation and experiment in CIR. Adapted from \citep{dumais2014understanding}.}
    \label{tab:evaluation_types}
\begin{tabularx}{\linewidth}{l|X|X|X}
\hline
 & & \textbf{Observation}  & \textbf{Experiment}  \\ 
\hline
 & \textbf{Lab study:} Bring in users for detailed instrumentation and observation. & Instrument, observe, and question the users, to understand their experience. & Test a hypothesis about conversational search in a controlled experiment. \\ \cline{2-4} 
\parbox[t]{2mm}{\multirow{3}{*}{\rotatebox[origin=c]{90}{\textbf{User-oriented}}}} & \textbf{Field study:} Observe real-world usage, but with extra instrumentation and/or surveys. & Develop new hypotheses about the search experience, by observing and questioning users. & Test a hypothesis about how users will respond, by changing their experience. \\ \cline{2-4} 
 & \textbf{Log study:} Analyze the behavior of a user population, without extra instrumentation or surveys. & Mine the logs of a CIR system for patterns that suggest a new hypothesis. & Test a hypothesis. A/B randomized controlled trial. \\ \hline
 & \textbf{End-to-end evaluation:} Evaluate the full functionality of a CIR system. & \cellcolor{lightgray} ~ & Compare systems, considering all system components and actions, in conversational context. \\ \cline{2-4} 
\parbox[t]{2mm}{\multirow{3}{*}{\rotatebox[origin=c]{90}{\textbf{System-oriented}}}}  & \textbf{Retrieval evaluation:} Evaluate a CIR component that retrieves results. & \cellcolor{lightgray} & Compare retrieval components, in conversational context, on a metric such as NDCG. \\ \cline{2-4} 
 & \textbf{Other component evaluation:} Evaluate components with non-retrieval functionality. & \cellcolor{lightgray} & Action selection. Asking clarifying questions. Sensitive content detection. Dialog state tracking.  \\ \hline
\end{tabularx}
\end{table}

\section{Forms of Evaluation}

Research in information retrieval has long embraced the fact that we are evaluating an interactive system with a human in the loop. The two main approaches for capturing the human element are summarized in Table~\ref{tab:evaluation_types} as system-oriented evaluation and user-oriented evaluation.

\emph{System-oriented evaluation} captures the user's requests and preferences in a fixed dataset. The dataset can be shared and reused. There can be datsets for retrieval components and non-retrieval components of the system, and potentially even action selection among both types of action. When using this form of evaluation, researchers focus on developing models and systems that best match the user preferences that were captured in the dataset. 

\emph{User-oriented evaluation} observes the interactions of a real user with the search system \citep{dumais2014understanding}. The users can be in a lab, which allows detailed instrumentation and interpretation of their use of the system. The users can be in a field study, where they do real-world tasks in the wild, but still can be probed for extra details through surveys or diary studies. If the conversational system has been deployed and is in use by real users, the user-oriented study can also be carried out based on log data. The logs capture user behavior, and can be collected in large quantity, but then a key challenge is how to interpret their behavior for use in understanding and evaluating the system. In each of these cases it is possible to carry out a randomized controlled experiment, where users are assigned a different experience to test a hypothesis, or to just carry out an observational study.

An important aspect of these evaluation methods is that they inform each other. In traditional information retrieval, for example, the lab study is done using a system that was probably trained using a system-oriented evaluation dataset. The dataset might be based on user-oriented data, if the queries were sampled from real logs or the dataset's definition of what makes a good response is based on insights from past user studies. In the lab study we may discover new insights into user behavior and preferences. Those insights can be fed back into the development of new system-oriented datasets. Building datasets that are grounded in reality makes us more confident that our results have external validity, meaning that the results in a research study are predictive of real-world performance. This process is intended to develop our understanding of the human element in CIR, with the hope that evaluation results are indicative of real-world performance of a CIR system.

Section~\ref{sec:eval_system} is an overview of system-oriented evaluation. 
Section~\ref{sec:eval_user} is an overview of user-oriented studies. By involving real users, these studies may yield insights that are difficult to capture with a fixed system-oriented experiment. Section~\ref{sec:eval_emerging} introduces some emerging topics in evaluation, such as detecting sensitive content that is offensive, abusive or hateful, when moderating the responses of a conversational system.

\section{System-oriented Evaluation}
\label{sec:eval_system}

System-oriented evaluation involves using a fixed dataset that captures the input and desired output of the retrieval system or one of its components. First we present examples of evaluation without conversational context. Then we show how these can be adapted to capture conversational context, and list a number of example datasets. Finally we describe system-oriented evaluation of non-retrieval components.

\subsection{Evaluating retrieval}
\label{sec:eval_system_retrieval}

For some components of a CIR system (Figure~\ref{fig:cir-system-arch}), system-oriented evaluation can be carried out by considering the current request without any context, as though it were a user's first request. Such evaluation can be useful because handing a single request is a capability that any retrieval system should have. Two examples of this kind of retrieval evaluation without context are a document search (action) module, which can be evaluated using an IR test collection dataset, and a result generation module, which can be evaluated using a reading comprehension dataset.

System-oriented evaluation of a document search component is the same as evaluation with a traditional information retrieval test collection. A test collection comprises a set of documents, a set of user queries, and ground truth labels indicating query-document relevance \citep{voorhees2005trec}. In a typical research test collection the document corpus is fixed, and the retrieval system is expected to produce a ranking of documents for each query, with the goal of having the most relevant documents at the top of the list. 

If there are multiple grades of relevance (irrelevant, partially relevant, relevant, highly relevant) and multiple items in the ranked list, there are a variety of metrics which can be used to generate a list-level measure of quality. Here we choose one well-known example, which is Normalized Discounted Cumulative Gain (NDCG). The metric is Cumulative because it sums over rank positions. The Discount is based on the rank $i$ of the document, giving more importance to documents at the top of the ranking, since users will see those first. The Gain is based on the relevance judgment the $i$th result $G(rel_i)$, indicating how valuable the document is for the user. This gives us NDCG:
\begin{equation*}
\mathrm{NDCG}@k = Z * \sum\limits_{i=1}^k \frac{G(rel_i)}{\log_2 (i+1)}
\end{equation*}
The Normalizer is a value $Z$ that means a perfect ranking, that is ranked in descending order of Gain, has $\mathrm{NDCG}@k = 1$. Like most IR metrics, NDCG rewards the system for retrieving results with the most valuable results at the highest ranks.

The magnitude of mean NDCG and other IR metrics can vary depending on the application setting and query sample. Rather than focusing on magnitude, studies usually calculate NDCG for two systems on the same set of queries. Their performance is typically compared using a paired statistical significance test, to compare mean NDCG.

System-oriented evaluation of a result generation component can be thought of as evaluation of a machine reading comprehension system (Chapter \ref{chp:c-mrc}), to see how well the system can generate or extract an answer from a retrieved paragraph of text. Take extractive readers (Section \ref{sec:extractive-readers}) as an example. Each instance in the evaluation considers a particular question and paragraph, capturing the user's required answer by identifying one or more spans of paragraph text that answer the question \citep{rajpurkar2016squad}. Identifying multiple spans is useful to identify the human level of agreement on the task, and to capture that there can be multiple possible correct spans. 

The system sees a different paragraph for each test instance, modeling a situation where some upstream retrieval module has already been applied to find a candidate paragraph. The system identifies a span of text, and it can be evaluated using two metrics
\begin{itemize}
    \item Exact match: The proportion of questions where the predicted span is an exact match for a ground truth span.
    \item F1 score: Calculates the F1 score, which is the harmonic mean of precision and recall. The precision is the proportion of predicted span tokens that are in the ground truth span, and recall is the proportion of ground truth span tokens that are in the predicted span. For a question with multiple ground truth spans we take the maximum F1.
\end{itemize}
This kind of evaluation is also used in a comparative fashion, for example to see if a new approach has (significantly) better performance than a baseline. 

These examples of system-oriented datasets, for document retrieval and result generation, have a common structure. They supply some content, a query, and some annotations indicating what content answers the query. It is then possible to evaluate two or more systems using the same fixed dataset, identifying which system had query responses that better match the ground truth annotation. The advantage of system-oriented evaluation is that it is possible to reuse the dataset across many different experiments and studies. One disadvantage is that it is not guaranteed that the dataset captures a valid real-world application. If the content, query or annotation fails to capture the requests and preferences of users on a real application, results on the dataset may tell us very little about the true quality of the systems being evaluated. Another disadvantage is that reuse of the dataset may lead to false conclusions via overfitting, where the same test data was used multiple times by the same researcher, leading the researcher to inadvertently make decisions using the test data. One way to avoid overfitting is to limit the number of iterations by having a leaderboard where each submission is publicly tracked, as in \cite{rajpurkar2016squad}. Another is to have a single-shot evaluation where the test annotations are not even made public until after all submissions are finalized, as in \cite{voorhees2005trec}. Another is to require that published studies evaluate the same approaches over multiple datasets without per-dataset iteration.

\subsection{Evaluating Retrieval in Conversational Context}
\label{sec:eval_system_end-to-end}

We have seen how to evaluate document retrieval and answer extraction, which could be actions in a CIR system, or sub-components of an action that combines retrieval and extraction. Improving these could lead to an overall improvement in the quality of CIR results. However, the datasets we considered so far are based on a single self-contained query request. In a conversational setting, the query request happens in the context of an ongoing multi-turn interaction. The CIR system can make use of the past requests and responses in generating its response for the current request, in a way that brings its response more in line with the current user needs, making the response more appropriate given the ongoing conversation.

This leads to CIR-specific datasets, which incorporate conversational context. One example of this is the TREC Conversational Assistant Track (CAsT) \citep{cast2019overview}, which is an IR test collection with a corpus, queries and labels. However, when we process the query in CAsT, we now know what happened in previous turns of the conversation. The labeling also takes this context into account. For example, if the current query is ``What is different compared to previous legislation?'' the previous query may be ``What is the purpose of GDPR?'' then both the retrieval system and the labeling must take into account the previous query. The system is rewarded for finding information on how GDPR compares to previous legislation.

Relevance judgments for the current query take this conversational context into account. This means that evaluation using a metric such as NDCG is evaluating the system's ability to make appropriate use of context when ranking. We note, although it evaluates the system in conversational context, the previous steps of the conversation are shared for all systems being evaluated, even though each system may have dealt with the previous steps differently. Each query is like a branching point, after some predetermined conversation steps, and NDCG tells us whether the first branching step had good results. This makes it off-policy evaluation. We can compare this to on-policy evaluation, where a system is deployed, and we test its ability to use its own past behavior to improve performance. One way of understanding such effects is to evaluate with real users, as we will describe later in the chapter.

The previous section described a typical non-conversational dataset as having three main aspects: The query, the content and the annotation of which content is correct for the query. This section describes how the query may be interpreted and annotated in a conversational context. The main difference is that we would now describe a dataset as having: query+context, content and annotation. The annotation, and potentially the content selection, can now take the context into account. Table~\ref{tab:evaluation_benchmarks} summarizes a number of benchmarks and other datasets in this way. The non-conversational IR test collection approach is exemplified by the MS MARCO passage dataset \citep{bajaj2016ms}. The corpus contains 8.8 million passages, the queries do not have any conversational context, and the annotations are passage-level relevance judgments with respect to a query. The answer extraction approach we described is SQuAD \citep{rajpurkar2016squad}, which has no special conversational context and is evaluated based on identifying correct spans from a paragraph. We also described TREC CaST, which is related to MS MARCO, but adds the conversational context. This makes it much more suitable for CIR evaluation, since we have argued that user revealment over multiple steps of conversation is an important property of a CIR process.

\begin{table}[]
    \caption{Summary of benchmarks and datasets, including those used in this book.}
    \label{tab:evaluation_benchmarks}
\small
\resizebox{\textwidth}{!}{
\begin{tabular}{llll}
\hline
\textbf{Dataset}                & \textbf{Annotation}              & \textbf{Content}              & \textbf{Query context}     \\
\hline
\multicolumn{4}{l}{\textbf{Finding answers}}                                                           \\
\hline
- SequentialQA \citep{iyyer2017search} & Table cells                & Wikipedia table         & Session     \\
- Complex SQA \citep{saha2018complex} & Varied        & Wikidata                & Session     \\
- ConvQuestions \citep{christmann2019look} & Entities & Wikidata                & Session     \\
- CoSQL \citep{yu2019cosql} & Tables and text         & Databases               & Session     \\
- SQuAD \citep{rajpurkar2016squad} & Span             & Paragraph               & None    \\
- MS MARCO \citep{bajaj2016ms} & Passage      & Passage corpus          & None    \\
- QuAC \citep{choi2018quac} & Passage                 & Section of wiki page    & Session     \\
- CoQA \citep{reddy2019coqa} & Answer \& Rationale    & Passage                 & Session     \\
- QBLink \citep{elgohary2018dataset}  & Short answer & Document corpus? & Session     \\
- Natural Questions \citep{natural-questions} & Long and short answers     & Document corpus         & None    \\
- TriviaQA \citep{joshi2017triviaqa} & Short answer + context & Document corpus & None    \\
- TREC CAsT \citep{dalton2019overciew} & Passage      & Passage corpus          & Session     \\
- OR-QuAC \citep{qu2020orqa} & Span                   & Passage corpus          & Session \\
\hline
\multicolumn{4}{l}{\textbf{Asking clarifying questions}}                                      \\
\hline
- CLARQUA \citep{xu2019asking} & Clarifying Q         & Entities                & None    \\
- Qulac \citep{aliannejadi2019asking} & Clarifying Q facets & Facets            & Session?    \\
- StackExchange \citep{rao2018learning} & Question ranking & Questions          & Post \\
- Amazon H+K \citep{rao2019answer} & Question & Product description  & Product description \\
- MIMICS \citep{zamani2020mimics} & Generate  Q+face & Reference data          & None       \\
- ShARC \citep{saeidi2018sharc} & Generate Q            & N/A                   & Passage+Q    \\
\hline
\multicolumn{4}{l}{\textbf{Dialog observation and generation}}                                         \\
\hline
- MultiWOZ \citep{budzianowski2018multiwoz} & Generated responses        & Reference data          & Session     \\
- TaskMaster \citep{byrne2019taskmaster}             & Generated responses        & Reference data          & Session     \\
- Raddle \citep{peng2020raddle} & Generated responses & Reference data & Session \\
- Frames \citep{asri2017frames} &    Generated responses                        & Reference data     & Session             \\
- MISC \citep{thomas2017misc} & \multicolumn{3}{c}{(Observing information seeking conversation)} \\
- SCS \citep{trippas2017scs} & \multicolumn{3}{c}{(Observing information seeking conversation)} \\
- CCPE-M \citep{radlinski2019coached} & \multicolumn{3}{c}{(Observing cold-start recommendation)}        \\
\hline
\multicolumn{4}{l}{\textbf{User prediction and simulation}}                                            \\
\hline
- MIMICS                 & Predict user response      & Labels. Clicks.         & Query       \\
\hline
\end{tabular}
}
\end{table}

The section of Table~\ref{tab:evaluation_benchmarks} about finding answers lists several other kinds of dataset that are worth noting, several of which are also described in other parts of the book. The content given in some cases is a paragraph of text, which means we assume some process has already taken place to find a paragraph of interest to the user. In other cases, the content is structured data, such as Wikipedia tables, Wikidata knowledge triples and databases. We also see cases where the content is a text corpus, so we are either searching a corpus of full documents or a corpus of shorter passages of text.

In all cases the annotated output is shorter than a full document, reflecting the goal of a conversational system to identify outputs that are short enough for presentation to the user on a small screen or through voice. The answer can be a passage of text such as a sentence, or it can even be a span of text that gives a direct answer. Some datasets identify both the passage and the direct answer. In a real conversational search system, users might be interested in several levels of granularity, for example highlighting a two-word answer on the screen, while also showing the containing passage and providing a link to the document that contains the passage. Users can check the document to justify the answer that was found, and in general may be more comfortable if the system provides easy access to such information.

\subsection{Evaluating Non-retrieval Components}
\label{sec:eval_system_other}

Although CIR components that find answers in conversational context play a key role, there are other CIR system components that have some impact on the quality of CIR results, which can be evaluated using other kinds of datasets.

Several components in the CIR system may be used to understand and augment the user's query. These could include components for named entity recognition, co-reference resolution, query completion and query rewriting. If these are being carried out in conversational context, they could also make use of a state tracker that stores current and past requests and responses. If the component augments the query before retrieval, then it can be evaluated using the retrieval-based evaluation in the previous section. Using TREC CAsT data, several studies employed methods that combine the current query with previous queries, with the goal that the augmented query will yield improved search results \citep{yang2019query,vakulenko2021question,voskarides2020query}. These methods for query expansion and query rewriting are described in more detail in Chapter~\ref{chp:CIR}, along with other methods that use recognition and resolution to better process the text of the query and the candidate answers.

A CIR system may generate clarifying questions. Different generation approaches and their evaluation are described in more detail in Chapter~\ref{chp:proconv}. One evaluation approach is to use data from online forums, such as StackExchange \citep{rao2018learning}. This identifies naturally-occurring clarifications that happend in a real human-to-human conversation. The assumption is that a CIR component that can answer such questions would be useful.

Another evaluation approach is to use logs of clarifying questions that occurred in real-world search system such as the Bing search engine  \citep{zamani2020mimics}. The paper uses such clarification data to propose a number of forms of evaluation. One is to evaluate the generation of clarifying questions. Since evaluation of text generation is difficult, the paper suggests a mixed approach that uses the search engine data for training, but then includes some human annotation in the final evaluation. Other question-level forms of evaluation are selecting the best clarifying question and choosing when to ask a clarifying question (action selection). Since clarifying questions can propose multiple choice answers, there are also answer-level forms of evaluation, such as ranking the potential answers and predicting user engagement at the answer level.

Clarification evaluation based on reference clarifications, from forums or log data, has the problem that the CIR system may generate an equally good clarification question that is different from the reference. One solution is to use human annotation, to identify which clarifying questions seem most appropriate and useful \citep{aliannejadi2019asking}. The approach was to collect clarifying questions relating to different facets of the query, based on crowdsourcing. The task of selecting a clarifying question can then be framed as a question retrieval problem.


Many IR systems make use of summarization (Chapter \ref{chp:qmds}), to give users their first view of some content, so they can decide whether to engage further the content. This is usually via query-focused summaries, meaning that the summary can change depending on the user's query, even if summarizaing the same content, for example the same document \citep{tombros1998advantages}. Since CIR is amenable to hands-free and screen-free interaction, summarization can play an important role. Summarization is related to answer extraction, as described in the previous section, but it can also be applied if the query cannot have a short answer, if the conversation is about some complex topic. This makes summarization much more difficult to evaluate, perhaps leading to evaluation against a reference summary using metrics such as BLEU or ROUGE. We note that there are a large number of studies of query-independent summarization, but relatively fewer studies of query-focused summarization, and studies of conversation-biased summarization, where the summary depends on the entire conversation context that has happened so far. One way of implementing such a summarizer would be to apply conversation-aware query rewriting, then generate a query-focused summary based on the rewritten query.



In general, in summarization and also generation of clarifying question, it is possible for the CIR system to do natural language generation. There are three main methods for evaluating such generation \citep{celikyilmaz2020evaluation}. One is to involve humans, asking them if the generated text is good. The main drawback is that the CIR system may generate new text that a human has never seen before, so additional human input is needed. The second method is to compare the generated texts to a reference text, using a metric based on simple similarity metrics such as string overlap. The reference text may be human-generated and/or human-annotated. The main drawback is that the CIR system may generate text that is too different from the references, but is still valid, which would lead it to being penalized. The third approach is to use machine learned metrics to play the role of the human judge. The main drawback would be if the metric does not do a perfect job of simulating the human.



\section{User-oriented Evaluation}
\label{sec:eval_user}

User-oriented observation and evaluation is necessary if we wish our CIR system to be grounded in the real world. Even system-oriented datasets benefit from being based on conversation traces from a real user interaction, bringing the dataset's distribution in line with real-world conditions. Even so, we could criticize a system-oriented dataset as being \emph{off-policy} evaluation, meaning that we are using traces from some past system to generate a dataset, which is then used to evaluate a different system. To see how a system performs for a real user in a multi-turn interaction, we need user-oriented evaluation. We also need user-oriented observation to understand user needs and scenarios, so we can feed that understanding into the design of all future forms of evaluation (including future user studies). We will now describe the forms of user-oriented evaluation that are summarized in Figure~\ref{tab:evaluation_types}.



\subsection{Lab Studies}

In observational lab studies, the goal is to understand the CIR process, by observing users interacting with a CIR system. In such studies, it is possible for a human to play the role of a CIR system \citep{trippas2018informing} or to use an operational CIR system \citep{kiseleva2016understanding} or use both in one study \citep{vtyurina2017exploring}. The human acting as a CIR system has access to a traditional IR system, and performs searches on behalf of the user, keeping track of their conversational context and potentially asking clarifying questions. The human CIR system may do a better job of understanding a long conversation and remembering the right context, revealing what users would like to do and the limitations of present-day CIR systems. This can include not only the the functionality of the CIR system but also how to achieve a more naturalistic conversation style \citep{thomas2018style}. 

In one observational user study, \citet{trippas2018informing} considered voice-only interaction between a user, who was performing a given search task, and a human intermediary acting as a CIR system, who was carrying out searches in a traditional IR system. One observation was that users would say multiple things in a single utterance, such as giving feedback and asking a question. Users would build a mental model of how to work with the intermediary, for example realizing they can guide the intermediary's use of querying and browsing search results. The intermediary could also understand the presentation preferences of the user, by for example learning that a certain user likes to hear a list of objects or entities, and asking the user whether they want a similar listing in their next task.

Users were observed to start out with various kinds of initial utterances, such as a query-like list of keywords, a natural language query, query plus instructions on what kind of results to retrieve, and others. Perhaps a CIR system should support all these query types. Users gave feedback on the quality of results, which the intermediary can take into account, so a CIR system could do this too.

\citet{kiseleva2016understanding} ran a lab study, where the design was based on observations from large-scale logs. The logs were from an intelligent system that combines the CIR functionality with the functionality of controlling the user's device. Based on observing the logs they identified three scenarios. First is device control, such as making a phone call or launching an app. Second is searching for a short answer or Web page. Third is searching that requires complex interaction, such as planning a holiday.

Their user study had 60 participants. For all three scenarios, they defined fixed tasks but avoided being too specific to maximize the variation in user responses. For example, the short answer task ``finding the score of the user's favorite sports team'' could lead to each participant entering quite a different query, depending on which team they think of. Participants were asked post-task whether they succeeded and whether they were satisfied. The key finding was that factors contributing to self-reported success differed by scenario. For device control, satisfaction and success were the same thing. For CIR scenarios, satisfaction was related more to effort than task completion, meaning that it is possible to complete the task but be unhappy with the process. Task-level satisfaction could not be reduced to query-level satisfaction.

\citet{vtyurina2017exploring} compared search using a text interface where the CIR system is a real system, a human expert or a human posing as a CIR system. They found that users do not have a bias against conversational systems, but the real-world CIR system was not up to the task of searching for complex information needs. Specifically the known human and human CIR system performed similarly on overall satisfaction, ability to find information and topical quiz success, while the automatic system performed significantly worse on all three counts. The automatic system would fail to maintain context and misunderstand partially-stated questions. The automatic system in this study would give a URL with each result, which the users preferred to the human and human CIR system, who did not. When dealing with the human, people felt uncomfortable asking too many questions. Feedback was found to be important.

This aligns with the user revealment, system revealment, mixed initiative, relevance feedback, etc. expected in the theoretical model of conversational search \citep{radlinski2017theoretical}.

\subsection{Scaling Up Evaluation}

Studies of CIR at scale, either in a panel form or with log analysis, will be an important future step. To explain the role of such studies, we can consider how panels and logs were used in related applications: Web search, spoken dialog systems, and social chatbots. These past examples can explain how larger-scale user-oriented evaluation may play a role in the future of CIR.

In Web search, a key early contribution was from \cite{broder2002taxonomy}, describing a taxonomy of Web search queries: Navigational, informational, and transactional. Looking at log traces of the AltaVista search engine, an earlier paper had already listed ``yahoo'' is one of the most common queries \citep{silverstein1999analysis}. Broder was also using AltaVista logs, and may have noticed that the clicks for that query were overwhelmingly on \url{yahoo.com}, indicating an intent to navigate to a specific site. This observation, of the queries and preferred responses of real users, would be an example of user understanding by behavior log observation (Table~\ref{tab:evaluation_types}).

The taxonomy is important because the navigational result of \url{yahoo.com} is different from an informational result, which would be a page with information about Yahoo such as a news article. A transactional result would be one related to buying or obtaining something. The query ``coca cola'' could have all three intents: (1) Informational intent, to find history, sales, advertising, stock price, and controversies; (2) Navigational intent, to visit \url{coca-cola.com}; and (3) Transactional intent, to buy Coca-Cola products. Logs play an important role to understand that these different kinds of intent can exist and also to understand which intent seems to be most likely for a given query. Based on Bing logs, clicks on the Coca-Cola homepage are ten times more common than clicks on the Wikipedia page for Coca-Cola, suggesting that the navigational intent is more important.

To study intents more systematically, the original study \citep{broder2002taxonomy} used surveys and also carried out manual annotation of a sample of queries. The finding was that navigational, informational and transactional queries are roughly 20\%, 50\% and 30\% of queries, respectively. Based on this it was possible to set up system-oriented evaluation that also included these types of intents, for example capturing that the desired output is Navigational in the TREC Web Track \citep{voorhees2005trec}. All of this suggests that log analysis of CIR systems, to understand usage patterns and gain insights into possible user intents, will be a crucial step to the future of the field. Bringing to light the kinds of CIR research systems and datasets that correctly reflect a real-world CIR system.

Another important step in user-oriented understanding is to set up a panel, to bridge the gap between user behavior logs, which provide implicit signals of success, and direct feedback from users on whether they succeeded. One study in Web search \citep{fox2005evaluating} set up a panel that recorded people's Web search behavior such as queries and clicks, then periodically also popped up a survey to ask about explicit satisfaction. This helped confirm important insights, such as the dwell time on a search result being related to its relevance, with a short dwell time sometimes indicating that the result was not relevant. In spoken dialog systems \citep{su2016line}, a system was described for requesting feedback only when a reward model is uncertain about user success, which could reduce the number of surveys necessary. The reward model from such a process may encode the same kinds of insights that would have been found by other means such as lab studies or more traditional panels.

After observational studies in the lab, running panels, and mining logs, we can begin to consider running large-scale experiments. Online randomized controlled trials split users into two groups and provide them with different versions of the system, logging their responses, to identify how user responses change. Such trials are run by many types of product such as search engines, recommenders, social networks, and online marketplaces. Authors from a broad range of such products \citep{gupta2019top} identified a number of important steps, including establishing an overall evaluation criterion (OEC). The OEC should align with long-term business goals, it should evaluate all parts of the product, encourage a diverse set of improvements, and be sufficiently sensitive and affordable.

In the area of social chatbots, a study with large-scale A/B experimentation is \citet{zhou2020design}. The paper describes the metric Conversation-turns Per Session (CPS). The more turns we see from a user in a session, the more engaged they are with the social chatbot. They observe that this metric is better used as a long-term metric. Specifically, if the goal is to have a short-term increase in CPS, then the system can just generate responses such as ``I don't understand, what do you mean?'', which could entice some users to respond, but would make them like the system less, and reduce CPS in the long term. Similarly, adding some new task completion functionality could reduce CPS in the short term, by helping users complete their task more quickly. However, by making the system more useful and trustworthy, this functionality is likely to increase future engagement in the long term.


Candidates for OEC in Web search were considered in \citet{chapelle2012large}, confirming that signals relating to clicks and dwell time are still valid. That study also demonstrated that higher quality Web results are associated with shorter sessions. The assumption is that users have a fixed task, and if the results are worse they need to try more queries to get the same results. The discrepancy with CPS, which rewards longer sessions, could be because Web search is often task-oriented, which is different from a conversational chatbot that encourages longer interaction. To resolve which approach works best for CIR, it would be possible to consider the long-term effects of different A/B hypotheses, and see which short-term CIR user behaviors (such as longer or shorter sessions) best correlate with long-term engagement.


\section{Emerging Forms of Evaluation}
\label{sec:eval_emerging}

This section describes two notable forms of evaluation that are currently emerging in the field: CIR user simulation and responsible CIR.

\paragraph{CIR user simulation.}
We have seen that evaluation using a fixed dataset can capture conversational context, although the CIR system is effectively jumping into a conversational context, where the previous steps were carried out by some other CIR system. Interactive studies with real users do not have this problem, but raise the issue of either running large-scale A/B tests, which is not available to every academic, or running a lab or field study, which is expensive to do. A third form of evaluation may emerge in the coming years, of evaluating with fixed data rather than a real user, but with a simulated user that can carry out multiple conversational steps.

Although it is not common to evaluate CIR using user simulation, in related areas of conversational AI there are already user simulation methods. As summarized by \cite{gao2019neural} simulators for dialog agents can be based on randomly generated scenarios. A scenario has two parts. One is the things that the user does not yet know, for example to book a movie ticket, the user wants to find out the ticket, theater and start time. The other is the user's constraints, that she wants to book for three people, tomorrow and for the movie ``batman vs superman''. The simulation can then be carried out using a rule-based approach, where the simulator represents the users ``state of mind'' using a stack data structure known as the user agenda. The simulation can also use a model-based approach, which still has the user's goal and constraints, but trains a sequence-to-sequence model for simulation. Simulators, particularly rule-based ones, can incorporate domain knowledge in the process. For example, the user must always have a number of people in mind, and that number of people should be greater than zero but not too large.

Taking the simulation approach from dialog systems, it is possible to adapt it to conversational recommendation \citep{zhang2020evaluating}. Building on the agenda-based approach, the simulated user maintains a stack of requests, and will select between pulling and pushing (replacing) on the stack depending on whether the user action has been met with an appropriate response. The user actions are disclose, reveal, inquire, navigate, note and complete. The user's preferences are modeled via a set of known preferred items and a knowledge graph that allows generalization to related items. Natural language understanding is based on entity linking to understand what the conversational recommendation system is talking about, and also based on the assumption that the user can understand which of the actions the agent is carrying out, based on some training data based on past agent behavior with action annotation. Natural language generation is based on a small set of templates.

It is likely that these simulation approaches for dialog systems and conversational recommendation can be adapted to CIR. The difference is the lack of structure, due to the free-form nature of IR information needs. The information need underlying a CIR session could be to book a movie ticket at a particular time for a particular number of people, like in the dialog system. It could also be to recommend a movie, given the preferences of a user across known movies and their knowledge graph attributes. The CIR system could help the user visit a movie booking site or search for documents about movie recommendations. But the CIR user's need may also be to learn about a movie director Stanley Kubrik's preparation process for a feature film, which would take on average four years. If there are a great variety of information needs, the CIR system needs to be open-domain, answering needs that do not fit into a dialog slot-filling paradigm, or a knowledge graph recommendation. We hope to see more progress in open-domain user simulation in the coming years.




\paragraph{Responsible CIR.}
Since search and recommendation systems can make the difference between users discovering content and never seeing it at all, recent papers have considered the ethical concerns of such retrieval systems. One consideration is whether the IR system is systematically under-representing certain types of content \citep{singh2018fairness,biega2018equity}, such as mostly showing male CEOs for the query ``ceo'', under-representing female CEOs. Another consideration is that the IR system may not perform equally well for all groups of users, perhaps providing lower quality search results to some group \citep{mehrotra2017auditing} because it is smaller and contributes less to the overall training data of the IR system. 

Such considerations are an emerging topic in online systems for search and recommendation, and should be also considered in the context of CIR. The problems may be similar in conversational and non-conversational systems. For example, the concerns of under-represented content and under-served users, described in the previous paragraph, are also likely to arise in conversational systems. In all types of system, we should be concerned with protecting users from potentially harmful results such as scams or disinformation. The Alexa Prize \citep{Ram2018Alexa}, focused on conversational assistants, identified a related set of inappropriate response types: Profanity, sexual responses, racially offensive responses, hate speech, insulting responses, and violent responses. 

In the Xiaoice social chatbot \citep{zhou2020design} a number of areas were identified for ongoing study. Privacy is a concern, since users reveal information about themselves to the conversational system. This is certainly true in Web search as well, where users reveal their thoughts to the engine in the form of queries, but could happen even more if the conversation with the agent elicits further details or even (in the case of a chatbot) is treated as a friend. Another concern is allowing Xiaoice to suggest not conversing or even refuse a conversation, either because it is late at night or because the topic is inappropriate. Another concern is that the system should make it clear what it can and cannot do, admitting that it is fallible, because users who trust the system too much could make a mistake. The overall concern is how all the possible impacts on human users can be correctly evaluated and optimized in a machine learning based system.



\chapter{Conversational Search}
\label{chp:CIR}

In this chapter we discuss {conversational document search}, which is to find relevant documents from document collections to meet user's information needs behind conversational search queries. 
It is often referred to in a short form as \textit{conversational search}, 
as \emph{documents} are the most common form of system output. Differing from ad hoc search~\citep{croft2010search}, conversational search systems need to take into account not only the input query but also conversational context when retrieving documents. 
The retrieved documents can either be directly present to a user as the ``ten blue links'' in a search result page or be used as inputs to search result generation components e.g., to produce a concise summary using the query-focused summarization module or to generate a direct answer to the input query using the machine comprehension module. These search result generation modules will be described in the later chapters.

This chapter is organized as follows. Section~\ref{sec:cir_overview} introduces the conversational search task. 
Section~\ref{sec:cir_benchmarks} presents public benchmarks.
Section~\ref{sec:cir_preliminary} reviews Transformed-based pre-trained language models which are widely used for developing the models and methods for conversational search and CIR in general.
Section~\ref{sec:convsearch_arch} presents a typical architecture of a conversational search system, describing its main components and the challenges faced when developing them.
Section~\ref{sec:cqu} describes contextual query understanding methods of converting a conversational search query and its dialog context to a \emph{de-contextualized} query which can be used alone (without conversational context) for ad hoc document retrieval.   
Section~\ref{sec:sparse_retrieval} describes sparse document retrieval methods that are based on the classic bag-of-words (BOW) representations of queries and documents.
Sections~\ref{sec:dense_retrieval} and~\ref{sec:conv_dense_retrieval} describe dense document retrieval methods for ad hoc search and conversational search, respectively.
Section~\ref{sec:doc_ranking} describes neural document ranking models.

\section{Task}
\label{sec:cir_overview}

The growing of conversational search in recent years is driven by two factors. The first is the ubiquitous accessibility of Internet. Now, people are doing almost everything on Web. As a result, search tasks become so complex that they often require intensive human-machine interactions. 
The second factor is the rapid adoption of a new generation of conversational assistants running on mobile and edge devices with small or no screens, such as Google Assistant, Siri, and Alexa. Compared to desktop and laptop computers, these devices rely more on speech-based human-machine interactions that contain voice search queries which are naturally more conversational. 

Conversational search presents both new technical challenges and new opportunities for search engine developers.
On the one hand, a conversational search system needs to understand users' search intent not only from a single query, but by taking into account conversational context. In addition, a CIR system has to present its search result, based on the retrieved documents of conversational search, in a very concise form, such as a direct answer to an input query or up to three search results, each or all in a short summary, as users have little patience to read or hear an answer consisting of more than three sentences. This requires the system to produce more accurate search result than a non-conversational IR system which can afford to produce ``ten blue links''.
On the other hand, the dialog interface of conversational search allows users to more easily clarify search intent and more effectively perform complex search tasks, such as exploratory search for investigation and learning, via multi-turn conversations.

In this chapter we describe conversational search as a component of a general CIR system, discussing the technical challenges it faces and the solutions developed recently. The other CIR components that use conversational search results as inputs are described in the later chapters: Chapter~\ref{chp:qmds} presents methods of summarizing conversational search results. Chapter~\ref{chp:c-mrc} presents machine comprehension methods of generating a direct answer to a user query based on conversational search results.

Although the specific setting of conversational search can vary slightly among different benchmarks, as will be described in Section \ref{sec:cir_benchmarks}, the conversational search task, in comparison to classical ad hoc search, can be characterized as follows:
\begin{itemize}
    \item The \textit{information needs} remain the same as that in classical ad hoc search.
    \item The \textit{user search intent} is present in a dialog session, which often includes multiple turns of natural language queries on the same topic or a set of related topics.
    \item The \textit{system output} is a set of relevant documents, same as that in classical search.
\end{itemize}

As an example, the conversational search task in the TREC Conversational Assistant Track (CAsT)~\citep{cast2019overview, cast2020overview} is defines as follows.
Given a series of natural conversational turns for a topic $\mathcal{T}$ with queries for each turn $\mathcal{T}=\{Q_1,...Q_i...,Q_n\}$, the task is to identify a set of relevance documents (or passages) $\mathcal{D}_i$ for each turn $Q_i$ to satisfy the information needs in turn $i$ with the context in turn $Q_{<i}=Q_1 : Q_{i-1}$. 

\begin{table}[t]
    \begin{subtable}[t]{\textwidth}

    \begin{tabular}{ll}
\hline
\multicolumn{2}{l}{\textbf{Title}: Acid Reflux} \\
\multicolumn{2}{l}{\textbf{Desc}: Information about Acidic Reflux, its cause and potential treatments.} \\ \hline
\textbf{Turn} & \textbf{Conversation Utterances} \\ \hline
1 &	What causes acidic reflux in the morning? \\
2 &	Does it have long term side effects?\\
3 &	What is the best OTC for it? \\
4 &	What are the side effects of long term PPI use? \\
5 &	Tell me about natural treatments \\
6 &	What foods cause it? \\
7 &	What ones reduce it?  \\
8 &	How does exercise affect it? \\
\hline
\end{tabular}
\caption{CAsT 2019 Topic 44} \label{tab:cast2019_q}
    \end{subtable}
    \vspace{0.3cm}
    
    \begin{subtable}[t]{\textwidth}
        \begin{tabular}{ll}
\hline
\multicolumn{2}{l}{\textbf{Title}: GDPR} \\
\multicolumn{2}{p{12cm}}{\textbf{Desc}:  learn about GDPR, the privacy issues in social networks, and the addiction of it.} \\ \hline
\textbf{Turn} & \textbf{Conversation Utterances} \\ \hline
1 &	What is the purpose of GDPR? \\
2 & What is different compared to previous legislation? \\
3 &	What are the privacy implications of those technologies? \\
4 &	Oh, IP addresses are considered PII? What is the full range of personal data? \\
5 &	How do big companies adapt to GDPR? \\
6 &	OK. Tell me about the privacy issues in social networks. \\
7 &	What do they get in return for their privacy? \\
8 &	What are the symptoms of that addiction? \\
\hline
\end{tabular}
\caption{CAsT 2020 Topic 91}
    \end{subtable}
\caption{Two example conversational search topics from TREC Conversational Assistance Track 2019 and 2020, respectively. Each topic includes a series of conversational queries.
Titles and Descriptions are meta data for reference and assessment, not to be used by the conversational search systems.
}
\label{tab:cast_q}
\end{table}

Table~\ref{tab:cast_q} shows two conversational search topics sampled from TREC CAsT 2019 and 2020.
Each topic consists of a title and a detailed description (Desc) which are used as meta-information for evaluation but are not supposed to be used by any automatic search systems to be evaluated, and a sequence of conversational queries. 

Compared to ad hoc search, conversational search is unique in search queries. As shown in Table~\ref{tab:cast_q}, the queries for each topic forms a natural conversation session, demonstrating typical dialog behaviors observed in human conversations. For example, a conversational search session typically starts with a query requesting a general introduction about the topic, followed by queries that ask for more information about the topic being discussed, forming an exploratory information seeking trajectory. Topic shift and return are also observed in Table~\ref{tab:cast_q} (b) where PII is discussed in Turn 4 before asking about GDPR again in Turn 5. 

Ellipsis phenomena are frequently encountered in conversational search. 
For example, the query ``what ones reduce it?'' (Turn 7 in Table~\ref{tab:cast_q} (a)) cannot lead to any meaningful search result without knowing what ``ones'' and ``it'' refer to, respectively.
Thus, the search system has to take into account not only the current query but also its conversational context, which includes previous user queries and system responses in the same search dialog session, to identify the user's search intent to retrieve relevant documents. 

Contextual query understanding and conversational document retrieval are the main research topics to be discussed in this chapter.
\section{Benchmarks}
\label{sec:cir_benchmarks}

This section describes in detail two widely used benchmarks of conversational search, the TREC CAsT  benchmark~\citep{cast2019overview, cast2020overview} and the OR-QuAC dataset~\citep{qu2020orqa}. 
They are representative examples of two benchmark construction approaches, TREC and crowdsourcing, respectively.
TREC CAsT is a standard TREC-style benchmark which is constructed by the IR community to evaluate IR systems, following the \emph{test collection evaluation paradigm}~\citep{cleverdon1966factors, voorhees2005trec}. ``At the core of this  methodology was the idea that live users could be removed from the evaluation loop, thus simplifying the evaluation and allowing researchers to run in vitro–style experiments in a laboratory with just their retrieval engine, a set of queries, a test collection, and a set of judgments (i.e., a list of relevant documents).'' 
OR-QuAC is an open-retrieval conversational question answering (QA) benchmark constructed via \emph{crowdsourcing}. 
It is initially curated using crowdsourcing for conversational QA and then extended for the conversational search task by deriving document-level relevance labels from the answer annotations~\citep{chen2017reading}. 
We end this section by briefly reviewing other benchmarks.

\subsection{TREC CAsT}
TREC CAsT is the result of an IR community's effort of constructing a reusable, standardized, and TREC-quality benchmark for conversational search. As of December 2021, the track has run for three years (in 2019, 2020, and 2021). 
Similar to other TREC tracks, CAsT is an instantiation of a vision shared by the IR community that conversational search is one of the most important themes for building next-generation IR systems~\citep{swirl2018}.

As the first step towards conversational search, CAsT focuses on the document retrieval task, as discussed in Section~\ref{sec:cir_overview}: to retrieve relevant documents (or passages) from a text corpus for a series of conversational queries.

\paragraph{Corpus.} 
The CAsT corpus is a combination of corpora from MS MARCO~\citep{bajaj2016ms} and TREC CAR~\citep{dietz2017trec}. MS MARCO includes an 8-million-passage corpus collected from Bing's QA system when it is originally released by Microsoft. Later, MS MARCO is augmented by a 3.3-million-document corpus constructed by fetching the full content of the URLs associated with those passages~\citep{craswell2019overview}.
The TREC CAR corpus is constructed for the Complex Answer Retrieval Track and includes around 30 million passages collected from Wikipedia. In CAsT 2019 and CAsT 2020, the passage corpora are used and the task is passage retrieval. In CAsT 2021, the task is document retrieval and the MS MARCO document corpus is used.

\paragraph{Information Needs.}
The CAsT organizers semi-manually constructed exploratory information needs (topics) from the combination of previous TREC topics, MS MARCO conversational sessions, and their own interests and experiences (e.g., they pick queries by themselves or using reference to derived pseudo  search sessions from Bing search logs)~\citep{dalton2019overciew}. 
\paragraph{Conversational Queries.} 
The queries are curated by the CAsT organizers manually. They start from a topic, write an initial query as the first round, and then grow a conversational trajectory with follow-up queries while interacting commercial Web search engines (Google and Bing) (in CAsT 2019) or a baseline search system (in CAsT 2020).
When curating the conversational trajectories, the organizers follow a shared guideline~\citep{cast2019overview} to mimic users' behavior when using a conversational search engine.
As a result, the conversational queries in CAsT reflect the combination of the CAsT organizers' vision of conversational search engines and the conversation behavior of live users of Web search engines.

\begin{table}   
\caption{The manual de-contextualized queries for CAsT 2019 Topic 44.}\label{tab:cast_manualq}
    \resizebox{\textwidth}{!}{
    \begin{tabular}{ll}
\hline
\multicolumn{2}{l}{\textbf{Title}: Acid Reflux} \\
\multicolumn{2}{l}{\textbf{Description}: Information about Acidic Refulx, its cause and potential treatments.} \\ \hline
\textbf{Turn} & \textbf{Conversation Utterances} \\ \hline
1 &	What causes acidic reflux in the morning?\\
2 &	Does acidic reflux in the morning have long term side effects? \\
3 &	What is the best OTC for acidic reflux in the morning? \\
4 &	What are the side effects of long term PPI use? \\
5 &	Tell me about natural treatments for acid reflux in the morning. \\
6 &	What foods cause acidic reflux in the morning? \\
7 &	What ones reduce acidic reflux in the morning? \\
8 &	How does exercise affect acidic reflux in the morning?\\
\hline
\end{tabular}
}
\end{table}

\paragraph{Manual Query Rewrites.} 
The conversational queries include various dialog phenomena including co-reference and omission as shown in Table~\ref{tab:cast_q}. 
These conversational queries require context information from previous rounds to be understood.
To facilitate assessment and research, CAsT also provides an annotated dataset of \textit{manually de-contextualized} queries. 
Table~\ref{tab:cast_manualq} shows the sequence of manually de-contextualized queries for CAsT 2019 Topic 44, 
where each query is rewritten by the organizers to contain all of the information required to represent the single turn of the underlying information need.

\paragraph{Relevance Labels.} 
The labels in CAsT are annotated by TREC accessors using the standard TREC \emph{pooling} method. 
The track participants submitted their system's document retrieval results for the given CAsT queries and corpus. The organizers merged the top-$K$ (10 to 15) ranked documents per submitted run to form the \emph{pool} of query-document pairs. TREC annotators accessed the relevance of the pooled pairs and the annotated labels were released as the official TREC benchmark. This TREC-style pooling method was developed to provide robust evaluation of IR systems and has been the standard in the IR community for the past several decades~\citep{voorhees2005trec}.

\paragraph{Advantages.} 
The TREC approach to building IR benchmarks has several advantages.
The information needs are derived from search sessions of real users.
The conversational query rounds are curated to mimic potential user behaviors in a future conversational search engine by IR experts. 
The queries are curated independently with the relevance accessing process, simulating live search scenarios where the retrieved documents of a query are unknown to a user when she forms the query.
The relevance labels allow robust evaluation of different IR systems
because the TREC-style pooling strategy ensures the reusability of TREC benchmarks.
TREC accessors are also well-known to provide high quality annotations.

\paragraph{Limitations.} 
The TREC setup also has some limitations.
The first is the small quantity of the relevance labels, as a trade-off to high quality. 
For example, TREC CAsT consists of in total 100 conversational search topics annotated, each with about ten queries. 
This makes TREC CAsT a few-shot setting as the number of labels is insufficient to reliably train any large-scale deep learning models (from scratch) in a supervised learning manner.
The other limitation is that the queries are curated by the CAsT organizers as a reflection of their visions of conversational search. 
As a result, the dataset presents properties and challenges that the organizers considered as important for future CIR systems, but their predictions could be biased.

\subsection{OR-QuAC}

Unlike TREC CAsT which is constructed semi-manually by IR experts, OR-QuAC is an open-retrieval conversational QA benchmark developed via crowdsourcing, as many other recently developed QA benchmarks~\citep[e.g.,][]{rajpurkar2016squad, choi2018quac, reddy2019coqa}.
OR-QuAC is designed for an open-retrieval setting where an IR system to be evaluated needs to retrieve passages from a collection, given a query and its conversational context~\citep{qu2020orqa}. 
OR-QuAC is an extension of the QuAC benchmark~\citep{choi2018quac}, which offers information-seeking conversations, to an open-retrieval setting by (1) creating a collection of over 11 million passages using the Wikipedia corpus that serves as the knowledge source of answering questions, and (2) generating relevance labels for query-passage pairs based on whether the passage contains the answer span of the query~\citep{chen17open}. 
In addition, OR-QuAC aggregates the CANARD dataset~\citep{elgohary2018dataset} which contains context-independent (de-contextualized) rewrites of QuAC queries.   

\paragraph{Conversational Queries.} 
The QuAC query-answer pairs are collected in a process that involves facilitating information-seeking conversations between two crowd workers, a teacher and a student, who discuss a section from a Wikipedia article about an entity~\citep{choi2018quac}. The student is permitted to see only the section’s title and the first paragraph of the article, while the teacher is additionally provided with full access to the section text.  The conversation starts with the student formulating a free-text question (query) from the limited information they have been given. The teacher is not allowed to answer with free text; instead, they must select a contiguous text span in the section. While this decision limits the way answers are generated, it makes evaluation simpler and more reliable. The conversation continues until (1) twelve questions are answered, (2) one of the workers decides to end the interaction, or (3) more than two unanswerable questions were asked.

\paragraph{Manual Query Rewrites.} 
The queries in QuAC also include ellipse phenomena which make them ambiguous and under-specified, similar to those in CAsT. 
~\cite{elgohary2019canard} recruit a group of crowd workers to manually rewrite the QuAC queries into \emph{de-contextulized} queries that include all the information necessary to understand the information needs. The resultant CANARD dataset includes manually de-contextualized queries for all QuAC's dev set and a sample of 4,873 conversation topics (34,956) from QuAC's training set.
Then, the QuAC dev set is used as a test set, as the QuAC test set is hidden, and the annotated training topics is split to train/dev with a 9:1 ratio.

\paragraph{Relevance Labels.} 
An open-retrieval benchmark needs document-level (or passage-level) relevance labels. These can be derived from the query-answer pairs of QuAC using the method proposed by \cite{chen2017reading}. Given a query-answer pair, a passage is labeled as relevant to the query if the answer span occurs in the passage.

\paragraph{Corpus.} 
The passage collection of OR-QuAC is constructed by splitting the Wikipedia documents using a standard passage breaker. A passage that appears in the section used in construction and includes the answer span is labeled as relevant to the corresponding query; the rest passages in the corpus are labeled irrelevant to the query. The train/dev/test split follows that of CANARD.

\paragraph{Advantages.} 
A significant advantage of the OR-QuAC benchmark is its quantity. 
As a dataset curated via crowdsourcing, OR-QuAC includes tens of thousands of crowd-sourced question-answer pairs and manually de-contextualized queries.
This is a very valuable resource to train large models for the designated open retrieval task and also to provide (weak) supervision data to develop models for other related tasks. 
For example, CANARD is often used as an additional (weakly supervised) training dataset for improving systems evaluated on TREC CAsT~\citep{cast2019overview}.

\paragraph{Limitations.} 
The major issue of OR-QuAC is data quality. The questions written by crowd workers are of lower quality than the TREC CAsT queries written by IR experts. This is expected because a crowd worker is unlikely to spend as much time as an IR expert in making interesting and challenging information seeking conversations for evaluating IR systems. 
Another issue is that the interactive information-seeking task designed for collecting conversational QA pairs is artificial and does not reflect the complexity of real-world conversational search scenarios. For example, it has been reported that the dialog trajectories in QuAC often follow a pattern that students start a dialog by asking a question about the beginning of the section before progressing to asking questions about the end~\citep{choi2018quac}. 
The setting also ensures that the answer span is in the same document section based on which the question is generated. 
These artifacts can be easily exploited by machine-learning models to produce good results only on the OR-QuAC benchmark, which are not generalizable to real-world settings.  
In addition, the automatic relevance labeling process is problematic. The process labels a passage as relevant to a query based on whether it contains the answer of the query. However, the passage that contains the answer span might not be truly relevant if it does not include enough evidence to answer the question. On the other hand, there could be many relevant passages that do not contain the answer span.

\begin{table}
 \centering
  \caption{Statistics of the CAsT-19 and OR-QuAC benchmarks.  }
  \label{tab:convsearch_dataset}
 \centering
    \begin{tabular}{l|r|rrr} 
    \hline
                  & \textbf{CAsT-19}                & \multicolumn{3}{c}{\textbf{OR-QuAC}}  \\
                     {\textbf{Statistics}}          & \textbf{Test}   & \textbf{Train}  & \textbf{Dev}   & \textbf{Test}        \\ 
    \hline
    \# Conversations           & 20                         & 4,383  & 490   & 771         \\
    \# Questions               & 173                        & 31,526 & 3,430 & 5,571       \\
    \# Labels                  & 29,571                         & 31,526  & 3,430  & 6,544        \\
    \# Avg. Question Tokens    & 6.1                          & 6.7    & 6.6   & 6.7         \\
    \# Avg. Questions / Conversation & 9.6                           & 7.2    & 7.0   & 7.2         \\
    \# Avg. Labels / Question  & 170.9                           & 1.0    & 1.0   & 1.2       \\
    \hline
    \# Documents               & \multicolumn{1}{c|}{38M}                        & \multicolumn{3}{c}{11M}  \\
    \hline
    \end{tabular}
\end{table}

\begin{table}
  \caption{Notable properties of TREC CAsT and OR-QuAC datasets.}
  \label{tab:convdataset_properties}
   \centering
  \begin{tabular}{l|p{4cm}|p{4cm}}
  \hline
       & \textbf{TREC CAsT} & \textbf{OR-QuAC} \\ \hline
Topic       & Curated by Experts with reference to web search & Wikipedia Entities\\ \hline
Query       & Written by Experts with Interaction to search systems & Crowd-Sourced via Question-Answering between Workers \\ \hline
Corpus       & MS MARCO and Wikipedia. Passages in 2019 and 2020; Documents in 2021 & Wikipedia Passages \\ \hline
Relevance Label & Pooled Track Systems and Accessed by TREC &  Post Constructed by Answer Containment \\ \hline
Manual Query  Rewrites & By Experts & By Crowd Source Workers\\ \hline
Amount of Labels  & Small Scale, hundreds of queries & Large Scale, tens of thousands of queries \\ \hline
Response Dependency & Included in 2020 and 2021, not in 2019 & Include dependency on previous answer spans \\ 
       \hline
  \end{tabular}
\end{table}

\bigskip

The statistics of the CAsT and OR-QuAC benchmarks are listed in Table~\ref{tab:convsearch_dataset}. Their notable properties are summarized in Table~\ref{tab:convdataset_properties}. Note that we discuss the limitations of these benchmarks not to discourage the use of them, but to raise the awareness of potential misjudging the effectiveness of the search methods being evaluated due to these limitations. For example, a technique may not work well on CAsT only because there is no sufficient training signals in CAsT; a technique that uses spurious information (e.g., the description of a CAsT topic, or the design constrain that all relevant passages of a OR-QuAC topic are from the same Wiki page section) could have shown artificially high performance on these benchmarks but it is unlikely to generalize well to real-world scenarios.
``All benchmarks are wrong, but some are useful''\footnote{\url{https://en.wikipedia.org/wiki/All\_models\_are\_wrong}}. 
It is important to understand the pros and cons of a benchmark before it can be correctly used to evaluate research progress.

\subsection{Other Related Resources}
Besides TREC CAsT and OR-QUAC, there are many other datasets and resources related to conversational search.
The first group of related resources are the open-domain QA benchmarks, which consist of single round questions, answer labels, and open retrieval labels. 
These benchmarks include the open-retrieval version of SQuAD~\citep{chen2017reading, lee2019latent}, TriviaQA~\citep{joshi2017triviaqa}, and Natural Questions~\citep{natural-questions}.
The second group of related resources are the ad hoc search benchmarks that include single-round queries and relevance documents or passages. 
The most notable one is MS MARCO~\citep{bajaj2016ms}, which is one of the largest search relevance benchmark.
Both open-domain QA and ad hoc search datasets are frequently used as additional (weak) supervision data in conversational search.
Recently, the QReCC dataset~\citep{anantha2020open} is released as yet another conversational search benchmark. It extends CAsT and QuAC with additional conversations constructed from Natural Questions and additional passages sampled from Common Crawl.

\section{Pre-Trained Language Models}
\label{sec:cir_preliminary}

Before discussing the methods of conversational search, we first introduce pre-trained language models (PLMs), which have been adopted in nearly all the components of CIR systems, including  conversational search. In fact, many recent advances in CIR are attributed to PLMs. 

Language model pre-training is an active research field. 
In this section we describe BERT~\citep{devlin2018bert}, a classic PLM based on the Transformer architecture~\citep{vaswani:17}, and the pre-training and fine-tuning framework that has been widely used to adapt a PLM to downstream applications. For more detailed surveys we refer to \citet{xu2021pre} and encourage readers to check more recent updates in the field.

\subsection{BERT: A Transformer-Based PLM}
A Transformer is a deep learning model that adopts the self-attention mechanism to produce contextualized representations. 
Transformer models have been proved effective in performing a wide range of natural language and computer vision tasks, while also being relatively stable and robust in large-scale training. They are the de facto neural architecture for PLMs.

In the rest of the chapter, we use the following equations to denote the text encoding process using a Transformer-based PLM,
\begin{align}
    X &\xrightarrow{\text{Transformer}} \mathbf{H}, \text{or} \\
    \mathbf{H} &= f_\text{Transformer}(X),
\end{align}
where an input token sequence $X=\{x_1,...,x_i,...,x_N\}$, and $x_i$ is a sub-word (e.g., a BPE token~\citep{sennrich2015neural}), encoded by a sequence of contextual dense vectors $\mathbf{H}=\{\mathbf{h}_1,...,\mathbf{h}_i,...,\mathbf{h}_N\}$.

Now, we describe the encoding process in detail. As an example, we consider BERT (Bidirectional Encoder Representations from Transformers), one of the most widely used Transformer-based PLMs. 
BERT makes use of the attention mechanism of Transformer to learn contextual relations between words (or sub-words) in a text. In its vanilla form, Transformer includes two separate modules — an encoder that reads the text input and a decoder that produces a prediction for the task. Since BERT is an autoencoding language model, only the encoder module is needed.

BERT performs encoding in two steps: (1) input embedding using an embedding layer, and (2) contextualized encoding using a stack of Transformer layers. 

\paragraph{Embedding Layer.} 

\begin{figure}[t] 
\centering 
\includegraphics[width=0.8\linewidth]{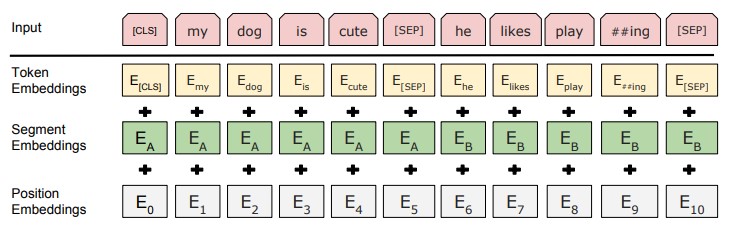}
\vspace{-1mm}
\caption{The BERT input embeddings are the sum of the token embeddings, the segmentation embeddings and the position embeddings. Figure credit: \cite{devlin2018bert}.} 
\label{fig:bert-input}
\vspace{0mm}
\end{figure}

The embedding layer converts the input sequence of discrete tokens to a sequence of continuous embedding vectors. 
As illustrated in Figure~\ref{fig:bert-input}, for a given token, its embedding vector is constructed by summing the corresponding token, segment, and position embeddings.
The Transformer in its vanilla form~\citep{vaswani:17} uses relative position embeddings while BERT uses absolute position embeddings, and DeBERTa, a recent variant to BERT, uses disentangled position embeddings~\citep{he2020deberta}.

\begin{figure}
    \centering
    \includegraphics[width=0.95\textwidth]{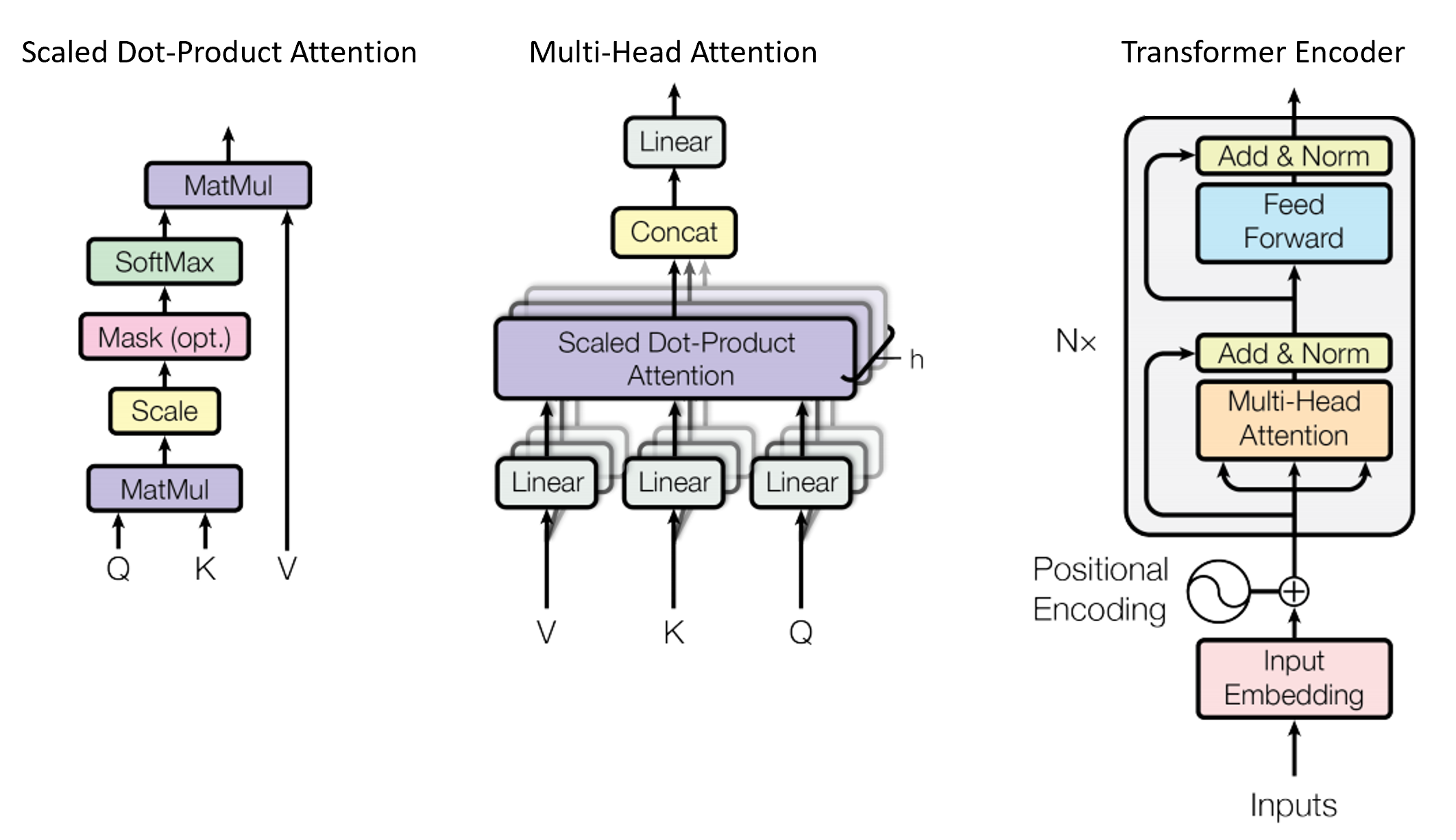}
    \caption{
    (Left) Scaled dot-product attention. 
    (Middle) Multi-head attention consists of several attention layers running in parallel. 
    (Right) The Transformer encoder. 
    Figure credit:~\cite{vaswani:17}.}
    \label{fig:Transformer}
\end{figure}

\paragraph{Transformer Layers.}

The embedding vectors are then fed into a stack of Transformer layers to produce contextual representations. 
Each Transformer layer has two sub-layers comprising a multi-head attention layer followed by a position-wise feed forward network, as depicted in Figure~\ref{fig:Transformer} (Right).

The attention function used by the Transformer is called \emph{Scaled Dot-Product Attention}, as illustrated in Figure~\ref{fig:Transformer} (Left). The input consists of queries and keys of dimension $d_k$, and values of dimension $d_v$.  In BERT, the embedding vectors of the input (sub-)words serve as the queries, keys and values, and we have $d_k = d_v = d$. To compute the attention functions on each input token, the queries, keys and values are packed together into matrices $\mathbf{Q}$, $\mathbf{K}$, $\mathbf{V}$. The attention matrix is computed as

\begin{equation} \label{eqn:ch3-transformer-attention}
\text{Attention}(\mathbf{Q},\mathbf{K},\mathbf{V}) = \text{softmax} \left( \frac{\mathbf{Q} \mathbf{K}^{\top}}{\sqrt{d_k}} \right) \mathbf{V},
\end{equation}
where $\sqrt{d_k}$ is the scaling factor of dot-product (multiplication) attention. 

Instead of performing a single attention function with $d$-dimensional keys, values and queries, \cite{vaswani:17} propose to perform \emph{Multi-Head Attention}, as illustrated in Figure~\ref{fig:Transformer} (Middle). 
It first linearly projects the queries, keys and values multiple times with different, learned linear projections to $d$ dimensions. Then, on each of these projected versions of queries, keys and values, the scaled dot-product attention function is performed in parallel, yielding $d$-dimensional output values. These are concatenated and once again projected, resulting in the final values.

\begin{align} \label{eqn:chp3-transformer-multi-heand-attention}
\text{MultiHead}(\mathbf{Q},\mathbf{K},\mathbf{V}) &= \text{Concat} \left(\text{head}_1,...,\text{head}_H \right) \mathbf{W}^O, 
\\
\text{head}_i &= \text{Attention}(\mathbf{Q}\mathbf{W}_i^Q, \mathbf{K}\mathbf{W}_i^K,  \mathbf{V}\mathbf{W}_i^V),
\end{align}
where $\mathbf{W}$’s are trainable parameter matrices.

\begin{figure}
    \centering
    \includegraphics[width=1\textwidth]{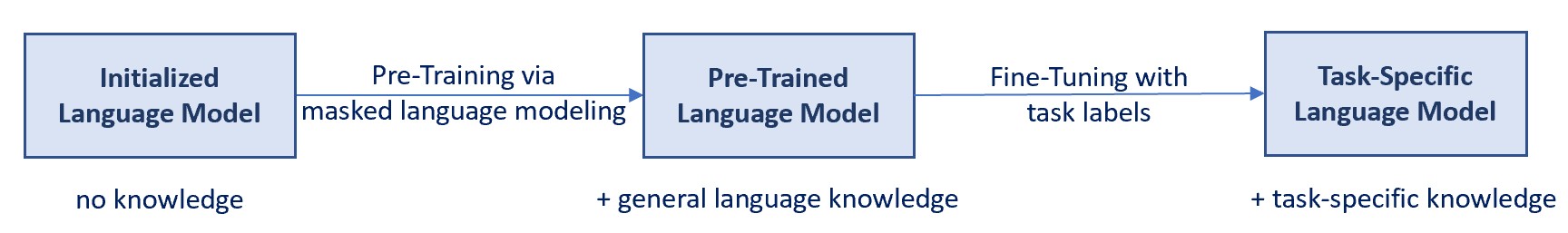}
    \caption{The pre-training and fine-tuning framework.}
    \label{fig:plm-frame}
\end{figure}

\subsection{The Pre-Training and Fine-Tuning Framework}
Training a deep Transformer model from scratch requires a large amount of labels that is only available in a few applications, such as machine translation where millions of parallel sentences can be used for training~\citep{vaswani:17}. A more efficient method is to first \emph{pre-train} a Transformer-based language model on raw text corpora via self-supervised learning ~\citep{devlin2018bert, liu2019roberta}, and then \emph{fine-tune} the pre-trained language model to downstream tasks with limited amounts of task-specific labels. 

Figure~\ref{fig:plm-frame} illustrates the 
\emph{pre-training and fine-tuning} framework. 
This is an instance of transfer learning, where the knowledge learned by the PLM for a language modeling task (e.g., masked language modeling) is transferred to perform a different downstream task.
In this section we review the masked language modeling task used for BERT pre-training~\citep{devlin2018bert}, and then discuss several common setups to fine-tune PLMs to downstream tasks.

\paragraph{Pre-training with Masked Language Modeling.}
The Masked Language Modeling (MLM) task can be illustrated as

\begin{align*}
    \texttt{[CLS]} x_1 \dots \texttt{[MASK]}_i \dots x_N \xrightarrow{\text{Transformer}} \mathbf{H} \xrightarrow{\text{MLM head}} P_{r{\text{MLM}}} (x_i | \mathbf{h}_i).
\end{align*}
Before feeding word sequences $X=\{x_1,...,x_N\}$ into BERT, 15\% of the words in each input sequence are replaced with a \texttt{[MASK]} token. The model then attempts to predict the original value of each masked word $x_i$, based on its contextual representation $\mathbf{h}_i$, which encodes context provided by the other, non-masked, words in the sequence. 
\texttt{[CLS]} is a special token, added at the beginning of an input sequence, to produce the representation of the entire sequence, which is useful for many downstream tasks.
Technically, the prediction of the output words requires adding an MLM head on top of the encoder output:
\begin{align*}
P_r (x | \mathbf{h}_i) = \mathrm{softmax}(\mathbf{W} \cdot \mathbf{h}_i) = \frac{\exp(\mathbf{W}_i \mathbf{h}_i)}{\sum_{x_t \in \mathcal{V}} \exp(\mathbf{W}_t \mathbf{h}_t)}.
\end{align*}
 This classification layer multiplies the output vectors $\mathbf{h}_i$ by a projection matrix $\mathbf{W}$, transforms them into the vocabulary dimension $|\mathcal{V}|$, and calculates the probability of each word in the vocabulary with softmax as

\begin{figure}
    \centering
    \includegraphics[width=0.70\textwidth]{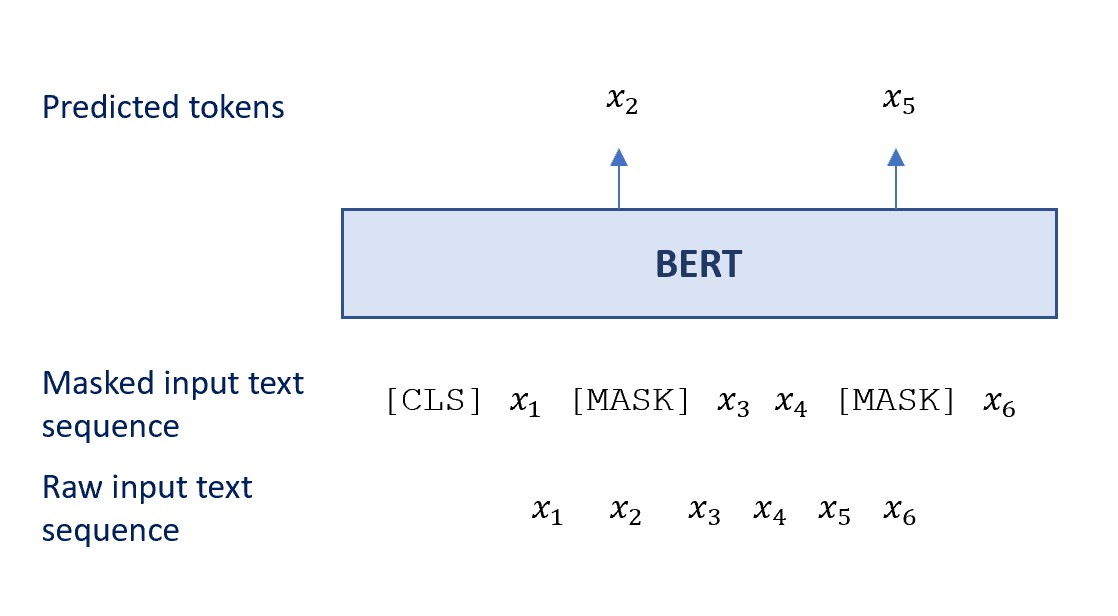}
    \caption{Pre-training BERT with Masked Language Modeling~\citep{devlin2018bert}.}
    \label{fig:BERT}
\end{figure}

Figure~\ref{fig:BERT} shows an example of pre-training BERT using MLM. Since no manual label is required, BERT can be trained on large amounts of raw text corpora. 
For example, the BERT base model, a 12-layer Transformer with 110 million parameters, is pre-trained on the Wikipedia and Google Book Corpora which amount to 16 GB words~\citep{devlin2018bert}. 
Recent results show that training a larger PLM on a larger dataset significantly improve model's performance, especially in few-shot or zero-shot settings where there is few or zero task labels in the downstream task~\citep{liu2019roberta, raffel2019exploring,brown2020language,roberts2020much,schick2020s}.

While the root reason why PLMs are so effective remains an open research topic, a recent study shows that BERT does capture linguistic information, representing the steps of the traditional NLP pipeline (e.g., POS tagging, parsing, NER, semantic roles, then coreference) in an interpretable and localizable way~\citep{tenney2019bert}.


\paragraph{Fine-Tuning for Downstream Tasks.}

\begin{figure}
         \includegraphics[width=1.0\textwidth]{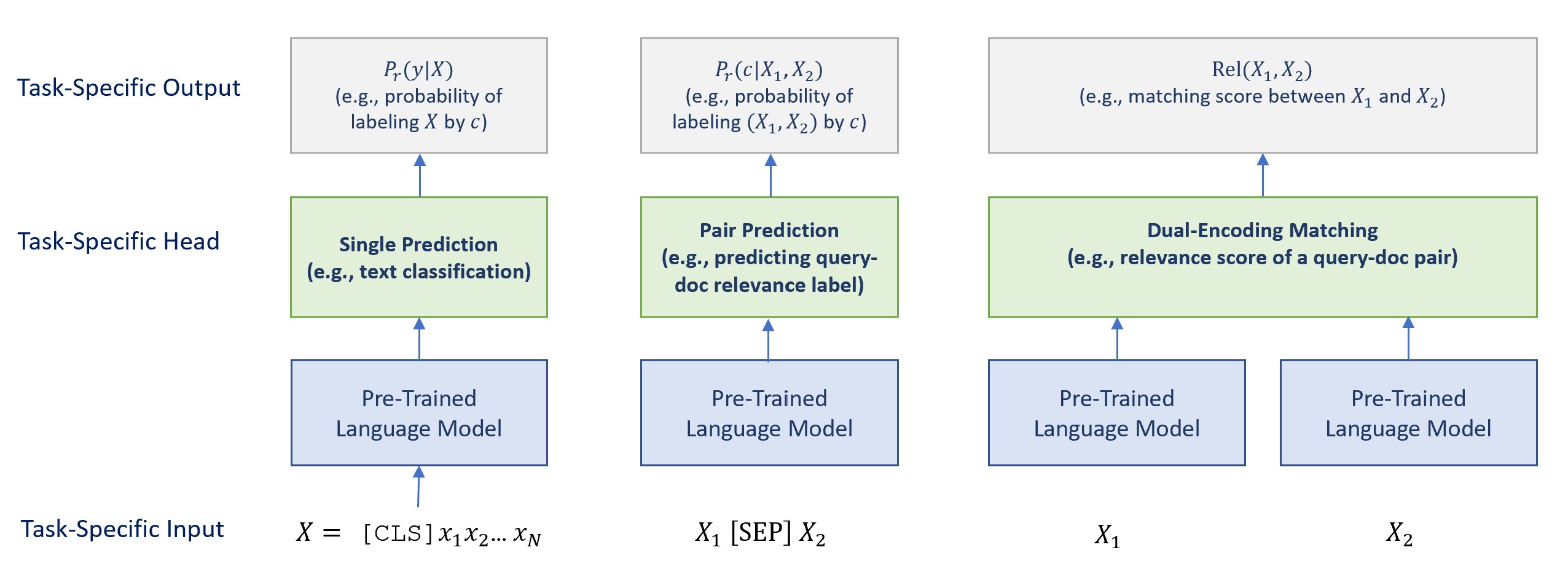}
            \caption{Three formulations of downstream tasks: single prediction (Left), pair prediction (Middle) and dual-encoding matching (Right).}
     \label{fig:plm_usage}
\end{figure}

A PLM can be adapted to downstream tasks by continually fine-tuning the parameters of the PLM and an added task-specific head using task labels, e.g., relevance labels in document search. 
Depending on the formulation of the target downstream task, we use different types of task-specific heads, and process training data accordingly.
Figure~\ref{fig:plm_usage} presents three commonly used formulations of downstream tasks for fine-tuning PLMs.

The first formulation is \textit{Single Prediction}, as illustrated in Figure~\ref{fig:plm_usage} (Left). It is often used to predict a label for a single text sequence. Take the text classification task as an example. The probability that  text $X$ is labeled as class $y$ (i.e., the positive or negative sentiment) is predicted by a logistic regression with softmax: 
\begin{align}
    P_r(y|X) = \text{softmax}(\mathbf{W} \cdot \mathbf{x}), \label{eq:single_predict}
\end{align}
where $\mathbf{x}$ is the contextual vector of $X$ (i.e., the contextual vector of the \texttt{[CLS]} token), and $\mathbf{W}$ is the task-specific parameter matrix.
Note that we can make prediction using the contextual vectors of other token, rather than \texttt{[CLS]}, in the input. For example, we might want to predict the POS tag of a word or a text span in extractive OA tasks.

The second formulation is \textit{Pair Prediction}, as illustrated in Figure~\ref{fig:plm_usage} (Middle). It is used to predict a label for a pair (or a set) of text sequences. An example is to predict the relevance label of a query-document pair. This can be achieved by feeding the concatenation of query string $X_1$, a \texttt{[SEP]} token, and document string $X_2$ to the model, and the relevance label $y$ is predicted using a logistic regression layer with softmax:
\begin{align}
    P_r(y|X_1, X_2) = \text{softmax}(\mathbf{W} \cdot \mathbf{x}),
    \label{eq:pair_predict}
\end{align}
where $\mathbf{x}$ is the contextual representation of the concatenated input sequence (i.e., the contextual vector of the \texttt{[CLS]} token), and $\mathbf{W}$ is the task-specific parameter matrix.
In neural IR, this pair prediction task is sometimes denoted as BERT$_\text{CAT}$ or Transformer$_\text{CAT}$.

The third formulation is \textit{Dual-Encoding Matching}, as illustrated in Figure~\ref{fig:plm_usage} (Right). This is often used in text similarity or matching tasks. For example, in document search we might want to compute the relevance score Rel(.) between query $X_1$ and each candidate document $X_2$ as
\begin{align}
    \text{Rel}(X_1, X_2) = \text{sim}(\mathbf{x}_1, \mathbf{x}_2;\theta),
    \label{eq:dual_encoding}
\end{align}
where $\mathbf{x}_1$ and $\mathbf{x}_2$ are the contextual representations of $X_1$ and $X_2$, generated by the two encoders (dual-encoding), respectively, and sim(.) is a similarity function which can be implemented using cosine similarity or a model parameterized by $\theta$.
Compared to the pair prediction formulation, dual-encoding is more often used for document retrieval where runtime efficiency is important. Dual-encoding allows documents and queries to be encoded separately so that we can \emph{pre-encode} all documents and the IR system only needs to encode an input query on the fly for retrieval. Dual-encoding is widely used for dense retrieval~\citep{huang2013learning, lee2019latent, karpukhin2020dense, xiong2020approximate}, which will be discussed in later sections. 
This formulation is also referred to as the representation-based model, dual-encoder, siamese-encoder, weak interaction model, deep structured similarity model etc.

The PLM fune-training follows the standard supervised learning procedure, where the parameters of the PLM and its added task-specific head are optimized w.r.t. the task loss, e.g., cross entropy for prediction, rank loss in relevance ranking, and NCE loss in dense retrieval. 

\section{System Architecture}
\label{sec:convsearch_arch}

\begin{figure}[t]
    \centering
    \includegraphics[width=0.95\linewidth]{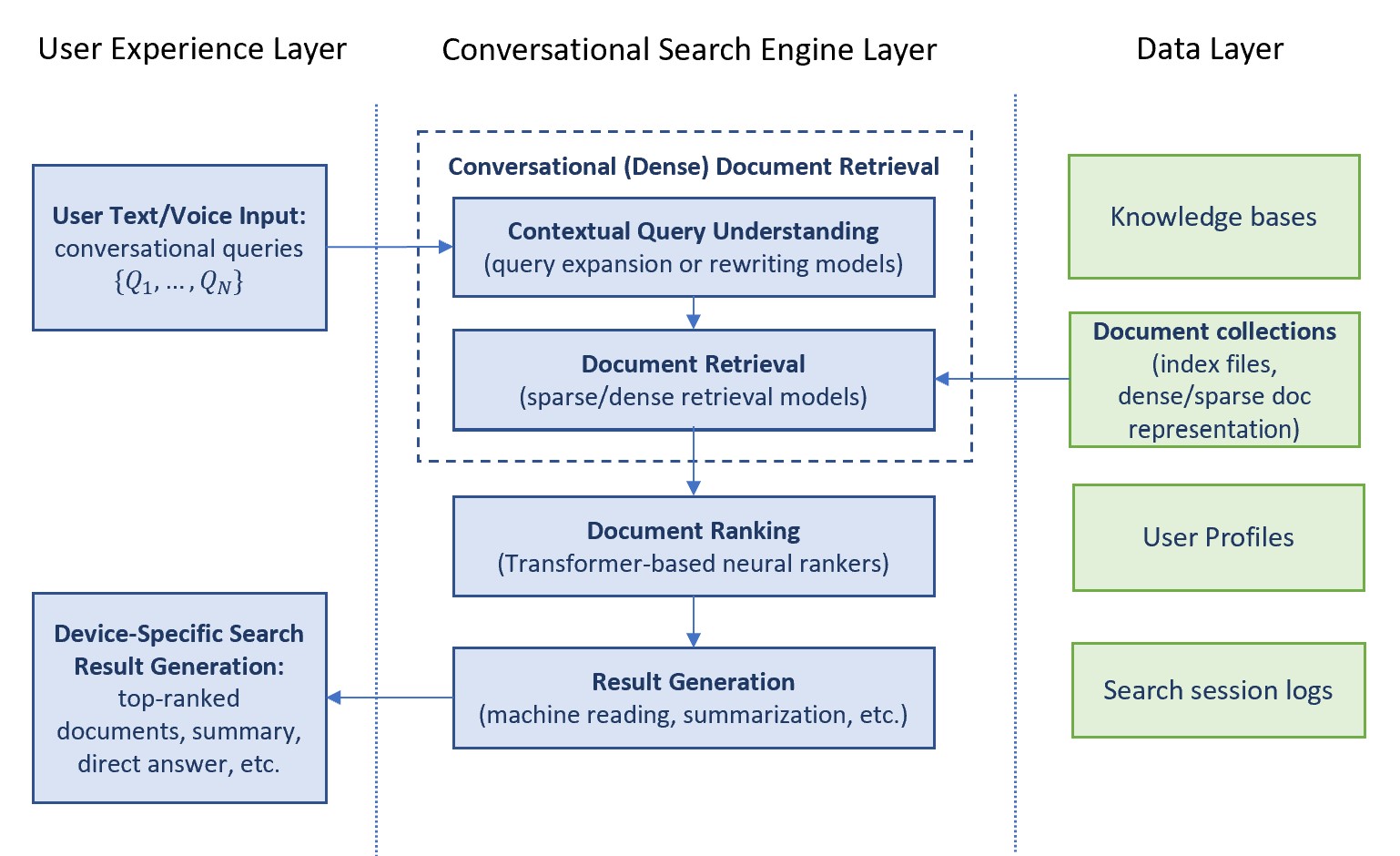}
    \caption{ A general architecture of conversational document search systems. \label{fig:document-search-arch}}
\end{figure}

A typical architecture of conversational search systems is illustrated in Figure~\ref{fig:document-search-arch}. It consists of a subset of the components of the CIR architecture in Figure~\ref{fig:cir-system-arch}.
Following the task setting described in Section~\ref{sec:cir_overview}, conversational search is performed as follows.
A user issues a series of conversational queries $\{Q_1,...,Q_N\}$ for a topic $\mathcal{T}$, where each query $Q$ is part of the conversational trajectory and its answer depends on not only $Q$ but its dialog history $\mathcal{H}$ which includes all its previous queries and their system responses.
The system returns for each query $Q$ a list of documents $\{D_1,...,D_M\}$, ranked by their relevance scores with respect to $Q$.
The search is conducted using three components: contextual query understanding, document retrieval, and document ranking. 
The result generation module will be described in Chapters~\ref{chp:qmds} and \ref{chp:c-mrc}. 

\paragraph{Contextual Query Understanding.} 
Since ellipsis phenomena are frequently encountered in conversational search, understanding a conversational query $Q$ relies on its context.  
Contextual query understanding conceptually acts as a query rewriter, which uses conversational context $\mathcal{H}$ to rewrite $Q \in \mathcal{T}$ to a de-contextualized query $\hat{Q}$ for document retrieval. 
The rewriter can be implemented as a separate module using query expansion methods or neural rewriting models (Section~\ref{sec:cqu}), or as part of the conversational dense retrieval module (Section~\ref{sec:conv_dense_retrieval}). 

\paragraph{Document Retrieval.} 
Identifying a short list of relevant documents of query $Q$ (or its de-contextualized form $\hat{Q}$) from a large document collection $\mathcal{D}$ is computationally expensive. For efficiency, a coarse-to-fine search strategy is widely adopted where the overall search process is divided into two phases, document \emph{retrieval} and \emph{ranking}, either of which may consist of multiple steps to perform coarse-to-fine document retrieval or (re-)ranking, respectively. 
Conceptually, document retrieval needs to examine each $D \in \mathcal{D}$ its relevance to $Q$. Since a brute-force method is prohibitively expensive,
it is necessary to construct an index of $\mathcal{D}$ for efficient retrieval. 
For example, sparse retrieval methods (Section~\ref{sec:sparse_retrieval}) often use the (weighted) inverted index to help find the top-$K$ relevant documents efficiently, as that in ad hoc search~\citep{croft2010search}.
Dense retrieval methods (Section~\ref{sec:dense_retrieval}, on the other hand, have to resort to efficient similarity search methods \citep{aumuller2017ann,FAISS,li2020improving} to find relevant documents in a continuous vector space.   

\paragraph{Document Ranking.} 
Given top-$K$ candidate documents retrieved from $\mathcal{D}$, where $K \ll |\mathcal{D}|$, we can afford to use sophisticated ranking models, such as Transformer-based PLMs, to (re-)rank these documents to return a short list of the most relevant documents to the user.   

\bigskip 

The rest of this chapter describes the models and methods recently developed for the aforementioned three components. 

\section{Contextual Query Understanding}
\label{sec:cqu}

The contextual query understanding (CQU) task is to reformulate a conversational query $Q$, given its conversational context $\mathcal{H}$, to a de-contextualized query $\hat{Q}$ that can be used alone (without $\mathcal{H}$) for document retrieval. So, $\hat{Q}$ is also called a \emph{standalone} query.  

This section presents three types of CQU methods based on heuristic query expansion, machine learning based query expansion, and neural query rewriting. 
All the methods we describe below are developed on the early TREC CAsT datasets, where turns in a conversational search session depend only on previous queries but not system responses, as shown in the examples in Table~\ref{tab:cast_q}. Thus, the conversational context of a query at the $i$-th dialog turn $Q_i$ contains only previous conversational queries, $\mathcal{H}_i = \{Q_1,…, Q_{i-1}\}$.

\subsection{Heuristic Query Expansion Methods}
\label{subsec:heuristic_qe}

Conversational query expansion methods~\citep[e.g.,][]{lin2020multi,yang2019query, voskarides2020query} generate a standalone query $\hat{Q}_i$ for a conversational query $Q_i$ by adding to $Q_i$ a set of query terms $\mathcal{S}$. The expansion terms are often selected from its dialog history $\mathcal{H}_i$ to optimize the retrieval result of $\hat{Q}_i$ as
\begin{align}
    \mathcal{S}^* &= \text{argmax}_{\mathcal{S} \subseteq \mathcal{H}_i} \text{Accuracy}(Q_i \cup \mathcal{S}), 
    \label{eq:resotarget}
\end{align}
where $\text{Accuracy}(Q_i \cup \mathcal{S})$ denotes the quality (accuracy) of the ad hoc retrieval results using the standalone query, which is $Q_i$ expanded with terms in $\mathcal{S}$.

Similar to query expansion in ad hoc search~\citep{croft2010search}, it is difficult, or often impossible, to directly optimize Equation~\ref{eq:resotarget}. The document search system, which is used to estimate $\text{Accuracy}(Q_i \cup \mathcal{S})$, is in general not differentiable. Thus, many approaches have been developed to obtain $\mathcal{S}^*$ approximately using either heuristics or machine learning methods. 

As an example of heuristic query expansion methods of CQU, we describe the Historical Query Expansion (HQE) method proposed by ~\cite{lin2020multi}. HQE is a very simple and effective method, leading to one of the top-performing systems in TREC CAsT 2019~\citep{yang2019query}. 
HQE was developed on the TREC CAsT dataset, taking advantage of two characteristics of conversational queries observed from the dataset.

\begin{enumerate}
    \item \text{Main topic and subtopic:} A conversational session is centered around a main topic, and the turns in the session dive deeper into subtopics, each of which only lasts a few turns. For example, in Table~\ref{tab:cast_q} (a), the main topic of the session is ``acid reflux'', Turns 3, 4 and 5 discuss the subtopics of ``OTC'', ``PPI'', and ``treatment'', respectively, while Turns 5 to 8 are related to ``treatments''. 
    \item \text{Degree of ambiguity:} The conversational queries can be labeled by degree of ambiguity of three categories. The first category includes queries with clear intent, which can be directly used as standalone queries, such as Turn 1 in Table~\ref{tab:cast_q} (a). The second contains those starting a subtopic (e.g., Turns 3, 4 and 5). The third contains ambiguous queries that continue a subtopic (e.g., Turns 6, 7, and 8).
\end{enumerate}

HQE performs CQU in three steps. For each search query $Q$ in a dialog session, (1) the main topic and subtopic keywords are extracted from $Q$; (2) the ambiguity of $Q$ is measured; and (3) a standalone query $\hat{Q}$ is formed by expanding $Q$ with the main topic and subtopic keywords extracted from its dialog history $\mathcal{H}$. These steps are performed using a keyword extractor and a query performance predictor.

\paragraph{Keyword Extractor (KE).}
KE computes the importance score of each query term $q$ with respect to the query (or topic) it occurs. \cite{lin2020multi} propose to approximate the importance score $R(q)$ using the retrieval score (e.g., BM25) of $q$'s highest-scoring document as
\begin{align}
    R(q) &= \max_{D \in \mathcal{D}} \text{BM25}(q, D).
\end{align}
The intuition behind this design is that the importance of a word can be judged from those documents that are (potentially) highly relevant to it. That is, if a word is representative of its relevant documents, it is likely a keyword of the topics of these documents \citep{lin2020multi}. 

\paragraph{Query Performance Predictor (QPP).}
QPP measures a query’s ambiguity. Assuming that the degree of query ambiguity is closely related to its ambiguity with respect to the collection of documents being searched. Following the design of KE, the ambiguity of a query $Q$ is scored as the retrieval score (e.g., BM25) of the query's highest-scoring document as 
\begin{align}
    \text{QPP}(Q) &= \max_{D \in \mathcal{D}} \text{BM25}(Q, D).
\end{align}
The higher the QPP score, the less ambiguous $Q$ is.

\bigskip

With KE and QPP defined above, HQE works in three steps:
\begin{enumerate}
    \item For each query $Q=\{q_1,...,q_N\}$ in a conversation session, HQE extracts topic (subtopic) keywords from $Q$ if $R(q) > R_{\text{topic}}$ ($R(q) > R_{\text{sub}}$), where $R_{\text{topic}}$ ($R_{\text{sub}}$) is a hyperparameter, and collects them in the keyword set $\mathcal{S}_{\text{topic}}$ ($\mathcal{S}_{\text{sub}}$).
    \item QPP measures the ambiguity of each query $Q$, and labels it as ambiguous if $\text{QPP}(Q) < \eta$, where $\eta$ is a hyperparameter.
    \item For all queries $Q_i$ in the session, except the first turn $Q_1$, HQE first rewrites $Q_i$ by concatenating it with the topic keyword set $\mathcal{S}_{\text{topic}}$ collected from its history $\mathcal{H}_i$. Moreover, if $Q_i$ is ambiguous, HQE further adds the subtopic keywords $\mathcal{S}_{\text{sub}}$ from the previous $M$ turns, where $M$ is a hyperparameter, assuming that subtopic keywords only last $M$ turns. The heuristic is based on the assumption that the first query in a session is always well-specified and that following queries belong to the second or the ambiguous category. 
\end{enumerate}

Although HQE performs well in TREC CAsT 2019, it is not clear whether the modeling assumptions,  based on the characteristics of the CAsT training data, are generalizable to real-world conversational search settings. In the next section, we describe another category of query expansion methods that more heavily utilize machine learning techniques.

\subsection{Machine Learning Based Query Expansion Methods}
\label{subsec:ml_qe}

Machine learning based query expansion methods tackle CQU of Equation~\ref{eq:resotarget} by defining a binary term classification problem: for each term appearing in dialog history $q \in \mathcal{H}_i$, decide whether to add it to the current query $Q_i$. That is, $\mathcal{S}^*$ consists of all the terms selected by a binary term classifier learned from training data ~\citep[e.g.,][]{voskarides2020query,cao2008selecting,xiong2015query}. 

In what follows, we describe in detail the method of Query Resolution by Term Classification (QuReTeC)~\citep{voskarides2020query} as an example. QuReTeC is composed of two parts: a BERT-based binary classifier to predict whether to add a term to the current query, and a distant supervision method of constructing term-level labels to train the classifier. 

\begin{figure}[t]
    \centering
    \includegraphics[width=1.0\linewidth]{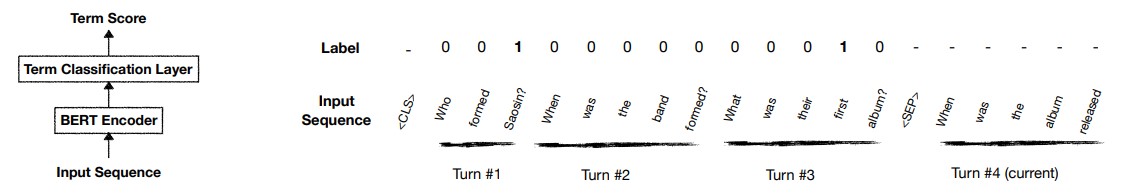}
    \caption{(Left) The QuReTec model architecture. (Right) An example input sequence and gold standard term labels (1: relevant, 0: non-relevant) for  QuReTeC, where the labels of \texttt{<CLS>}, \texttt{<SEP>} and the current turn terms are masked out. Figure credit:~\cite{voskarides2020query}}
    \label{fig:qeretec_eg}
\end{figure}

\subsubsection*{Term Classification.}

QuReTeC uses a BERT-based classifier, as shown in Figure~\ref{fig:qeretec_eg} (Left), which is composed of three components: an input sequence, a BERT encoder, and a term classification layer.
\begin{enumerate}
    \item The \emph{input sequence} consists of all the terms in dialog history $\mathcal{H}_i$ and the current turn $Q_i$. A \texttt{[CLS]} token is added to the beginning of the sequence, and a \texttt{[SEP]} token is added between the terms in $\mathcal{H}_i$ and the terms in $Q_i$. Figure~\ref{fig:qeretec_eg} (Right) shows an example input sequence and the gold standard term labels.  
    \item The \emph{BERT encoder} first represents the input terms with WordPiece embeddings using a 30K vocabulary, and then applies multiple Transformer layers to output a contextual representation (encoding vector) for each term.
    \item The \emph{term classification layer}, which consists of a dropout layer, a linear layer and a sigmoid function, is applied on top of the contextual representation of the first sub-token of each term to output a scalar for each term, indicating how likely the term should be added to $\mathcal{S}^*$. 
\end{enumerate}

\subsubsection*{Generating Distance Supervision.}

To train QuReTeC using the standard binary cross entropy loss, we need a dataset containing a gold binary label for each $q \in \mathcal{H}_i$. Since such labels are in general not available, we have to resort to distant supervision methods~\citep{mintz2009distant} to derive these labels from document-level relevance labels, which are widely available, based on how a term influences the document retrieval accuracy with respect to a query when the term is used to expand the query~\citep{cao2008selecting}.

Specifically for QuReTeC, given a query $Q_i$ and its labeled relevant document set $\mathcal{D}^+$, each term in its dialog history $q \in \mathcal{H}_i$ is labeled as position if $q$ appears in $\mathcal{D}^+$ but not in $Q_i$. 

Although the above procedure is noisy and can result in adding
terms to $Q_i$ that are non-relevant, or adding too few relevant terms $Q_i$, \citet{voskarides2020query} report good performance on the TREC CAsT 2019 dataset, where the majority of expansion terms in the oracle manual de-contextualized queries (see the example in Table~\ref{tab:cast_manualq}) are indeed from the previous query turns. However, it remains to be validated whether the observation and good performance hold on other conversational search benchmarks or in real-world scenarios.

\subsection{Neural Query Rewriting}
\label{subsec:neural_qr}

The query rewriting approach to CQU uses a neural natural language generation (NLG) model to generate standalone query $\hat{Q}_i$, using the conversational query $Q_i$ and its dialog history $\mathcal{H}_i$. In the TREC CAsT setting, we have $\mathcal{H}_i = \{ Q_1,...,Q_{i-1} \}$. Thus, the approach can be written as 
\begin{align}
    \{Q_1,...,Q_{i-1}\}, Q_i \xrightarrow{\text{NLG}(\theta)} \hat{Q}_i,
\end{align}
where the query rewriter $\text{NLG}(\theta)$ is often implemented using a Transformer-based auto-regressive language model, parameterized by $\theta$.
For example, \cite{vakulenko2021question} propose to use a fine-tuned GPT-2 model~\citep{radford2019language} for query rewriting as
\begin{align}
    \hat{Q}_i &= \text{GPT-2}(\texttt{[CLS]} \circ Q_1 \circ \texttt{[SEP]} \circ Q_2 ; \texttt{[SEP]}....\texttt{[SEP]}Q_{i} \circ \texttt{[GO]}),
    \label{eq:query_rewriter}
\end{align}
where $\circ$ is the concatenation operator, the input sequence starts with a \texttt{[CLS]} token and ends with a \texttt{[GO]} token, and the output standalone query $\hat{Q}_i$ is generated token by token using a beam search decoder or a greedy decoder. 

Fine-tuning a pre-trained GPT-2 model for query rewriting requires large amounts of gold query rewrites $\hat{Q}^*$ as training data. However, the TREC CAsT datasets consists of only a few hundred manually generated query rewrites, raising a concern that such a small amount of training data is likely to lead to overfitting when used to fine-tune deep neural networks like GPT-2. A common approach is to use the crowd sourced query rewrites in CANARD~\citep{elgohary2019canard} as additional training data.
It has been reported that the neural query writer based on fine-tuned GPT-2 is simple and effective. It leads to one of the best performers in TREC CAsT 2019~\citep{cast2019overview} and later becomes a standard solution adopted by many participants in TREC CAsT 2020~\citep{cast2020overview}. The neural query rewriter is also easy to be combined with other conversational query understanding methods ~\citep[e.g.,][]{lin2020multi} to achieve better IR results.

However, in real-world IR scenarios, large amounts of gold query writes are not available. In what follows, we describe a method of training the neural query rewriter in a few-shot learning setting.

\subsubsection*{Training Data Generation via Rules and Self-Supervised Learning}

\begin{figure}[t]
    \centering
    \includegraphics[width=1.0\linewidth]{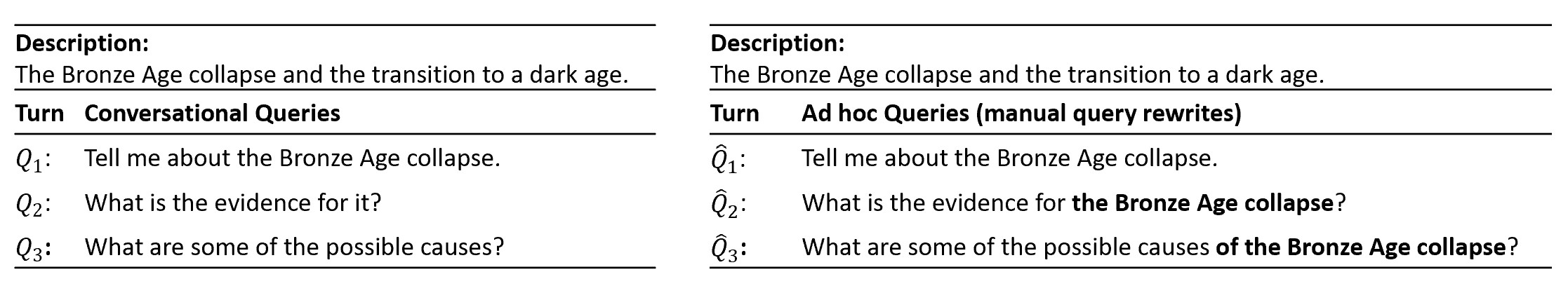}
    \caption{A conversational search session (Left) and an ad hoc search session (Right), adapted from~\cite{Yu2020Rewriter}}.
    \label{fig:conv_ad_hoc_session}
\end{figure}

\citet{Yu2020Rewriter} propose two methods, based on rules and self-supervised learning, respectively, to generate weak-supervision data using large amounts of ad hoc search sessions to fine-tune GPT-2 for conversational query rewriting. Compared to conversational search session, ad hoc search sessions are more available in large quantities as ad hoc search is more mature and the applications are more widely deployed. As illustrated in Figure~\ref{fig:conv_ad_hoc_session}, ad hoc search sessions can be viewed as pseudo target query
rewrites $\mathcal{\hat{H}} = \{\hat{Q}_1,...,\hat{Q}_n\}$, and they can be converted to conversation-like sessions $\mathcal{{H}} = \{{Q}_1,...,{Q}_n\}$ using a \emph{query simplifier}. Then, $(\mathcal{\hat{H}},\mathcal{H})$ pairs can serve as weak supervision to approximate gold query rewrites for fine-tuning GPT-2 for query rewriting. The query simplifier can be implemented using rule-based or machine learning approaches.

The rule-based approach uses two simple rules to mimic two ellipsis phenomena in conversations, omission and coreference, to simplify query $\hat{Q}_i$ given its session context $\mathcal{\hat{H}}_i$ as follows:
\begin{itemize}
    \item Omission: A noun phrase is omitted if it occurs after a preposition and appears in previous queries;
    \item Coreference: previously appeared singular and plural noun phrases are respectively replaced with pronouns.
\end{itemize}

The machine learning approach fine-tunes a pre-trained GPT-2 model for query simplification using a handful manual query rewrites. The model works similarly to the neural query rewriter of Equation~\ref{eq:query_rewriter}, but generates $Q$ from $\hat{Q}$'s. 
\begin{align}
    {Q}_i &= \text{GPT-2}(\texttt{[CLS]} \circ \hat{Q}_1 \circ \texttt{[SEP]} \circ \hat{Q}_2 \circ \texttt{[SEP]}....\texttt{[SEP]}\hat{Q}_{i} \circ \texttt{[GO]}),
    \label{eq:query_simplifier}
\end{align}
The query simplifier can then be applied to the ad hoc search sessions (e.g., MS MARCO) to generate more conversation-like sessions.
\cite{Yu2020Rewriter} find that it is easier to train a query simplifier (Equation~\ref{eq:query_simplifier}) that learns to omit information from an ad hoc query using session context than to train a query rewriter (Equation~\ref{eq:query_rewriter}) that learns to recover the omitted information. Empirically, the weak-supervision data generated using the machine learning approach significantly boosts the performance of the GPT-2 based query rewriter on the TREC CAsT benchmarks.

\section{Sparse Document Retrieval}
\label{sec:sparse_retrieval}

The de-contextualized query produced by the CQU module (Section~\ref{sec:cqu}) can be used as a standalone query to retrieve documents using any ad hoc retrieval system. Depending on how queries (and documents) are represented, retrieval systems can be grouped into two categories: \emph{sparse retrieval} systems that use sparse vectors to represent queries (and documents), and \emph{dense retrieval} systems that use dense vector representations. Sparse retrieval is described in this section, and dense retrieval the next section.

To retrieve a set of documents that are relevant to  query $Q$ from document collection $\mathcal{D}$, a sparse retrieval system first represents $Q$ and each $D \in \mathcal{D}$ using sparse vectors, computes the relevance score between query vector $\mathbf{q}$ and each document vector $\mathbf{d}$, and then selects top-$K$ most relevant documents.

The most commonly used query and document vectors are based on bag-of-words (BOW) representations,
\begin{align}
    \mathbf{q} &= \{(x, w) | \forall x \in Q\} \\
    \mathbf{d} &= \{(x, w) | \forall x \in D\},
    \label{eq:bow}
\end{align}
where $\mathbf{q}$ are $\mathbf{d}$ are high-dimensional sparse vectors, where each dimension corresponds to a term $x$ in a pre-defined vocabulary, and the term weight $w$ is none-zero only if the term appears in the query or document. The term weight can be binary, term frequency counts, or machine learned~\citep{croft2010search}.

Then, the query-document relevance score can be computed using a retrieval function, and top-$K$ relevant documents are returned. One of the most popular retrieval functions is BM25~\citep{robertson2009probabilistic}:
\begin{align}
    \text{BM25}(Q, D) &= \sum_{x\in Q} \text{TF}(x, Q)\text{IDF}(x, \mathcal{D}) \cdot 
    \frac { \text{TF}(x, D) \cdot (k_1 + 1)} 
    {\text{TF}(x, D) + k_1 \cdot (1-b+b \cdot \frac{|D|} {L})}, 
    \label{eq:bm25}
\end{align}
where $k_1$ and $b$ are hyperparameters, and the function scores a query-document pair using three sets of sufficient statistics:
\begin{enumerate}
    \item Term Frequency (TF) is the frequency of a term appearing in the query $\text{TF}(x, Q)$ and the document $\text{TF}(x, D)$. The higher the TF, the more important the term is.
    \item Invert Document Frequency (IDF) is the inverse of the fraction of documents in $\mathcal{D}$ containing this term $\text{IDF}(x, \mathcal{D})$. The higher the IDF, the more informative the term is.
    \item The ratio of the document length $|D|$ and the average document length $L$ in $\mathcal{D}$ is used to normalize the TF-IDF score so that short documents will score better than long documents given they both have the same number of term matches.
\end{enumerate}

BM25 is originally derived from the Binary Independence Model~\citep{robertson1976relevance}, which assumes that terms are mutually independent so that the relevance score of a query-document pair is a weighted combination of the matching score of each query term with respect to the document. This allows efficient document retrieval using data structures such as inverted index. 
BM25 is also sometimes referred to as a \emph{family} of BOW scoring functions with slightly different components and parameters. The function of Equation~\ref{eq:bm25} is one of the most prominent instantiations. We refer the readers to ~\cite{croft2010search} for a comprehensive review of sparse retrieval and indexing methods.
\section{Dense Document Retrieval}
\label{sec:dense_retrieval}

Sparse retrieval methods are based on lexical matching. Although simple and computationally efficient, lexical matching can be inaccurate due to the fact that a concept is often expressed using different vocabularies and language styles in documents and queries. For example, semantically similar words (e.g., ``automobile' and ``vehicle'') are not viewed as a match by BM25. In addition, sparse retrieval functions, such as BM25, are not trainable, and thus cannot leverage large amounts of search logs to improve retrieval performance.

Dense retrieval methods can effectively address these issues. The use of dense vector representations for retrieval has a long history since Latent Semantic Analysis~\citep{deerwester1990indexing}. Recently, dense retrieval methods that are based on deep learning models trained on labeled query-document pairs have become popular~\citep[e.g.,][]{huang2013learning,shen2014latent,karpukhin2020dense}. This section describes the dual-encoder architecture which is widely used for neural dense retrieval models (Section~\ref{subsec:dual_encoder_arch}), how dense retrieval models are applied efficiently to retrieve relevant documents over a large document collection using approximate nearest neighbor search (Section~\ref{subsec:ann_search}), and the way dense retrieval models are trained using search logs (Section~\ref{subsec:dr_model_training}).

\subsection{The Dual-Encoder Architecture}
\label{subsec:dual_encoder_arch}

\begin{figure}[t]
    \centering
    \includegraphics[width=0.6\linewidth]{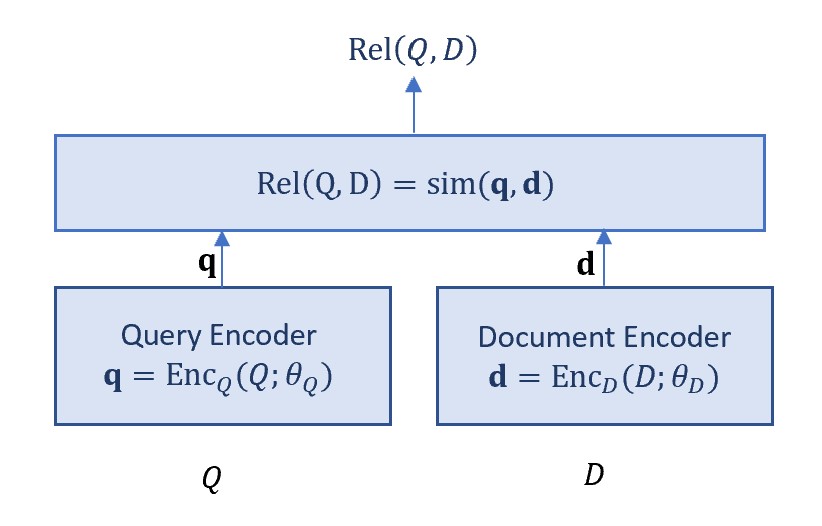}
    \caption{The dual-encoder architecture of neural dense retrieval models.}
    \label{fig:dual-encoder-dr}
\end{figure}

The dual-encoder architecture of neural dense retrieval models~\citep{huang2013learning,shen2014latent,karpukhin2020dense} is shown in Figure~\ref{fig:dual-encoder-dr}.

Dense retrieval works by using two neural encoders (e.g., a pair of fine-tuned BERT models, as shown in Figure~\ref{fig:plm_usage} (Right)). 
One, called document encoder $\text{Enc}_D(;\theta_D)$ parameterized by $\theta_{D}$, encodes document $D$ into a dense vector $\mathbf{d}$. The other, called query encoder $\text{Enc}_Q$, encodes query $Q$ into a dense vector $\mathbf{q}$. 
Using a BERT encoder as an example, the input is the text sequence of $Q$ (or $D$), appending the \texttt{[CLS]} token in the beginning of the input sequence, and the output vector of \texttt{[CLS]} produced by BERT can be used as a dense vector representation of $Q$ (or $D$).  

Then, the relevance score of $D$ given $Q$ can be computed as similarity between the two dense vectors,
\begin{equation} \label{eq:dual_enc}
\begin{aligned}
    \text{Rel}(Q, D ; \theta) &= \text{sim} (\mathbf{q}, \mathbf{d}) \\ 
    &= \text{sim}(\text{Enc}_Q(Q;\theta_Q), \text{Enc}_D(D;\theta_D)),
\end{aligned}
\end{equation}

where the similarity function can be implemented using dot product, L2 distance, cosine similarity, etc.

During training, we feed query-document pairs into the model, and the model parameters $\theta = \{\theta_Q, \theta_D\}$ are optimized to maximize the relevance score between a query and its relevant documents, as to be detailed in Section~\ref{subsec:dr_model_training}.

At indexing time, all documents in the collection $\mathcal{D}$ need to be encoded into dense vectors using $\text{Enc}_D$. The vectors are then stored and indexed. 

At runtime, only the input query needs to be encoded into $\mathbf{q}$ on the fly using $\text{Enc}_Q$. Next, $\mathbf{q}$ is compared against the already indexed document vectors, and top-$K$ most relevant documents, estimated using Equation~\ref{eq:dual_enc}, are returned as the retrieval result.

Unlike sparser retrieval where efficient retrieval can be achieved using inverted index, a dense retrieval process, which requires computing the relevance score of the input query vector against every document vector $\mathbf{d} \in \mathcal{D}$, could be prohibitively expensive for a large document collection. Thus, efficient dense retrieval relies on specific designs of indexing strategies, similarity search algorithms, and large-scale computer infrastructures.

\subsection{Approximate Nearest Neighbor Search}
\label{subsec:ann_search}

Formally, the nearest neighbor similarity search problem is defined as follows. Given query vector $\mathbf{q}$ and the collection of $N$ document vectors $\mathcal{D} = \{\mathbf{d}_1,...,\mathbf{d}_N\}$, we search the $K$ nearest neighbors of $\mathbf{q}$ in terms of a similarity (or distance) function as
\begin{align}
    K\text{-argmax}_{\mathbf{d} \in \mathcal{D}} \text{sim}(\mathbf{q}, \mathbf{d}).
    \label{eq:ann}
\end{align}
Approximate nearest neighbor (ANN) search relaxes the guarantee of exactness for efficiency by vector compression and by only searching a subset of $\mathcal{D}$ for each query. Searching a larger subset increases both accuracy and latency. We review some commonly used ANN methods, following closely the descriptions in~\cite{li2020improving,FAISS}.

\paragraph{Vector Compression.}
The first source of approximation comes from compressed vectors. The most popular vector compression methods includes binary codes~\citep{gong2012iterative,he2013k}, and vector quantization methods~\citep{jegou2010product,pauleve2010locality}. These methods have the desirable property that searching neighbors does not require reconstructing the vectors. For example, in a vector quantization method, a vector is first reduced by
principal component analysis dimension reduction and then is subsequently quantized. Although the process introduces similarity approximation error, it often results in orders of magnitude of compression, significantly improving efficiency in vector storage and similarity calculation. 

\begin{figure}[t]
    \centering
    \includegraphics[width=1.0\linewidth]{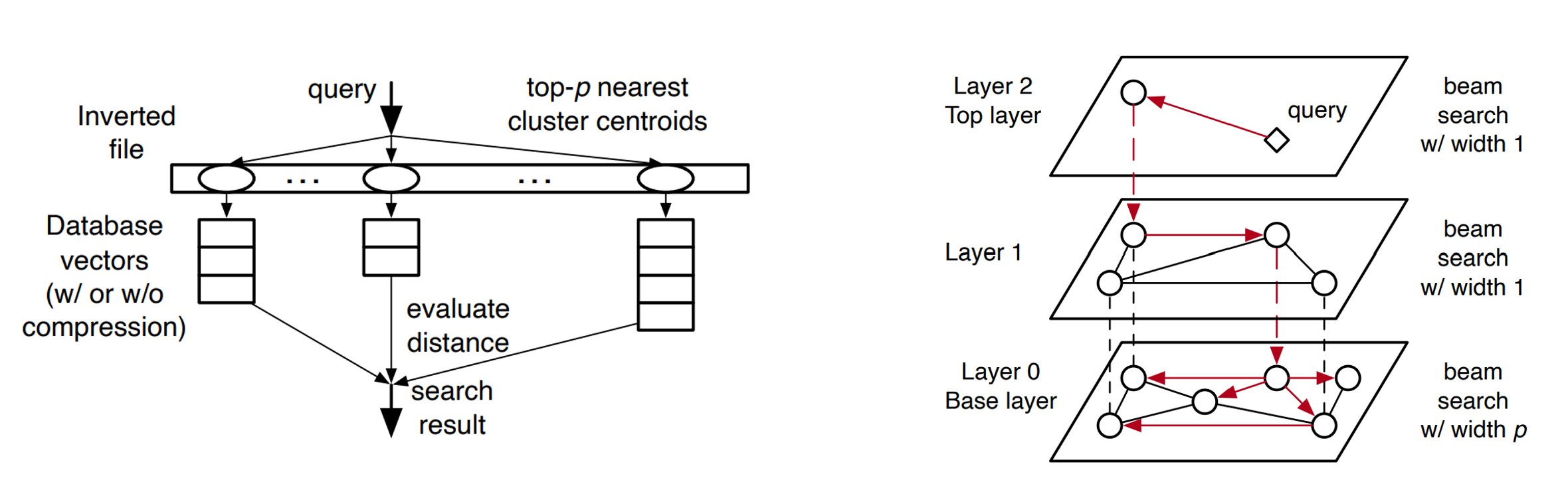}
    \caption{Two state-of-the-art ANN indexing methods: the IVF index (Left) and the (three-layer) HNSW index (Right). Figure credit: \cite{li2020improving}}.
    \label{fig:ann-index}
\end{figure}

\paragraph{ANN Indices.}
The second source of approximation comes from ANN indices that restrict the distance evaluations to a subset of document vectors. Two state-of-the-art methods are inverted file index (e.g., IVF [31]) and graph-based indexing (e.g., HNSW [41]).
As illustrated in Figure~\ref{fig:ann-index} (Left), the IVF index groups document vectors into different clusters. When building the index, a list of cluster centroids is learned via K-means clustering, and each document vector is assigned to the cluster with the closest centroid. During searching, the
index first computes the similarity between the query and all cluster centroids, then evaluates the document vectors belonging to the top-$p$ nearest clusters. A larger $p$ increases both accuracy (good) and search latency (bad). 
Hierarchical Navigable Small World Graphs (HNSW) is a graph-based indexing approach. As illustrated in Figure~\ref{fig:ann-index} (Right), HSNW includes multiple layers of proximity graphs. The top layer contains a single node and the base layer all document vectors. Each intermediate layer contains a subset of document vectors covered by the next lower layer. During indexing, document vectors are inserted one by one into multiple layers from the base layer up to a certain layer determined by an exponentially decaying probability distribution. At each insertion, the newly-inserted vector is connected to at most a fixed number of nearest nodes previously inserted to the same graph to create an approximate KNN-graph. In addition, nodes from different clusters are selected to be connected to improve the global
graph connectivity. At runtime, given query $\mathbf{q}$, beam search is performed at each layer, starting from the top layer, to identify nearest neighbor of $\mathbf{q}$ among the top-$p$ best candidate nodes in a coarse-to-fine manner. Like in IVF, a larger $p$ increases accuracy and search latency.  

\paragraph{When to Stop Search.}
While state-of-the-art ANN approaches use fixed configurations that apply the same termination condition (the size of subset to search) for all queries, \citet{li2020improving} point out that the number of dense vectors that need to be searched to find top-$K$ nearest neighbors varies widely among queries, and that the intermediate search result (after a certain amount of search steps) provides important information for estimating how much more search should be performed.  To achieve a better trade-off between latency and accuracy, they propose an approach that adaptively determines search termination conditions for individual queries. A set of gradient boosting decision tree classifiers are trained to predict when to stop searching. Applying these classifiers achieves the same accuracy with much less total amount of search compared to the fixed configurations.

\paragraph{Open-Source ANN libraries.} There are several open-source libraries that provide implementations for ANN search. The most popular one is perhaps FAISS \citep{FAISS}. It contains algorithms that search in sets of vectors of any size, up to the ones that possibly do not fit in RAM. FAISS is written in C++ with complete wrappers for Python/numpy. Some of the most useful algorithms are implemented on the GPU.

\subsection{Model Training}
\label{subsec:dr_model_training}

Let $\theta=\{ \theta_Q, \theta_D \}$ be the parameters of the neural dense retrieval model of Equation~\ref{eq:dual_enc}. $\theta$ is learned to identify the most effective query and document representations for document retrieval. Consider a query $Q$ and two candidate documents $D^+$ and $D^-$, where $D^+$ is more relevant than $D^1$ to $Q$. Let $\text{Rel}(Q,D; \theta)$, as defined in Equation~\ref{eq:dual_enc}, be the relevance score of $D$ given $Q$. We want to maximize $\Delta = \text{Rel}(Q,D^+; \theta) - \text{Rel}(Q,D^-; \theta)$. We do so by minimizing a smooth pair-wise rank loss \citep{huang2013learning} defined as
\begin{align}
    \mathcal{L}(\Delta; \theta) &= \log (1 + \exp (-\gamma \Delta)), \label{eq:rank_loss_1}
\end{align}
where $\gamma$ is a scaling factor,
or more commonly the margin-based loss~\citep{liu2009learning} defined as
\begin{align}
    \mathcal{L}(\Delta; \theta) &= [ \lambda - \Delta ]_{+}, \label{eq:rank_loss_2}
\end{align}
where $[x]_{+} := \max(0, x)$, and $\lambda$ is the margin hyperparameter.

The quality of the trained model depends on the quality and quantity of the training data. One common approach is to collect weak relevance labels in large quantities from search click logs. For example, \cite{huang2013learning} treat all the documents that have been clicked more than 5 times by users for query $Q$ as its relevant (positive) documents $D^+$, and randomly sample a set of documents from the document collection $\mathcal{D}$ as irrelevant (negative) documents $D^-$.
However, such randomly sampled $D^-$'s are often so irrelevant to $Q$ (i.e., resulting in $\Delta \gg 0$ in Equations~\ref{eq:rank_loss_1} and~\ref{eq:rank_loss_2}) that they do not provide useful information for model training~\citep{karpukhin2020dense}. 
\cite{xiong2020approximate} provides a theoretical analysis on the impact of negative sampling for dense retrieval model training using the variance reduction and negative sampling framework~\citep{alain2015variance, katharopoulos2018not}. The analysis shows that randomly sampling negatives from $\mathcal{D}$ is unlikely an effective importance sampling strategy and often yields inefficient learning or slow convergence.

One remedy to the problem is to sample \emph{challenging} negative documents that have high ranking scores assigned by a sparse ranking model such as BM25~\citep{karpukhin2020dense,ding2020rocketqa}.
The drawback of this approach is that the training signals provided by the sparse model helps improve the performance of the target dense retrieval model by forcing it to \emph{mimic} the sparse model. Thus, the trained dense model is unlikely to suppress or compensate the sparse model~\citep{luan2020sparsedense}.

To address this drawback, \cite{xiong2020approximate} propose a self-training method where the negative documents are sampled by the dense retrieval model being optimized. Since the dense retrieval model is updated in each mini-batch, the self-training follows an asynchronous iterative process ~\citep{guu2020realm}, where a set of GPUs are reserved to recalculate the dense vectors for the documents in $\mathcal{D}$ using the document encoder trained in the last iteration, and then update the ANN indices for negative sampling.

\section{Conversational Dense Document Retrieval}
\label{sec:conv_dense_retrieval}

The document retrieval modules, described in the last two sections, take as input a standalone query $\hat{Q}$, which is rewritten from a conversational query $Q$ given its dialog history $\mathcal{H}$ by the CQU module (Section~\ref{sec:cqu}). This \emph{pipeline} approach is appealing because we can re-use well-developed ad hoc document retrieval models for conversational search. But the drawback of this approach is that the CQU module cannot be optimized directly for retrieval performance, e.g., using labeled query-document pairs, because query rewriting is a discrete step which makes the gradient-based training algorithms not applicable. Moreover, developing the CQU module independently of the retrieval module requires gold query rewrites, which are much more difficult to collect in large amounts than labeled query-document pairs which can be obtained from search click logs. 

This section describes conversational dense retrieval (ConvDR) models which take conversational query $Q$ and its dialog history $\mathcal{H}$ as input, and retrieve top-$K$ relevant documents without using a separate CQU module. This \emph{integrated} approach is motivated by the fact that dense retrieval models search relevant documents using a dense vector query, which can encode both the query and its conversational context.

\cite{qu2020orqa} propose one of the first ConvDR models by extending the dual-encoder-based dense retrieval model of Equation~\ref{eq:dual_enc}. Formally, at the $i$-th turn, the ConvDR model retrieves top-$K$ relevant documents via
\begin{align}
    K\text{-argmax}_{\mathbf{d} \in \mathcal{D}} \text{sim}(\mathbf{q}, \mathbf{d})
\end{align}
where
\begin{align}
    \mathbf{q} &= \text{Enc}_{\hat{Q}}(Q_i, \mathcal{H}_i ; \theta_{\hat{Q}})
    \label{eq:conv_dr} \\
    \mathbf{d} &= \text{Enc}_{D}(D, \theta_D),
\end{align}

Note that the query encoder encodes both $Q_i$ and $\mathcal{H}_i$ to $\mathbf{q}$, taking the roles of both CQU and the ad hoc query encoder of Equation~\ref{eq:dual_enc}. Therefore, the ConvDR model can be trained end-to-end to directly optimize document retrieval performance on query-context-document $(\{Q, \mathcal{H}\} , D)$ pairs in the same way the ad hoc dense retrieval model is trained, as described in Section~\ref{subsec:dr_model_training}.

\cite{qu2020orqa} show that with sufficient training data (i.e., query-context-answer pairs), ConvDR significantly outperforms the pipeline approach on the open-retrieval conversational QA (OR-QuAC) task.

However, unlike the case of ad hoc search where labeled query-document pairs can be extracted from search click logs, conversational search logs are not available in large quantities because conversational search engines have not been widely deployed. Next, we will describe a method of training ConvDR models in a few-shot learning setting.

\subsection{ConvDR Model Trainining}
\label{subsec:convdr_model_training}

\cite{yu2021convdr} propose a few-shot learning method to train a ConvDR model using the teacher-student knowledge distillation framework~\citep{hinton2015distilling}, where the query encoder of ConvDR (student) learns to produce the output of the query encoder of a well-trained ad hoc dense retrieval model (teacher).
Specifically, the method assumes that we have 
\begin{enumerate}
    \item A few hundred gold query rewrites, $(\{Q,\mathcal{H}\}, \hat{Q})$, where $\hat{Q}$ is the standalone query, manually rewritten based on search query $Q$ and its conversational context $\mathcal{H}$, as in the CANARD dataset~\citep{elgohary2019canard};
    \item A few labeled query-context-document pairs, $(\{Q,\mathcal{H}\},D^+)$ (or $(\{Q,\mathcal{H}\},D^-)$), where $D^+$ (or $D^-$) is labeled as relevant (irrelevant) to $Q$ and its conversational context $\mathcal{H}$; and
    \item A well-trained ad hoc dense retrieval model, e.g., ANCE \citep{xiong2020approximate}, which uses the dual-encoder architecture.
\end{enumerate}

The ANCE document encoder is used directly as the document encoder of ConvDR. The query encoder of ConvDR, as in Equation~\ref{eq:conv_dr}, is trained using multi-task learning, where the parameters $\theta_{\hat{Q}}$ are optimized by minimizing a combination of two loss functions, one for each tasks.

\begin{align}
    \mathcal{L}(\theta_{\hat{Q}}) = \mathcal{L}_{\text{KD}}(\theta_{\hat{Q}}) + \mathcal{L}_{\text{rank}}(\theta_{\hat{Q}})    
\end{align}

$\mathcal{L}_{\text{KD}}$ is the loss of the knowledge distillation (KD) task, where $\text{Enc}_{\hat{Q}}$ (student) tries to mimic the ANCE query encoder $\text{Enc}_{Q}^{\text{ANCE}}$ (teacher) as much as possible. The underlying assumption is that since the information need behind the standalone query $\hat{Q}$ is identical to that of the conversational query $Q$ in its dialog context $\mathcal{H}$, their dense representations should be the same. This can be achieved by minimizing the mean square error (MSE) between the produced query representations. On a single training sample $(\{Q,\mathcal{H}\}, \hat{Q})$, $\mathcal{L}_{\text{KD}}$ is defined as

\begin{align}
    \mathcal{L}_{\text{KD}}(\theta_{\hat{Q}}) := \text{MSE} \left( \text{Enc}_{\hat{Q}}(Q, \mathcal{H}; \theta_{\hat{Q}}), \text{Enc}_{Q}^{\text{ANCE}}(\hat{Q}) \right)
\end{align}

The second task is the pair-wise ranking task as described in Section~\ref{subsec:dr_model_training}. $\mathcal{L}_{\text{rank}}(\theta_{\hat{Q}})$ is the loss defined in Equation~\ref{eq:rank_loss_1} or~\ref{eq:rank_loss_2}.
\section{Document Ranking}
\label{sec:doc_ranking}

As the retrieval module retrieves only a few hundred candidate documents from a large document collection, the document ranking step can afford to use more sophisticated models than those used for document retrieval to (re-)rank the retrieved candidates.

A common approach to document ranking is to learn a ranking model $f(Q,D; \theta)$, parameterized by $\theta$ and trained using relevance labels, to score each retrieved candidate document $D$ with respect to input query $Q$.  The ranking model can be a feature-based statistical classifier such as SVM~\citep{joachims2002optimizing} and Boosted Tree~\citep{wu2010adapting}, or a neural network model~\citep{mitra2018introduction}. Recently, ranking models based on pre-trained language models (PLMs) have achieved SOTA on many IR benchmarks, and have become the new standard solution to document ranking~\citep{lin2020pretrained}. In what follows, we present a BERT-based document ranker as an example. 

\cite{nogueira2019passage} cast document ranking as a \emph{pair prediction} task that predicts a document to be relevant or not for an input query, and fine-tune a pre-trained BERT model using the pair prediction formulation, as illustrated in Figure~\ref{fig:plm_usage} (Middle). Specifically, we feed the concatenation of query $Q$, a \texttt{[SEP]} token, and document $D$ to the ranker, and the relevance label $y \in \{1, 0\}$ is predicted using a logistic regression layer with softmax as
\begin{align}
    P_r(y|Q, D) = \text{softmax}(\mathbf{W} \cdot \mathbf{x}),
    \label{eq:bert-doc-ranker}
\end{align}
where $\mathbf{x}$ is the contextual representation of the concatenated input sequence (i.e., the contextual vector of the \texttt{[CLS]} token, produced by BERT), and $\mathbf{W}$ is the parameter matrix of the logistic regression layer.

Note that while the BERT-based ranker uses the pair prediction formulation for fine-tuning, the PLM-based dense retrievers described in Sections~\ref{sec:dense_retrieval} and \ref{sec:conv_dense_retrieval} use the dual-encoding formulation, as illustrated in Figure~\ref{fig:plm_usage} (Right). The different design choices are due to the efficient-effectiveness trade-off in different settings. 

The pair prediction formulation allows the model to explicitly capture the term-level matches between $Q$ and $D$, as its self-attention mechanism of BERT is applied on all pairs of query and document terms. The dual-encoder model, however, has to compress all information of $Q$ or $D$ to two dense vectors, respectively, and performs query-document matching at the document (or query) level, which is coarser than the term-level matching. \cite{mitra2018introduction,xiong2017knrm} report that the ranking models based on term-level matching are more effective than the models based on document-level matching for IR.

Although pair prediction models are widely used for document ranking due to their effectiveness, they are not feasible for dense retrieval which requires to score all the documents in a large document collection. Dual-encoder models, on the other hand, are widely used for document retrieval due to their efficiency. They allow all documents to be pre-encoded so that only the input query needs to be encoded on the fly. Then the relevance scores can be computed efficiently e.g., using dot product, and the retrieval is support by efficient ANN algorithms.

\chapter{Query-Focused Multi-Document Summarization}
\label{chp:qmds}

Query-focused multi-document summarization (QMDS) aims at producing a concise and fluent summary of a set of documents that are returned by the document search module in response to a specific input query. 
QMDS is one of the key components of the result generation module of a CIR system, as illustrated in Figure~\ref{fig:cir-system-arch}. 

Compared to classical automatic text summarization tasks (e.g., single-document or multi-document summarization) which have been comprehensively studied for decades \citep{mani1999advances}, dating back to Luhn's work at IBM in the 1950's \citep{luhn1958automatic}, QMDS had received much less attention. 
Only in the last 5 to 6 years, have we observed growing interests in QMDS research driven by various CIR and conversational recommendation applications that are being deployed on the devices with no or small screens (e.g., voice search and mobile search) where a concise answer to a query is more desirable than a traditional SERP.

This chapter is organized as follows. 
Section~\ref{sec:qmds-task} introduces the QMDS task and reviews public QMDS datasets. 
Section\ref{sec:text-summarization-methods} presents an overview of text summarization methods, including extractive and abstractive methods, with or without using neural network models.
Extractive methods produce a summary by concatenating sentences or paragraphs selected from the source documents. Abstractive methods generate a concise summary that captures salient information of the source documents. 
Section~\ref{sec:qmds-methods} describes neural approaches to QMDS, based on extractive and abstractive neural models, respectively.
Section~\ref{sec:sum-eval} discusses how we evaluate the factuality of the summaries generated by abstractive summarizers. 

\section{Task and Datasets}
\label{sec:qmds-task}

\begin{figure}[t] 
\centering 
\includegraphics[width=0.80\linewidth]{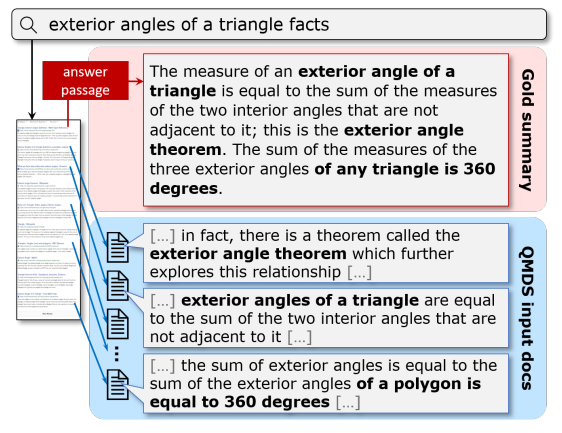}
\vspace{0mm}
\caption{A QMDS sample from the QMDS-IR dataset \citep{pasunuru2021data}, illustrating how a query-focused summary (gold summary) of a set of retrieved documents help a user to answer her input query without going over the individual documents. Figure Credit: \cite{pasunuru2021data}.} 
\label{fig:qmds-example} 
\vspace{0mm}
\end{figure}

Aiming to create a short summary from a set of documents that answers a specific query \citep{dang2005overview}, QMDS is becoming imperative in various CIR and recommendation scenarios (e.g., voice assistant and mobile search) where the search result needs to be presented as a concise summary such that a user can easily judge whether it contains the information she is looking for without going over the long list of returned items, and decides what to do next, e.g., fetching the answer and closing the search session, issuing a more specific query, or switching to a related topic.
Unlike classical text summarization which summarizes important information in the sources document(s), QMDS generates a summary that highlights only the query-focused aspect of the source documents, as illustrated by the example in Figure~\ref{fig:qmds-example}.

Formally, the QMDS task in the context of CIR is defined as follows.
Given input query $Q$, dialog history $\mathcal{H}$, and a set of $N$ documents $\mathcal{D} = \{D_1,...,D_N\}$ retrieved by a conversational search module based on $Q$ and $\mathcal{H}$, a QMDS model needs to generate a summary $Y$ of $\mathcal{D}$ that answers $Q$.  

A series of text summarization tasks have been evaluated in the Document Understanding Conference (DUC). The \emph{main} and \emph{update} tasks evaluated in DUC 2007 \footnote{https://duc.nist.gov/duc2007/tasks.html\#pilot}, for example, are conceptually identical to QMDS as defined above.
The main task is, given query $Q$ and a set of 25 relevant documents $\mathcal{D}$, to generate a 250-word summary $Y$ of the documents that answers $Q$.
The update task, in addition, takes into account conversational context $\mathcal{H}$ when generating the summary. The task is to produce short (100 words) multi-document update summarizes $Y$ of newswire articles $\mathcal{D}$, regarding a given DUC topic $Q$, under the assumption that the user has already read a set of earlier articles included in $\mathcal{H}$. The purpose of each updated summary is to inform the user of new information about the topic. 

For a typical conversational search session, the main task corresponds to summarizing the search result with respect to the first query whereas the update task corresponds to summarizing the search results with respect to follow-up queries. 

To perform the evaluation, DUC organizers provides for each task a small test set that is manually generated (e.g., 40 queries for the main task and 10 queries for the update task). The lack of large-scale high-quality dataset has been the biggest hurdle to be overcome to advance the QMDS research.

Although it is challenging to collect QMDS data in large quantities e.g., by mining Web search session logs, researchers have proposed various approaches to generating \emph{simulated} QMDS datasets by leveraging existing document collections and Web search APIs
\citep[e.g.,][]{liu2018generating,pasunuru2021data,kulkarni2020aquamuse,kulkarni2021comsum,zhao2021qbsum}.
Below, we review, as examples, three such datasets that are publicly available.

\paragraph{WikiSum.} 
This dataset consists of approximately 2.2 M $(Q, \mathcal{D}, Y)$ triples \citep{liu2018generating}, and is developed based on a subset of English Wikipedia articles.  
The authors view Wikipedia as a collection of summaries on various topics or queries given by their title, e.g., ``Canada'' or ``Artificial Intelligence'', and the source documents to be summarized can be viewed as all reputable documents on the Web or books. 
Thus, each $(Q, \mathcal{D}, Y)$ in WikiSum is constructed using a Wikipedia article as follows. $Q$ is the title of the article, $\mathcal{D}$ is a collection of non-Wikipedia reference documents, and $Y$ is the first section, or lead, of the Wikipedia article, conditioned on reference text. 
WikiSum has been used in several QMAD studies. 
Although WikiSum is a simulated QMAS dataset and the results on WikiSum have to be taken with a grain of salt since neither the queries nor the summaries are natural  \citep{pasunuru2021data}, the dataset is among the most widely used ones for QMAS research.

\paragraph{QMDS-CNN.}
This dataset consists of approximately 312K $(Q, \mathcal{D}, Y)$ triples \citep{pasunuru2021data}, and is developed using the CNN / Daily Mail (DM) dataset which is commonly used for single-document summarization \citep{hermann2015teaching}.
Each $(Q, \mathcal{D}, Y)$ triple is generated from a news article. $Q$ is the title of the article, $Y$ is the single-document summary of the article provided in the original CNN/DM dataset. $\mathcal{D}$ consists of documents generated using two methods. First, each news article is chunked into around 20 small paragraphs, and each paragraph is treated as $D \in \mathcal{D}$ if it is part of the news article. Second, a BM25 search engine is used to retrieve for each query (or title) top-4 paragraphs from all the paragraphs that do not belong to the news article.
The authors argue that QMDS-CNN is a better dataset for QMDS research than WikiSum because (1) summaries are of high-quality because they are manually generated, and are query-focused since they reflect the title; and (2) each summary contains information from multiple documents by design.  
The only main downsize is that the queries are not real but simulated by titles. The simulation might not be desirable since titles are generated by the authors who also create the documents while queries are issued by users who have no idea about the documents. 

\paragraph{QMDS-IR}
This dataset consists of queries issued by real search engine users, thus addressing the problem of WikiSum and QMDS-CNN.  The dataset consists of 102K $(Q, \mathcal{D}, Y)$ triples, and each is constructed as follows~\citep{pasunuru2021data}.
First, queries that have natural language answers returned and clicked by users are randomly sampled from Bing search logs in the United States. 
Second, for each query $Q$, we collect its top-10 ranked documents returned by the Bing search engine. The Bing QA system also generates for each returned document an answer passage.
Third, one document is randomly picked among the top-10 documents, and its answer passage is used as summary $Y$, and the rest 9 documents form $\mathcal{D}$.
The QMDS task on this dataset is to recover the query-focused answer passage using the other 9 documents retrieved by the search engine.
To the best of our knowledge, the task defined on QMDS-IR is by far the closest approximation to the QMDS scenario in CIR. A sample of $(Q, \mathcal{D}, Y)$ from this dataset is shown in Figure~\ref{fig:qmds-example}. 

\section{An Overview of Text Summarization Methods}
\label{sec:text-summarization-methods}

\citet{radev2002introduction} define a summary as ``a text that is produced from one or more texts, that conveys important information in the original text(s), and that is no longer than half of the original text(s) and usually, significantly less than that.''
Text summarization is the task of producing a concise and fluent summary while preserving key information content and overall meaning of the original documents. It has been applied in a wide range of applications, taking different forms. For example, QMAD is a special form of the task where search engines generate summaries (or snippets) as the preview of the documents retrieved to answer a specific user query \citep{gao2020standard}.  

In general, there are two approaches to text summarization: \emph{extraction} and \emph{abstraction}. 
Extractive summarization methods produce summaries by choosing a subset of the sentences in the original document(s). This contrasts with abstractive summarization methods, where the information in the text is rephrased. 

\subsection{Extractive Summarizers}
\label{subsec:extractive-summarizers}

The extractive approach has been dominating the field due to its simplicity and effectiveness. In a recent survey, \citet{allahyari2017text} point out that most extractive summarizers take three steps to generate a summary: sentence representation, scoring and selection.

\subsubsection*{Sentence Representation}

Each sentence in $\mathcal{D}$ is represented as a feature vector. A classical approach is to use the BOW (Bag-Of-Words) representation, where each element in the vector corresponds to a topic word in a pre-defined vocabulary, and the value of the element is the weight of the topic word assigned using a scoring model, such as TF-IDF, BM25, log-likelihood ratio, and so on.  
The feature vector can be enriched by incorporating indicative information such as sentence length and position in the document.
In neural approaches, the feature vector is produced using sentence encoders implemented using various neural network models. 
For example, \citet{cheng2016neural} employ a recurrent neural network (RNN) or a convolutional neural network (CNN) to encode the words in a sentence to obtain a vector representation of the sentence.  
\citet{liu2019text} utilize pre-trained language models (e.g., BERT) to encode sentences.
A recent survey on this topic is \citet{gao2020standard}.  

\subsubsection*{Sentence Scoring}

Each sentence is assigned an importance score, indicating how well the sentence represents the main idea of the original documents. This can be achieved by first encoding the document(s) using a vector, and then computing the cosine similarity between the sentence vector and the document vector as the importance score.
Document encoding can be achieved by either aggregating the vectors of all the sentences in the document, or simply treating the document as a long sentence and applying the same sentence encoder.
More advanced methods also take into account inter-sentence dependencies and the discourse structure of documents \citep{erkan2004lexrank}, and introduce machine learning methods to estimate the scores on training data. We will present an example in detail in the next section.

\subsubsection*{Summary Sentence Selection}

The summarizer needs to select a subset of sentences to form the summary, subject to a pre-set length limit. A naïve method uses a greedy algorithm to select top-$K$ most important sentences. But this might lead to high redundancy, i.e., a set of important but similar sentences is selected. 

A better strategy is to consider both the importance and novelty of a sentence. A classic implementation of the strategy is the MMR (maximal marginal relevance) method \citep{carbonell1998use}. At each iteration, MMR selects one sentence from $\mathcal{D}$ and includes it in the summary $Y$ until a length limit is reached. The selected sentence $S_i$ is the most important one among the remaining sentences and it has the least content overlap with the current summary. As in Equation~\ref{eqn:mmr} below, $\mathrm{sim1}(S_i , \mathcal{D})$ measures the similarity of the sentence $S_i$ to the documents, indicating the sentence importance. $\max_{S_j \in Y} \mathrm{sim2}(S_i , S_j)$ measures the maximum similarity of the sentence $S_i$ to each of the summary sentences, acting as a proxy of redundancy. $\lambda$ is a balancing factor.

\begin{equation}
\mathrm{MMR} \triangleq \arg \max_{S_{i} \in \mathcal{D} \backslash Y} 
\left[
\lambda \mathrm{sim1}(S_i, \mathcal{D}) - (1-\lambda)\max_{S_j \in Y}\mathrm{sim2}(S_i, S_j)
\right]
\label{eqn:mmr}
\end{equation}

\subsection{Abstractive Summarizers}
\label{subsec:abstractive-summarizers}

Compared to the extractive approach where a summary is constructed using extracted sentences, abstractive summarizers paraphrase the idea of the original documents in a new form, and have a potential to generate more concise and coherent summaries.
However, developing effective abstractive summarizers is harder since we have to deal with problems, like semantic representation, inference and natural language generation, which are considered more challenging than sentence extraction. Therefore, early abstractive methods underperform extractive methods, and often rely on an extractive pre-processor to produce the abstract of text \citep{berg2011jointly,knight2000statistics}.

Recently, the abstractive approach is getting more attentions in the research community by applying modern neural language models, such as feed-forward language models (FFLMs) \citep[e.g.,][]{rush2015neural}, recurrent neural networks (RNNs) \citep[e.g.,][]{nallapati2016abstractive,chopra2016abstractive} and Transformers \citep[e.g.,][]{liu2018generating,liu2019hierarchical}, to generate summaries conditioned on source documents.

\subsubsection*{FFLMs}

\citet{rush2015neural} present the first widely used neural abstractive summarizer. The summarizer utilizes a beam search algorithm to generate a summary word-by-word based on an attend-based feed-forward neural language model as

\begin{equation}
\begin{aligned}
P_r(y_{i+1} | \mathbf{Y}_{C}, \mathcal{D}; \theta) &\propto \exp{(\mathbf{Vh} + \mathbf{W} \cdot  \text{Enc}(\mathbf{Y}_C, \mathcal{D}))}, \\
\mathbf{h} &= \tanh \left( \mathbf{U} [\mathbf{E}\mathbf{y}_{i-C_1},...,\mathbf{E}\mathbf{y}_{i}] \right),
\label{eqn:fflm4sum}
\end{aligned}
\end{equation}
where $y_{i+1}$ is next summary word to be predicted, $\mathcal{D}$ is the set of the source documents, $\mathbf{Y}_{C} \triangleq \mathbf{y}_{[i-C+1,...,i]} $ is a sequence of indicator vectors that represent the $C$ generated summary words immediately before $y_{i+1}$.

The model is parameterized as a standard feed-forward neural network language model \citep{bengio2003neural} with an additional contextual encoder $\text{Enc}$.  
The model parameters are $\theta = (\mathbf{E}, \mathbf{U}, \mathbf{V}, \mathbf{W})$ where $\mathbf{E} \in \mathbb{R}^{D \times V}$ is a word embedding matrix, $\mathbf{U} \in \mathbb{R}^{(CD) \times H}$, $\mathbf{V} \in \mathbb{R}^{V \times H}$, $\mathbf{W} \in \mathbb{R}^{V \times H}$ are weight matrices, $4 \times D$ is the size of the word embeddings, and $\mathbf{h}$ is a hidden layer of size $H$. 
The contextual encoder \text{Enc} returns a vector of size $H$ representing $(\mathbf{Y}_{C}, \mathcal{D})$. \text{Enc} can be implemented using any sentence encoders developed for extractive summarizers as described in Section~\ref{subsec:extractive-summarizers}. \citet{rush2015neural}   shows that an attention-based encoder works the best. 

\subsubsection*{RNNs}

\citet{nallapati2016abstractive} and \citet{chopra2016abstractive} have independently developed the first set of neural abstractive summarizers implemented using attentional encoder-decoder RNNs (or sequence-to-sequence models), which are originally developed for machine translation \citep{bahdanau2014neural,sutskever2014sequence}.  In these summarizers, a RNN decoder generates a summary of input $\mathcal{D}$ which is encoded by an attention-based encoder to ensure that the decoder focuses on the appropriate input words at each step of generation:

\begin{equation}
P_r(Y | \mathcal{D}) = \prod_{i=1}^{N} P_r(y_i | y_{1},...,y_{i-1}, \mathcal{D}; \theta).
\label{eqn:rnn4sum}
\end{equation}

\begin{figure}[t] 
\centering 
\includegraphics[width=0.99\linewidth]{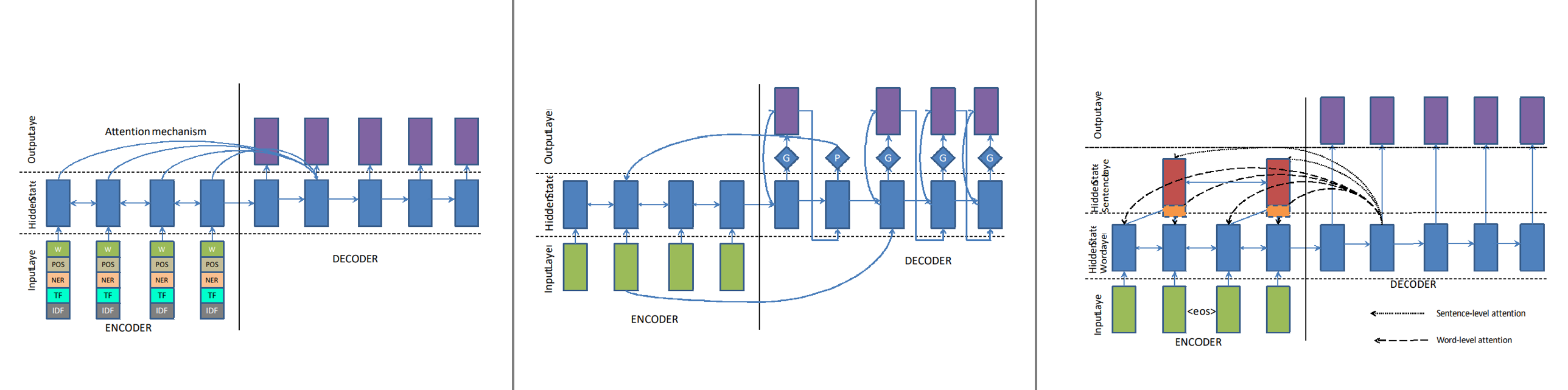}
\vspace{0mm}
\caption{Three RNN models for abstractive summarization \citep{nallapati2016abstractive}. 
(Left) A feature-rich encoder captures rich linguistic features of source documents, including parts-of-speech (POS) tags, named-entity (NE) tags, and TF and IDF statistics of words. 
(Middle) A switching generator-pointer decoder can generates OOV words, such as named entities that are unseen in training data. When the switch shows `G’, the traditional generator is used to produce a word, and when it shows `P’, the pointer network is activated to copy the word from one of the source document positions.
(Right) A hierarchical encoder with hierarchical attention identifies both keywords and key sentences in source documents that are important to generate the summary. The attention weights at the word level, represented by the dashed arrows are re-weighted by the corresponding sentence level attention weights, represented by the dotted arrows.
} 
\label{fig:rnn4sum} 
\vspace{0mm}
\end{figure}

\citet{nallapati2016abstractive} propose three novel modeling techniques to address the unique challenges of text summarization. 
First, a feature-rich encoder is developed to identify key concepts and entities that are crucial to capture the main idea and content of the source documents. As illustrated in Figure~\ref{fig:rnn4sum} (Left), in addition to word embeddings, the encoder captures rich linguistic features, including parts-of-speech (POS) tags, named-entity (NE) tags, and TF and IDF statistics of words.

The second is the use of a switching generator-pointer decoder to allow the summarizer to generate OOV words, such as named entities that are unseen in training data. As illustrated in Figure~\ref{fig:rnn4sum} (Middle), the decoder is equipped with a switch that decides between using the generator (G) or a point (P) for each summary word to be generated. If the switch is on, the word is in the vocabulary and can be produced by the generator as usual. Otherwise, the decoder generates a pointer to a word-position in the source documents, and the word at that position is copied into the summary.

Third, a hierarchical attention mechanism is utilized to identify both keywords and key sentences in source documents that are important to generate the summary. As illustrated in Figure~\ref{fig:rnn4sum} (Right), the attention mechanism operates at the word level and sentence level, and the word-level attention scores are re-weighted by the corresponding sentence-level attention scores, implying that the importance of a word increases if it also occurs in an important sentence.  

Combining all these techniques, \citet{nallapati2016abstractive} show that the encoder-decoder RNNs significantly outperform the model of \citet{rush2015neural} on the DUC benchmark.

\subsubsection*{Transformers}

\citet{liu2018generating,liu2019hierarchical} propose an abstractive multi-document summarizer, which also uses the encoder-decoder architecture, but based on a multi-layer Transformer model \citep{vaswani:17}. The effectiveness of the summarizer is validated empirically on the WikiSum dataset. 

\begin{figure}[t] 
\centering 
\includegraphics[width=0.99\linewidth]{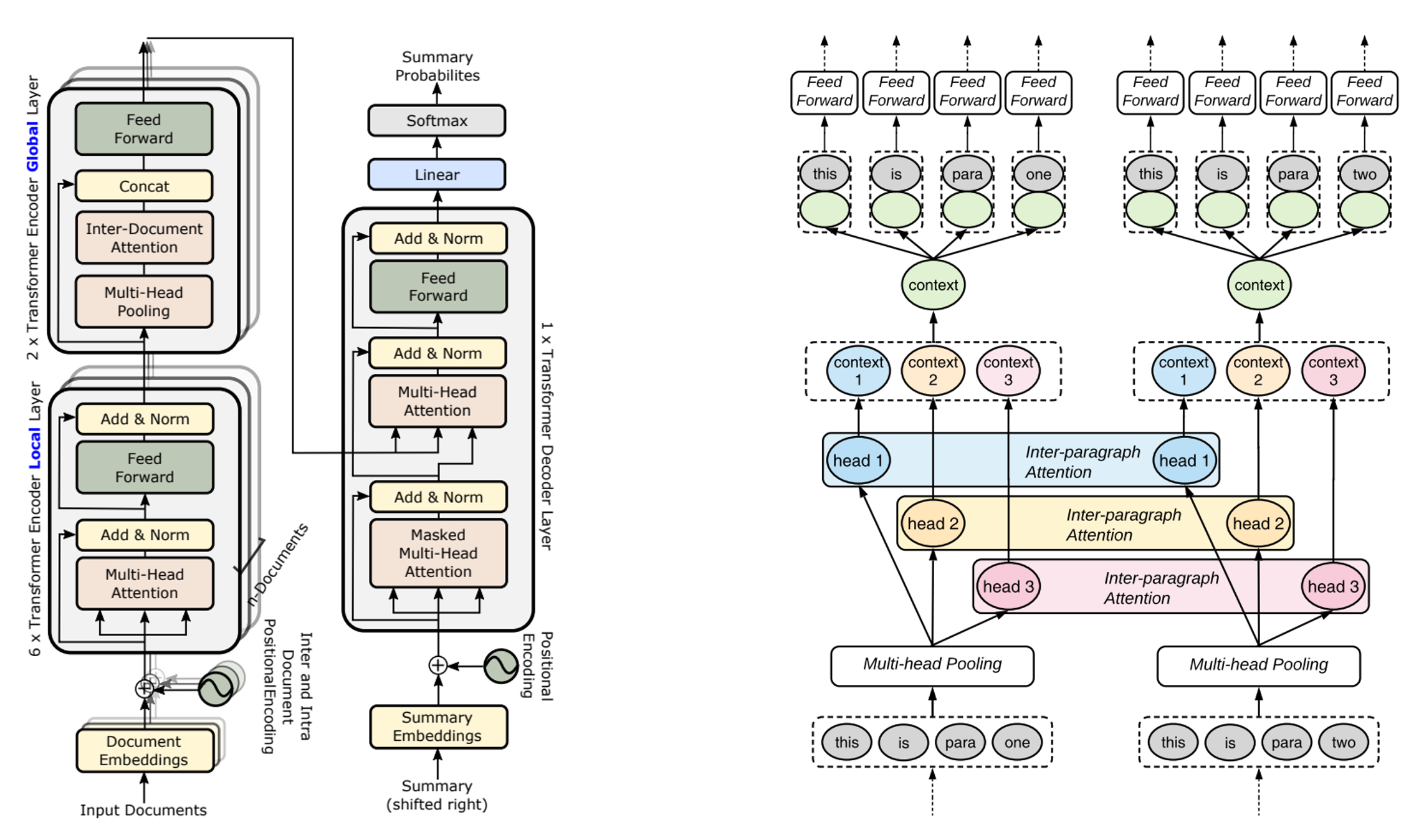}
\vspace{0mm}
\caption{A hierarchical Transformer for abstractive multi-document summarization \citep{liu2019hierarchical}.
(Left) The summarizer consists of four components: the document embedding layer, local Transformer layer, global transformer layer, and decoder. Figure credit: \citet{pasunuru2021data}.
(Right) A global transformer layer. Different colors indicate different heads in multi-head pooling and inter-paragraph attention. Figure credit: \citet{liu2019hierarchical}.
} 
\label{fig:transformer4sum} 
\vspace{0mm}
\end{figure}

As illustrated in Figure~\ref{fig:transformer4sum} (Left), the summarizer consists of four components: the document embedding layer, local Transformer layer, global transformer layer, and decoder.

\paragraph {Document Embedding Layer.} Each token in each document is mapped to a vector. Unlike RNNs, Transformer does not retain sequence information of input, thus a position embedding vector is needed for each word in input. For multi-document summarization, each token has two positions, the rank of the document the token occurs and the position of the token in the document.

\paragraph {Local Transformer Layer.} The vanilla Transformer layer \citep{vaswani:17} is used to encode contextual information for tokens within each document.

\paragraph {Global Transformer Layer.} This layer encodes the inter-document context information. As shown in Figure~\ref{fig:transformer4sum} (Right), first, a multi-head pooling operation is applied to each paragraph (or document). Different heads encode paragraphs with different attention weights. Then, for each head, an inter-paragraph attention mechanism is applied to collect for each paragraph its dependencies on other paragraphs by self-attention, generating an inter-document context vector. Finally, these context vectors are concatenated, linearly transformed, added to the vector of each token, and fed to a feed-forward layer, updating the representation of each token with global information.

\paragraph {Decoder.} A vanilla Transformer is utilized to generate a summary token by token while attending to the source input. Beam search and a length penalty are used in the decoding process to generate fluent and short summaries.

\bigskip

There are many other works on abstractive summarization. We name a few here.
Abstractive summarizers are known to generate fluent summaries but be weak in retaining the main idea of source documents. Therefore, many researchers try to introduce a content selection mechanism so that the summary is generated conditioned on only the most important content selected.  
\citet{zhou2017selective} extend the sequence-to-sequence framework with a selective mechanism that allows the decoder to generate a summary based on a selective portion of the source documents that is the most important to retain the main idea. 
Similarly, \citet{chen2018fast} propose an abstractive summarizer that first selects salient sentences and then rewrites them abstractively (via compression and paraphrasing) to generate a concise summary.
\citet{gehrmann2018bottom}  propose a data-efficient content selector to over-select phrases in a source document that should be part of the summary. The selector is used in a bottom-up attention step to constrain the model to using only the most likely phrases. 

\citet{see2017get} propose a hybrid pointer-generator network that can copy words from the source text via pointing, which aids accurate reproduction of information, while retaining the ability to produce novel words through the generator. This work is a significant extension to \citet{nallapati2016abstractive} shown in Figure~\ref{fig:rnn4sum} (Middle) since the mixture of copy mechanism and the language model is used to generate all summary words, not just the OOV words.

\citet{celikyilmaz2018deep} present deep communicating agents, which use an encoder-decoder architecture, to address the challenges of representing a long text (e.g., a concatenation of multiple documents) for abstractive summarization. The task of encoding a long text is divided across multiple collaborating encoders (agents), each encoding a subsection of the input text. These encoders are connected to a single decoder, trained end-to-end using reinforcement learning to generate a focused and coherent summary.

\section{QMDS Methods}
\label{sec:qmds-methods}
QMDS methods can be grouped into two categories, extractive and abstractive methods, which are the extensions of extractive and abstractive text summarizers, respectively, as described in Section \ref{sec:text-summarization-methods}.

\subsection{Extractive Methods}
\label{subsec:extractive-qmds-methods}

Extractive QMDS methods extend the extractive approach described in Section \ref{subsec:extractive-summarizers}. Most of these methods also take three steps to generate a summary: sentence representation, scoring and selection. The main difference is that in each step the input query $Q$, in addition to source document set $\mathcal{D}$, is also considered.   

\subsubsection*{Sentence Representation}
We can simply treat query $Q$ as a document, and then use the same methods of Section \ref{subsec:extractive-summarizers} to represent each sentence, including the sentences in $Q$, as a feature vector.

\subsubsection*{Sentence Scoring}
The importance of each sentence $S$ in $\mathcal{D}$ depends on how relevant $S$ is to query $Q$, $\text{sim1}(S, Q)$ and how well $S$ represents the main idea of $\mathcal{D}$, $\mathrm{sim2}(S, \mathcal{D})$ as
\begin{equation}
\mathrm{sim}(S,(Q,\mathcal{D})) \triangleq 
\mu \mathrm{sim1}(S, Q) + (1-\mu)\mathrm{sim2}(S,\mathcal{D})
\label{eqn:sent-score}
\end{equation}
where $\mu \in (0,1)$ controls the extent to which query relevance information influences sentence importance.

To take into account the inter-sentence dependencies and the discourse structure of documents, many extractive summarizers \citep[e.g.][]{wan2008using,xu2020coarse} use the well-known LexRank algorithm \citep{erkan2004lexrank} to compute sentence importance based on the concept of eigenvector centrality in a graph representation of sentences. 
LexRank builds for $\mathcal{D}$ a graph $\mathcal{G}=(\mathcal{V},\mathcal{E})$ where nodes $\mathcal{V}$ are sentences in $\mathcal{D}$, $\mathcal{E}$ are undirected edges and are weighted according to matrix $\mathbf{E}$, where each element $\mathbf{E}_{i,j}$ is the transition probability from vertex $i$ to $j$. 
Similar to Equation~\ref{eqn:sent-score}, $\mathbf{E}_{i,j}$ needs to take into account both the inter-sentence similarity and the relevance to $Q$, and can be computed as 

\begin{equation}
\begin{aligned}
&\mathbf{E}_{i,j} = \mu P_r(S_i, Q) + (1 - \mu) P_r(S_i, S_j) \\
&P_r(S_i, Q) \propto \mathrm{sim1}(S_i, Q) \\
&P_r(S_i, S_j) \propto \mathrm{sim2}(S_i, S_j).
\label{eqn:sent-score-lexrank}
\end{aligned}
\end{equation}
The importance of each node is computed using random walk on $\mathcal{G}$. Specifically, a Markov chain is run with $\mathcal{E}$ on $\mathcal{G}$ until it converges to stationary distribution 

\begin{equation}
\mathbf{p} = \mathbf{E}^{\top} \mathbf{p}, 
\label{eqn:stat-dist-lexrank}
\end{equation}
where each element of eigenvector $\mathbf{p}$ denotes the importance of a sentence. 

\subsubsection*{Summary Sentence Selection}
The summary can be constructed by selecting sentences from $\mathcal{D}$ based on their importance and novelty scores, similar to the MMR method of Equation~\ref{eqn:mmr}, where for QMDS $\mathrm{sim1}(.)$ is replaced by $\mathrm{sim}(S,(Q,\mathcal{D}))$ of Equation~\ref{eqn:sent-score} or the elements of $\mathbf{p}$ in Equation~\ref{eqn:stat-dist-lexrank}.

\subsubsection*{Coarse-to-Fine QMDS}

Unlike traditional text summarization tasks where $\mathcal{D}$ is pre-selected, in QMDS the size and number of source documents in $\mathcal{D}$ can be very large, depending on the retrieval result of an input query. Therefore, encoding all sentences in $\mathcal{D}$ to generate a summary can be so computationally expensive as to lead to a noticeable delay in system responses, thus hurting the user experience of human-machine interactions during conversational information retrieval.

\begin{figure}[t] 
\centering 
\includegraphics[width=0.9\linewidth]{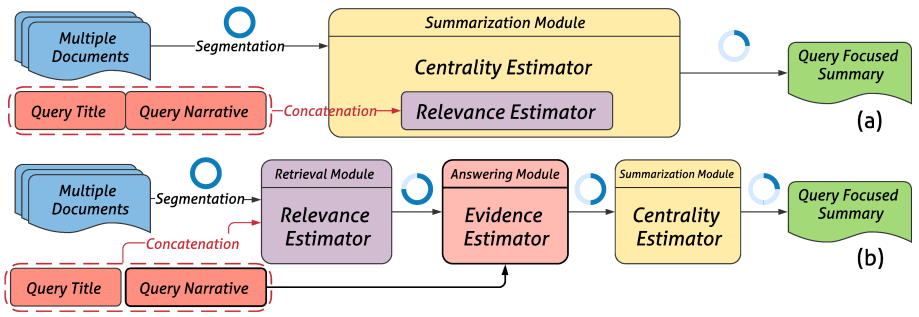}
\vspace{0mm}
\caption{Classic (a) and coarse-to-fine modeling framework for QMDS  \citep{xu2020coarse}.
The classic methods score all sentences in source documents simultaneously.
The coarse-to-fine approach employs progressively more accurate modules for estimating whether a sentence is relevant to a query (relevance estimator), likely to contain answers to the query (evidence estimator), and should included in the summary (centrality estimator). 
The blue circles indicate a coarse-to-fine estimation process from source documents to final summaries where modules gradually discard sentences.
Figure credit: \citet{xu2020coarse}.
} 
\label{fig:coarse-to-fine-qmds} 
\vspace{0mm}
\end{figure}

\citet{xu2020coarse} propose a coarse-to-fine QMDS modeling framework to address the challenge. As shown in Figure~\ref{fig:coarse-to-fine-qmds}(b), the approach employs progressively more accurate modules for estimating whether a sentence is relevant to a query (relevance estimator), likely to contain answers to the query (evidence estimator), and should be included in the summary (centrality estimator). The authors argue that compared to classic methods where all sentences are encoded and scored simultaneously (Figure~\ref{fig:coarse-to-fine-qmds}(a)), the coarse-to-fine approach is more efficient, especially for highly interactive CIR scenarios, since at each step the model processes a decreasing number of sentences and the computational cost does not increase with the size of the original input $\mathcal{D}$. 

Another advantage of this approach is that the modules (i.e., estimators) can be independently developed by leveraging training data if available. For example, estimating the relationship between the query and the candidate sentences can be formulated as an extractive QA task (to be described in Section~\ref{sec:textqa-task-datasets}), which aims to detect the answer span for a query from a set of pre-selected sentences \citep{yao2013answer,yang2015wikiqa} or text passages \citep{rajpurkar2018know,nguyen2016ms}. Thus, the evidence estimator can be trained by leveraging distant supervision signals from existing QA datasets, including WikiQA \citep{yang2015wikiqa}, TrecQA \citep{yao2013answer} and SQuAD 2.0 \citep{rajpurkar2018know}.

\subsection{Abstractive Methods}
\label{subsec:abstractive-qmds-methods}

Abstractive QMDS methods are the extensions of the abstractive summarizers described in Section~\ref{subsec:abstractive-summarizers} to address two challenges of QMDS: how to efficiently encode document set $\mathcal{D}$, and how to incorporate query $Q$ for summary generation.

\begin{figure}[t] 
\centering 
\includegraphics[width=0.8\linewidth]{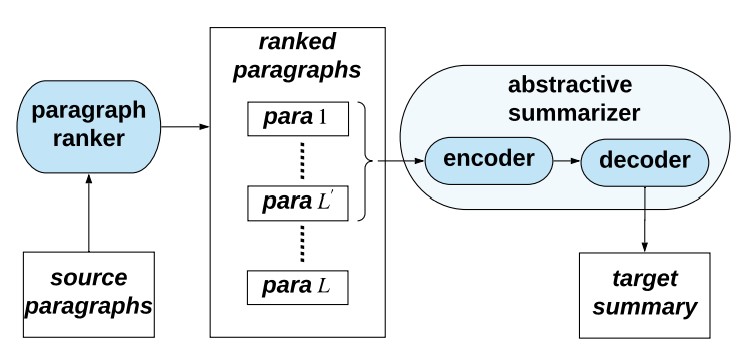}
\vspace{0mm}
\caption{The two-stage method of query-focused multi-document summarization. $L$ source paragraphs (documents) are first ranked and the $L'$-best ones serve as input to an abstractive summarizer (i.e., an encoder-decoder model) which generates the target summary.
Figure credit: \citet{liu2019hierarchical}.
} 
\label{fig:two-stage-qmds} 
\vspace{0mm}
\end{figure}

Since the amount of text in $\mathcal{D}$ can be very large as discussed earlier, it is infeasible to train an end-to-end abstractive model given the memory constraints of current hardware and latency constraints of interactive CIR scenarios. Therefore, a two-stage approach is often used. As illustrated in Figure~\ref{fig:two-stage-qmds}, in the first stage, a subset of $\mathcal{D}$ is coarsely selected using an extractive summarizer, such as the ones described in Section~\ref{subsec:extractive-qmds-methods}; and in the second stage, an abstractive summarizer (e.g., based on encoder-decoder models) generates the target summary conditioning on the extraction (e.g., the $L'$-best paragraphs). This two-stage process simulates how humans summarize multiple long documents: first highlight pertinent information, and then conditionally generate the summary based on the highlights.

A simple method of incorporating $Q$ for summary generation is to append $Q$ to the top-ranked document before sending it to the encoder of an abstractive summarizer, such as the hierarchical Transformer-based summarizer illustrated in Figure~\ref{fig:transformer4sum}, which is referred to as \emph{baseline summarizer} below. However, some recent studies show that it is beneficial to treat $Q$ differently from $D \in \mathcal{D}$. 

\begin{figure}[t] 
\centering 
\includegraphics[width=0.7\linewidth]{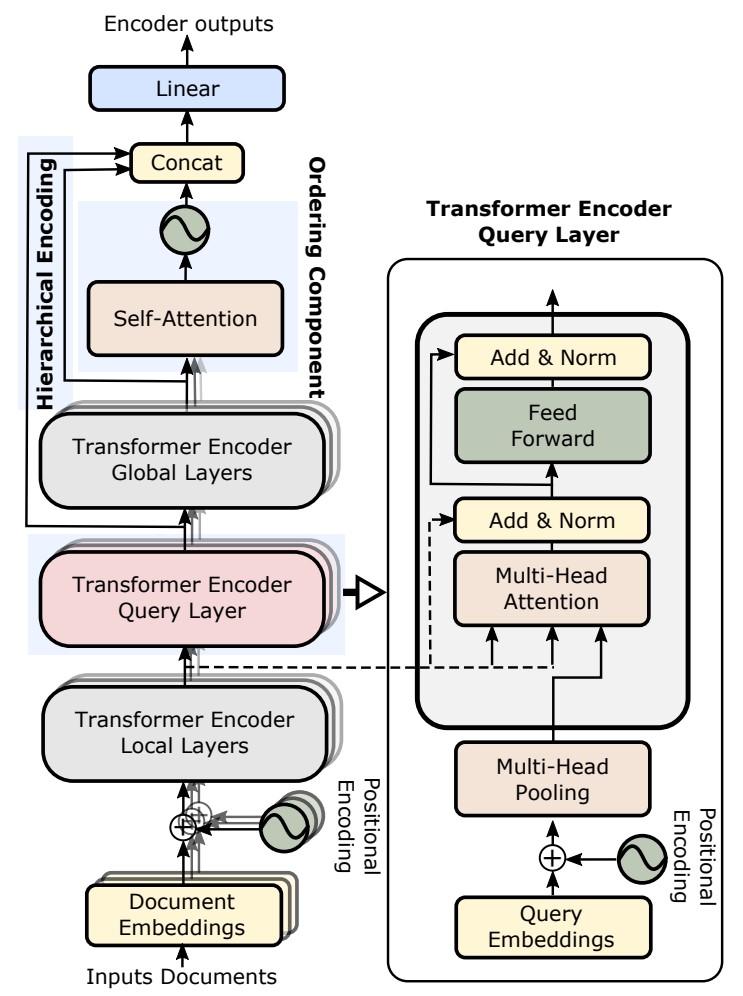}
\vspace{0mm}
\caption{The HEROSumm encoder \citep{pasunuru2021data}, which extends of the encoder of the baseline summarizer \citep{liu2019hierarchical} by introducing two components to incorporate query for summary generation: a query Transformer layer for query encoding and a document ordering component for ranking documents based on their relevance to query. Figure credit: \cite{pasunuru2021data}.
} 
\label{fig:herosumm-encoder} 
\vspace{0mm}
\end{figure}

As an example, we describe the HEROSumm (HiErarchical queRy focused Order-aware multi-document) model \citep{pasunuru2021data} which extends the baseline summarizer \citep{liu2019hierarchical}.
Like the baseline summarizer, HEROSumm uses a encoder-decoder architecture based on a hierarchical Transformer model. 
HEROSumm differs from the baseline in that the former introduces two new components in its encoder modules to incorporate $Q$ for summary generation: a query encoder, and a document ordering component that ranks documents in $\mathcal{D}$ based on their relevance to $Q$. The encoder of HEROSumm is illustrated in Figure~\ref{fig:herosumm-encoder}. HEROSumm consists of the following components.

\paragraph{Document Embedding Layer.}
Each token in each document is mapped to a vector. Compared to the baseline summarizer, the only difference is that the position embedding vector of each word encodes only the position of the token in the document while the rank of the document is randomly assigned and will be updated by the document ordering component.

\paragraph{Document Ordering Component.}
The self-attention mechanism is used to compute the relevance (attention) score $R_i$ between query and each document $D_i \in \mathcal{D}$, indicating how much information in $D_i$ needs to be take into account when generating the summary. $R$ is encoded in the position embedding vector as

\begin{equation}
\begin{aligned}
\mathrm{PE}_{D_i,2j} &= \sin{(R_i / 1000 ^ {2j / d_{model}})}\\
\mathrm{PE}_{D_i,2j+1} &= \cos{(R_i / 1000 ^ {2j / d_{model}})}
\label{eqn:herosumm-ordering}
\end{aligned}
\end{equation}
where $\mathrm{PE}(D_i,2j)$ is the $i$-th dimensional position embedding of $D_i$, and $d_{model}$ is the size of model's hidden layers. The position encoding is inserted in the last layer of the HEROSumm encoder, as illustrated in Figure~\ref{fig:herosumm-encoder}.

\paragraph{Local Transformer Layer.}
Like the baseline summarizer, the vanilla Transformer layer \citep{vaswani:17} is used to encode intra-document contextual information for tokens within each document.

\paragraph{Global Transformer Layer.}
Like the baseline summarizer, this layer encodes the inter-document contextual information, as illustrated in Figure~\ref{fig:transformer4sum} (Right).

\paragraph{Query Transformer Layer.}
This layer sits between the local and global Transformer layers, as illustrated in Figure~\ref{fig:herosumm-encoder} (with an enlarged view provided on the right of the figure). It encodes two query-context vectors. One captures dependencies between the input query and tokens within each document. The other captures dependencies between the query and each document (or paragraph) in $\mathcal{D}$. Finally, these query-context vectors, together with the intra- and intern-document context vectors, are fed to the decoder to generate a summary.   

\paragraph{Decoder.}
The decoder of HEROSumm uses a standard decoder Transformer \citep{vaswani:17}, as that in the baseline summarizer. There are studies of improving the decoder using the idea of MMR to avoid generating repeated phrases and increase the query-relevance and diversity of summaries. For example, \citet{nema2017diversity} propose a neural-based Seq2Seq framework which ensures that context vectors in attention mechanism are orthogonal to each other. Specifically, to alleviate the problem of repeating phrases in the summary, successive context vectors are treated as a sequence and a modified LSTM cell is used to compute the new state at each decoding time step. In decoding steps, the attention mechanism is used to focus on different portions of the query at different time steps.

\section{Factuality Metrics for Summarization Evaluation}
\label{sec:sum-eval}

\begin{figure}[t] 
\centering 
\includegraphics[width=0.68\linewidth]{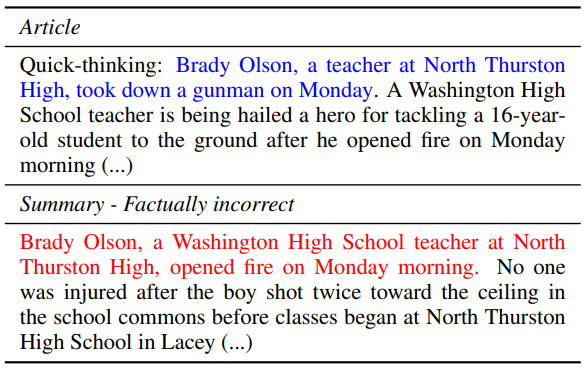}
\vspace{0mm}
\caption{Example of a factually incorrect summary generated by an abstractive summarizer. (Top) Source document. (Bottom) Model-generated summary. Figure credit:  \citet{kryscinski2019neural}.
} 
\label{fig:sum-eval-example} 
\vspace{0mm}
\end{figure}

Automatic evaluation of summarizers, especially the abstractive summarizers, is challenging \citep{kryscinski2019neural}. A common approach is to collect for each test case multiple human generated summaries as references, then measure the $N$-gram overlap between the machine-generated summaries and the references using metrics such as ROUGE \citep{lin2004rouge}.
However, this approach fails to evaluate the factuality of machine-generated summaries. According to \citet{kryscinski2019neural}, 30\% of summaries generated by abstractive summarizers contain factual inconsistencies, caused by distorted or fabricated facts about the source documents, as illustrated by the example in Figure~\ref{fig:sum-eval-example}.

Developing factuality metrics for evaluating abstractive summarizers and other text generation models becomes an increasingly important topic. 
Recently proposed factuality metrics can be grouped into two categories, based on proxy evaluation tasks and extracted facts, respectively.

\begin{figure}[t] 
\centering 
\includegraphics[width=0.99\linewidth]{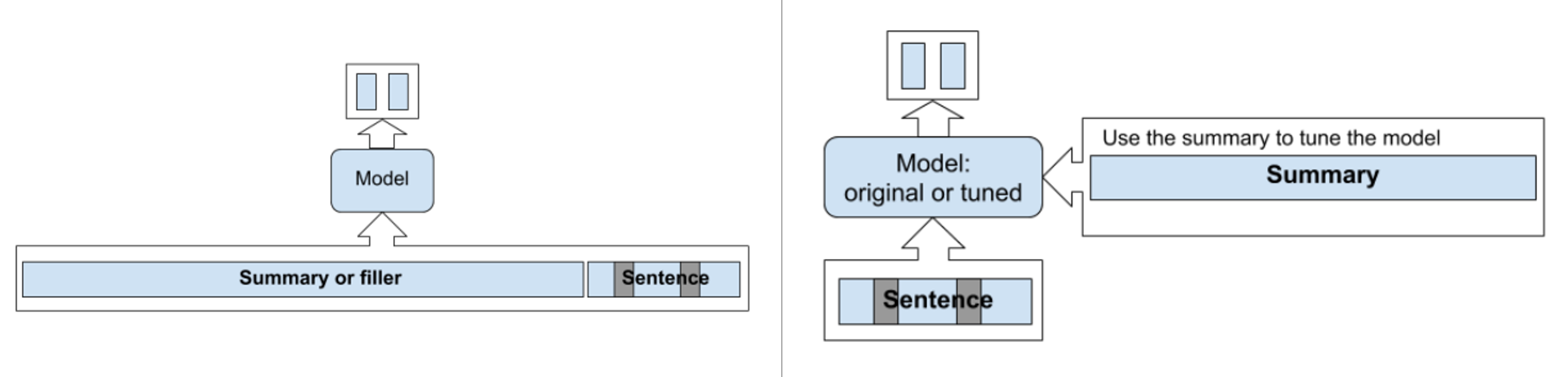}
\vspace{0mm}
\caption{Two cloze-based metrics for evaluating the factuality of model-generated summaries \citep{vasilyev2020fill}.
(Left) BLANC-help is defined by the difference in accuracy of two reconstructions of masked tokens: with summary vs. filler concatenated in front of the sentence with masked tokens. The model input is a summary (or filler) + sentence with masked (grey) tokens. The output is the unmasked tokens.
(Right) BLANC-tune is defined by the difference in accuracy of two reconstructions of masked tokens: with model tuned on the summary vs. with the original model. Both models are given the same input: a sentence with masked (grey) tokens. Each model outputs the unmasked tokens.
Figure credit: \citet{vasilyev2020fill}.
} 
\label{fig:cloze-based-sum-eval} 
\vspace{0mm}
\end{figure}

The first category uses proxy evaluation tasks, such as question answering (QA), and a masked token prediction cloze task.  
Given $(\mathcal{D}, Y)$ where $Y$ is a model-generated summary of $\mathcal{D}$, the QA-based metrics assign a quality score to $Y$ by measuring how accurate, in terms of precision/recall/F-score, a QA system can answer questions generated from $\mathcal{D}$ based on $Y$, or vice versa. For example, the SummaQA metric \citep{scialom2019answers} answers questions derived from $\mathcal{D}$ based on $Y$, whereas the FEQA metric \citep{durmus2020feqa} answers questions from $Y$ based on $\mathcal{D}$.
The cloze-based metrics assign the quality of a summarizer by measuring the performance boost gained by a pre-trained language model with access to its generated summaries while performing the token prediction task on the source documents \citep{vasilyev2020fill}. As illustrated by the two examples in Figure~\ref{fig:cloze-based-sum-eval}, the BLANC-help metric (Left) is defined by the difference in accuracy of two reconstructions of masked tokens: with summary versus filler concatenated in front of the sentence with masked tokens, where the model input is a summary (or filler) and sentence with masked (grey) tokens, and the output is the unmasked tokens.  The BLANC-tune metric (Right) is defined by the difference in accuracy of two reconstructions of masked tokens: with model tuned on the summary versus with the original model, where both models are given the same input consisting of a sentence and its masked (grey) tokens, and each model outputs the unmasked tokens.

The second category extracts facts from the source documents or the reference summaries, and compute how much a machine-generated summary contains consistent facts.
For example, \citet{zhang2019evaluating} proposes to adopt information extraction tools, such as OpenIE, to extract two sets of facts from a machine-generated summary and its references, respectively, and check the consistency of the two sets by computing their semantic relevance score, precision, recall and F1-score.
In \citet{chang2021webqa}, salient keywords (facts) are extracted manually from the source documents as a by-product of generating the reference answers. The factuality of a machine-generated answer is scored by recall, i.e., how many salient keywords are contained in the generated answer.  

Despite these efforts, evaluating the quality of machine-generated text remains an open research topic. 
A recent comparative study of factuality evaluation metrics is \citet{gabriel2020go}.
A recent survey of evaluation methods of natural language generation systems is \citet{celikyilmaz2020evaluation}.

\chapter{Conversational Machine Comprehension}
\label{chp:c-mrc}
 
This chapter discusses neural approaches to conversational machine comprehension (CMC). 
A CMC module, which is often referred to as \emph{reader}, generates a direct answer to a user query based on query-relevant documents retrieved by the document search module (Chapter~\ref{chp:CIR}). It is a component of the result generation module of a CIR system, as shown in Figure~\ref{fig:cir-system-arch}

Equipped with CMC, a CIR system can be used as a conversational open-domain question answering (QA) system. 
Compared to a (conversational) document search system, a QA system is much easier to use in mobile devices in that they provide concise, direct answers to user queries, as opposed to be a list of document links. To simplify the terminology, a QA engine and a machine comprehension module (or a reader) are exchangeable terms in this chapter.


This chapter starts with an introduction to the CMC task (Section~\ref{sec:textqa-task-datasets}). Using the CoQA \citep{reddy2019coqa} and QuAC \citep{choi2018quac} benchmarks as examples, we discuss the dialog features, such as topic shift, drill down, clarification, which the CMC module needs to handle.  
Then, we describe two categories of machine comprehension models, known as \emph{extractive} and \emph{generative} readers, originally developed for sing-turn QA tasks. 
While extractive readers (Section~\ref{sec:extractive-readers}) select a text span in the retrieved documents as the answer to the user query,
generative readers (Section~\ref{sec:generative-readers}) can generate answer strings which might not be contained in the retrieved documents.
These two types of readers can be combined to produce better QA results (Section~\ref{sec:hybrid-readers}).
We then examine approaches to extending these readers to use conversational context for CMC (Section~\ref{sec:cmc-readers}). 

\section{Task and Datasets}
\label{sec:textqa-task-datasets}

\begin{figure}[t] 
\centering 
\includegraphics[width=1.0\linewidth]{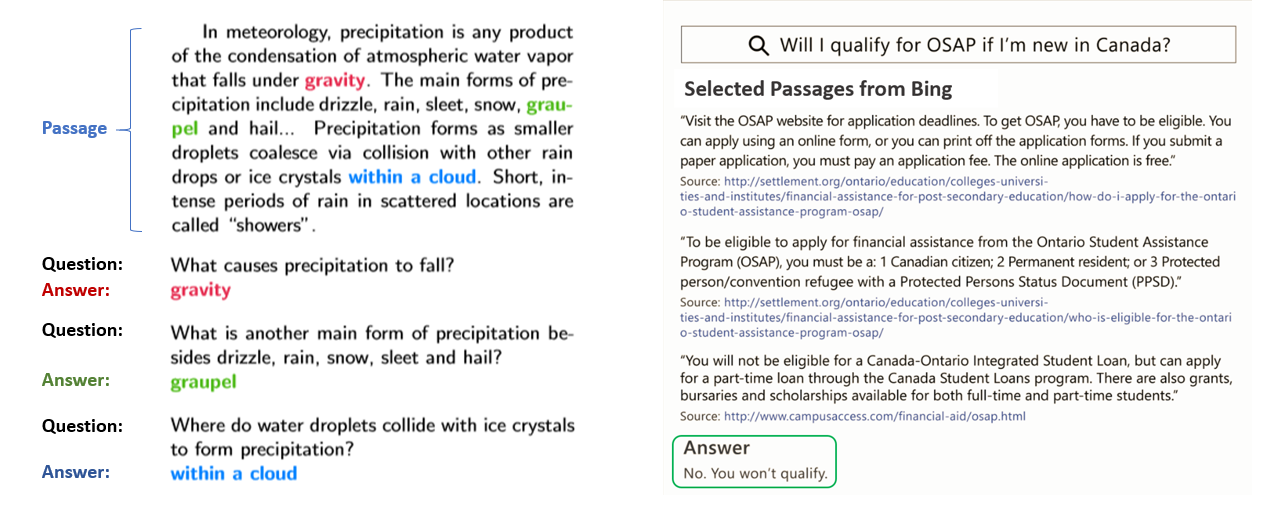}
\vspace{-2mm}
\caption{The examples from two types of MC task. (Left) An extractive QA example in the SQuAD dataset, adapted from \citet{rajpurkar2016squad}. Each of the answers is a text span in the passage. (Right) An abstractive QA example in the MS MARCO dataset, adapted from \citet{nguyen2016ms}. The answer, if there is one, is synthesized from the given passages.} 
\label{fig:two-mrc-datasets}
\vspace{0mm}
\end{figure}

\begin{figure}[t] 
\centering 
\includegraphics[width=1.0\linewidth]{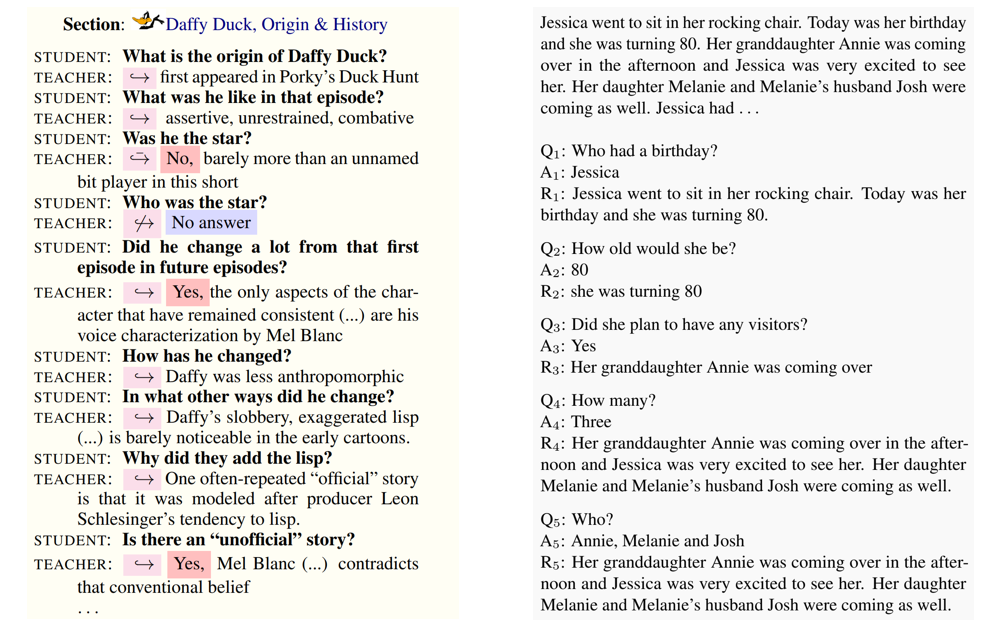}
\vspace{-2mm}
\caption{The examples from two conversational QA datasets. (Left) A QA dialogue example in the QuAC dataset.  The student, who does not see the passage (section text), asks questions. The teacher provides answers in the form of text spans and dialogue acts. These acts include (1) whether the student should $\hookrightarrow$, could 
$ \bar{\hookrightarrow} $, 
or should not 
$ \not\hookrightarrow $ 
ask a follow-up; (2) affirmation (Yes / No), and, when appropriate, (3) No answer. Figure credit: \citet{choi2018quac}. (Right) A QA dialogue example in the CoQA dataset.  Each dialogue turn contains a question ($\text{Q}_i$), an answer ($\text{A}_i$) and a rationale ($\text{R}_i$) that supports the answer. Figure credit: \citet{reddy2018coqa}.} 
\label{fig:two-conversational-mrc-datasets}
\vspace{0mm}
\end{figure}

Machine comprehension (MC) tasks defined in the research community can be categorized along two dimensions. Depending on how the answers are produced, MC tasks can be
(1) \emph{extractive}, where answer is a text span in the given documents (or passages), such as the SQuAD task (Figure~\ref{fig:two-mrc-datasets} (Left)) \citep{rajpurkar2016squad}, and
(2) \emph{abstractive}, where the answer is free-form and needs to be synthesized from the given passages, such as the MS MARCO benchmark (Figure~\ref{fig:two-mrc-datasets} (Left)) \citep{nguyen2016ms}.
Depending on whether conversational context is used, MC tasks can be \emph{single-turn}, such as SQuAD and MS MARCO, or \emph{conversational}, such as CoQA and QuAC (Figure~\ref{fig:two-conversational-mrc-datasets}).

Without loss of generality, the CMC task is defined as follows.
Given an input question $Q$, a set of passages $\mathcal{P}$ (which are retrieved by the document search module with respect to $Q$ in CIR), and dialog history $\mathcal{H}$,
the CMC reader needs to produce the answer $A$ to $Q$. 

Figure~\ref{fig:two-conversational-mrc-datasets} (Left) shows an information seeking QA dialog in the QuAC dataset. The conversation is student-driven and centered around a short evidence passage. The student, who does not see the passage, asks questions. The teacher provides answers in the form of text spans and dialog acts. These acts include
\begin{itemize} 
    \item Whether the student should, could, or should not ask a follow-up question,
    \item Yes/No affirmation, and when appropriate,
    \item No answer.
\end{itemize}

Figure~\ref{fig:two-conversational-mrc-datasets} (Right) shows an example conversational QA session in CoQA. It is a conversation between two humans who are reading a passage. One is acting as a questioner and the other an answerer. Answering every question, except the first one, depends on the dialog history. For example, “she” in Q$_2$ refers to the entity ``Jessica'' in A$_1$.  To ensure the naturalness of answers in a conversation, the answers are not restricted to text spans in the given passage (as in an extractive QA task) but are free-form text. For evaluation, the dataset provides for each answer a text span from the passage as a rationale to the answer. For instance, the answer to Q$_4$ (how many) is ``Three’’ which is not a text span while its rationale spans across multiple sentences.  

For both datasets, the main evaluation metric is macro-average F1 score of word overlap between the prediction and the references, after removing stopwords.

\cite{yatskar2018qualitative} studies the dialog behaviors and abstractive behaviors of the CoQA and QuAC datasets, and summarizes the unique challenges that CMC models need to handle. 
There are five dialog behaviors:
\begin{enumerate}
    \item \text{Topic Shift:} A question about something discussed in the dialog history (e.g., ``Q: how does he try to take over?... Q: where do they live?'').
    \item \text{Drill Down:} A request for more information about a topic being discussed (e.g., ``A: the Sherpas call Mount Everest Chomolungma. Q: is Mt. Everest a holy site for them?'').
    \item \text{Topic Return:} Asking about a topic again after it had been shifted away from.
    \item \text{Clarification:} Reformulating a question that had previously been asked.
    \item \text{Definition:} Asking what is meant by a term.
\end{enumerate}

Yatstar finds that although the topic shift and drill down behaviors are well represented in both datasets, 
neither dataset has a significant number of returns to previous topics, clarifications, or definitional interactions. Therefore, the community needs a more representative CMC dataset.

Yatakar also observes that abstractive answers are necessary for the following cases in the two datasets.
\begin{enumerate}
    \item \text{Yes/No:} Questions whose answers are ``yes'' or ``no''.
    \item \text{Coref:} Co-reference is added to previously mentioned entities in either context or question (e.g., ``Q: how was France's economy in the late 2000's? A: \textbf{it} entered the recession.'')
    \item \text{Count:} Counting how many entities of some type were mentioned.
    \item \text{Fluency:} Adding a preposition, changing the form of a word, or merging two non-contiguous spans (e.g., ``Q: how did he get away? A: \emph{by} foot.'')
\end{enumerate}

Beside Yes/No questions, QuAC contains no other abstractive phenomena while CoQA contains only a small number of predominately insertions, often at the beginning of an extracted answer span, for co-reference or other fluency improvements. Therefore, state-of-the-art extractive readers can be trivially adapted to both tasks, as we will detail later. Therefore, the community needs a more representative abstractive QA dataset.

In addition to CoQA and QuAC, we list below four related public benchmarks that provide complementary features regarding dialog and abstractive behaviors.

\paragraph{QBLink~\citep{elgohary2018dataset}.} 
QBLink is a dataset for sequential question answering where the questioner asks multiple related questions about the same concept one-by-one. After each question, the answerer provides an answer before the next question is asked. The dataset is designed to evaluate the ability of question-answering systems to leverage the additional context in the form of a history of previously asked questions and answers.  The dataset consists of 18,644 sequences (56,000 question-answer pairs). Each sequence starts with a lead-in that defines the topic of the questions, followed up three questions and answers.
    
\paragraph{Qulac~\citep{aliannejadi2019asking}.} 
This is the first dataset dedicated for clarifying questions in an IR setting. The dataset is built on top of the TREC Web Track 2009-2012 dataset and consists of 10K QA pairs for 198 topics with 762 facets. Each topic is coupled with a facet. Thus, the same question would receive a different answer based on user's information need (facet).  

\paragraph{Natural Questions (NQ)~\citep{natural-questions}.} 
Although NQ is a single-turn QA dataset, it is the among the most popular large-scale datasets for developing any end-to-end open-domain QA system which consists of the document search module and the machine comprehension module. 
    
\paragraph{TriviaQA~\citep{joshi2017triviaqa}.} 
Like NQ, TriviaQA is a single-turn QA dataset widely used for developing open-domain QA systems. It contains trivia question-answer pairs that were scraped from the web. Different from NQ, the questions in TriviaQA are written with known answers in mind. TriviaQA is known to be challenging because it contains complex and compositional questions which require cross-sentence reasoning to find answers.

\section{Extractive Readers}
\label{sec:extractive-readers}

This section describes neural machine comprehension readers for single-turn extractive MC. The next two sections describe how these readers are extended to abstractive MC and conversational MC, respectively.


Extractive MC can be cast as a classification problem. Given a question $Q=\{q_1,...,q_I\}$ and a (set of) passage(s) $P=\{p_1,...,p_J\}$, the reader needs to classify each passage word as start, end, or outside of the answer span, and pick the most probable span as the answer.

A typical architecture of neural MC models for single-turn extractive MC consists of two modules. 
The first is a question-passage encoder. The output of the encoder is a sequence of integrated context vectors, each for one passage word, encoding information of its word embedding, passage context and question context. 
The second module is an answer predictor. It takes the integrated context vectors as input, predicts for each word its probabilities of being the start or the end of the answer span, respectively, and picks the most probable span.

Early extractive readers commonly use sequence models (such as Recurrent Neural Networks) with attention for their encoders. After the Transformer~\citep{vaswani:17} is proposed and widely used to develop pre-trained language models (such as BERT~\citep{devlin2018bert}), the extractive readers that use pre-trained Transformer models as encoders become state of the art.
In what follows, we use two examples to illustrate how these extractive readers are implemented. 
The first is the Bi-Directional Attention Flow (BiDAF) model~\citep{seo2016bidirectional}, which is one of the most popular baseline readers that use sequence models as encoders. 
The second is the state-of-the-art extractive reader based on BERT~\citep{devlin2018bert}. 

\subsection{BiDAF}
\label{subsec:bidaf}

\begin{figure}[t] 
\centering 
\includegraphics[width=1.0\linewidth]{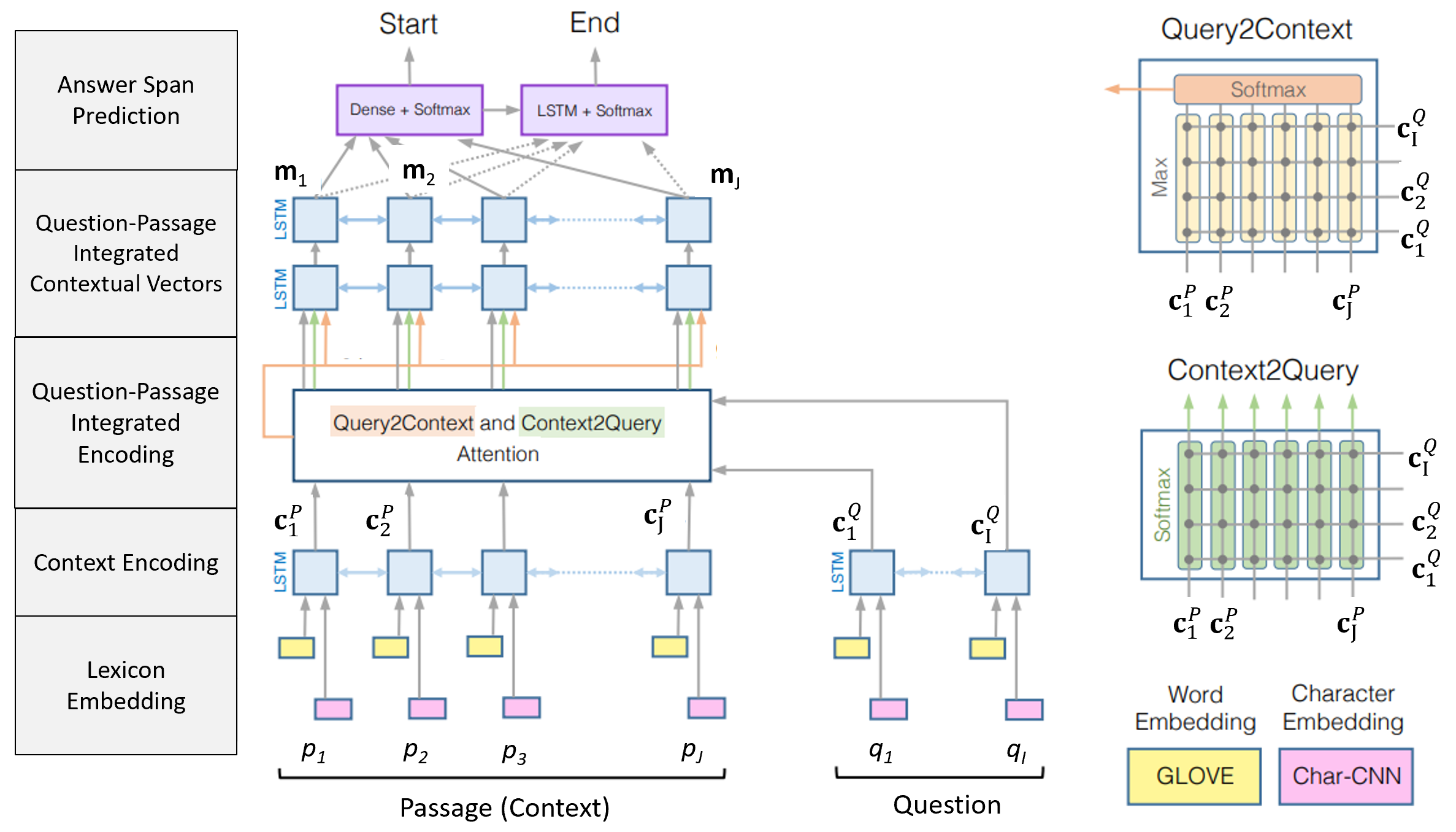}
\vspace{-1mm}
\caption{The Bi-Directional Attention Flow model for extractive QA, adapted from \citet{seo2016bidirectional}.} 
\label{fig:bidaf}
\vspace{0mm}
\end{figure}

BiDAF first encodes the input question $Q$ and passage $P$ into a set of integrated context vectors $\mathbf{M}$ using three components (i.e., lexicon embedding, context encoding, and question-passage integrated encoding), and then predicts the answer span based on $\mathbf{M}$, as illustrated in Figure~\ref{fig:bidaf}. 

\paragraph{Lexicon Embedding.}

This component extracts information from a question of length $I$, $Q=\{q_1,...,q_I\}$, and a passage of length $J$, $P=\{p_1,...,p_J\}$, at the word level and normalizes for lexical variants. It typically maps each word to a dense vector using a pre-trained word embedding model, such as word2vec~\citep{mikolov2013distributed} or GloVe~\citep{pennington2014glove}, such that semantically similar words are mapped to the vectors that are close to each other in neural space. 
Word embedding can be enhanced by concatenating each word embedding vector with other linguistic embeddings such as those derived from Part-Of-Speech (POS) tags, and named entities etc.
To tackle the out-of-vocabulary problem, character-level word representations or sub-word representations are used, such as CharCNN~\citep{kim2016character}, 
FastText~\citep{bojanowski2017enriching}, 
Byte-Pair Encoding (BPE)~\citep{sennrich2016neural} and 
WordPiece~\citep{wu16google}.
Given $Q$ and $P$, the word embeddings for the tokens in $Q$ is a matrix $\mathbf{E}^Q \in \mathbb{R}^{d \times I}$ and 
tokens in $P$ is $\mathbf{E}^P \in \mathbb{R}^{d \times J}$, where $d$ is the dimension of the word embedding vectors.

\paragraph{Context Encoding.}

This component uses contextual cues from surrounding words to refine the word embeddings. As a result, the same word maps to different vectors in a neural space depending on its context, such as ``bank'' in ``bank of a river'' vs. ``bank of American''. Context encoding is typically achieved by using Recurrent Neural Networks (RNNs). As shown in Figure~\ref{fig:bidaf}, BiDAF uses two Long Short-Term Memory (LSTM) networks in both directions (i.e., left-to-right and right-to-left), respectively, and concatenate the outputs of the two LSTMs to obtain a matrix $\mathbf{C}^Q \in \mathbb{R}^{2d \times I}$ as the contextual representation of $Q$, and a matrix $\mathbf{C}^P \in \mathbb{R}^{2d \times J}$ as the contextual representation of $P$.

\paragraph{Question-Passage Integrated Encoding.}

This component couples the question and passage vectors and produces the integrated contextual matrix $\mathbf{M}$, which consists of a set of question-aware feature vectors, one for each passage word, over which the answer span is predicted. The integration is achieved by summarizing information from both $\mathbf{C}^Q$ and $\mathbf{C}^P$ via the \emph{attention} process as follows:   

\begin{enumerate}
    \item Compute an attention score, which signifies relevance of question words to each passage word: $s_{ij}=\text{sim}(\mathbf{c}_i^Q,\mathbf{c}_j^P; \theta_s) \in \mathbb{R}$ for each $\mathbf{c}_i^Q$ in $\mathbf{C}^Q$, where $\text{sim}(;{\theta_s})$ is the similarity function e.g., a bilinear model, parameterized by $\theta_s$.
    \item Compute the normalized attention weights through softmax: $\alpha_{ij}=\exp{(s_{ij})} / \sum_k \exp{(s_{kj})}$.
    \item Summarize information for each passage word via $\mathbf{\hat{c}}_j^P=\sum_i \alpha_{ij} \mathbf{c}_i^Q$ to obtain a matrix $\mathbf{\hat{C}}_P \in \mathbb{R}^{2d \times J}$ as the question-aware representation of $P$.
\end{enumerate}
BiDAF computes attentions in two directions: from passage to question $\mathbf{\hat{C}}^Q$ as well as from question to passage $\mathbf{\hat{C}}^P$. Then, BiDAF generates $\mathbf{M}$ by first concatenating three matrices $\mathbf{C}^P$, $\mathbf{\hat{C}}^P$ and $\mathbf{\hat{C}}^Q$, and then using a bidirectional LSTM to encode for each word its contextual information with respect to the passage and question.

\paragraph{Answer Span Prediction.}

The probability distribution of the start index over the entire passage is obtained by a logistic regression with softmax:

\begin{equation} \label{eqn:bidaf-start-index}
P_{r}^{(start)}(j) = \text{softmax} (\mathbf{W}^{(start)} \mathbf{M}_{:j}),
\end{equation}
where $\mathbf{M}_{:j}$ is the $j$-th column vector (corresponding to the $j$-th passage word), $\mathbf{W}^{(start)}$ is the parameter matrix. 
Similarly, the probability distribution of the end index over the entire passage is computed as
\begin{equation} \label{eqn:bidaf-end-index}
P_{r}^{(end)}(j) = \text{softmax} (\mathbf{W}^{(end)} \mathbf{M}_{:j}).
\end{equation}
The probability of an answer span $(i,j)$ is
\begin{equation} \label{eqn:bidaf-span-index}
P_{r}^{span}(i,j) = P_{r}^{start}(i) P_{r}^{end}(j).
\end{equation}


\paragraph{Training.}

The training loss is defined as the sum of the negative log probabilities of the true start and end indices by the predicted distributions, averaged over all training samples:

\begin{equation} \label{eqn:mrc-loss}
\mathcal{L}(\theta) = - \frac{1}{|\mathcal{D}|} \sum_{n=1}^{|\mathcal{D}|} \left( \log \left( {P_{r}^{(start)}} (y^{(start)}_n) \right) + \log \left( {P_{r}^{(end)}} (y^{(end)}_n) \right) \right),
\end{equation}
where $\theta$ is the set of trainable model parameters, $\mathcal{D}$ is the training set, $y^{(start)}_n$ and $y^{(end)}_n$ are the true start and end of the answer span of the $n$-th training sample, respectively.

\subsection{BERT-Based Readers}
\label{subsec:bert-based-reader}

Most of the state-of-the-art extractive readers use a Transformer-based pre-trained language model, such as BERT~\citep{devlin2018bert} and its variants like RoBERTa~\citep{liu2019roberta} and DeBERTa~\citep{he2020deberta}, as the encoder.

\begin{figure}[t] 
\centering 
\includegraphics[width=1.0\linewidth]{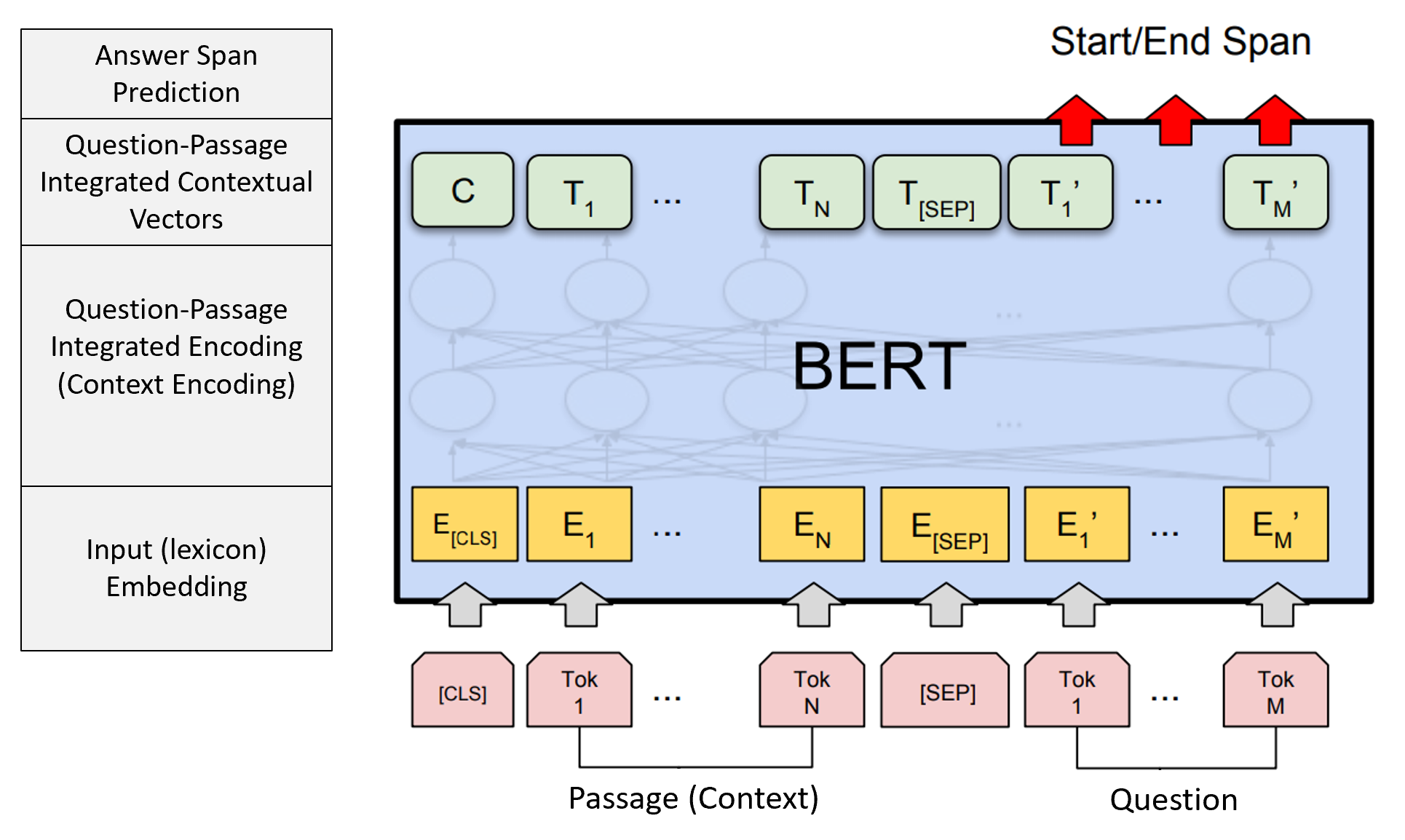}
\vspace{-1mm}
\caption{The BERT-based reader for extractive MC, adapted from \citet{devlin2018bert}.} 
\label{fig:bert4mrc}
\vspace{0mm}
\end{figure}

As illustrated in Figure~\ref{fig:bert4mrc}, a BERT-based reader takes the word embeddings (or, more preciously, sub-word embeddings such as WordPiece) of the concatenated question-passage sequence as input, and uses the build-in attention mechanism of the Transformer to compute for every word in question and passage an attention score to encode the influence each word has on another. In the output of the encoder, each word in passage is represented using an integrated contextual vector that incorporates contextual information from other passage words and question words. These integrated contextual vectors are then fed into a logistic regression layer with softmax for answer span prediction as Equations \ref{eqn:bidaf-start-index} to \ref{eqn:bidaf-span-index}.

The build-in attention mechanism of the Transformer is described in detail in Section~\ref{sec:cir_preliminary} and Figure~\ref{fig:Transformer}, following~\cite{vaswani:17}.

\paragraph{Training.}
BERT-based readers are trained in two stages. 
In the first stage, the BERT encoder is pre-trained on large amounts of raw text data by predicting masked words conditioned on their context. The learning is self-supervised and does not need any labels.  Hugging Face\footnote{https://huggingface.co} maintains a rich repository of pre-trained language models (PLMs), including the BERT models and their variants, which are developed for various NLP tasks. However, these PLMs are trained on general-domain text corpora (e.g., Web). If the target domain is dramatically different from general domain, we might consider adapting a general-domain PLM using in-domain data by continual pre-training the PLM. For domains with abundant unlabeled text, such as biomedicine, pre-training domain-specific language models from scratch might also be a good choice~\citep{gu2020domain}.
In the second stage, the answer span predictor is added on top of the BERT encoder, following the single or pair prediction formulation as illustrated in Figure~\ref{fig:plm_usage}, and both the encoder and the answer predictor are fine-tuned using the MC training data.

\section{Generative Readers}
\label{sec:generative-readers}

While an extractive reader extracts contiguous spans from the retrieved passages as the answer, generative readers sequentially decode the answer string which might not be contained in the retrieved passages. A typical generative reader is implemented as a sequence-to-sequence model that generates answer $A$ conditioned on retrieved a (set of) passage(s) $P$ and question $Q$.

\begin{figure}[t] 
\centering 
\includegraphics[width=1.0\linewidth]{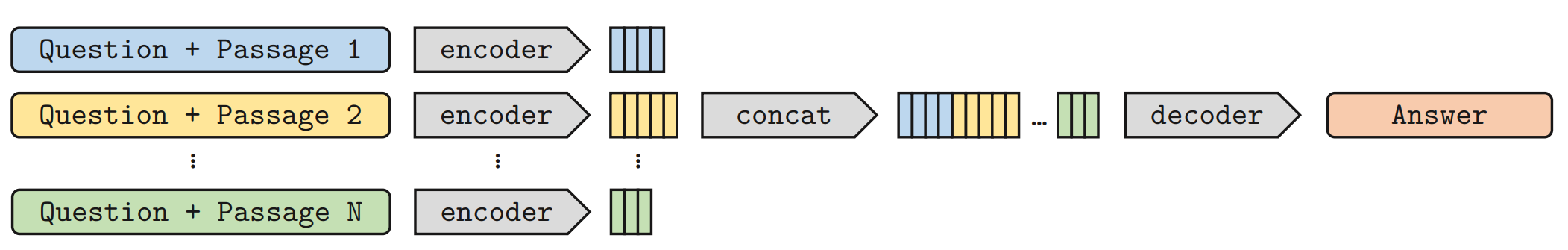}
\vspace{-1mm}
\caption{The Fusion-In-Decoder (FID) model. Figure credit: \citet{izacard2020leveraging}} 
\label{fig:fusion-in-decoder}
\vspace{0mm}
\end{figure}

We describe the Fusion-In-Decoder (FID) model \citep{izacard2020leveraging} as an example. 
FID is developed for open-domain QA, and is based on a pre-trained sequence-to-sequence model, T5~\citep{raffel2019exploring}. As illustrated in Figure~\ref{fig:fusion-in-decoder}, given question $Q$ and a set of $N$ retrieved passages $\mathcal{P} = \{P_1,...,P_N\}$, the encoder, which is similar to what is described in Section~\ref{sec:extractive-readers}, encodes each $(Q,P_n)$ pair independently, and produces an integrated contextual vector representation for each token, $\mathbf{m}^n_j$ for the $j$-th token of the $n$-th pair. The decoder then performs attention over the concatenation of the representations of all the retrieved passages, and generate the answer string.

Let $\{Q, \mathcal{P} \}$ denote the input of the question and all retrieved passage, and $A=\{a_1,...,a_K\}$ the answer string consisting of $K$ tokens. The generative reader is trained to minimize a sequence-to-sequence loss for a given $(\{Q, \mathcal{P} \},A)$,

\begin{equation} \label{eqn:fid}
\mathcal{L}(\{Q, \mathcal{P} \},A;\theta) = -\sum_k^K\log P_r(a_k|\{Q, \mathcal{P} \}, a_{1:k-1};\theta),
\end{equation}
where $\theta$ is the model parameter. During inference, a greedy decoding can be used to produce the answer.

FID differs from other generative readers \citep[e.g.,][]{min2020ambigqa,lewis2020retrieval} in that it processes passages independently in the encoder but jointly in the decoder. As pointed out by \cite{izacard2020leveraging}, the computational cost of FID grows linearly with the number of passages, instead of quadratically since the encoder of FID only performs self-attention over one context at a time. On the other hand, conditioning on all passage encodings jointly allows the decoder to effectively aggregate evidence from multiple passages. 
\section{Hybrid Readers}
\label{sec:hybrid-readers}

Since extractive and generative readers adopt different answer inference strategies, a hybrid extractive-generative reader can be a better option for open-domain QA. \cite{cheng2021unitedqa} show that answers produced by those two types of reader are complementary to each other. Thus, they propose a simple ensemble approach, known as UnitedQA, for combining the predictions from extractive and generative readers. Denote the predicted answer strings from $M$ extractive and $N$ generative readers as 
$A^{E}_{1}, ..., A^{E}_{M}$ and $A^{G}_{1}, ..., A^{G}_{N}$, respectively.
The hybrid prediction $A^*$ is obtained by
\begin{align}  \label{eqn:united-qa}
    A^* = \arg\max_{A \in \mathcal{A}} \tau \sum_{m=1}^{M} \mathbbm{1}(A, A^E_{m}) +  \delta \sum_{n=1}^{N} \mathbbm{1}(A, A^G_{n}),
\end{align}
where $\mathcal{A}$ is the set of all predicted strings, $\mathbbm{1}(A, A^\prime)$ is an indicator function and $\tau=0.6$, $\delta=0.4$. 
UnitedQA sets new state-of-the-art on two representative open-domain QA benchmarks, NaturalQuestions and TriviaQA, when the paper is published. 

\section{Conversational Machine Comprehension Readers}
\label{sec:cmc-readers}

The Conversational Machine Comprehension (CMC) readers that we  discuss in this section are developed on the CoQA and QuQA datasets. A recent literature review of CMC readers is \cite{gupta2020conversational}.

Since some important dialog features, such as topic return, clarification, definition, are not well-represented in the two datasets, as discussed in Section \ref{sec:textqa-task-datasets}, these CMC readers are designed primarily to answer the input (current) question based on the given passage and dialog history, but not to take other types of system actions such as recommending a new topic or asking clarifying questions.
Thus, 
these CMC readers do not use a dialog policy for system action selection. 
In Chapter~\ref{chp:c-kbqa}, we will describe how a conversational knowledge base QA module learns a dialog policy to select system actions.

Specifically, the CMC task on the CoQA and QuQA datasets is defined as follows. 
Given a passage $P$ and dialog history $\mathcal{H} = \{ H_1,...,H_{N-1} \} $ where each dialog turn is a question-answer pair: $H_n = (Q_n, A_n)$, the CMC reader needs to generate answer $A_N$ to input question $Q_N$. 

Compared to the single-turn readers described in Sections \ref{sec:extractive-readers} and \ref{sec:generative-readers}, CMC readers need to make use of dialog history $\mathcal{H}$, in addition to $P$, to answer the input question. 
Dialog history is useful because the answer to $Q_N$ often depends on the previous answers and questions in the same dialog session, as demonstrated in the dialog behaviors of drill down and topic return, and because ellipsis phenomena are frequently encountered in dialog sessions (e.g., in Figure \ref{fig:two-conversational-mrc-datasets} (Right), "she" in Q2 refers to "Jessica" in A1).

CMC readers extend single-turn readers by 
adding a history selection module to select a subset of $\mathcal{H}$ that are useful to answer the input question, and by 
extending the encoding module to encode not only the passage and input question, but also the (selected) dialog history. The answer module (i.e., the answer predictor of an extractive reader, or the answer decoder of a generative reader) remains the same.

\subsection{History Selection}
\label{subsec:history-selection}

There are two reasons for selecting a subset of history rather than using the whole history for CMC. 
The first is due to computational cost. 
A dialog history can be arbitrarily long. Encoding a long history is computationally expensive. For example, many Transformer-based readers (such as BERT-based readers) take a word sequence up to 512 tokens as input.
The second reason is that not all dialog turns in the history are useful for answering the input question. As the topics of conversation are likely to shift, it is more desirable to base the answer to be generated on the relevant subset of the history. 

Many CMC readers use heuristics to select dialog history. 
\cite{ju2019technical,choi2018quac,yatskar2018qualitative} simply truncate the history to retain the $M$ immediate turns based on a pre-defined length limit.
\cite{zhu2018sdnet,ohsugi2019simple} use the last two turns based on their empirical study which shows that including more turns does not improve performance.

\begin{figure}[t] 
\centering 
\includegraphics[width=0.8\linewidth]{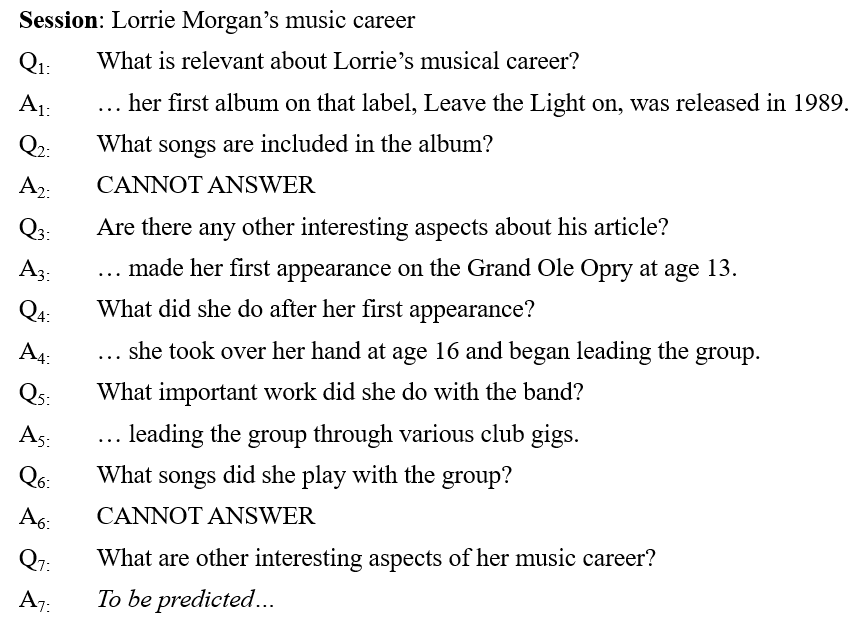}
\vspace{0mm}
\caption{An example of a conversational question answering session from QuAC \citep{choi2018quac}. Q2, Q4, Q5 and Q6 are closely related to their immediate previous turns. Q7 is related to a remote question Q1. Q3 does not follow Q2 but shifts to a new topic. Figure credit: \citet{qu2019attentive}} 
\label{fig:ham-example}
\vspace{0mm}
\end{figure}

The history attention mechanism (HAM) \citep{qu2019attentive} is among a very few methods that take a model-based approach to selecting history on-the-fly based on the input question. 
HAM is motivated by the observation that previous dialog turns in a dialog session are of different importance for answering the input question, and the importance does not necessarily depend on their relative position. As illustrated in the example in Figure~\ref{fig:ham-example}, while Q2, Q4, Q5 and Q6 are closely related to their immediate previous turns, Q7 is related to a remote question Q1, and Q3 does not follow Q2 but shifts to a new topic. 
The HAM module computes for each turn in dialog history a weight based on how useful it is for answering the input question. Then, the history used for generating the answer to the input question is represented as a weighted sum of the vector representations of previous dialog turns. 

\subsection{History Encoding}
\label{subsec:history-encoding}

The methods of encoding history for CMC can be grouped into two categories. 
The first category of methods simply append the selected history to the input question, thus converting the CMC problem into a single-turn MC problem \citep[e.g.,][]{ju2019technical,zhu2018sdnet}. 
There are some variants where history is first encoded into feature vectors before being appended. 
For example, \cite{choi2018quac} augment BiDAF to encode dialog context from previous $M$ QA pairs as follows. First, the input question embedding is extended by appending a feature vector that encodes dialog turn number. Then, the passage embedding is extended by appending a feature vector that encodes the locations of the previous answer spans.

The second category uses the FLOW-based approach \citep[e.g.,][]{huang2018flowqa,yeh2019flowdelta,chen2019graphflow}. 
The FLOW mechanism is proposed to incorporate intermediate representations generated during the process of answering previous questions. Compared to the approaches that simply append previous QA pairs to the input question as input, FLOW integrates the latent semantics of dialog history more deeply. The first FLOW-based CMC reader, called FlowQA \citep{huang2018flowqa}, achieves state of the art on CoQA and QuAC when it is published. 

\begin{figure}[t] 
\centering 
\includegraphics[width=0.95\linewidth]{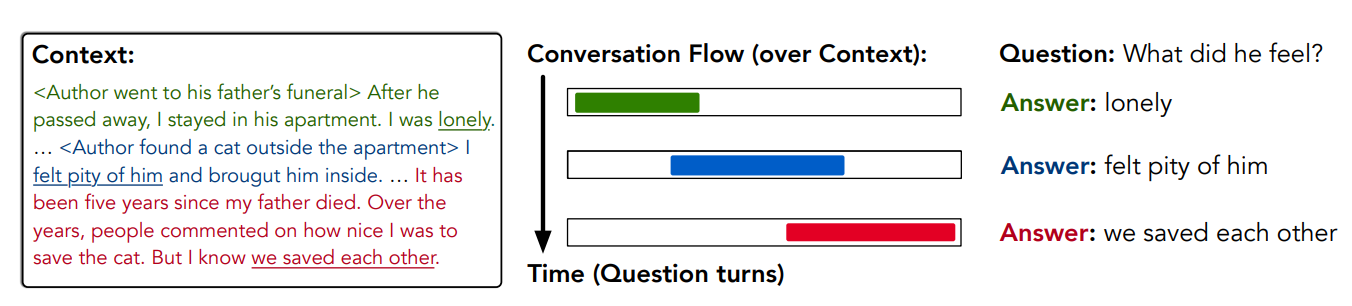}
\vspace{0mm}
\caption{An illustration of the conversation flow. As the current topic changes over time, the answer to the same question changes accordingly. Figure credit: \citet{huang2018flowqa}} 
\label{fig:flowqa}
\vspace{0mm}
\end{figure}

FLOW is designed to capture how a dialog flows, which can be illustrated using the example in Figure~\ref{fig:flowqa}. As the conversation progresses, the topic being discussed changes over time. The answer to the same question varies depending on the current topic. For example, the answer to the question ``what did he feel'' is ``lonely'' when the topic being discussed is the author's father's funeral. But the answer becomes ``we saved each other'' when the dialog flows to a different topic regarding five years after the death of the author's father. 
Thus, modeling how a dialog flows is crucial to the performance of a CMC reader. 

\cite{huang2018flowqa} define FLOW as a sequence of latent topic representations computed based on question-passage context. 
Take the BiDAF model as an example, such topic representations are the integrated contextual vectors of dialog turns.
A naïve implementation of FLOW is to read the context in order and process the dialog turns sequentially, then pass the output hidden vectors from each integration layer during the $(M-1)$-th dialog turn to the corresponding integration layer for $Q_M$. This process is highly nonparallel. Thus, training a CMC reader on large amounts of training dialogs can be prohibitively expensive. 

\begin{figure}[t] 
\centering 
\includegraphics[width=0.95\linewidth]{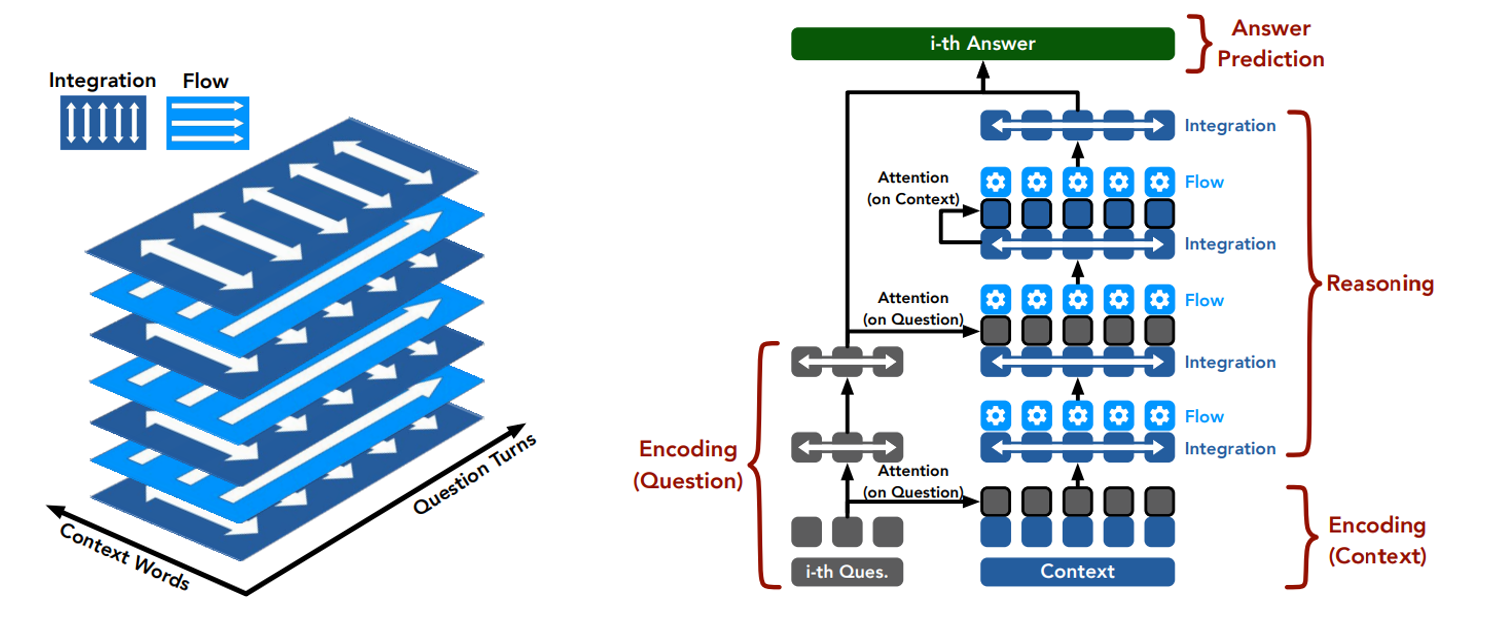}
\vspace{0mm}
\caption{The integration flow (IF) mechanism and the architecture of FlowQA. (Left) The IF layer performs reasoning in parallel for each question. (Right) FlowQA employs multiple IF layers on top of the context encoding, inter-weaved with attention (first on question, then on context itself), to generate the history-aware integrated contextual vectors. Figure credit: \citet{huang2018flowqa}}
\label{fig:flowqa-arch}
\vspace{0mm}
\end{figure}

\cite{huang2018flowqa} propose the Integration Flow (IF) mechanism to improve the training efficiency.  
As illustrated in Figure~\ref{fig:flowqa-arch} (Left), an IF layer performs reasoning in parallel for each question, and then refines the reasoning results across different dialog turns. 
Given the input question and previous questions, each represented using their integrated contextual vectors,
the IF layer generates a history-aware integrated contextual vector for each passage word using a LSTM.
So, the entire integrated contextual vectors for answering previous questions can be used to answer the input question.
As illustrated in Figure~\ref{fig:flowqa-arch} (Right), FlowQA employs multiple IF layers on top of contextual encoding, inter-weaved with attention (first on question, then on context itself), to generate the history-aware integrated contextual vectors. 
The answer is predicted on the integrated context vectors. 

\begin{figure}[t] 
\centering 
\includegraphics[width=0.7\linewidth]{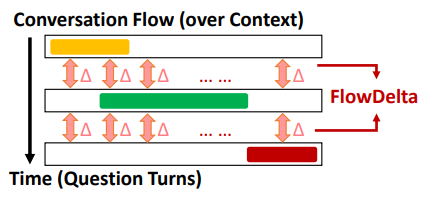}
\vspace{0mm}
\caption{Illustration of the flow information gain modeled by the FlowDelta mechanism. Figure credit: \citet{yeh2019flowdelta}.}
\label{fig:flowdelta}
\vspace{0mm}
\end{figure}

There have been extensions of FlowQA, achieving even better results. We present two examples below.
FlowDelta \citep{yeh2019flowdelta} explicitly models the information gain through the dialog flow in order to allow the CMC reader to focus on more informative cues, as illustrated in Figure~\ref{fig:flowdelta}.

\cite{chen2019graphflow} argue that FlowQA is not quite natural since it does not mimic how humans perform reasoning. 
Using IF layers, FlowQA first processes sequentially in passage, in parallel of question turns and then processes sequentially in question turns, in parallel of passage words.  However, humans typically do not first perform reasoning in parallel for each question, and then refine the reasoning results across different turns.  Doing so is sub-optimal because the results of previous reasoning processes are not incorporated into the current reasoning process. Therefore, \cite{chen2019graphflow} propose an improved FLOW-based CMC reader, GraphFLOW, that captures dialog flow using a graph neural network.  GraphFlow dynamically constructs a question-specific context graph from passage text at each turn, then models the temporal dependencies in the sequence of context graphs. 

\bigskip

\begin{figure}[t] 
\centering 
\includegraphics[width=0.95\linewidth]{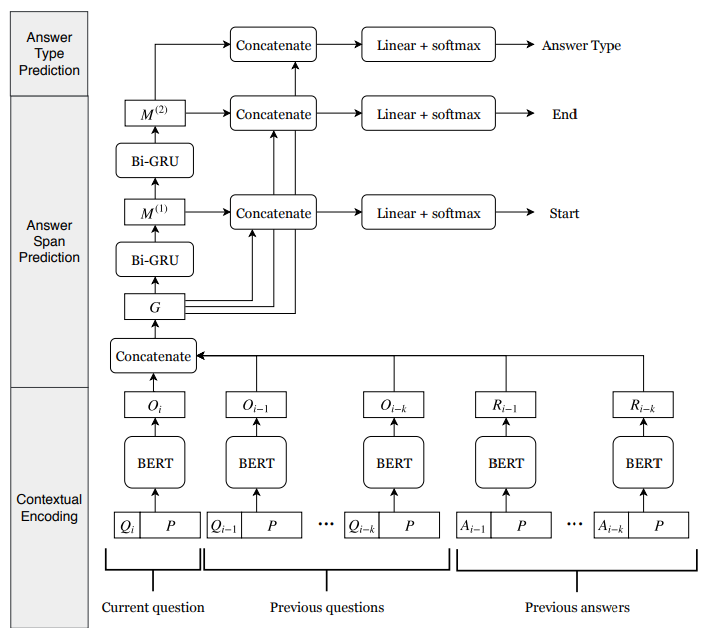}
\vspace{0mm}
\caption{A CMC reader that uses BERT to encode dialog history. Figure credit: \citet{ohsugi2019simple}.}
\label{fig:bert-based-cmc-reader}
\vspace{0mm}
\end{figure}

The FLOW mechanism is originally proposed for sequence-model-based readers. It also inspires the development of several BERT-based CMC readers that use the integrated context vectors of dialog history for answer prediction.
\cite{ohsugi2019simple} use BERT to encode for each question and answer a contextual representation (i.e., a sequence of vectors) conditioned on the passage. As illustrated in Figure~\ref{fig:bert-based-cmc-reader}, given a history consisting of $K$ QA pairs and the input question, there are $2K+1$ resultant contextual representations. Then, a bidirectional RNN model runs over the concatenation of the these vector sequences for answer span prediction. 

\cite{qu2019attentive} propose a BERT-based CMC reader that uses HAM to select and encode dialog history context in three steps. 
In the first step, each dialog turn in dialog history, together with the passage and the input question, is encoded into a context vector using BERT.  For example, we can construct for each dialog turn an instance sequence, which is a concatenation of the token \texttt{[CLS]}, dialog turn $H$, input query $Q$ and passage $P$. We feed the instance into BERT, and take the output representation of \texttt{[CLS]} as the integrated contextual vector of the instance, denoted by $\mathbf{h}$.  
Let $\mathbf{h}_k (k = 1,...,K)$ be the integrated contextual vector of the instance constructed using dialog turn $H_k$. Note that the instance for $H_K$ is simply the input question $Q$. 
In the second step, a single-layer feed-forward network is used to compute the weight of each dialog turn with respect to the input question as

\begin{equation} \label{eqn:ham1}
\alpha_{k} = \text{softmax}(\mathbf{W} \cdot \mathbf{h}_k),
\end{equation}
where $\mathbf{W}$ is the projection matrix.
In the third step, dialog history is incorporated into the input-question-passage integrated contextual vector $\mathbf{h}_K$ via history attention to form the history-aware integrated contextual vector $\hat{\mathbf{h}}_K$ as

\begin{equation} \label{eqn:ham2}
\hat{\mathbf{h}}_{k} = \sum_{k=1...K} \alpha_k \mathbf{h}_k.
\end{equation}
The answer span is predicted on $\hat{\mathbf{h}}_K$.

 

\chapter{Conversational QA over Knowledge Bases}
\label{chp:c-kbqa}
 
Techniques and methods developed for Conversational Question Answering over Knowledge Bases (C-KBQA) are fundamental to the knowledge base search module of a CIR system, as shown in Figure~\ref{fig:cir-system-arch}. Unlike the machine comprehension module (Chapter~\ref{chp:c-mrc}), which has to rely on the documents retrieved by the document search module to generate answers, the knowledge base search module can be used as a standalone system, known as a C-KBQA system, which generates the answer to an input query by reasoning over its knowledge base. 
Many task-oriented dialog systems are implemented as instances of a C-KBQA system~\citep{gao2019neural}. These systems consist of many dialog components that have counterparts in a CIR system as in Figure~\ref{fig:cir-system-arch}, such as contextual query understanding, dialog policy, dialog state tracker, clarifying question generation, and response generation.  

This chapter describes C-KBQA as a special case of CIR, where the answers to input queries are generated not from a collection of indexed documents, but from a (set of) structured knowledge base(s).
Section~\ref{sec:knowledge-bases-and-questions} introduces the C-KBQA task, and describes the forms of knowledge bases and public benchmarks. 
Section~\ref{sec:system-overview} gives an overview of the architecture of a typical C-KBQA system that is based on semantic parsing. 
The next four sections (Sections~\ref{sec:semantic-parsing},  \ref{sec:dialog-state-tracking}, \ref{sec:dialog-policy}, and \ref{sec:response-generation}) describe each of the four main modules of a semantic-parser-based C-KBQA system, respectively, including a semantic parser, a dialog state tracker, a dialog policy, and a response generator. 
Section~\ref{sec:gtg4cqa} discusses a unitary (non-modular) C-KBQA system that is based on a Transformer-based language model which unifies the C-KBQA modules and can generate answers grounded in input queries, dialog states and knowledge.
 
\section{Knowledge Bases and Questions}
\label{sec:knowledge-bases-and-questions}

Many C-KBQA systems are developed as a natural language interface to access their knowledge bases (KBs), which are typically implemented as relational databases or knowledge graphs.

\begin{figure}[t]
\centering 
\includegraphics[width=0.9\linewidth]{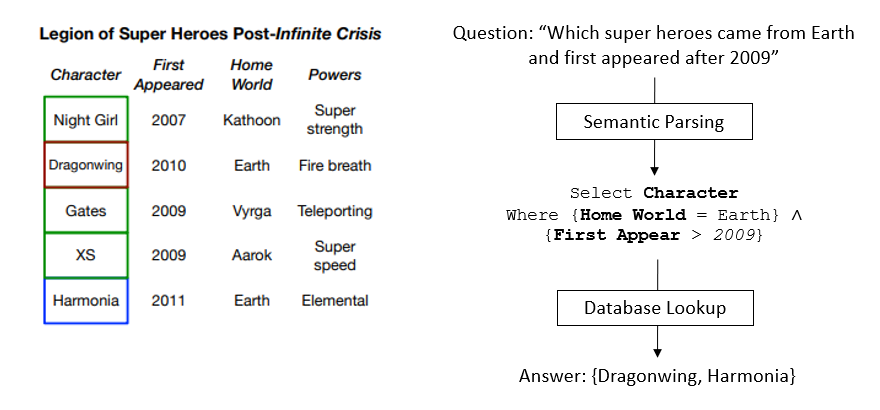}
\vspace{-2mm}
\caption{A semantic-parsing-based question answering system over a relational database. (Left) A sub-table of the relational database related to the comics \texttt{Legion of Super Heroes Post-Infinite Crisis}. (Right) A question, its semantic parse, and the answer. Figures adapted from \citet{iyyer2017search}} 
\label{fig:kbqa-db-example} 
\vspace{0mm}
\end{figure}

A relational database (DB) is also known as an entity-centric KB. 
It stores a list of entities, each associated with a set of properties. 
Figure~\ref{fig:kbqa-db-example} (Left) shows a table in a relational database that contains a list of characters who appear in the comics \texttt{Legion of Super Heroes} (entities), and for each character a set of properties such as \texttt{first-appeared}, \texttt{home-world}, \texttt{powers}.

\begin{figure}[t]
\centering 
\includegraphics[width=1\linewidth]{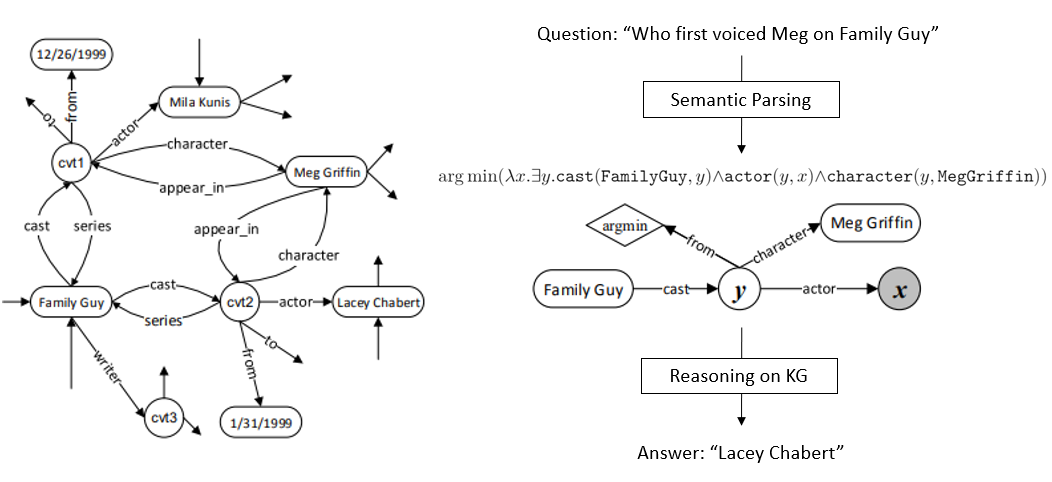}
\vspace{-2mm}
\caption{A semantic-parsing-based question answering system over a knowledge graph. (Left) A sub-graph of Freebase related to the TV show \texttt{Family Guy}. (Right) A question, its logical form in $\lambda$-calculus and query graph, and the answer. Figures adapted from \citet{yih2015semantic}.} 
\label{fig:kbqa-kg-example} 
\vspace{0mm}
\end{figure}

A knowledge graph (KG) consists of a collection of subject-relation-object triples $(s,r,o)$ where $s, o\in\mathcal{E}$ are entities and $r\in\mathcal{R}$ is a relation (or predicate). The ``knowledge graph'' is named after its graphical representation, i.e., the entities are nodes and the relations are the directed edges that link the nodes.
Figure~\ref{fig:kbqa-kg-example} (Left) shows a small sub-graph of Freebase \citep{bollacker2008freebase} related to the TV show \texttt{Family Guy}. Nodes include some names, dates and special Compound Value Type (CVT) entities.\footnote{CVT is not a real-world entity, but is used to collect multiple fields of an event or a special relationship.} A directed edge describes the relation between two entities, labeled by a relation.

The primary difference between KGs and relational DBs is that the former work with paths while the latter work with sets.  
In a KG, the relations are stored at the individual record level, while in a relational DB the structure is defined at a higher level (i.e., the table schema). This has important ramifications. 
First, a relational DB is much faster than a KG when operating on large numbers of records. In a KG, each record needs to be examined individually against an input question to determine the structure of the data in order to locate the sub-graph in which answer nodes reside, as illustrated in Figure~\ref{fig:kbqa-kg-example} (Right). 
In a relational DB, the schema of each table is pre-defined. Thus, an input question can be readily parsed to a SQL-like query based on the pre-defined schema, which is then executed on the DB to retrieve answer records, as illustrated in Figure~\ref{fig:kbqa-db-example} (Right).
Second, relational DBs use less storage space, because it is not necessary to store all those relationships for each record or entity.

Due to these differences, KGs and relational DBs are applied in different settings. 
KGs are mainly used to store Web-scale open-domain KBs, such as Freebase \citep{bollacker2008freebase} and DBPedia \citep{auer2007dbpedia}, where there are many variations in relations (or predicates) among entities, and users can issue complex questions which need to be parsed to map to different paths in the knowledge graph to retrieve the answer nodes.  
Relational DBs, on the other hand, are widely used to store task-specific KBs where the attributes of entities can be pre-defined. For example, the KBs of many task-oriented dialog systems,
such as the ones designed for movie ticket booking or table reservation in a restaurant, are stored using relational DBs. 

The QA sessions in Figures~\ref{fig:kbqa-db-example} and \ref{fig:kbqa-kg-example} are \emph{single-turn}. 
But it is not reasonable to assume that users can always compose in one shot a complicated, compositional natural language question that can uniquely identify the answer in the KB. 
As discussed in Section \ref{sec:how-people-search}, many information seeking tasks, such as exploratory search, require an intensive human-machine interaction over a long-term iterative sensemaking process. Thus, a multi-turn C-KBQA system is more desirable because they allow users to query a KB interactively without composing complicated questions.
For example, \citet{iyyer2017search} show that users rarely issue complicated questions like 
\[
\text{``Which super heroes came from earth and first appeared after 2009?''} 
\]
as in Figure~\ref{fig:kbqa-db-example} (Right). Instead, the same intent is more naturally expressed through a sequence of simpler questions as:
\begin{enumerate} 
    \item ``Who are the super heroes?''
    \item ``Which of them come from earth?''
    \item ``Of those, who appears after 2009?''
\end{enumerate}

\begin{figure}[t]
\centering 
\includegraphics[width=1.0\linewidth]{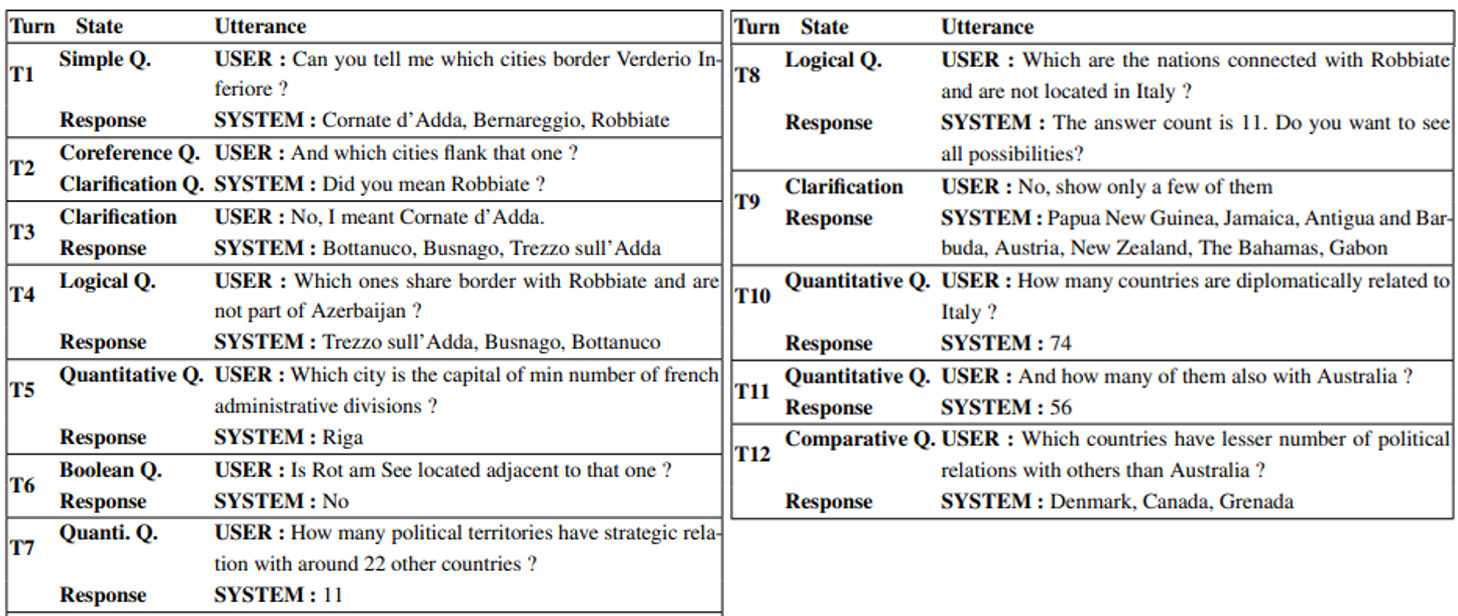}
\vspace{-2mm}
\caption{A sample conversational QA session from the CSQA dataset \cite{saha2018complex}.} 
\label{fig:csqa-example} 
\vspace{0mm}
\end{figure}

\citet{saha2018complex} present a dataset that consists of 200K QA sessions for the task of Complex Sequence Question Answering (CSQA). Figure~\ref{fig:csqa-example} shows a sample conversational QA session from the dataset. The session contains a sequence of questions.  
Some are simple questions that can be answered from a single subject-relation-object triple in the knowledge graph, such as the question in T1.
Some are complex questions that require logical, comparative and quantitative reasoning over a larger sub-graph of the knowledge graph, such as the questions in T4 and T5.
We also note some characteristics of a conversation session. All question-answer pairs in the same session are related, and the session contains typical elements of a dialog such as co-reference, ellipses, clarifications, and confirmations.

CSQA combines two sub-tasks: 
(1) answering factoid questions through complex reasoning over a large-scale KB, and (2) learning to converse through a sequence of coherent QA pairs.
To handle these tasks, a C-KBQA system is designed as a task-oriented dialog system as in Figure~\ref{fig:modular-task-bot-arch} that have the capabilities of
\begin{enumerate} 
    \item parsing input natural language questions,
    \item using conversational context to resolve co-references and ellipsis in user utterances, 
    \item deciding whether and how to ask clarifying questions based on a dialog policy, and 
    \item retrieving relevant paths in the KB to answer questions.
\end{enumerate}


\subsection{Open Benchmarks}
\label{subsec:kbqa-open-benchmarsk}

Open benchmarks have been the key to achieving progress in many AI tasks including dialog and information retrieval. Although C-KBQA is a relatively nascent research problem, some open benchmarks have already been developed.

    \paragraph{Sequential Question Answering (SQA) \citep{iyyer2017search}.} The dataset is collected via crowdsourcing by leveraging WikiTableQuestions (WTQ) \citep{pasupat2015compositional}, which contains highly compositional questions associated with HTML tables from Wikipedia. Each crowdsourcing task contains a long, complex question originally from WTQ as the question intent. The workers are asked to compose a sequence of simpler questions that lead to the final intent; an example of this process is shown in Figure~\ref{fig:kbqa-db-example} (Left). Each sequence forms a dialog session. The dataset consists of 2,022 sessions with an average of 2.9 questions per session.  
    \paragraph{Complex Sequential Question Answering (CSQA)~\citep{saha2018complex}.} This dataset contains 200K dialog sessions with a total of 1.6M dialog turns. CSQA is more challenging than SQA in that many questions in the CSQA sessions are complex and require reasoning over a large sub-graph of the knowledge graph, as illustrated in the examples in Figure~\ref{fig:csqa-example}. 
    \paragraph{ConvQuestions~\citep{christmann2019look}.} The dataset contains 11,200 conversations compiled by Turkers on five domains over Wikidata: books, movies, soccer, music, and TV series. Similar to CSQA, the dataset presents a variety of complex question phenomena like comparisons, aggregations, compositionality, and temporal reasoning. Answers are grounded in Wikidata entities to allow an easy comparison of different methods. Some questions are not answerable by Wikidata alone, but requiring seeking information in the open Web. 
    \paragraph{CoSQL~\citep{yu2019cosql}.} This a corpus for building cross-domain Conversational text-to-SQL systems. It is the dialog version of the Spider~\citep{yu2018spider} and SParC~\citep{yu2019sparc} tasks. CoSQL consists of 30K dialog turns and 10K annotated SQL queries, obtained from a Wizard-of-Oz collection of 3K conversations querying 200 complex databases spanning 138 domains. Each conversation simulates a real-world DB query scenario with a crowd worker as a user exploring the DB and a SQL expert retrieving answers with SQL, clarifying ambiguous questions, or otherwise informing of unanswerable questions.
    \paragraph{CLAQUA~\citep{xu2019asking}.} This dataset is developed for evaluating the system's capability of asking clarification questions for C-KBQA. The dataset supports the evaluation of three sub-tasks: identifying whether clarification is needed given a question, generating a clarification question, and predicting answers base on external user feedback.

\section{System Overview}
\label{sec:system-overview}

\begin{figure}[t] 
\centering 
\includegraphics[width=0.75\linewidth]{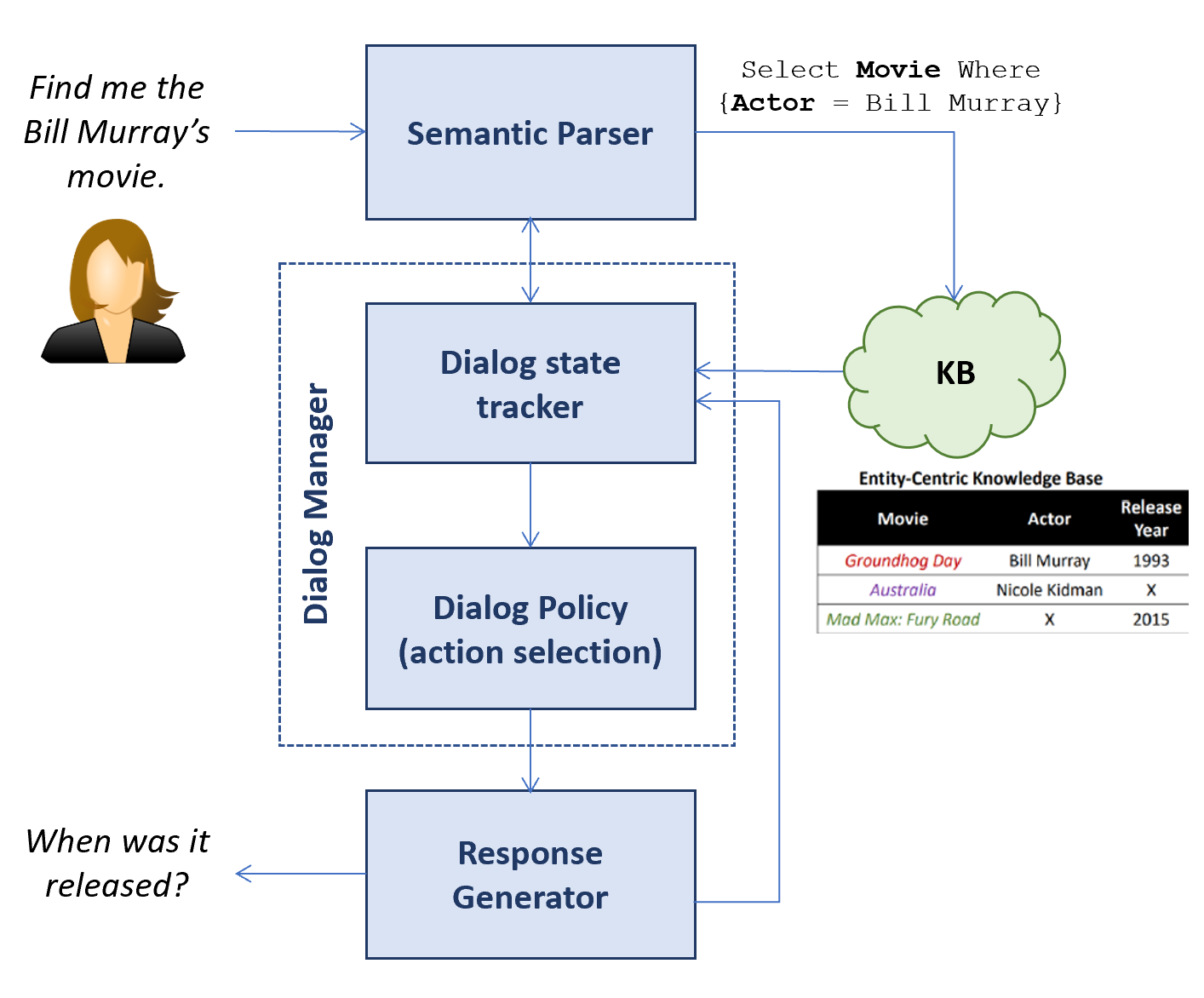}
\vspace{-2mm}
\caption{A typical architecture of C-KBQA systems.} 
\label{fig:c-kbqa-arch} 
\vspace{0mm}
\end{figure}

A typical architecture of C-KBQA systems is illustrated in Figure~\ref{fig:c-kbqa-arch}.
It is composed of the following components
\begin{enumerate}
    \item \text{Semantic parser:} It maps an input user question (or user utterance) and the dialog context to a meaning representation (or formal query) in formal query languages like SQL (Figure~\ref{fig:kbqa-db-example}), $\lambda$-calculus (Figure~\ref{fig:kbqa-kg-example}) and SPARQL, which would be executed on the KB to fetch the answer. These formal query languages have a well-defined grammar, allowing for complex answer retrieval and generation involving local operands (e.g., disjunction, conjunction, and negation), aggregation functions (e.g., grouping and counting), filtering based on conditions, etc., and are supported by many popular KBs.
    \item \text{Dialog manager:} It consists of (1) a \emph{dialog state tracker} that updates the dialog state to maintain information (e.g., entities, predicates and action sequences) from conversation history, and (2) a \emph{dialog policy} that selects the next system action (e.g., ask clarifying or navigation questions, return the answer, reject the question since there is no answer, greet, terminate the conversation, etc.)
    \item \text{Response generator:} This component converts the system action selected by the dialog policy to a natural language response. 
\end{enumerate}

The C-KBQA architecture shares a lot of similarities with the architectures of task-oriented dialog systems (Figure~\ref{fig:modular-task-bot-arch}) and CIR systems (Figure~\ref{fig:cir-system-arch}).
One difference is that C-KBQA relies on structured KBs, not document collections, for generating answers. Another noticeable difference is that the query understanding module (which is the NLU component in Figure~\ref{fig:modular-task-bot-arch} and the contextual query understanding module in Figure~\ref{fig:cir-system-arch}) is implemented as a semantic parser in C-KBQA. 
The semantic parser takes both a user utterance and the dialog state (which represents the dialog history) as input to generate a formal query to execute on the KB to produce the answer.  
The parsed formal query and its retrieved result are then used to update the dialog state.  
Note that the retrieved result is not always presentable to users. For example, if a user question is not specific enough, its retrieved result may contain many entities (e.g., there could many movies starred by ``Bill Murray'' in Figure~\ref{fig:c-kbqa-arch}). In such a case, the system has to ask a clarifying question (e.g., ``When was it released?'') to make the question more specific. This is particularly important for QA applications on devices with small or no screen.

In the next four sections, we describe each of the components, illustrated using examples.

\section{Semantic Parsing}
\label{sec:semantic-parsing}

This section starts with an overview of semantic parsing methods based on recent literature surveys  \citep[e.g.,][]{chakraborty2019introduction,kamath2018survey}, then, as a case study, describes in detail a dynamic neural semantic parser (DynSP) \citep{iyyer2017search}, and reviews methods of improving DynSP or similar neural semantic parsers using pre-trained language models. 

\subsection{Overview}
KBQA is typically cast as a semantic parsing problem. 
Given a KB $\mathcal{K}$ and a formal query language $\mathcal{F}$, a semantic parser converts an input natural language question $Q$ to a formal query $F\in\mathcal{F}$ that not only can execute over $\mathcal{K}$ to return the correct answer(s) $A$ but also accurately captures the meaning of $Q$.

Many state-of-the-art semantic parsers are implemented using neural network models trained in either a fully supervised setting where the training data consists of question-parse $(Q,F)$ pairs, or a weakly supervised setting where the training data consists of question-answer $(Q,A)$ pairs.  Since it is expensive to generate $(Q,F)$ pairs in large quantities, the weakly supervised setting prevails in real-world applications. However, this presents a challenge to model training since we need to avoid spurious formal queries which often hurt generalization. These spurious formal queries coincidentally execute over a KB to return correct answers but do not capture the meaning of the original questions.

Neural network based semantic parsers can be grouped into three categories: translation, classification and ranking \citep{chakraborty2019introduction}.

\paragraph{Translation models.} We view semantic parsing as a machine translation problem where $Q$ is translated to $F$. A popular approach is to treat $Q$ and $F$ as sequences of symbols, and apply neural sequence-to-sequence models, which are originally developed for machine translation~\citep{sutskever2014sequence}. 
However, training these models in a weakly supervised setting is challenging, and there is no guarantee that the formal queries generated by these translation models are always grammatical and readily executable over the KB. 

\paragraph{Classification models.} The structure of a formal query can be of arbitrary size and complexity given the input questions. However, for some simple questions, we can assume a fixed structure for $F$. For example, the factoid questions, such as T1 in Figure~\ref{fig:kbqa-kg-example}, can be answered from a single subject-relation-object $(s,r,o)$ triple in the KB. The corresponding formal query thus consists of a subject entity $s$ and a relation $r$, and the answer is the missing object entities. Thus, semantic parsing can be performed using text classification, i.e., predicting $s$ and $r$ from natural language question $Q$. 
    
\paragraph{Ranking models.} Ranking models are the most widely used among the three categories. Instead of using any pre-defined fixed structure, a search procedure is employed to find a suitable set of candidate formal queries (with different structures), and some ranking models are used to compare these candidates (e.g., by estimating how much they are semantically similar to the input question $Q$) and to select the best-ranked one.  The neural semantic parser to be studied next belongs to this category.

\subsection{A Dynamic Neural Semantic Parser}
The description of the Dynamic Neural Semantic Parser (DynSP) follows closely that in \citet{iyyer2017search}.
Consider the example in Figure~\ref{fig:kbqa-db-example}. 
Given a user question ``Which super heroes came from Earth and first appeared after 2009?'' and a table of a relational DB that stores the entities and their associated properties, the formal query generated by DynSP is
\[
\small
\texttt{SELECT character WHERE \{home-world = ``Earth''\} $\wedge$ \{first-appeared > ``2009''\}}
\]
and its returned result is
\[
\texttt{\{``Dragonwing'', ``Harmonia''\}}.  
\]
The formal query in this example is a SQL-like query, which consists of a \texttt{SELECT} statement that is associated with the name of the answer column, and zero or more conditions, each containing a condition column and an operator (=, <, >) and arguments, which enforce additional constraints on which cells in the answer column can be chosen.

\begin{figure}[t] 
\centering 
\includegraphics[width=0.6\linewidth]{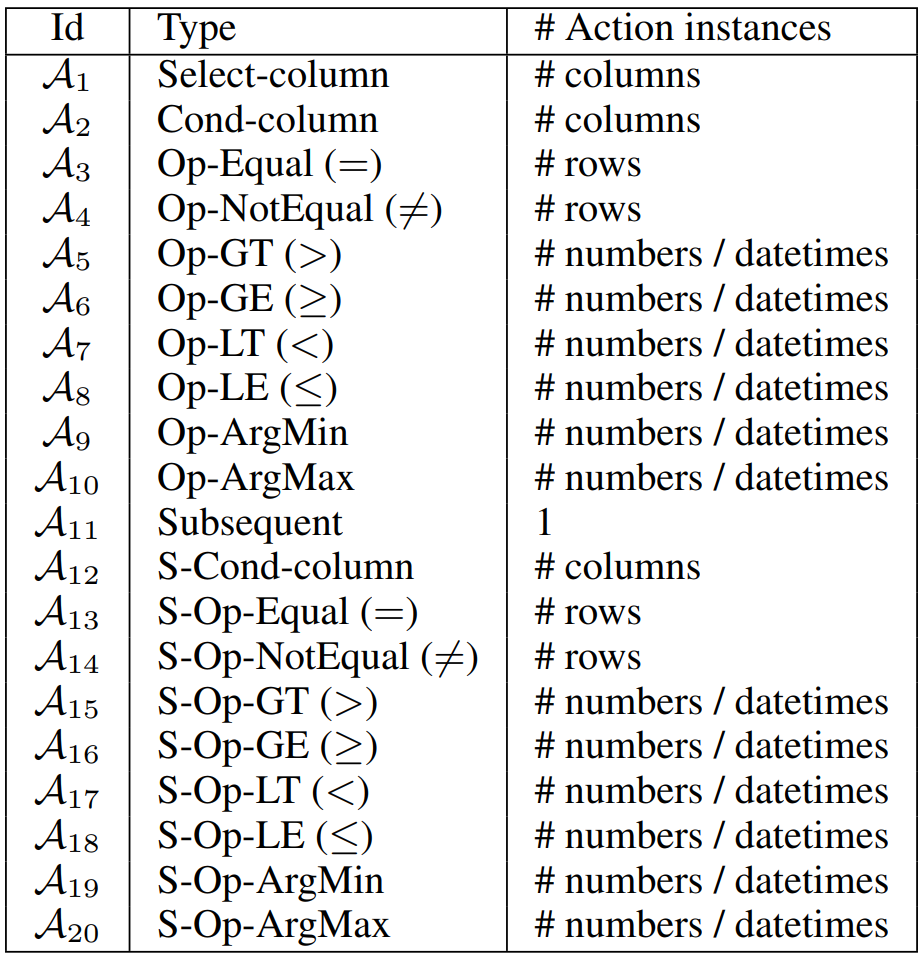}
\vspace{-1mm}
\caption{Types of actions and the number of action instances in each type, defined in \cite{iyyer2017search}.} 
\label{fig:dynsp-actions} 
\vspace{0mm}
\end{figure}

DynSP formulates semantic parsing as a state-action search problem, where a state $s$ is defined as an action sequence representing a complete or partial parse, and an action $a$ (of an action type $\mathcal{A}$, as exemplified in Figure~\ref{fig:dynsp-actions}) is an operation to extend a parse. Parsing is cast as a process of searching an end state with the highest score.

DynSP is inspired by STAGG, a search-based semantic parser \citep{yih2015semantic} and the dynamic neural module network (DNMN) \citep{andreas2016learning}.
Like STAGG, DynSP pipelines a sequence of modules as search progresses; but these modules are implemented using neural networks, which enables end-to-end training as in DNMN. 
Note that in DynSP the network structure is not predetermined, but are constructed dynamically as the parsing procedure explores the state space. 
Figure~\ref{fig:dynsp-actions} shows the types of actions and the number of action instances in each type, defined in \cite{iyyer2017search}. 
Consider the example in Figure~\ref{fig:kbqa-db-example}, one action sequence that represents the parser is 
\begin{enumerate} [noitemsep]
    \item ($a_1$) select-column \texttt{character}, 
    \item ($a_2$) cond-column \texttt{home-world},
    \item ($a_3$) op-equal \texttt{``Earth''}, 
    \item ($a_2$) cond-column \texttt{first-appeared}, 
    \item ($a_5$) op-gt \texttt{``2009''}.
\end{enumerate}

The follow-up questions in conversational QA can be handled using a preamble statement \texttt{SUBSEQUENT} as shown in Figure~\ref{fig:dynsp-actions}.
A subsequent statement contains only conditions because it adds constraints to the semantic parse of the previous question. 
For example, the corresponding formal query of the follow-up question ``Which of them breathes fir?'' is
\[
\texttt{SUBSEQUENT WHERE \{powers = ``fire breath''\}}. 
\]
The answer to the question is 
\[
\texttt{\{``Dragonwing''\}}, 
\]
which is a subset of the previous answer.

\begin{figure}[t] 
\centering 
\includegraphics[width=0.6\linewidth]{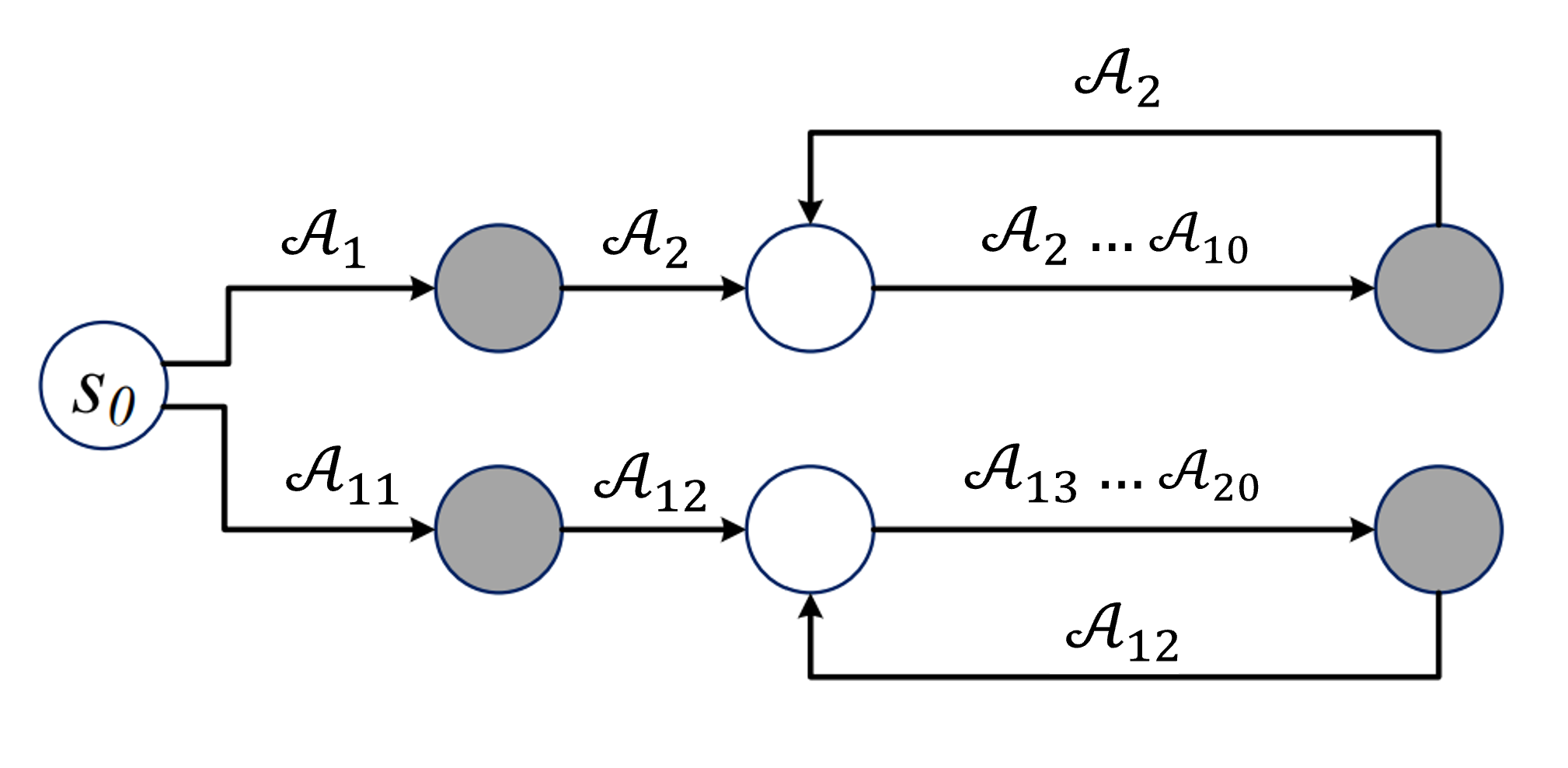}
\vspace{-1mm}
\caption{Possible action transitions based on their types (see Figure~\ref{fig:dynsp-actions}), where shaded circles are end states \cite{iyyer2017search}.} 
\label{fig:dynsp-fsm} 
\vspace{0mm}
\end{figure}

Since many states represent semantically equivalent parses, to prune the search space, the actions that can be taken for each state are pre-defined, as shown in Figure~\ref{fig:dynsp-fsm}. 
Then, beam search is used to find an end state with the highest score in the space using the state value function as: 

\begin{equation}
V(s_t ; \theta) = V(s_{t-1} ; \theta) + \pi(s_{t-1}, a_t ; \theta), V(s_0 ; \theta) = 0 
\label{eq:dynsp}
\end{equation}
where $s_t$ is a state consisting of an action sequence $a_1,...a_t$, and $\pi(s,a ; \theta)$ is the policy function that scores action $a$ given state $s$.

Equation~\ref{eq:dynsp} shows that the state value function can be decomposed as a sequence of policy functions, each implemented using a neural network. 
Therefore, the state value function is a state-specific neural network parameterized by $\theta$ which can be trained end-to-end. 
Conceptually, these neural networks measure semantic similarities among the words in the question and tables (either the column names or cells). 
Take the select-column action type ($\mathcal{A}_1$) as an example. The corresponding policy function computes the matching score between question $Q=\{ q_1,...,q_I \}$ and column $C= \{ c_1,...,c_J \} $. 
Let $\mathbf{q}_i$ be the embedding of question word $q_i$ and $\mathbf{c}_j$ the embedding of word $c_j$ in the target column name. The matching score can be computed as
\begin{equation}
\frac{1}{J} \sum_{j=1...J} \max_{i=1...I} {\mathbf{q}_i^{\intercal}} \mathbf{c}_j 
\label{eq:dynsp1}
\end{equation}
where the word embeddings can be generated using a pre-trained language model, as described in Section~\ref{subsec:using-pre-trained-language-models}.

\citet{iyyer2017search} propose to learn $\theta$ using weakly supervised learning on query-answer pairs without ground-truth parses. 
This is challenging because the supervision signals (rewards) are delayed and sparse. For example, whether the generated parse is correct or not is only known after a \emph{complete} parse is generated and the KB lookup answer is returned. 
The authors use an approximate reward, which is dense, for training. 
A \emph{partial} parse is converted to a query to search the KB, and the overlap of its answer with the gold answer is used as the training supervision. A higher overlap indicates a better partial parse. 

Let $A(s)$ be the answer retrieved by executing the parse represented by state $s$, and $A^*$ be the gold answer of a question $q$. The approximated reward of $s$ is defined as
\begin{equation}
R(s,A^*) = \frac{|A(s) \cap A^*|}{|A(s) \cup A^*|} 
\label{eq:dynsp-reward}
\end{equation}

The model parameters $\theta$ are updated in such a way that the state value function $V_\theta$ behaves similarly to the reward $R$. This is achieved by minimizing the following loss for every state $s$ and its approximated reference state $s^*$
\begin{equation}
\mathcal{L}(s) = (V(s ; \theta) - V(s^* ; \theta)) - (R(s) - R(s^*)) 
\label{eq:dynsp-loss}
\end{equation}
The model update algorithm is summarized in Algorithm~\ref{alg:dynsp-model-update}. It picks a training sample $(x,A^*)$, where $x$ represents the table and the question.
The approximate reward $R$ is defined by $A^*$, where $\mathcal{S}(x)$ is the set of end states for $x$. 
Line 2 finds the best approximated reference state $s^*$ and 
Line 3 finds the most violated state $\hat{s}$, both relying on beam search guided by the approximated reward $R(x,A^*)$. 
Line 4 updates model parameters by computing the loss in Equation~\ref{eq:dynsp-loss}.

\begin{algorithm}[H]
\caption{DynSP model parameters update \cite{iyyer2017search}}
\label{alg:dynsp-model-update}
\SetAlgoLined
\KwResult{Updated model parameters $\theta$}
\For {each labeled sample $(x,A^*)$} {
  $s^* \gets \arg\max_{s \in \mathcal{S}(x)} R(s,A^*)$ \;
  $\hat{s} \gets \arg\max_{s \in \mathcal{S}(x)} V_\theta(s) - R(s,A^*)$ \;
  update $\theta$ by minimizing $\max(\mathcal{L}(\hat{s}),0)$ \;
}
\end{algorithm}


\subsection{Using Pre-Trained Language Models}
\label{subsec:using-pre-trained-language-models}

Semantic parsing requires to link the entity mentions and predicates in a user utterance to the entities and relations (or column names) in a KB, respectively. The linking task is challenging due to the diversity of natural language. The same concept can be presented using different expressions in user questions and KBs.  
Consider the example of Figure~\ref{fig:kbqa-db-example}, the predicate ``came from'' in the user question needs to link to the column name \texttt{home-world} of the table. 
In the example of Figure~\ref{fig:kbqa-kg-example}, the question “Who first voiced Meg on Family Guy?” needs to link to a sub-graph which consists of two entity nodes of \texttt{MegGriffin} and \texttt{FamilyGuy}, and a set of relation edges including \texttt{cast}, \texttt{actor} and \texttt{character}.

Semantic parsers, such as DynSP, deal with the linking task using neural language models. These models map natural language expressions into vectors in a hidden semantic space where the vectors of the expressions that are semantically similar are close to each other.  As Equation~\ref{eq:dynsp} shows, words in the question and the target column name are encoded into vectors using a pre-trained language model for matching. Thus, the performance of semantic parsers relies on the quality of the language model they use. The state-of-the-art neural language models  (e.g., BERT \citep{devlin2018bert}, RoBERTa \citep{liu2019roberta}, and UniLM \citep{dong2019unified}) are typically pre-trained on large amounts of text data and then fine-tuned to various downstream NLP tasks.

We now describe a pre-trained language model, SCORE~\citep{score2021}, which is developed for conversational semantic parsing. SCORE achieves new state of the art on several conversational semantic parsing benchmarks when the paper is published. SCORE is initialized using RoBERTa which is pre-trained using the Masked Language Modeling (MLM) objective on Web corpora. Then, SCORE is continuously pre-trained on \emph{synthesized} conversational semantic parsing data using three pre-training objectives.

\begin{figure}[t] 
\centering 
\includegraphics[width=1.0\linewidth]{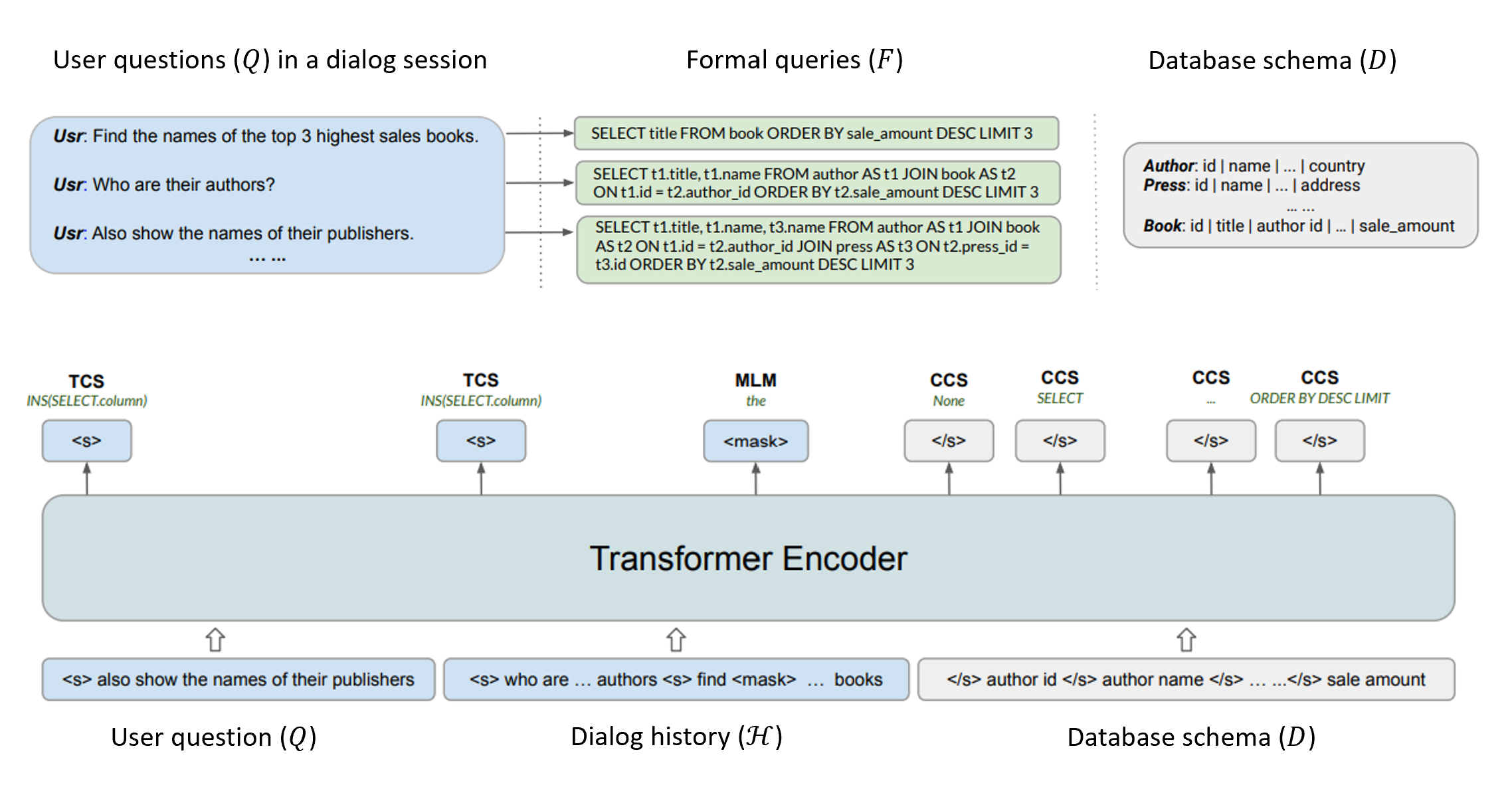}
\vspace{-1mm}
\caption{(Top) Training samples of a conversational semantic parsing task. Each sample consists of user question $Q$, dialog history $H$, database schema $D$ and formal query $F$. (Bottom) Pre-training SCORE using multi-task learning with three tasks, MLM (Mask Language Modeling), CCS (Column Contextual Semantics) and TSC (Turn Contextual Switch). The gold labels of TSC and CCS for each training sample are derived from the target formal query $F$. Figure credit: \cite{score2021}.} 
\label{fig:score-csp} 
\vspace{0mm}
\end{figure}

As shown in Figure~\ref{fig:score-csp}, each training sample consists of user question $Q$, dialog history $\mathcal{H}$, database schema $D$, and target formal query $F$.  
The dialog session used for training is synthesized as follows.
Given $D$ and a sampled single-turn question SQL template, the values (e.g., column names, cell values, and SQL operations) for typed slots in the template are randomly sampled to form a synthesized formal query $F$, and a context-free grammar is used to generate the corresponding user question $Q$ of $F$. 
Follow-up questions can be generated by editing the generated $F$ and applying the context-free grammar.
\cite{score2021} show that using synthesized data allows SCORE to easily adapt to new tasks in few-shot settings where there is few manually-generated task labels.

SCORE is a multi-layer Transformer model. To train the model, we need to present the input $(Q, \mathcal{H}, D)$ as a word sequence. We do so by first concatenating all column names in $D$ as a text sequence where column names are separated by a special token \texttt{</s>}, and then concatenating $Q$, $\mathcal{H}$ and $D$, separated by a special token \texttt{<s>}.  Since it is expensive to get manually labeled data, SCORE is trained on large amounts of synthesized data which are automatically generated using the database schema and the grammars of the formal language.

SCORE is continuously pre-trained using multi-task learning. In addition to MLM, SCORE pre-training also uses two auxiliary tasks designed for conversational semantic parsing, namely Column Contextual Semantics and Turn Contextual Switch.

\paragraph{Column Context Semantics (CCS).}  
CCS is intended to capture the alignment between $Q$ and $F$. $F$ can be decomposed into a set of operations on columns and tables, such as \texttt{SELECT} and \texttt{WHERE} for SQL queries. For each column (or table), CCS predicts how likely an operation should be performed on the column (or table) given the input $(Q, \mathcal{H}, D)$. In Figure~\ref{fig:kbqa-db-example}, given the input, \texttt{SELECT} is performed on the column \texttt{character}, and \texttt{WHERE} is performed on the columns \texttt{home-world} and \texttt{first-appeared}. 
As shown in Figure~\ref{fig:score-csp}, we use the encoded contextual vector of the special token \texttt{</s>} right before each column or table name, denoted by $\mathbf{e}$, to predict its corresponding operation $o$ by a logistic regression with softmax:
\begin{equation}
P_r(o|\mathbf{e}) = \text{softmax}(\mathbf{W}_{\text{CCS}} \cdot \mathbf{e}),
\label{eq:score-ccs}
\end{equation}
where $\mathbf{W}_{\text{CCS}}$ is a learned parameter matrix. 
The CCS loss is defined as cross-entropy with respect to the gold CCS labels that are derived from $F$:
\begin{equation}
- \sum_o \mathbbm{1}(\mathbf{e},o) \log(P_r(o|\mathbf{e})),
\label{eq:score-ccs-loss}
\end{equation}
where $\mathbbm{1}(\mathbf{e},o)$ is the binary indicator (0 or 1), indicating whether the correct operation $o$ is predicted on the column or table whose name is encoded by $\mathbf{e}$.

\paragraph{Turn Contextual Switch (TCS).} 
TSC is intended to capture how a formal query $F$ is generated or changed with the conversation flow. Consider the example in Figure~\ref{fig:kbqa-db-example}, if the user asks a follow-up question “which of them breathes fire?” then $F$ needs to be changed by inserting a new \texttt{WHERE} condition (i.e., the corresponding TCS label is \texttt{INT(WHERE)}).  Given the formal query language such as SQL, there is only a limited number of ways $F$ can be changed, as defined by the action types in Figure~\ref{fig:dynsp-actions}.  TSC predicts for each user turn in a conversation session, whether a particular type of action needs to take to change $F$ or create a new formal query. 
As shown in Figure~\ref{fig:score-csp}, we use the encoded contextual vector of the special token \texttt{<s>} right before each turn, denoted by $\mathbf{e}$, to predict action $a$ that is needed to take to change the formal query. Similar to CCS, we use a logistic regression model with softmax for prediction:
\begin{equation}
P_r(a|\mathbf{e}) = \text{softmax}(\mathbf{W}_{\text{TCS}} \cdot \mathbf{e}),
\label{eq:score-tcs}
\end{equation}
where $\mathbf{W}_{\text{TCS}}$ is a learned parameter matrix. 
The TCS loss is defined as cross-entropy with respect to the gold TCS labels derived from $F$:
\begin{equation}
- \sum_a \mathbbm{1}(\mathbf{e},a) \log(P_r(a|\mathbf{e})),
\label{eq:score-tcs-loss}
\end{equation}
where $\mathbbm{1}(\mathbf{e},a)$ is the binary indicator, indicating whether the correct action $a$ is predicted given the dialog turn encoded by $\mathbf{e}$.

\subsection{C-KBQA Approaches without Semantic Parsing}
\label{subsec:kbqa-wo-parser}

There have been approaches to C-KBQA without using a semantic parser to produce the intermediate formal queries. We briefly describe two examples below. 
\cite{muller2019answering} present an approach to answering conversational questions on structured data without logical forms (formal queries). They encode KB as graphs using a graph neural network, and select answers from the encoded graph using a pointer network. 
    
\cite{christmann2019look} propose CONVEX, an unsupervised method that can answer incomplete questions over a KG by maintaining conversation context using entities and predicates seen so far and automatically inferring missing or ambiguous pieces for follow-up questions.

These approaches are motivated by the observation that it is always difficult to collect large amounts of task-specific labeled data for learning semantic parsers for new tasks. 
However, after large-scale pre-trained language models are applied for semantic parsing, as described in Section~\ref{subsec:using-pre-trained-language-models}, high-quality semantic parsers can be effectively developed for new tasks with few task labels or using synthesized data~\citep{score2021}. 
It is interesting to see whether the pre-trained language models can also be successfully applied to the C-KBQA approaches without semantic parsing. 

\section{Dialog State Tracking}
\label{sec:dialog-state-tracking}

A dialog manager is composed of a dialog state tracker that maintains the dialog state and a dialog policy which selects the next system action. This section describes dialog state tracking, and the next section the dialog policy.

\begin{figure}[t] 
\centering 
\includegraphics[width=0.99\linewidth]{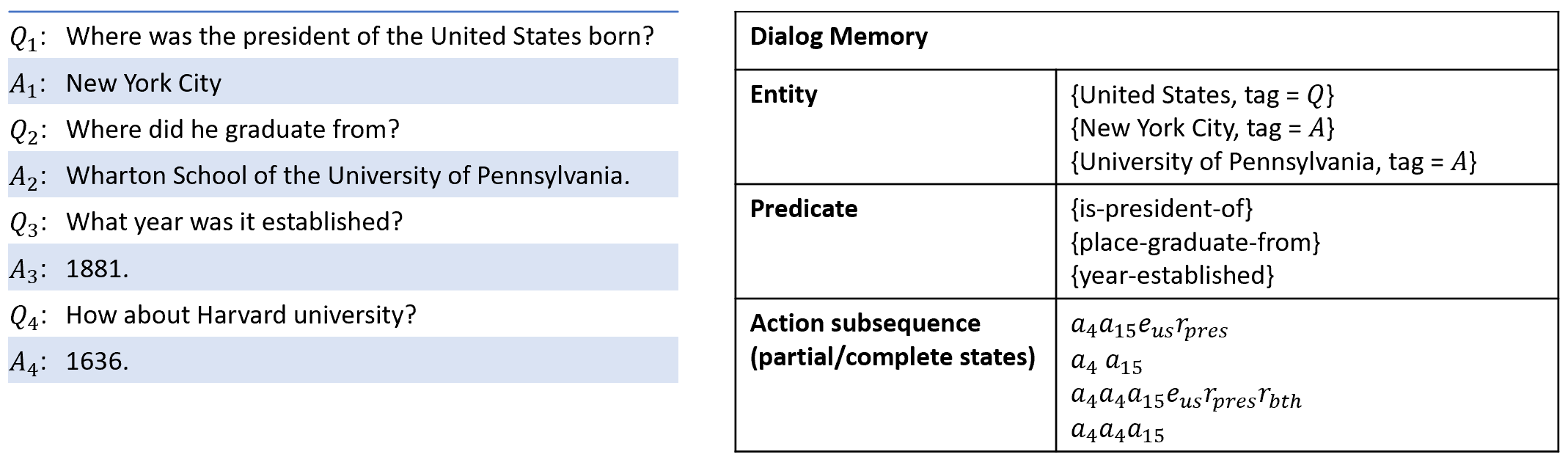}
\vspace{-1mm}
\caption{An example of conversational question answering session (Left) and its dialog memory (Right). Adapted from \cite{guo2018dialog}.} 
\label{fig:kbqa-dst-example} 
\vspace{0mm}
\end{figure}

The type of follow-up questions, which DynSP can deal with using the subsequent statement, is only sufficient to very simple cases where the answer to a follow-up question is a subset of the answer to its previous questions.
In many real-world applications, parsing user questions in dialog sessions where ellipsis phenomena are frequently encountered is a far more challenging task than what DynSP can handle.
Consider the example in Figure~\ref{fig:kbqa-dst-example} (Left). The ellipsis of the entity ``he'' in ${Q}_2$ refers to ``president of the United States'' in ${Q}_1$. The pronoun ``it'' in ${Q}_3$ refers to the answer ${A}_2$. In ${Q}_4$, the ellipsis of the predicate \texttt{year-established} comes from ${Q}_3$.
Thus, it is necessary to explicitly track the dialog state which contains all contextual information to understand what the user is looking for at the current turn of conversation.  

\citet{guo2018dialog} present a dialog state tracking method for generating a formal query $F$ from a user utterance $Q$ and dialog history (context) $\mathcal{H}$.
Given $(Q, \mathcal{H})$, the dialog state is extracted from $\mathcal{H}$ and stored in a dialog memory. As illustrated in Figure~\ref{fig:kbqa-dst-example} (Right), the dialog memory contains three types of information.
\begin{itemize}
    \item \textbf{Entities} are extracted from $\mathcal{H}$ and stored in the dialog memory to help resolve co-references when ellipsis of entities occurs in $Q$.  Entities can be extracted from previous questions and answers, as indicated using ``tags'' in Figure~\ref{fig:kbqa-dst-example} (Right).
    \item \textbf{Predicates} in $\mathcal{H}$ are recorded to deal with the questions where predicate ellipsis occurs. For example, ${Q}_4$ does not contain the predicate \texttt{year-established} as it is in ${Q}_3$.
    \item \textbf{Action subsequences} are partial or complete parses which can be reused in parsing follow-up questions in a similar way to that of DynSP.
\end{itemize}

\begin{figure}[t] 
\centering 
\includegraphics[width=0.95\linewidth]{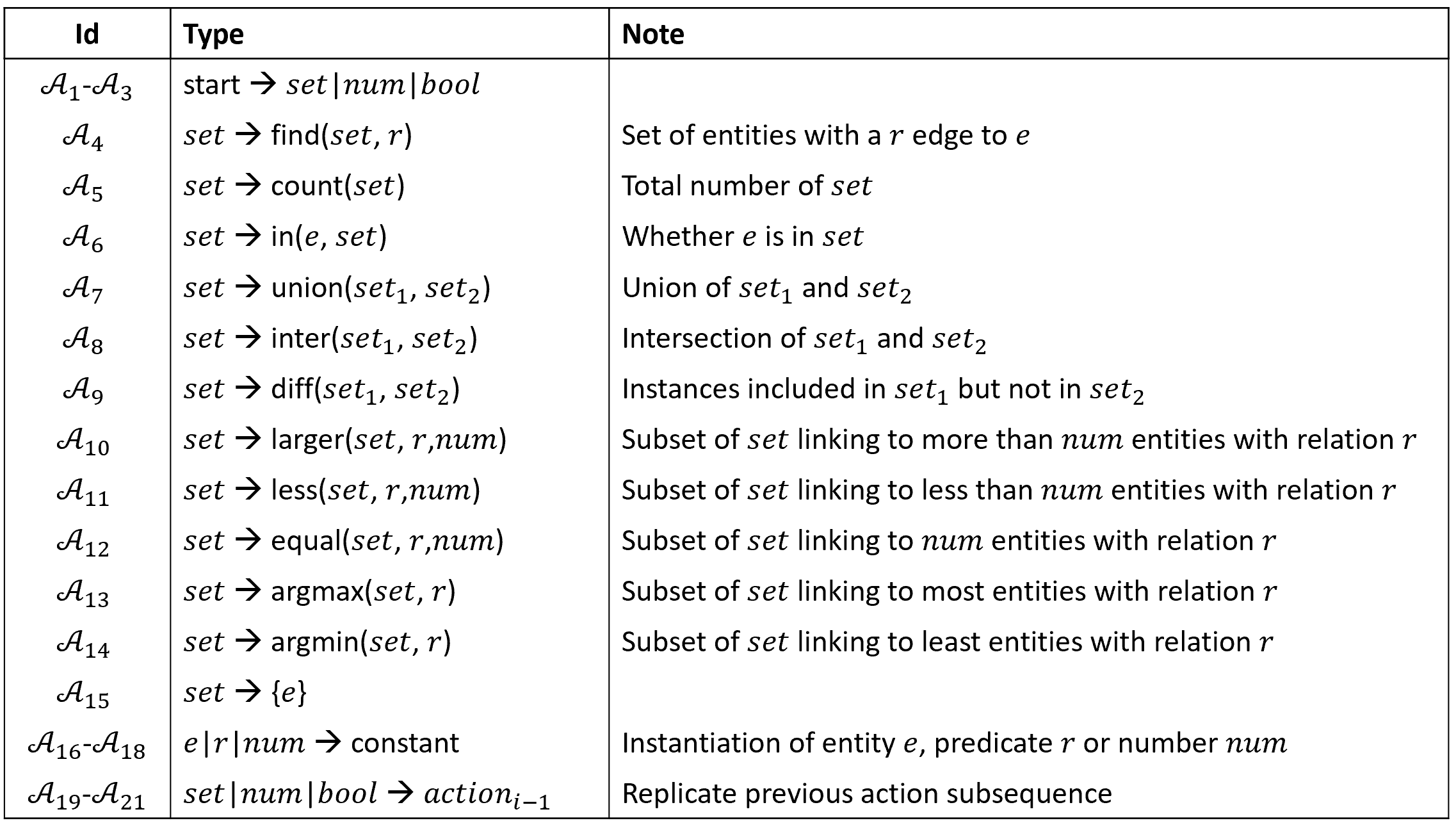}
\vspace{-2mm}
\caption{Types of actions used in \citet{guo2018dialog} for generating formal queries.} 
\label{fig:kbqa-dst-actions} 
\vspace{0mm}
\end{figure}

The semantic parser of \citet{guo2018dialog} is similar to DynSP. Given input $(Q, \mathcal{H})$, the parser searches for a sequence of actions to form $F$. Figure~\ref{fig:kbqa-dst-actions} lists the action types used by the parser, where $\mathcal{A}_1$ to $\mathcal{A}_{15}$ are used for context-independent parsing, which are conceptually similar to $\mathcal{A}_1$ to $\mathcal{A}_{10}$ defined for DynSP in Figure~\ref{fig:dynsp-actions}. Note that the actions defined in Figure ~\ref{fig:kbqa-dst-actions} operate on a knowledge graph while the actions of DynSP on tables in a relational database.
To leverage the dialog state for parsing, two types of actions are defined, as illustrated in Figure~\ref{fig:kbqa-dst-actions}.
\begin{itemize}
    \item \textbf{Instantiated actions} ($\mathcal{A}_{16}$ to $\mathcal{A}_{18}$) can access the dialog memory to identify related entities and predicates that are missing in $Q$ in the current dialog turn.
    \item \textbf{Replication actions} ($\mathcal{A}_{19}$ to $\mathcal{A}_{21}$) can choose and copy a previous action. These actions cover \texttt{SUBSEQUENT} used in DynSP as a special case. 
\end{itemize}

The contextual semantic parsing methods described so far require extending a context-independent parser to be context-dependent in order to leverage dialog history information (dialog state) for generating formal queries.  
In what follows, we turn to discuss an alternative approach based on \emph{contextual question rewriting}, where we do not need to develop a context-dependent semantic parser. These methods can be viewed as the query writing methods of  contextual query understanding, as described in Section~\ref{sec:cqu}. While the methods in Section~\ref{sec:cqu} are developed on the TREC CAsT dataset where $\mathcal{H}$ consists of only previous queries, the methods to be described below are developed for C-KBQA tasks where $\mathcal{H}$ consists of previous user inputs and system responses.

\subsection{Contextual Question Rewriting}
\label{subsec:cqr}

\begin{figure}[t] 
\centering 
\includegraphics[width=0.99\linewidth]{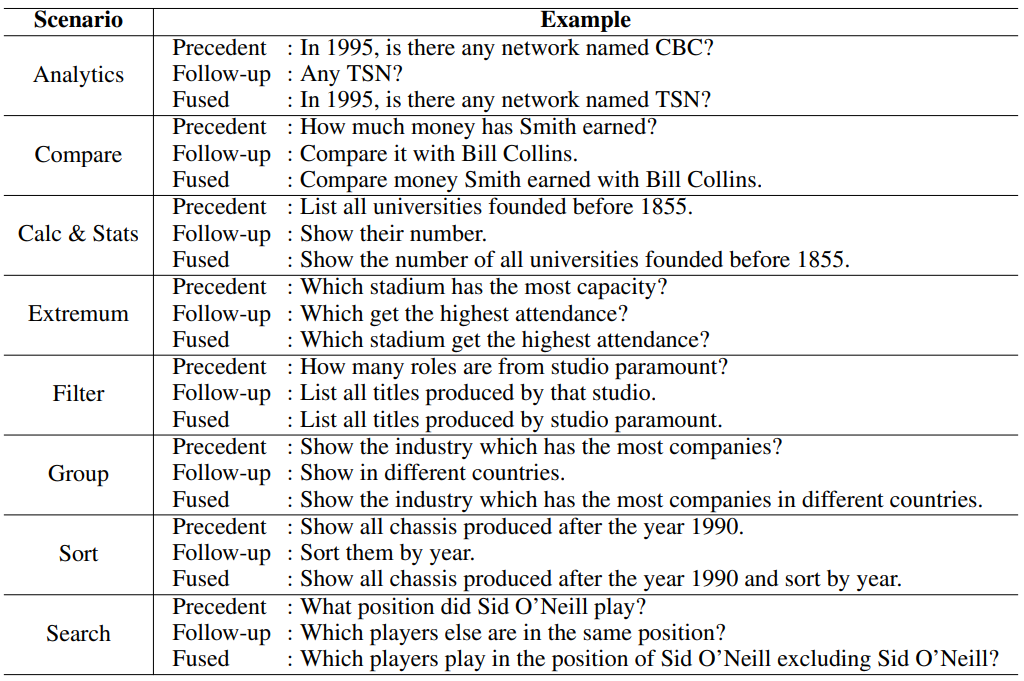}
\vspace{-1mm}
\caption{Some typical follow-up scenarios in conversational question answering \citep{liu2019fanda}.} 
\label{fig:follow-up-scenarios} 
\vspace{0mm}
\end{figure}

\citet{liu2019fanda} present a comprehensive follow-up question analysis and construct a dataset consisting of 1000 examples, covering some typical follow-up scenarios in conversational QA, as shown in Figure~\ref{fig:follow-up-scenarios}, where each example is a triple consisting of a context-independent question (precedent), a follow-up question, and a reformulated follow-up question (fused) that is fused with context from its precedent question.  
Liu et al. argue that whereas it is challenging to build a context-aware semantic parser to deal with a wide variety of follow-up scenarios, it is effective to rewrite a follow-up question to a de-contextualized question (e.g., the fused questions in Figure~\ref{fig:follow-up-scenarios}) by fusing context from its previous questions or dialog history and use a context-independent semantic parser to parse the rewritten question.

Such a contextual question rewriting (CQR) approach, which combines existing semantic parsing and query rewriting methods, has been widely used in many commercial systems, such as Bing \citep{gao2019neural} and Microsoft's XiaoIce system \citep{zhou2020design}.  In what follows, we describe three categories of the CQR models.

\begin{figure}[t] 
\centering 
\includegraphics[width=0.99\linewidth]{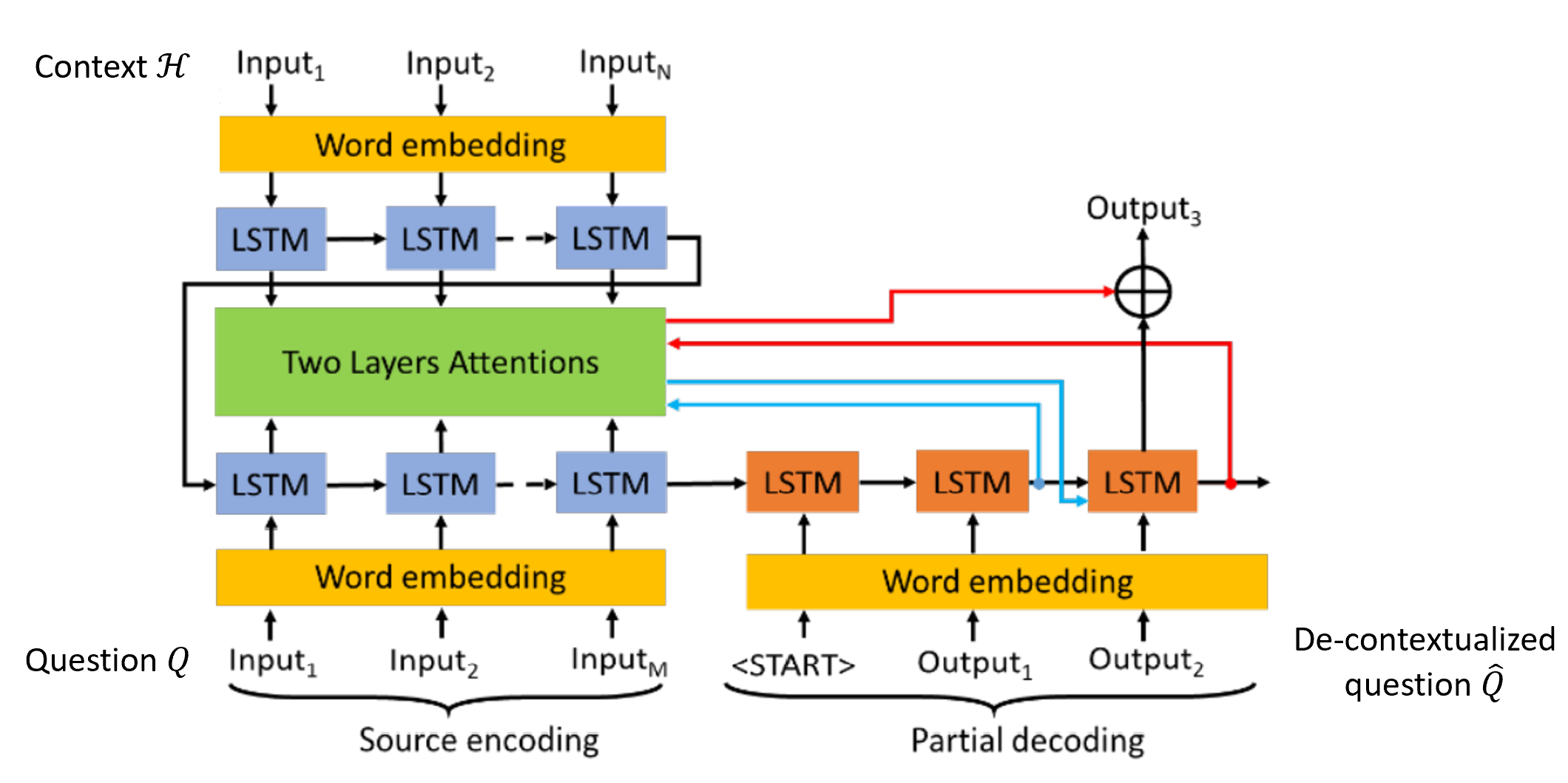}
\vspace{-1mm}
\caption{A sequence-to-sequence model for contextual question rewriting. Figure credit: \cite{ren18towards}.} 
\label{fig:s2s-cqr-model-2} 
\vspace{0mm}
\end{figure}

\paragraph{Sequence-to-Sequence Models.} 
\citet{ren2018conversational} formulate CQR as a sequence-to-sequence (s2s) problem, and have developed four s2s models that rewrite follow-up question $Q$ to context-independent question $\hat{Q}$ by fusing context (dialog history) $\mathcal{H}$ from its previous question. These models differ in how $\mathcal{H}$ is fused for question rewriting.
\begin{enumerate}
  \item Model-1 simply concatenates $Q$ and $\mathcal{H}$ as input to a standard LSTM-based s2s model.
  \item Model-2 uses two separate encoders to encode $Q$ and $\mathcal{H}$, respectively, and then uses a two-layer attention mechanism to fuse $Q$ and $\mathcal{H}$ into the decoder to generate $\hat{Q}$. The model is illustrated in Figure~\ref{fig:s2s-cqr-model-2}.
  \item Model-3 fuses $\mathcal{H}$ into $Q$ during encoding using an attention mechanism, then feeds the context-embedded representation of $Q$ into the decoder to generate $\hat{Q}$.
  \item Model-4 combines Model-2 and Model-3, and is the best performer among the four models as reported in \citet{ren18towards}.
\end{enumerate}

\begin{figure}[t] 
\centering 
\includegraphics[width=0.8\linewidth]{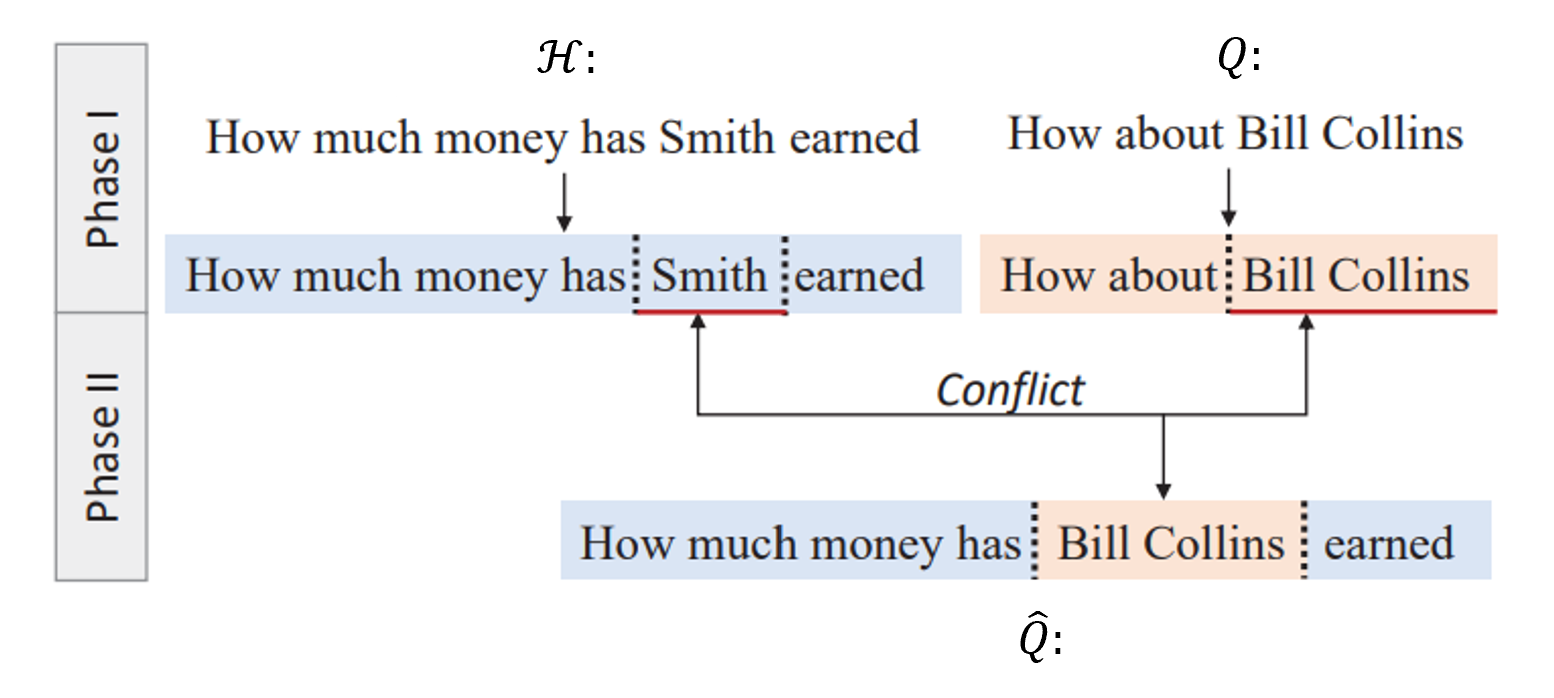}
\vspace{-1mm}
\caption{The two-phase split-and-recombine approach to contextual question rewriting. Figure credit: \cite{liu2019split}.} 
\label{fig:split-and-recombine-cqr} 
\vspace{0mm}
\end{figure}

\paragraph{Split-and-Recombine.}
\citet{liu2019split} present a split-and-recombine method of question rewriting. As illustrated in Figure~\ref{fig:split-and-recombine-cqr}, given input dialog history $\mathcal{H}$ and follow-up question $Q$, the reformulated de-contextualized question $\hat{Q}$ is generated in two phases.
\begin{enumerate}
  \item Split: $\mathcal{H}$ and $Q$ are split into a sequence of spans $Z$ according to model $P_r(Z|Q, \mathcal{H} ; \theta_{s})$
  \item Recombine: the spans are recombined to form $\hat{Q}$ according to model $P_r(\hat{Q}|Z ; \theta_r )$. Recombining is performed by identifying and resolving \emph{conflict} pairs. For example, ``Smith'' and ``Bill Collins'' form a conflict pair because both are named entities and take the same linguistic role in $Q$ and $\mathcal{H}$, respectively. Thus, ``Smith'' needs to be replaced with ``Bill Collins'' in $Q$ to form $\hat{Q}$.
\end{enumerate}

The model parameters $(\theta_{s}, \theta_{r})$ are trained on $(Q, \mathcal{H}, \hat{Q})$ pairs. Since the spans $Z$ are not available in training data, the REINFORCE algorithm \citep{williams92simple} is used for model training. Let $R(Z, \hat{Q}^*)$ be the function (or reward) to measure the quality of $Z$ with respect to the reference $\hat{Q}^*$ as
\begin{equation}
R(Z, \hat{Q}^*) = \sum_{\hat{Q} \in \mathcal{\hat{Q}}} P_r(\hat{Q}|Z ; \theta_{r}) r(\hat{Q},\hat{Q}^{*}),
\label{eq:split-recombine-1}
\end{equation}
where $\mathcal{\hat{Q}}$ is the set of all possible $\hat{Q}$ given $Z$ and $r(\hat{Q}, \hat{Q}^*)$ measures the similarity between $\hat{Q}$ and $\hat{Q}^{*}$. 
Note that \citet{liu2019split} use a process to deterministically generate $\hat{Q}$ from $Z$. So $\theta_r = \emptyset$. 
Then, the final objective function is defined as 
\begin{equation}
\mathbbm{E}_{(Q,\mathcal{H},\hat{Q}^{*}) \sim \mathcal{D}} \left[  \sum_{Z \in \mathcal{Z}} P_r(Z|Q,\mathcal{H} ; \theta_{s}) R(Z,\hat{Q}^{*}) \right] ,
\label{eq:split-recombine-2}
\end{equation}
where $\mathcal{Z}$ represents the set of all possible ways to split $(Q, \mathcal{H})$, and the summation can be approximated by samples generated from the probability distribution defined by $P_r(Z|Q,\mathcal{H}; \theta_{s})$. 

\paragraph{XiaoIce's CQR Component.}
The XiaoIce system \citep{zhou2020design} rewrites questions using the long conversation history which might span a few days or months. To do so, it requires to explicitly track the dialog history to identify and store named entity mentions and predicates from previous conversation turns produced by the user and the system. XiaoIce’s CQR component consists of three modules. It rewrites $Q$ to $\hat{Q}$ using $\mathcal{H}$ in three steps.
\begin{enumerate}
    \item Named entity identification: This module labels all entity mentions in the conversation, link them to the entities stored in the working memory of the state tracker, and store new entities in the working memory.
    \item Co-reference resolution: This module replaces all pronouns with their corresponding entity names.
    \item Sentence completion: If $Q$ is not a complete sentence due to the ellipsis of the predicate, it is completed by inserting the corresponding verb phrase of the previous utterance.
\end{enumerate}

\begin{figure}[t] 
\centering 
\includegraphics[width=0.9\linewidth]{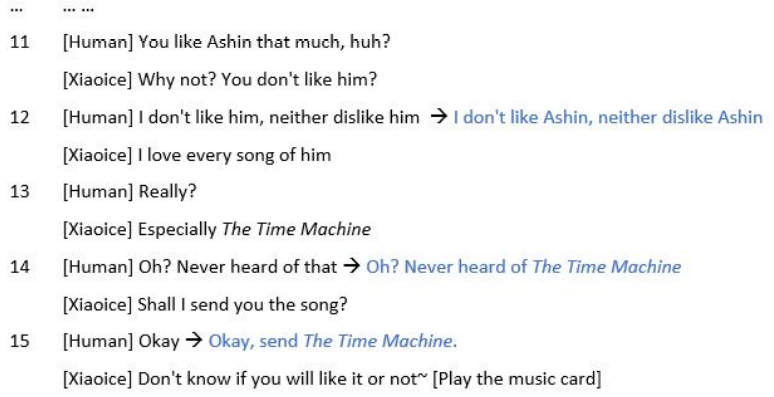}
\vspace{-1mm}
\caption{XiaoIce’s contextual question rewriting component rewrites user questions into de-contextualized questions as indicated by the arrows. Figure credit: \cite{zhou2020design}.} 
\label{fig:xiaoice-cqr} 
\vspace{0mm}
\end{figure}

As shown in Figure~\ref{fig:xiaoice-cqr}, XiaoIce rewrites user questions to include necessary context, for example, replacing “him” in Turn 12 with “Ashin,” “that” with “The Time Machine” in Turn 14, and adding “send The Time Machine” in Turn 15.

\section{Dialog Policy}
\label{sec:dialog-policy}

The tracked dialog state, consisting of a representation of conversational context and a representation of KB retrieval results, is passed to the dialog policy to select the next system action.
In this section we first present a case study of a set of dialog policies for a movie-on-demand system~\citep{li17end,dhingra17towards}, showing how these policies are developed and their impact on the dialog system performance. Then, we introduce dialog acts and discuss dialog policy optimization methods using reinforcement learning.

\subsection{A Case Study}
\label{subsec:case-study}

\begin{figure}[t] 
\centering 
\includegraphics[width=0.7\linewidth]{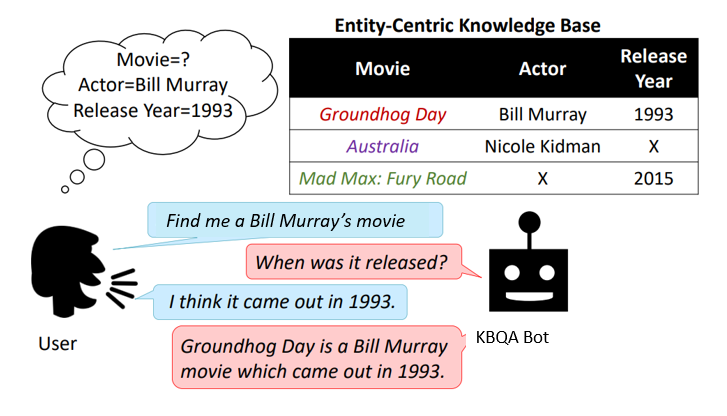}
\vspace{-1mm}
\caption{An interaction between a user and a C-KBQA system for the movie-on-demand task \citep{dhingra17towards}.} 
\label{fig:c-kbqa-case-study} 
\vspace{0mm}
\end{figure}

Consider a movie-on-demand dialog system as illustrated in Figure~\ref{fig:c-kbqa-case-study}. The system is an instance of C-KBQA systems that helps users navigate a KB in search of an entity (movie). The system needs to parse a user question and its context into a formal query to retrieve the requested movie(s) from the DB, which can be viewed as an entity-centric KB consisting of entity-attribute-value triples.  Based on the retrieved result (e.g., the number of entities that match the user goal), the system chooses either to present the result to the user if there is a small number of matched movies, or ask the user a clarifying question to provide more specific information. If the latter is chosen, the system also needs to select what information to ask first (i.e., movie attributes such as director, rating, genre, release year). In Figure~\ref{fig:c-kbqa-case-study}, the system chooses to ask \texttt{release-year} since ``Bill Murray'' stars in many movies. 

\begin{figure}[t] 
\centering 
\includegraphics[width=0.9\linewidth]{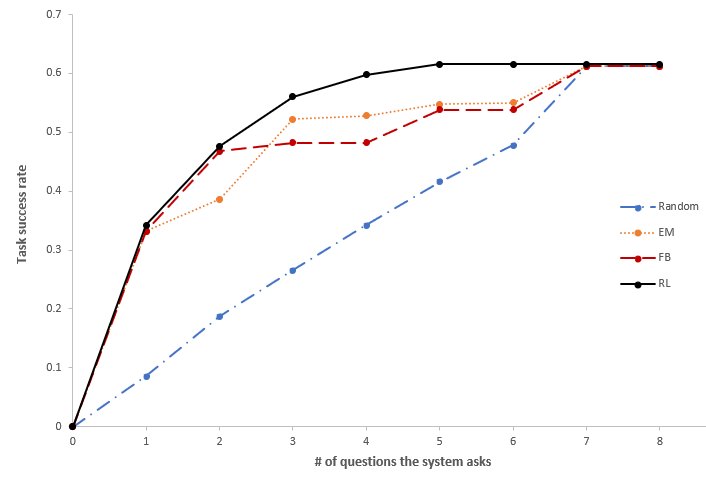}
\vspace{-1mm}
\caption{A comparison of four dialog policies of selecting a movie attribute (e.g., actor, director) to form the next clarifying question in a movie-on-demand dialog system. A random policy selects a movie attribute at random. An EM policy always asks for the value of the attribute with maximum entropy over the remaining entries in the knowledge base. A FB policy always asks for the value of the attribute that is the most frequently used in user queries among the remaining attributes. An RL (reinforcement learning) policy is learned in an online fashion using the data collected through user-system interactions.} 
\label{fig:c-kbqa-case-study-pol} 
\vspace{0mm}
\end{figure}

The performance of the system depends to a large degree upon whether the system can ask clarifying questions in an \emph{optimal} order. Let us measure the performance of a dialog system in task success rate -- the fraction of dialog sessions that successfully solves the user’s problem, e.g., finding the right movie. Figure~\ref{fig:c-kbqa-case-study-pol} compares four dialog policies of selecting a movie attribute to form the next clarifying question.

\begin{enumerate}
    \item \text{Random:} The random policy picks at random a movie attribute whose value is not provided by the user to form a clarifying question. As shown in Figure~\ref{fig:c-kbqa-case-study-pol}, as expected, the task success rate increases linearly with the number of questions being asked. 
    \item \text{EM:} The entropy-minimization (EM) policy \citep{wu2015probabilistic} suggests that the system always asks for the value of the attribute with maximum entropy over the remaining entries in the KB. This policy is proved optimal in the absence of language understanding errors. It works better than the random policy. However, it does not take into account the fact that some questions are easy for users to answer whereas others are not. For example, the system could ask users to provide the movie release ID which is unique to each movie but is often unknown to regular users.
    \item \text{FB:} The frequency-based (FB) policy suggests that the system always asks for the value of the attribute that is the most frequently used in user queries among the remaining attributes. Such information can be mined from query logs of commercial search engines. This policy is motivated by the observation that the more frequently an attribute (e.g., movie titles or actors) is used for querying movies by searchers, the more likely a user knows the value of the attribute. This is an example that a dialog policy is learned to mimic human policies (i.e., how human users query movies) using supervised learning.  In our study, the FB policy performs similarly to the EM policy, and works much better than the random policy.
    \item \text{RL:} The policy optimized using reinforcement learning (RL) is the best performer. The policy is learned in an online fashion using the data collected through user-system interactions, and thus combines the strengths of both the EM and FB policies. Optimizing dialog polices using RL is an active research topic, which will be discussed next.
\end{enumerate}

\subsection{Dialog Acts}
\label{subsec:dialog-acts}

The types of action a dialog system can perform are typically specified using dialog acts (DAs). 
A DA represents the meaning of an utterance at the level of illocutionary force~\citep{austin1975things}.  
A DA is defined as an intent which may have slots or slot-value pairs:
\[
\texttt{intent(slot$_1$=value$_1$,...,slot$_N$=value$_N$)}
\]
Commonly used intents include \texttt{inform}, \texttt{request}, \texttt{confirm} certain information, among others~\citep{stolcke2000dialogue}.
These intents are shared among dialog systems designed for different tasks. Slots, however, are task-specific as they are often defined based on the schema of the task-specific KB of the dialog system.

Consider the example in Figure~\ref{fig:c-kbqa-case-study}. Slots such as \texttt{movie}, \texttt{actor}, and \texttt{release-year}, are defined for the movie-on-demand task. 
The first system response in the example dialog  
\[
\text{``When was it released?''}
\]
is to request about a certain slot:
\[
\texttt{request(released-year),}
\]
while the second response
\[
\text{“Groundhog Day is a Bill Murray movie which came out in 1993.”}
\]
is to inform the matched movie: 
\[
\small
\texttt{inform(movie=``Groundhog Day'', actor=``Bill Murray'', released-year=``1993'').}
\]
The task of dialog policy optimizing is to learn to select the best action for each dialog state to maximize user satisfaction.   

\subsection{Reinforcement Learning for Policy Optimization}
\label{subsec:rl-policy}

A dialog policy is a function $\pi$ that maps dialog state $s$ to system action $a$.  
Supervised policy learning requires large amounts of labeled training data in the form of $(s, a)$ pairs, which are difficult to collect for complex dialog tasks, where the state-action space is large, and the environment may change over time.
For the same reasons, it is often impossible to manually specify a good policy \emph{a priori}. 
Thus, the best practice is to first build an initial policy using rules or supervised learning if training data is available, as a warm start, and then improve the policy online via interacting with (simulated) users.

A task-oriented dialog can be formulated as a decision-making process under the RL framework~\citep{young2013pomdp,gao2019neural}. The system navigates in a Markov Decision Process (MDP), interacting with its environment (e.g., users and task-specific KBs) over a sequence of discrete steps. At each step $t$, the system observes the current dialog state $s_t$, chooses an action $a$ according to a policy $\pi$, and then receives a reward $r_t$ and observes a new state $s_{t+1}$, continuing the cycle until the episode terminates. The goal of RL is to find the optimal policy to maximize expected rewards. 

Assume that the policy is implemented using a neural network parameterized by $\theta$, and $\pi(s;\theta)$ is a distribution over actions. The objective of RL is to learn $\theta$ so as to maximize the expected long-term reward it gets in a dialog of $T$ turns,

\begin{equation}
J(\theta) = \mathbbm{E} \left[\sum_{t=1}^T \gamma^{t-1} r_t | a_t \sim \pi(s_t;\theta)\right],
\label{eq:rl-loss}
\end{equation}
where $\gamma$ is a discount factor, and reward $r$ is defined according to task success rate, e.g., a successful dialog corresponds to a reward of 30, a failure to a reward of -10, and we assess a per turn penalty of -1 to encourage pithy exchanges.

Assuming that it is possible to estimate the gradient from collected dialogs, stochastic gradient ascent can be used to maximize $J$ as
\begin{equation}
\theta \leftarrow \theta + \eta \nabla_\theta J(\theta)\,, \label{eq:rl-pg}
\end{equation}
where $\eta$ is the learning rate.

One such algorithm, known as REINFORCE~\citep{williams92simple}, estimates the gradient as follows. 
Given a $T$-turn dialog generated by $\pi(\cdot;\theta)$. That is, the system action at every $t$ is sampled as $a_t \sim \pi(s_t;\theta)$ . Then, a stochastic gradient based on this single dialog is given by
\begin{equation}
\nabla_\theta J(\theta) = \sum_{t=1}^{T-1} \gamma^{t-1} \left(\nabla_\theta \log \pi(a_t|s_t;\theta) \sum_{h=t}^T \gamma^{h-t}r_h\right)\,. 
\label{eq:reinforce}
\end{equation}
REINFORCE suffers from high variance in practice, as its gradient estimate depends directly on the sum of rewards along the entire trajectory (dialog in our case) \citep{sutton18reinforcement}.  
Its variance may be reduced by using an estimated value function of the current policy, often referred to as the critic in actor-critic algorithms~\citep{sutton00policy,konda00actor}:
\begin{equation}
\nabla_\theta J(\theta) = \sum_{t=1}^{T-1} \gamma^{t-1} \left(\nabla_\theta \log \pi(a_t|s_t;\theta) \tilde{Q}(s_t,a_t,h) \right)\,, 
\label{eq:actor-critic}
\end{equation}
where $\tilde{Q}(s,a,h)$ is an estimated value function for the current policy $\pi(s;\theta)$ that is used to approximate $\sum_{h=t}^{T} \gamma^{h-t} r_h$ in Equation~\ref{eq:reinforce}.  
$\tilde{Q}(s,a,h)$ measures the average discounted long-term reward by first select $a$ and then following policy $\pi$ thereafter. The value can be learned by standard temporal difference methods \citep{sutton18reinforcement}. 
There have been many studies on the methods of computing the gradient $\nabla_\theta J$ more effectively than 
Equation~\ref{eq:actor-critic}. 
Interested readers can refer to a few related works and the references therein for further details
~\citep{kakade02natural,peters05natural,
schulman15trust,schulman15high,
mnih16asynchronous,gu17qprop,
dai18boosting,liu18action}.

RL allows a dialog system to learn how to respond in an environment which is different from the one where training data is collected. This is desirable since after we deploy a system to serve users, there is often a need over time to adapt to the changing environment (e.g., due to the update of the KB with new entity attributes). Unlike supervise learning, RL provides a solution for a dialog system to adapt without a teacher but from data collected by directly interacting with users in an initially unknown environment. 
While in supervised learning the system learns to mimic human responses by following human (teacher) examples explicitly presented in the labeled training data, in RL the system learns how to respond by exploring the state-action space and collecting reward signals by interacting with users. Thus, RL can potentially learn a policy better than human policies.

However, these advantages of RL come with a cost. Learning a good dialog policy from scratch against human users requires collecting many human-system interactions. This is not only prohibitively expensive but also often incurs real-world costs for failures. 
In what follows, we review three categories of methods proposed to reduce the cost.

\paragraph{Warm-Start Policy.} 
The RL process can be significantly sped up by restricting the policy search using expert-generated dialogs~\citep{henderson08hybrid} or teacher advice~\citep{chen17agent}.

\begin{figure}[t] 
\centering 
\includegraphics[width=0.99\linewidth]{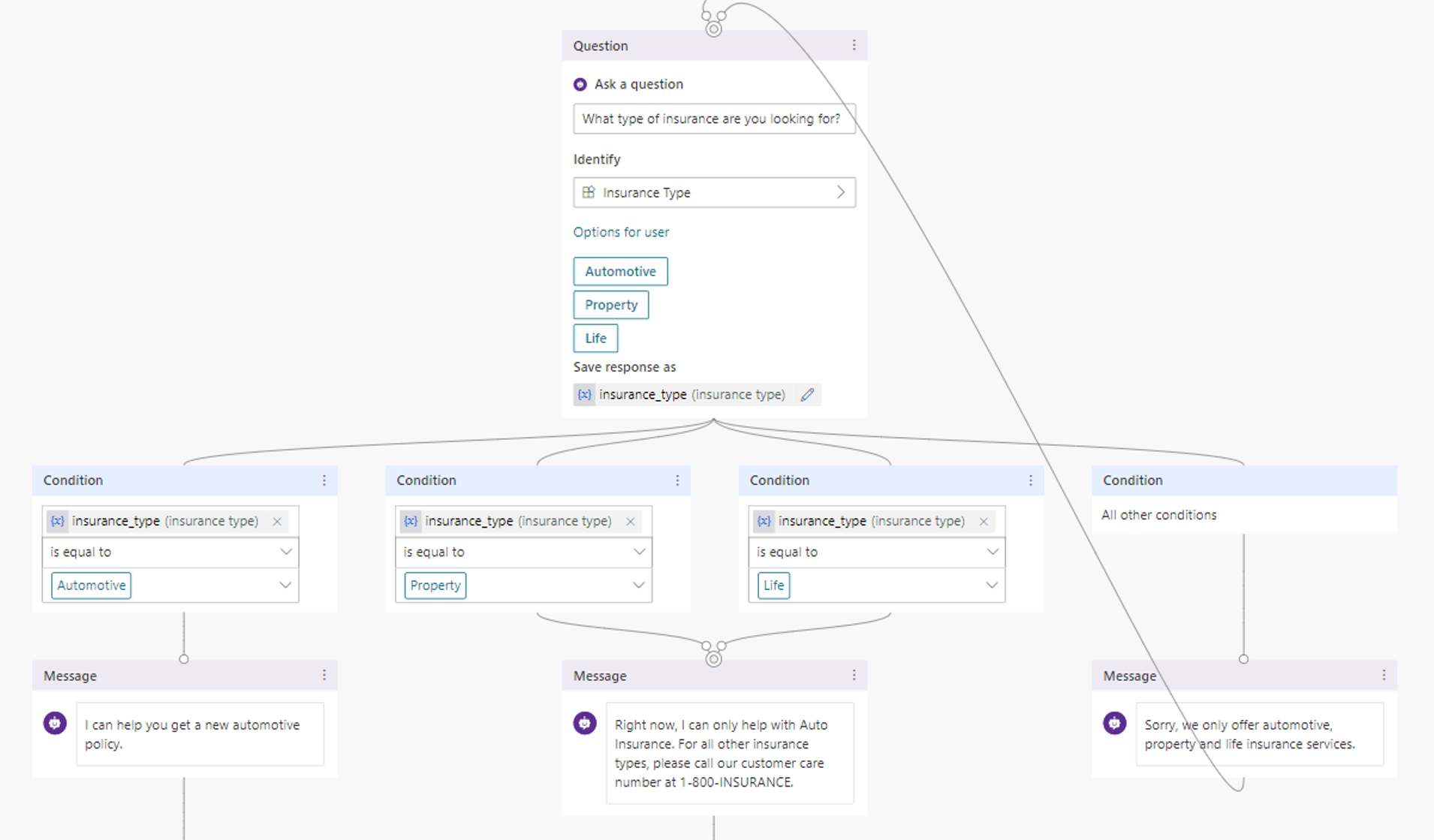}
\vspace{-1mm}
\caption{An example of a dialog flow specification using Microsoft's Power Virtual Agents.} 
\label{fig:pva} 
\vspace{0mm}
\end{figure}

Almost all commercial tools for building dialog systems,
including 
Google's Dialog Flow\footnote{{https://dialogflow.com/}},
Microsoft's Power Virtual Agents (PVA)\footnote{{https://powervirtualagents.microsoft.com/}}, 
Facebook's Wit.ai\footnote{{https://wit.ai/}}, 
Amazon's Lex\footnote{{https://aws.amazon.com/lex/}}, and 
IBM's Watson Assistant\footnote{{https://www.ibm.com/watson/}}, 
provide dialog composers that allow dialog authors (experts) to manually compose dialog policies.
For example,  Microsoft’s PVA expresses a dialog policy as a finite-state machine, with nodes representing dialog actions and arcs corresponding to states.
Figure~\ref{fig:pva} illustrates an example of a graphical dialog flow specification, where dialog authors need to explicitly specify dialog states (e.g., conditions), and for each state system actions (e.g., messages).  
However, PVA can only handle simple dialog tasks where the number of dialog states and actions is limited. The dialog flow can grow quickly to be too complex to manage as the task complexity increases or `off-track'' dialog paths have to be dealt with to improve the robustness of task bots.

An alternative approach is to use imitation learning (also known as behavioral cloning) to mimic an expert-provided policy.  A popular option is to use supervised learning to directly learn the expert's action in a state; see \cite{su16continuously,dhingra17towards,williams2017hybrid,liu17iterative} for a few recent examples. 
\citet{li14temporal} turn imitation learning into an induced reinforcement learning problem, and then applied an off-the-shelf RL algorithm to learn the expert's policy.

\paragraph{Efficient Exploration.}
Without a teacher, an RL agent learns from data collected by interacting with an initially unknown environment.  
In general, the agent has to try new actions in novel states in order to discover potentially better policies. 
Hence, it has to strike a good trade-off between exploitation (choosing good actions to maximize reward, based on information collected thus far) and exploration (choosing novel actions to discover potentially better alternatives), leading to the need for efficient exploration~\citep{sutton18reinforcement}.  In the context of dialog policy learning, the implication is that the policy learner actively tries new ways to converse with a user, in the hope of discovering a better policy in the long run. A recent survey is~\citet{gao2019neural}. 

One basic exploration strategy is known as Boltzmann exploration. 
For example, since $\pi(s;\theta)$ in Equation~\ref{eq:rl-loss} is the probability distribution over all actions, we can view it as a stochastic policy that allows the system to explore the state space without always taking the same action, thus handling the exploration-exploitation trade-off without hard coding it.   
In many problems where the environment does not change much, this simple approach is effective. 

We now describe a general-purpose exploration strategy developed particularly for dialog systems that need to adapt in a changing environment.

After a C-KBQA system is deployed to serve users, there may be a need over time to add more data types (new tables or entity attributes) to the KB. This problem, referred to as \emph{domain extension}~\citep{gasic14incremental}, makes exploration even more challenging: the system needs to explicitly quantify the uncertainty in its parameters for attributes/tables, so as to explore new ones more aggressively while avoiding exploring those that have already been learned.  \citet{lipton18bbq} approach the problem using a Bayesian-by-Backprop variant of Deep Q-Network (DQN)~\citep{mnih15human}.  
DQN learns a neural-network-based Q-function $Q(s,a;\theta)$ parameterized by $\theta$. Since Q-function measures the maximum expected reward for each $(s, a)$ pair, if the optimal Q-function is available, the optimal policy can be determined by $\pi(s;\theta) = \text{argmax}_a Q(s,a;\theta)$.   

The model proposed in \cite{lipton18bbq}, called BBQ, is identical to DQN, except that it maintains a posterior distribution $q$ over the network weights $\mathbf{w}=(w_1,\ldots,w_N)$. For computational convenience, $q$ is a multivariate Gaussian distribution with diagonal covariance, parameterized by $\theta=\{(\mu_i,\rho_i)\}_{i=1}^N$, where weight $w_i$ has a Gaussian posterior distribution, $\mathcal{N}(\mu_i, \sigma_i^2)$ and $\sigma_i = \log(1+\exp(\rho_i))$.  The posterior information leads to a natural exploration strategy, inspired by Thompson Sampling~\citep{thompson33likelihood,chapelle12empirical,russo18tutorial}.  When selecting actions, the agent simply draws a random weight $\tilde{\mathbf{w}} \sim q$, and then selects the action with the highest value output by the network.  Experiments show that BBQ explores more efficiently than state-of-the-art baselines for dialogue domain extension.

The BBQ model is updated as follows. Given observed transitions $\mathcal{H} = \{(s, a, r, s')\}$, one uses the target network 
to compute the target values $y$ for each $(s,a)$ in $\mathcal{H}$, resulting in the set $\mathcal{D} = \{(x,y)\}$, where $x=(s,a)$ and $y$ may be computed as in DQN~\citep{mnih15human}.  Then, parameter $\theta$ is updated to represent the posterior distribution of weights.  Since the exact posterior is not Gaussian any more, and thus not representable by BBQ, it is approximated as follows: $\theta$ is chosen by minimizing the \emph{variational free energy}~\citep{hinton1993keeping}, the KL-divergence between the variational approximation $q(\mathbf{w}|\theta)$
and the posterior $p(\mathbf{w}|\mathcal{D})$:
\begin{eqnarray*}
\theta^* &=& \operatorname{argmin}_{\theta} \operatorname{KL} 
[q(\mathbf{w}|\theta) || p(\mathbf{w}| \mathcal{D})] \\
&=& \operatorname{argmin}_{\theta} \Big\{ \operatorname{KL}[q(\mathbf{w}|\theta) || p(\mathbf{w})]
- \mathbf{E}_{q(\mathbf{w}|\theta)}
[\log p(\mathcal{D}|\mathbf{w})] \Big\} \,.
\label{eqn:var-free-energy}
\end{eqnarray*}
In other words, the new parameter $\theta$ is chosen so that the new Gaussian distribution is closest to the posterior measured by KL-divergence.

\paragraph{Integration of Planning and Learning.}

\begin{figure}
\centering
\includegraphics[width=0.95\linewidth]{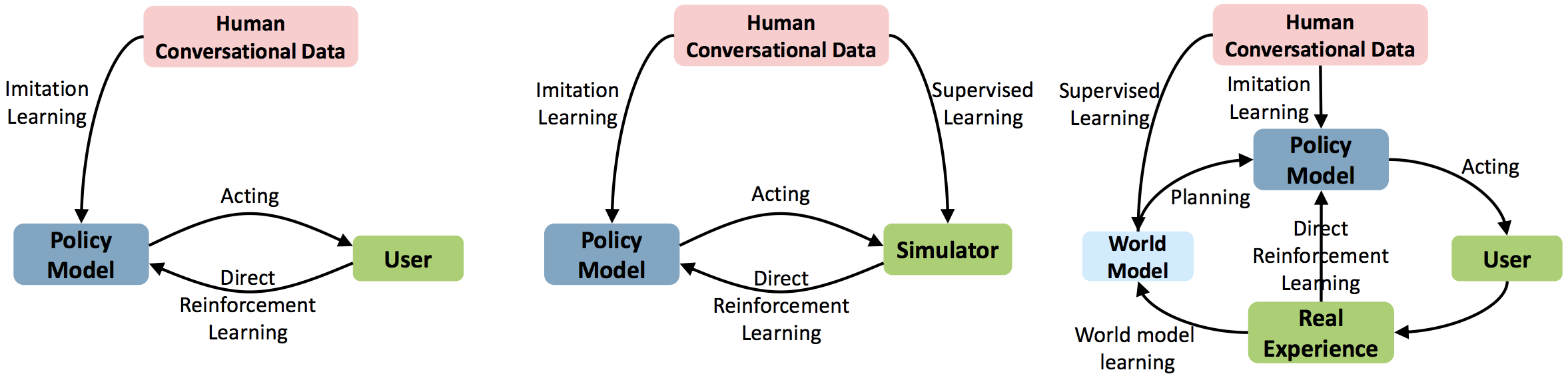}
\caption{Three strategies for optimizing dialog policies based on reinforcement learning (RL) \citep{peng2018deep} by interacting with human users (Left), user simulators (Middle), and both human users and user simulators (Right), respectively.} 
\label{fig:deep-dyna}
\end{figure}

Optimizing a dialog system against human users requires many interactions between the system and humans and often incurs real-world costs for failures (Figure \ref{fig:deep-dyna} (Left)).
User simulators provide an inexpensive alternative to RL-based policy optimization (Figure \ref{fig:deep-dyna} (Middle)). The user simulators, in theory, do not incur any real-world cost and can provide unlimited simulated experience for RL. But user simulators usually lack the conversational complexity of human interlocutors, and the trained task bot is inevitably affected by biases in the design of the simulator. \cite{dhingra17towards} demonstrates a significant discrepancy in a simulator-trained task bot when evaluated with simulators and with real users.

Inspired by the Dyna-Q framework \citep{sutton1990integrated}, \cite{peng2018deep} propose Deep Dyna-Q (DDQ) to deal with large-scale RL problems with deep learning models. As shown in Figure \ref{fig:deep-dyna} (Right), DDQ allows a bot to optimize the policy by interacting with both human users and user simulators. Training of DDQ consists of three parts:
\begin{itemize}
\item{\text{Direct reinforcement
learning}: the dialogue system interacts with a human user, collects real dialogs and improves the policy by either imitation learning or reinforcement learning;}
\item{\text{World model learning}: the world model (user simulator) is refined using real dialogs collected by direct reinforcement learning;}
\item{\text{Planning}: the dialog policy is improved against user simulators by reinforcement learning.}
\end{itemize}
Human-in-the-loop experiments show that DDQ is able to efficiently improve the dialog policy by interacting with real users, which is important for deploying dialog systems in practice \citep{peng2018deep}.

Several variants of DDQ are proposed to improve the learning efficiency. 
\cite{su2018discriminative} propose the Discriminative Deep Dyna-Q (D3Q) that is inspired by generative adversarial networks to better balance samples from human users and user simulators. Specifically, it incorporates a discriminator which is trained to differentiate experiences of user simulators from those of human users. During the planning step, a simulated experience is used for policy training only when it appears to be a human user experience according to the discriminator.
Similarly, \cite{wu2019switch} extend DDQ by integrating a \emph{switcher} that automatically determines whether to use a real or simulated experience for policy training.
\cite{zhang2019budgeted} propose a Budget-Conscious Scheduling-based Deep Dyna-Q (BCS-DDQ) to best utilize a fixed, small number of human interactions (budget) for task-oriented dialog policy learning. They extend DDQ by incorporating budget conscious scheduling to control the budget and improve DDQ’s sample efficiency by leveraging active learning and human teaching.

\section{Response Generation}
\label{sec:response-generation}


A response generator converts system actions in dialog acts (DAs) to natural language responses. 
It is an important component that affects naturalness of a conversational system, and thus users' experience.
The response should be \emph{adequate} to represent semantic DAs, and \emph{fluent} to engage users' attention.

Existing methods for response generation can be grouped into two major categories, template-based and corpus-based methods.

\paragraph{Template-Based Methods.} 
These methods require domain experts to handcraft templates for each task, and the system fills in slot-values afterward \citep{cheyer2014method,langkilde1998generation}. Thus, the produced responses are often adequate to contain the required semantic information, but not always fluent, hurting user experience.

\paragraph{Corpus-Based Models.} 
Statistical language models such as neural networks 
learn to generate fluent responses via training from DA-labeled corpora. 
One canonical model, known as Semantically Conditioned LSTM (SC-LSTM) \citep{wen2015semantically,chen2019semantically}, encodes dialog acts with one-hot representations and uses it as an extra feature to inform the sentence generation process. Despite its good performance on simple domains, it suffers from the scalability issue. SC-LSTM requires large amounts of task-specific annotated data which is not available for many tasks in real-world applications, especially for more complex tasks where the number of DAs grows exponentially with the number of slots. 
To address the issue, there is a growing interest in leveraging large-scale pre-trained models, such as GPT-2~\citep{radford2019language}, for controllable text generation~\citep{keskar2019ctrl}.

\bigskip

In what follows, we describe in detail a Semantically Conditioned GPT (SC-GPT) model that combines the strengths of GPT-2 and SC-LSTM for response generation. Our description follows closely \cite{peng2020few}.

\subsection{SC-GPT}


\begin{figure}
\centering
\includegraphics[width=0.90\linewidth]{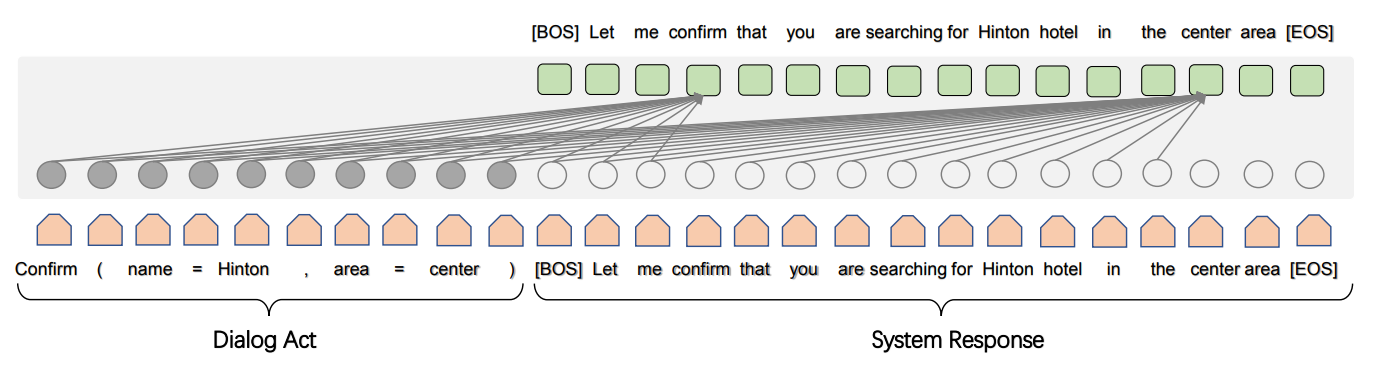}
\caption{Illustration of SC-GPT~\citep{peng2020few}. In this example, SC-GPT generates a new word token (e.g., ``confirm'' or ``center'') by attending the entire dialog act and word tokens on the left within the response. } 
\label{fig:sc-gpt}
\end{figure}

SC-GPT is an auto-regressive neural language model, which is implemented as a multi-layer Transformer, parameterized by $\theta$. 
As illustrated in Figure~\ref{fig:sc-gpt},  
SC-GPT generates natural language response $Y = \{y_1,...,y_T\}$ conditioned on dialog act $a$ as  
\begin{equation}
    P_r(Y|a ; \theta) = \prod_{t=1}^{T} P_r(y_t | y_{<t}, a ; \theta)
    \label{eq:sc-gpt}
\end{equation}
where $y_{<t}$ indicates all tokens before $t$. The model parameters are trained in the following three stages: (1) plain text pre-training, (2) DA-controlled pre-training, and (3) task-specific fine-tuning.

\paragraph{Plain Text Pre-Training.} 
Big neural language models pre-trained on large amounts of text corpora usually generalize well to new domains and tasks. 
Inspired by this, SC-GPT is initialized using the GPT-2 model, which
is pre-trained on the massive OpenWebText corpus~\citep{radford2019language}. 
Given text prompts, GPT-2 can often generate very fluent text.

\paragraph{DA-Controlled Pre-training.} 
To ensure the generated response being semantically controlled by DA, the GPT-2 model is continuously pre-trained on large amounts of annotated DA-response pairs $\mathcal{D}=\{(a_n, y_n)\}_{n=1}^{N}$, which have been developed in the research community, including the MultiWOZ dataset~\citep{budzianowski2018multiwoz} and the Frame corpus~\citep{asri2017frames}.
As illustrated in Figure~\ref{fig:sc-gpt}, each training pair is pre-processed into a sequence of tokens by concatenating $Y$ and $a$, and is fed into GPT-2. Model parameters $\theta$ are updated to maximize the log-likelihood of the conditional probability in Equation~\ref{eq:sc-gpt} over training samples: 
\begin{equation}
    \mathcal{L} (\theta) =  \sum_{n=1}^{|\mathcal{D}|} \sum_{t=1}^{T_n} \log P_r(y_{t,n} | y_{<t,n}, a_n ; \theta)
    \label{eq:sc-gpt-2}
\end{equation}

\paragraph{Fine-Tuning.} 
For a new task, its DAs often contains intents or slot-value pairs that are unseen in the training dataset collected for DA-controlled pre-training. 
SC-GPT needs to be fine-tuned to adapt to the new task. 
The fine-tuning follows the same procedure of DA-controlled pre-training, but uses only a few dozens of task-specific DA-labeled training samples.  

\cite{peng2020few} report that SC-GPT achieves new state-of-the-art on the public benchmark of dialog response generation, significantly outperforming all the other corpus-based methods including GPT-2 and SC-LSTM. 
In contrast to GPT-2 that generates fluent but not tightly DA-controlled text, SC-GPT can convert DAs into responses that not only are fluent but contain adequate intent and slot-value information specified by the DAs.
Compared to SC-LSTM, SC-GPT generalizes significantly more fluent text due to the pre-training on massive text corpora and DA-labeled datasets.

\section{Grounded Text Generation for C-KBQA}
\label{sec:gtg4cqa}

The C-KBQA systems described above use the modular architecture of Figure 
\ref{fig:c-kbqa-arch}, where each module (e.g., semantic parser, state tracker, dialog policy, response generator) needs to be developed or trained individually. 
Recently, there is a growing interest in developing unitry systems which directly generate natural language responses (e.g., answers) given a user input using a Grounded Text Generation (GTG) model.
Unlike a regular neural language model used for natural language generation (such as GPT-2), GTG is a stateful decision making model.
Specifically, GTG is a hybrid model which uses a large-scale Transformer neural network \citep{vaswani:17} as its backbone, combined with symbol-manipulation modules for knowledge base inference, to generate responses grounded in dialog state and real-world knowledge for completing tasks (e.g., C-KBQA).

This section presents a hybrid approach based on GTG to building robust C-KBQA systems at scale.
This approach is being developed contemporaneously by multiple research teams \citep[e.g.,][]{peng2020soloist,hosseini2020simple,Ham2020e2e,gao2020robust}. 
It combines the strengths of the classical modular approach (Figure~\ref{fig:c-kbqa-arch}) and the large-scale pre-trained neural language models. 
Our description follows closely \cite{peng2020soloist,gao2020robust}.

The hybrid approach aims to develop \emph{at scale} robust C-KBQA systems (or task bots) whose behaviors are \emph{interpretable} and \emph{controllable}. 
To ensure interpretability and controllability, we follow the classical modular approach to designing a task bot as a pipeline system (Figure~\ref{fig:modular-task-bot-arch}) where the output and input of each module in the pipeline is represented using symbols (such as slot-value pairs and templates) that are human comprehensible. 
To ensure scalability, the pipeline system is implemented using a GTG model which is equipped with an internal cognitive model of keeping track of dialog states, and can generate responses grounded in dialog states and a task-specific knowledge base. 
GTG can be pre-trained on open-domain datasets, and adapted to building bots for completing specific tasks with limited numbers of task labels.

\subsection{GTG for Task-Oriented Dialog}
\label{subsec:gtg4dialog}

\begin{figure}[t] 
\centering 
\includegraphics[width=0.99\linewidth]{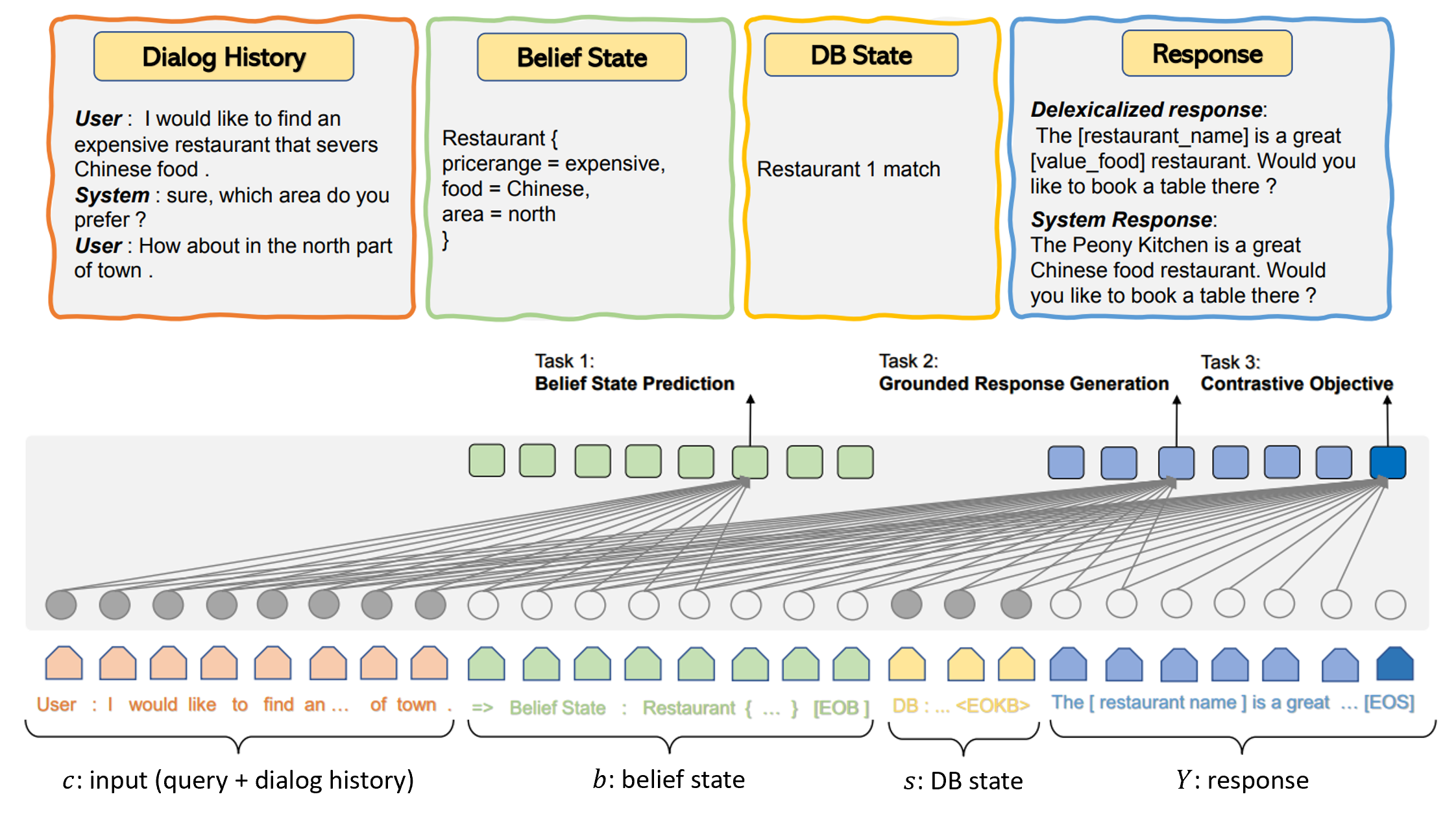}
\vspace{0mm}
\caption{GTG is an auto-regressive language model trained using multi-task learning. (Top) An example of a dialog turn. (Bottom) Multi-task learning of GTG using a dialog turn represented as a single text sequence in input and output. Adapted from \cite{peng2020soloist}.} 
\label{fig:gtg-mtl} 
\vspace{0mm}
\end{figure}

Consider a multi-turn task-oriented dialog, as illustrated in Figure~\ref{fig:gtg-mtl}. At each turn, GTG generates a natural language response $Y$ in the following steps. 

\paragraph{State tracking.} 
Let input $c$ denote an input query and its dialog history, GTG generates a dialog brief state $b$, 
\begin{equation}
b=\text{GTG}(c)        
\label{eq:dst}
\end{equation}
where $b$ is a list of tuples recording values for slots in a domain (e.g., \texttt{(restaurant, pricerange=expensive)}). 
In this step, GTG obtains the belief state directly from input, combining the functions of natural language understanding (NLU) and dialog state tracking (DST).

\paragraph{Knowledge base lookup.}
The belief state is used to query a task-specific knowledge base to retrieve the entities that satisfy the conditions specified in the belief state (e.g., the restaurants that meet the requirements on price, food type and location) as 
\begin{equation}
s=\text{GTG}(b)        
\label{eq:lookup}
\end{equation}
where $s$ is the database state, which includes the number of returned entities which is used later for response selection, and the attributes of each returned entity which are used later to lexicalize the response. In \cite{peng2020soloist}, database lookup is implemented using a keyword-matching-based lookup API. But this can also be implemented using neuro-symbolic retrieval models \citep[e.g.,][]{iyyer2017search}, as described in Section~\ref{sec:semantic-parsing}.

\paragraph{Grounded response generation.}
GTG generates a delexicalized response $Y$ grounded in dialog belief state $b$, task-specific knowledge $s$ and input $c$,
\begin{equation}
Y=\text{GTG}(c, b, s)        
\label{eq:grg}
\end{equation}
where $Y$ is delexicalized such that it contains only generic slot tokens (such as \texttt{[restaurant-name]} and \texttt{[value-food]}) rather than specific slot values. 
The final system response in natural language can be obtained by lexicalizing $Y$ using information from $b$ and $s$. 
Compared to lexicalized responses, letting GTG generate delexicalized responses makes GTG more generic and its training more sample-efficient.  

The above generation process suggests that GTG can be viewed as an auto-regressive model, which can be implemented e.g., using a multi-layer Transformer neural network, parameterized by $\theta$. 
Representing each dialog turn as $x = (c, b, s, Y)$, the joint probability $P_r(x; \theta)$ can be factorized as:
\begin{equation}
P_r(x ; \theta) = P_r(c; \theta) P_r(b|c ; \theta) P_r(s|b ; \theta) P_r(Y|c,b,s; \theta)        
\label{eq:gtg}
\end{equation}
where
$P_r(c; \theta)$ is the probability of the dialog history which can be computed recursively using Equation~\ref{eq:gtg},
$P_r(b|c; \theta)$ is the probability of belief state prediction as Equation~\ref{eq:dst},
$P_r(s|b ; \theta)=1$ since according to Equation~\ref{eq:lookup}, the database state $s$ is obtained using a deterministic database lookup process given dialog belief state $b$, and
$P_r(Y|c,b,s ; \theta)$ is the probability of grounded response generation as Equation~\ref{eq:grg}. 

\subsection{GTG Training}
\label{subsec:gtg-training}

GTG is pre-trained and fine-tuned for the completion of individual tasks in three stages.
\begin{enumerate}
    \item {\bf Language pre-training} via self-supervised learning. The GTG model learns the primary skills of understanding and generating natural language on large amounts of raw text. The pre-trained model encodes text as a sequence of symbolic tokens without grounding them in real-world concepts \citep{lucy2017distributional} and generates text by searching word co-occurrence void of meaning \citep{bisk2020experience}. Therefore, the model might generate fluent responses that are not useful to achieve any specific goals.
    \item {\bf Task-grounded pre-training} via supervised learning. GTG learns the primary skills of task completion, including tracking dialog belief states and user intents to build internal cognition models, knowledge base lookup (via semantic parsing), and deciding how to respond to complete a task. The pre-trained model grounds language words in perceptions and real-world concepts, and generates responses in terms of causality. 
    \item {\bf Task-specific fine-tuning} via supervised learning and reinforcement learning. GTG learns to adapt itself, in a fully embodied and social context, to complete specific tasks using a hybrid learning framework based on reinforcement learning and machine teaching, which is a form of supervised learning where training samples are generated by human teachers interacting with the system, as detailed in \cite{gao2020robust}, so that (1) the responses generated by GTG are grounded in task-specific rules and knowledge base that are encoded into the system via coding and machine learning, (2) the responses are optimized for task completion, and (3) the primary skills are constantly improved task by task so that the bot can adapt to new tasks more easily. 
\end{enumerate}

In what follows, we describe in detail task-grounded pre-training.
The model parameters $\theta$ are optimized on training data $\mathcal{D}=\{ x_n\}_{n=1}^{N}$ using multi-task learning \citep{peng2020soloist}, as illustrated in Figure~\ref{fig:gtg-mtl}, where a dialog turn is presented using a simple text format, with special delimiter tokens to indicate the types of different segments, (e.g., \texttt{[EOB]} indicates the end of the belief state segment). 
In addition to belief prediction and grounded response generation, whose losses are defined as
\begin{equation}
\mathcal{L}_{\text{dst}} = -\log(P_r(b|c ; \theta)) 
\label{eq:loss-dst}
\end{equation}
and
\begin{equation}
\mathcal{L}_{\text{rgr}} = - \log(P_r(Y|c,b,s ; \theta)),
\label{eq:loss-rgr}
\end{equation}
the task of contrastive learning is added to promote the matched items (positive samples $x$), while driving down mismatches (negative samples $x^{\prime}$). Specifically, a set of negative samples are sampled from $x$ by replacing some items in $x$ with probability 50\% with different items randomly sampled from the dataset $\mathcal{D}$. 
Since the the special token $\texttt{[EOS]}$ attends all tokens in the sequence, the output feature on $\texttt{[EOS]}$ is the fused representation of all items. We apply a binary classifier on top of the feature to predict whether the items of the sequence are matched $(g=1)$ or mismatched $(g=0)$:

\begin{equation}
\mathcal{L}_{\text{c}} = - g \log(P_r(x ; \theta)) - (1-g) \log(1-P_r(x^{\prime} ; \theta)). 
\label{eq:loss-cl}
\end{equation}

Thus, for a training dataset $\mathcal{D}=\{x_n\}_{n=1}^N$ consisting of $N$ dialog turns (regardless of whether they are from the same dialog sessions or not),  the full training objective for multi-task learning is
\begin{equation}
\mathcal{L}_{\theta}(\mathcal{D}) = \sum_{n=1}^{N}
(\mathcal{L}_{\text{dst}}(x_n) +
\mathcal{L}_{\text{rgr}}(x_n) +
\mathcal{L}_{\text{c}}(x_n)).
\label{eq:loss-mtl}
\end{equation}

The models proposed in \cite{Ham2020e2e,hosseini2020simple} differ from the GTG model described above (which follows \cite{peng2020soloist}) in that the response generation in \cite{Ham2020e2e,hosseini2020simple} is performed in two steps, following the classical pipeline architecture (Figure~\ref{fig:modular-task-bot-arch}) where action selection using dialog policy (POL) and natural language generation (NLG) are performed by two separate modules.
Instead of generating $Y$ directly from $(c, b, s)$ as in Equation~\ref{eq:grg}, system action $a$ is first generated based on $(c, b, s)$, and delexicalized response $Y$ is generated based on $(c, b, s, a)$ as:

\begin{equation}
a=\text{GTG}(c,b,s)        
\label{eq:gra}
\end{equation}
and
\begin{equation}
Y=\text{GTG}(c,b,s,a).        
\label{eq:nlg}
\end{equation}

\cite{peng2020soloist} argue that combining POL and NLG is critical to make the approach scalable. 
By separating POL and NLG, the system actions need to be labeled for training. 
However, system actions are task-specific labels, and are not easy to be collected in large amounts. 
The only labels needed for training GTG, on the other hand, are belief states, which are not only relatively task-independent but similar to named entity annotations that can be collected in large quantities much more easily, e.g., by transforming Wikipedia documents \citep{nothman2008transforming}.

Therefore, while the models of \cite{Ham2020e2e,hosseini2020simple} need to be trained on labeled, task-specific dialog data, one for each task, GTG can be pre-trained on open-domain datasets and then fine-tuned for specific tasks using much less task labels. For example, in \cite{peng2020soloist} the parameters of GTG model are initialized using GPT-2, and then pre-trained on large heterogeneous dialog corpora using multi-task learning according to the loss of Equation~\ref{eq:loss-mtl}.

\subsection{Remarks on Continual Learning for Conversational Systems}
\label{subsec:continual-learning}

The capability of continual learning is important for a conversational system to continuously improve its performance and adapt to a changing environment after its deployment.  
The environment could change due to various reasons. There may be a need over time to extend the domain of the target task by adding intents and slots \citep[e.g.,][]{lipton18bbq, gavsic2014incremental}, to serve a new group of users whose behaviors are different from the users based on which the system is originally designed and trained, or to update knowledge bases which might include incomplete or erroneous information.
Below, we review three primary continual learning methods in GTG: continual machine teaching, continual learning from user interactions, and updating knowledge for task completion.

\begin{figure}[t] 
\centering 
\includegraphics[width=0.99\linewidth]{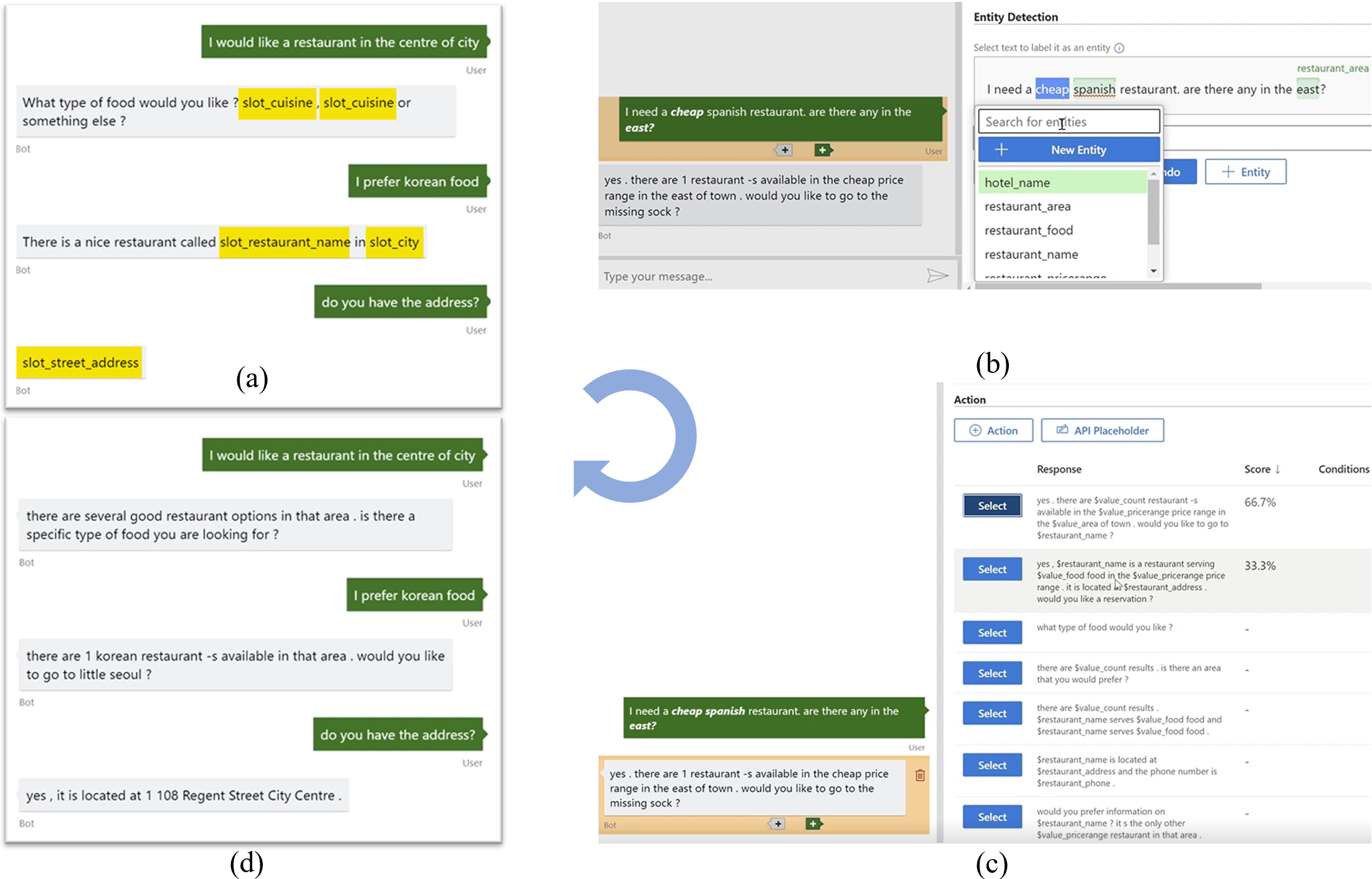}
\vspace{-1mm}
\caption{Illustration of the machine teaching process using a restaurant booking task as an example. (a) shows a conversation between a user and a pre-trained GTG. Dialog authors then incrementally review the logged dialogs, fine-tune the GTG by labeling slots (b) and correcting responses (c). (d) is the same conversation shown in (a) with the fine-tuned GTG.} 
\label{fig:cl-example} 
\vspace{0mm}
\end{figure}

\paragraph{Continual Machine Teaching.}
This is achieved by using machine teaching tools, such as conversation learning (CL) \citep{shukla2020conversation}. After the deployment of a system, CL stores user-system dialog sessions in the logs and selects a set of representative (failed) dialog sessions for dialog authors to teach the system to correct its behaviors so that the system can successfully complete the same or similar tasks next time. Figure \ref{fig:cl-example} illustrates the machine teaching process using CL. 
It is also crucial to avoid catastrophic forgetting \citep{french1999catastrophic,kirkpatrick2017overcoming}
in which the system improved using new dialog logs forgets how to deal with previously successfully handled tasks. This can be prevented by storing a collection of dialog sessions generated by previous versions of the system that are considered representative (the regression dataset). By always re-training the system using the regression dataset we can ensure that the bot learns new skills without forgetting previously learned skills.  
But how to select dialog sessions for the regression set and how to use the regression set for machine teaching are open research problems. One solution is to store a machine-learned generative model that can re-generate a regression set for system training and evaluation, instead of storing large amounts of previous dialog sessions.

\paragraph{Continual Learning from User Interactions.}
This can be achieved using reinforcement learning (RL), as described in Section~\ref{subsec:rl-policy}. However, current RL methods require a large amount of user-system interaction data, which may have too high a cost in real-world settings. 
To improve the sample efficiency in exploration, three main strategies are considered in the research community \citep{gao2019neural}. 
The first is to encourage the system to explore dialog states which are less frequently visited, and thus cause more uncertainty for the system in deciding how to respond \citep{lipton18bbq,singh2010intrinsically,gasic2010gaussian,pathak2017curiosity}. The second, known as model-based RL, is to use an environment model for efficient exploration. In task-oriented dialog, the environment model is a user simulator, which can be used alone (Figure~\ref{fig:deep-dyna} (Middle)) or 
together with human users (Figure~\ref{fig:deep-dyna} (Right)) for dialog policy learning. 
The third strategy is decomposing dialog tasks into smaller subtasks or skills that are easier to learn \citep{peng17composite,budzianowski2017sub,casanueva18feudal,tang18subgoal}. Ideally, we want to identify a set of basic skills that are common across many tasks. After learning these basic skills, a dialog system can learn to complete more complex tasks more easily via hierarchical reinforcement learning \citep{barto2003recent}.

\paragraph{Updating Knowledge.}
The knowledge base does not always contain all the information required for task completion. \cite{shen2016implicit} present a C-KBQA system with a \emph{shared memory} to store a compact version of the knowledge base. During model training, whenever the system fails to answer a question because some related information is missing in the knowledge base, the shared memory is updated to incorporate new information. Such a shared memory can be viewed as a long-term memory of the system that stores knowledge continuously learned from experiences.
\chapter{Proactive Human-Machine Conversations} 
\label{chp:proconv}

The conversational format provides a CIR system the opportunity to take a more active role when interacting with users. 
In addition to \emph{passively} responding to user queries, 
the CIR system can also \textit{proactively} lead the conversation to assist users in performing complicated information seek.

This chapter discusses methods and techniques 
that aim to equip a CIR system with the capability of proactively assisting a user to clarify her search intent by asking clarifying questions (Section~\ref{sec:clari_q}),  guiding the user to explore the topic being discussed by suggesting useful queries (Section~\ref{sec:suggest_q}), and recommending the user new topics in the hope of producing more fruitful search results (Section~\ref{sec:shift_topic}).

\section{Asking Clarifying Questions}
\label{sec:clari_q}

\begin{figure}[t]
    \centering
    \includegraphics[width=0.80\linewidth]{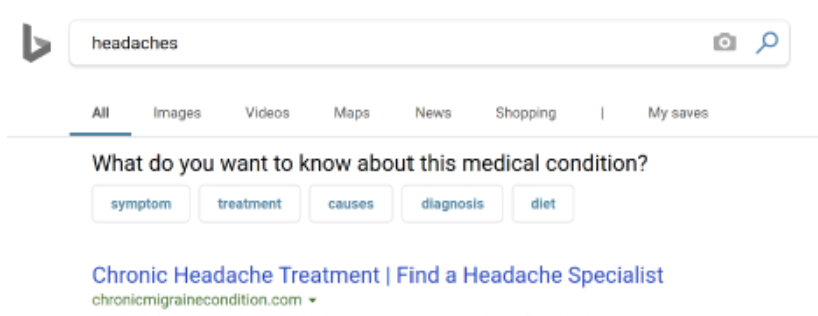}
    \caption{An example of clarifying questions in web search. Figure credit: \cite{zamani2020generating}.} \label{fig:clariQplane}
\end{figure}

Search queries are often succinct, and the underlying search intent may be ambiguous. As shown in Figure~\ref{fig:clariQplane}, a user may issue the query ``headaches'' to look for a treatment or study the symptom and causes for diagnosis.  To address this issue, commercial search engines, such as Bing and Google, diversify the search result and present retrieved documents relevant to multiple intents of the query in the SERPs. For search scenarios with \emph{limited bandwidth} interfaces, such as voice search and mobile search, asking the user a question to clarify her information need is a much more effective alternative approach. 

\citet{zamani2020generating} validate that asking clarifying questions is of significance in IR by performing a large scale user study, where a group of users are asked to use the Web search engine with an interface of using clarifying questions, as shown in Figure~\ref{fig:clariQplane}. The study concludes that ``users enjoy seeing clarifying questions, not only because of their \emph{functional} benefits, but also due to their \emph{emotional} benefits.''  In other words, a CIR system that asks clarifying questions looks more intelligent, and is easier to win users' trust.

As pointed out by \citet{aliannejadi2020convai3}, the two main research questions (RQ) on clarifying questions for a CIR system are 
\begin{itemize}
    \item RQ1: When to ask clarifying questions during conversations?
    \item RQ2: How to select or generate the clarifying questions?
\end{itemize}

RQ1 implies that a good CIR system would need to strike a balance between asking too many questions and providing irrelevant answers.
RQ2 has been studied in two different settings, where the clarifying question needs to be \emph{selected} from a pre-set question bank that contains all possible clarifying questions, or \emph{generated} on the fly. In what follows, we describe two methods, one for each setting.  

\subsection{Question Selection Methods}
\label{subsec:question-selection-methods}

The clarifying question selection setting is adopted by the ClariQ challenge at Search-oriented Conversational AI (SCAI) workshop at EMNLP 2020 \citep{aliannejadi2020convai3}. The challenge asks participants to develop CIR systems that can identify that a user question is ambiguous (RQ1), and, instead of trying to answer it directly, ask a good clarifying question (RQ2).  


The ClariQ dataset consists of 

\begin{itemize}
    \item A set of user requests (or topics) ${Q}$ in the conversational form (e.g. ``What is Fickle Creek Farm?'') with a label reflects if clarification is needed ranged from 1 to 4; 
    \item A question bank which contains possible clarifying questions collected for the user queries via crowdsourcing \citep{aliannejadi2019asking} (e.g. ``Do you want to know the location of fickle creek farm?''); and
    \item A set of user answers, one for each question (e.g. ``No, I want to find out where can I purchase fickle creek farm products.''), which are generated via crowdsourcing. The answer $A$ to a clarification question $Q'$ can be used to measure how much asking $Q'$ can help improve the search result, thus providing supervision for training question selection models.
\end{itemize}

\bigskip

The CIR system to be evaluated on the ClariQ dataset needs to perform the following two tasks: 
\begin{enumerate}
    \item \text{Unclear question identification}: Given a user request, returns a score from 1 to 4 indicating the necessity of asking clarifying questions, and 
    \item \text{Clarifying question selection}: Given a user request which needs clarification, select the most suitable clarifying question from the question bank.
\end{enumerate}

\cite{ou2020clarifying} have developed a question selection system that won the ClariQ challenge. The architecture of the system is shown in Figure~\ref{fig:context-query-understanding-example}.  It consists of the following three modules. 

\begin{figure}[t] 
\centering 
\includegraphics[width=0.99\linewidth]{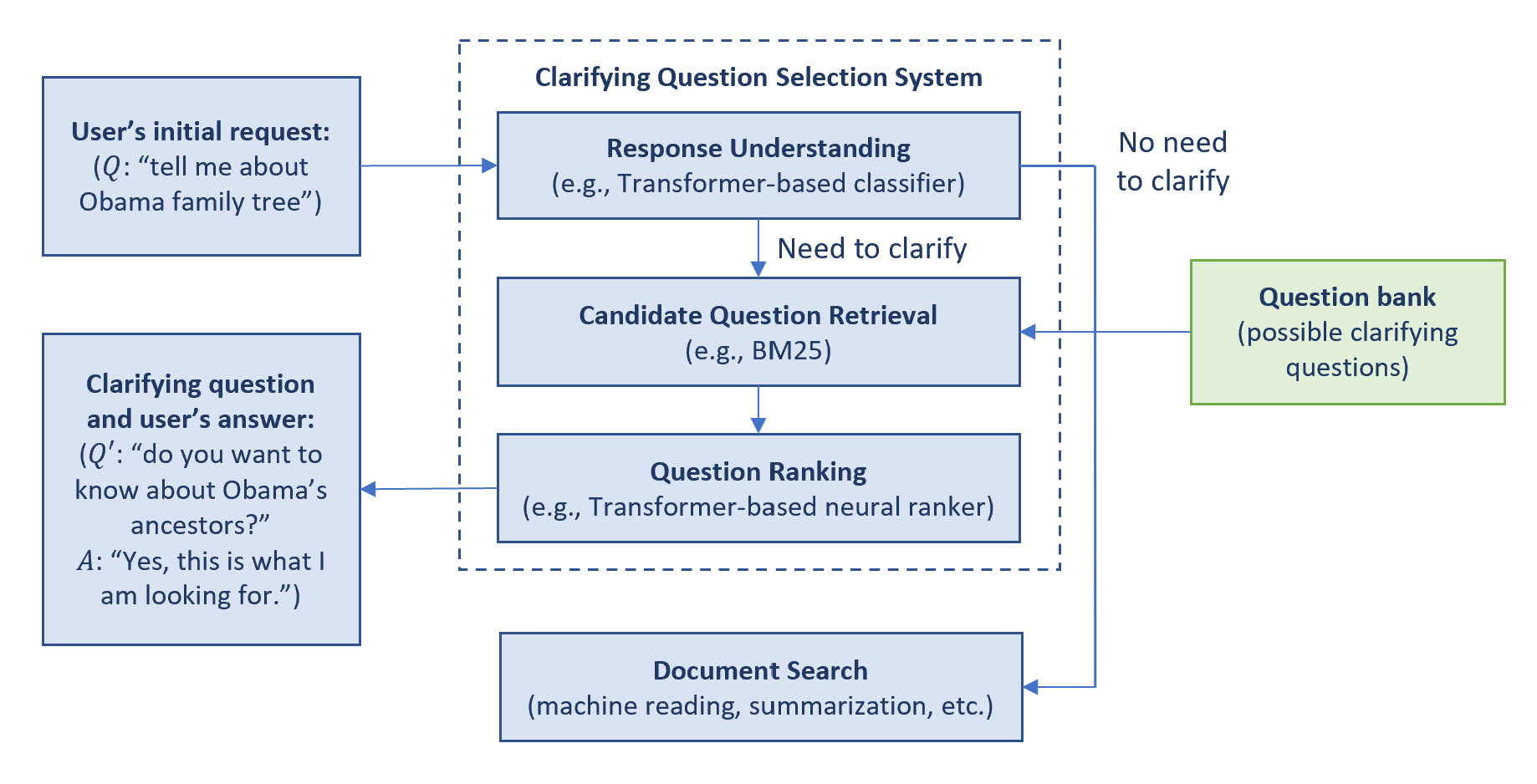}
\caption{The architecture of the clarifying question selection system of \cite{ou2020clarifying}.} 
\label{fig:question-selection-system} 
\vspace{0mm}
\end{figure}

\paragraph{Response Understanding.}
This module takes, as input, the concatenation of user request $Q$ and its dialog history $\mathcal{H}$, which consists of the last clarifying question and user answer, and determine if clarification is needed. The module is implemented using a pre-trained language model, ELECTRA \citep{clark2020electra}, which is fine-tuned on training data with clarification labels. Following the single prediction formulation of Figure~\ref{fig:plm_usage} (Left), the model predicts the clarification label $y \in \{1,0\}$ as 
$P_r(y|Q, \mathcal{H}) = \mathrm{softmax} (\mathrm{FC}(\mathbf{x}; \theta_1))$, 
where $\mathbf{x}$ is the representation of the input produced by ELECTRA, and $\mathrm{FC}(;\theta_1)$ is the task-specific head implemented as a 2-layer fully-connected feed-forward network and parameterized by $\theta_1$.

\paragraph{Candidate Question Retrieval.}
This module uses BM25 to retrieve, from the question bank, a set of candidate clarifying questions that are relevant to $Q$ and $\mathcal{H}$.

\paragraph{Question Ranking.}
This module scores all candidate questions, and returns the top-scored one. Each candidate question $Q'$ is scored by a ranking model, which is an ELECTRA model fine-tuned on supervision derived by using the collected answers to the clarifying questions in the question bank. Following the pair prediction formulation of Figure~\ref{fig:plm_usage} (Middle), the model scores $Q'$ as $\mathrm{FC}(\mathbf{x}, \theta_2)$, where $\mathbf{x}$ is the contextual representation of the concatenation of $Q$, $\mathcal{H}$, and $Q'$, produced by ELECTRA, and $\mathrm{FC}(;\theta_2)$ is the task-specific head, implemented as a 2-layer fully-connected feed-forward network, parameterized by $\theta_2$.

The question selection setting is widely used in many recent studies~\citep[e.g.,][]{rao2018learning,rao2019answer,aliannejadi2019asking}. But such a setting is only valid for task-specific dialog scenarios, such as the movie-on-demand dialog system described in Section~\ref{subsec:case-study}, where all of the possible clarifying questions can be derived from the schema of the movie-entity KB. For open-domain search tasks, however, a pre-defined question bank is often not available, and the clarifying questions need to be generated on the fly, as we will discuss next.

\subsection{Question Generation Methods}
\label{subsec:question-generation-methods}

This section describes a set of clarifying question generation models, following closely the description of \cite{zamani2020generating}.

\cite{zamani2020generating} perform a large-scale query log analysis using the data collected by the Bing search engine to identify a taxonomy of different clarification types required in open-domain IR. The taxonomy leads to a small number of question templates. 
Then, a rule-based model is proposed for generating clarifying questions. The model selects and fills out question templates using query aspects and query entity type information.
Furthermore, two neural question generation models are proposed. They are trained using supervised and reinforcement learning on the weak supervision data generated by the rule-based model.

\paragraph{Taxonomy.}
The taxonomy is derived by analyzing large amounts of query specification sessions collected from the Bing query log. Query specifications are a particular type of query reformulations that indicate the types of clarifications required for open-domain IR. For example, when the query ``trec'' is followed by a more specific query ``trec conference'', it shows whether and how the underlying search intent of ``trec'' can be clarified. 
Four types of clarifications are identified from the query specification sessions.

\begin{enumerate}
    \item Disambiguation: Some queries could refer to different concepts and a clarification question could help disambiguate the search intent. For example, ``trec'' can refer to ``Text Retrieval Conference'' or ``Texas Real Estate Commission''.
    \item Preference: Some queries are not ambiguous, but a clarification question can help identify a more precise information needs. For example, the search intent underlying the query ``sneakers'' can be clarified by asking whether the sneakers are for ``women'' or ``kids''. A user searching for ``apartment'' may be interested in ``renting'' or ``buying''.
    \item Topic: If the topic of the user's query is too broad, the system can ask for more information about the exact need of the user.
    \item Comparison: Comparing an entity with another one may help the user find the information they need. For example, the system may ask a user wanting to purchase a ``gaming console'' to compare ``xbox'' with ``play station''. 
\end{enumerate}

\paragraph{RTC: A Rule-Based Template Completion Model.}
Based on the taxonomy, five question templates are produced to cover most clarification types: 

\begin{enumerate}
    \item What do you want to know about \texttt{QUERY}?
    \item What do you want to know about this \texttt{QUERY-ENTITY-TYPE}?
    \item What \texttt{ASPECT-ENTITY-TYPE} are you looking for?
    \item Whom are you looking for?
    \item Whom are you shopping for?
\end{enumerate}

The RTC model works as follows. For each query $Q$, three types of information are extracted: (1) the word string of $Q$ (\texttt{QUERY}), (2) the entity type of $Q$ (\texttt{QUERY-ENTITY-TYPE}), and (3) the entity type for the majority aspect of $Q$ (\texttt{ASPECT-ENTITY-TYPE}).
An aspect of $Q$ is a term or phrase $Q'$ which can be concatenated with $Q$ to form a more specific query in the same query specification session. For example, given that ``trec'' is followed by ``trec conference'', we have $Q=\text{``trec''}$ and $Q' = \text{``conference''}$. When RTC extracts \texttt{ASPECT-ENTITY-TYPE}, the aspects of $Q$ are generated by a probabilistic Query Aspects Generation model (e.g., a neural or N-gram language model) trained on the logged query specifications. 


Then, a question template is selected using the following rules.
\begin{itemize}
    \item If \texttt{ASPECT-ENTITY-TYPE} is related to personal information, such as gender or age, and if \texttt{QUERY-ENTITY-TYPE} is a product (or related), Template 5 is chosen. If \texttt{QUERY-ENTITY-TYPE} is not a product, Template 4 is chosen.
    \item If no template is selected and if \texttt{ASPECT-ENTITY-TYPE} is not null, Template 3 is chosen.
    \item If still no template is selected and if \texttt{QUERY-ENTITY-TYPE} is not null, Template 2 is chosen.
    \item If still no template is selected, Template is chosen.
\end{itemize}

Despite its simplicity, RTC often generates appropriate clarifying questions.

\paragraph{QLM: Question Likelihood Maximization.}
The QLM model is a neural question generation model trained on the clarifying questions generated by the RTC model as a weak supervision data. It is expected to generalize the observed training set and perform better than RTC. 
 
\begin{figure}[t] 
\centering 
\includegraphics[width=0.90\linewidth]{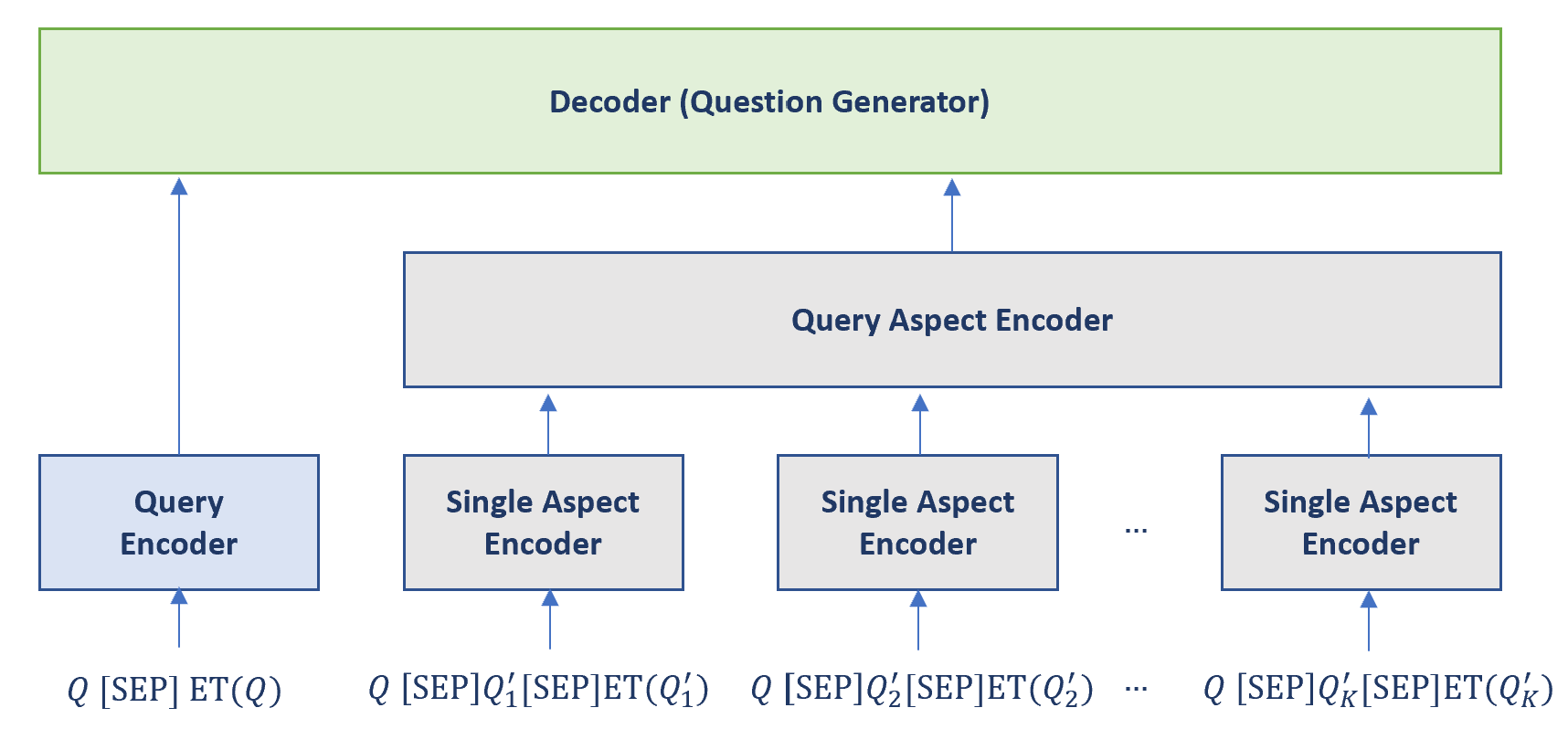}
\caption{The architecture of the QLM model for clarifying question generation \citep{zamani2020generating}.} 
\label{fig:question-generation-model} 
\vspace{0mm}
\end{figure} 
 
QLM uses a encoder-decoder architecture. 
As illustrated in Figure~\ref{fig:question-generation-model}, QLM uses a hierarchical encoder that consists of three components.

\begin{itemize}
    \item The \emph{query encoder} encodes $Q$ and its entity type $\mathrm{ET}(Q)$ (as well as dialog history $\mathcal{H}$ if available) into a $d$-dimensional dense vector.
    \item The \emph{single aspect encoder} encodes each predicted query aspect $Q'$ and its entity type $\mathrm{ET}(Q')$ into a vector.
    \item The \emph{query aspects encoder} takes the vectors of the top $K$ query aspects (sorted based on their probability computed by the Query Aspects Generation model) and computes a high-dimensional representation.
\end{itemize}

The QLM decoder takes the outputs of the query encoder and query aspect encoder and generates the clarifying question word by word. The encoder and decoder can be implemented using LSTMs or Transformers.

\paragraph{QCM: Query Clarification Maximization.} The QLM model is optimized to maximize the likelihood of generating questions observed in training data. Thus, it tends to generate common questions, which might not be effective to clarify search intent for IR. 
To address the issue, the QCM model is proposed to generate clarifying questions by maximizing a clarification utility function.
The model is obtained by continually training the QLM model to maximize the expected clarification utility \citep{rao2018learning,rao2019answer}, defined as
\begin{equation}
\mathbb{E}_{Q' \sim P_r(Q'|Q,\mathcal{H}; \theta)} [\mathbb{U}(Q',Q)]. 
\label{eq:clariq_utility}
\end{equation}
where $Q'$ is the clarifying question generated by the QCM model parameterized by $\theta$, and $\mathbb{U}(Q',Q)$ is a utility function that measures how much the clarifying question $Q'$ clarifies the information need of user when submitting the query $Q$. 
Since the utility function is often not differentiable \citep[e.g.,][]{rao2018learning,zamani2020generating}, the mixed incremental cross-entropy REINFORCE algorithm (MIXER) \citep{ranzato2015sequence} is used for training the QCM model. 

\cite{zamani2020generating} show that the QCM model can generate clarifying questions that are not only as fluent as the those in the training data but more useful for helping clarify search intents and improving search results than the questions generated by the QLM model.

\section{Suggesting Useful Questions}
\label{sec:suggest_q}

The methods described in the last section allow a CIR system to take a proactive role in conversations by asking clarifying questions to better understand a user's information needs. In this section we describe another category of methods that allows the CIR system to proactively lead users to more engaging experiences by suggesting interesting, informative, and useful follow-up questions.

Question suggestion methods have been widely used in social chatbots \citep[e.g.,][]{zhou2020design}, which are designed for building long-term emotional connections with users. These questions often make a user stay longer and more engaged in the conversation, allowing the chatbot to learn more information about the user (e.g., user preference). For example, the user study of the Sounding Board system \citep{fang2018sounding}, which won the first Alexa challenge, demonstrates that the \emph{conversation recommendation} function is crucial in maintaining user's interest in conversing with the system. Thus, the function has been widely adopted by other chatbots in later Alexa challenges \citep[e.g.,][]{yu2019gunrock}.  

The focus of this section is not on social chatbots, but on the methods of suggesting useful questions for information seeking. We start by presenting the \emph{People Also Ask} (PAA) function, a question suggestion feature provided by mainstream commercial search engines such as Google and Bing (Section~\ref{subsec:paa}), discuss the usefulness metric that goes beyond relevance and measures whether the machine-generated suggestions can further the users’ information need by bringing them to a productive next state of search (Section~\ref{subsec:usefulness-metric}), and describe two conversational question suggestion methods (Section ~\ref{subsec:question-suggestion-systems}).  Our description follows closely \cite{rosset2020leading}.

\subsection{Question Suggestions in Commercial Search Engines}
\label{subsec:paa}

As commercial search engines are getting significantly better in powering natural language based search, many users, however, still tend to issue single-turn keyword queries after years of experience with search engine's failures on supporting multi-turn natural language questions. 

To encourage users to ask more conversational questions, many commercial search engines (e.g., Google and Bing) have been presenting the \emph{People Also Ask} (PAA) feature, which aims to proactively engage users in conversation-like experiences by suggesting natural language questions. Instead of addressing a user’s current information needs, PAA suggests questions users would be interested in for the next step of their inquiry. 

\begin{figure}
    \centering
    \includegraphics[width=0.7\textwidth]{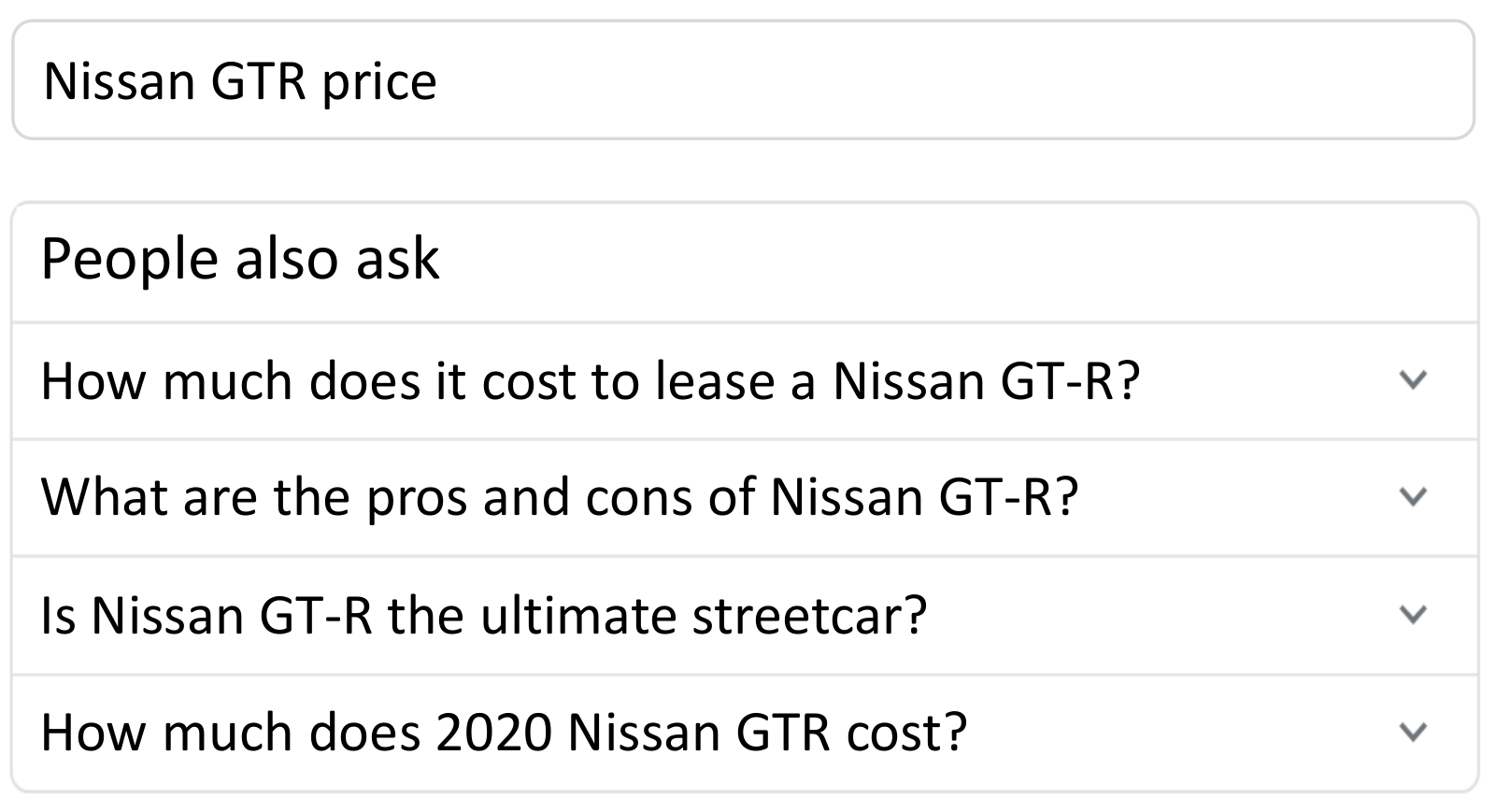}
    \caption{An example of question suggestions for user query ``Nissan GTR price'' in the People Also Ask feature of a commercial search engine. Figure credit: \cite{rosset2020leading}.}
    \label{fig:paa}
\end{figure}

Figure~\ref{fig:paa} shows an example of the PAA suggestion panel from Bing. Users can click on a suggested question to reveal its answer and the PAA panel can expand with more question suggestions based on the clicked question. 
The design of PAA is based on the assumption that the current information need is often addressed in organic results, thus one way of creating a conversational experience is to provide interesting question suggestions that a user is likely to  follow to continue the search. For example, the targeted question suggestions for ``Nissan GTR Price'' (Figure~\ref{fig:paa}) include those that help user complete a task (``leasing deals''), weigh options (``pros and cons about GTR''), explore a related topic (``ultimate streetcar''), or learn more details (``2020 GTR''). These question suggestions are \emph{proactive} in that they lead the user to an interactive search experience with related and diverse future search results.

\subsection{From Related to Useful Suggestions}
\label{subsec:usefulness-metric}

Relevance is a natural offline quality metric for conversational question suggestions. It measures whether a suggestion is on topic and related to the query. An off-topic suggestion is unlikely to provide a user a meaningful conversational search experience. However, as shown in \cite{rosset2020leading}, the offline relevance metric does not correlate well with user preferences in question suggestion observed in an online A/B test. A suggested question that is more relevant to a user query is not always preferred by the user, e.g., a suggestion that simply paraphrases the query is certainly relevant but does not provide any new information. A good suggestion should be not only relevant but also \emph{useful} in that it helps further the users’ information need by bringing them to a fruitful next state in their exploration.

\cite{rosset2020leading} propose a new \emph{usefulness} metric, which goes beyond relevance and measures whether the suggestion for a query brings real value to a user, e.g., new information she needs, the next step to complete a task, or helping her ask the right questions to explore a topic. \cite{rosset2020leading} define the guidelines of the usefulness metric, label query-suggestion suggestion paired sampled from Bing search logs, as exemplified in Table~\ref{tab:usefulness_eg}, and report five failure modes as categories of not useful which help diagnose the common challenges of CIR systems.

\begin{table}[t]
    \caption{Examples of (query, question suggestion) pairs in different usefulness/non-usefulness categorizes~\citep{rosset2020leading}. Labels are provided by human annotators.}
    \label{tab:usefulness_eg}
    \centering
    \resizebox{1\textwidth}{!}{
    \begin{tabular}{l|l|l}
    \hline
    \textbf{Query}     &  \textbf{Question Suggestion} & \textbf{Usefulness Label} \\ \hline
     used washer and dry                &  Can I store a washer and dryer in the garage ? & Misses Intent       \\  
 best questions to ask interviewer  &  What should I ask in an interview ?            & Dup. w/ Q       \\  
 medicaid expansion                 &  Did Florida accept Medicaid expansion ?        & Too Specific        \\  
 verizon yahoo purchase             &  Who bought out Yahoo ?                         & Prequel             \\ 
 jaundice in newborns               &  How to tell if your newborn has jaundice ?     & Dup. w/ Ans.        \\ 
 jonestown massacre                 &  What was in the Kool-Aid at Jonestown ?        & Useful              \\ 
 affirmative action                 &  Who does affirmative action benefit ?          & Useful              \\ 
 best hair clippers                 &  What clippers do barbers use ?                 & Useful \\     
\hline         
    \end{tabular}
    }

\end{table}

\begin{itemize}
    \item \text{Misses Intent:} This category includes suggestions which are completely off-topic, poorly formatted, or not forming cohesive natural language questions. 
    \item \text{Too Specific:} This includes suggestions that only apply to a narrow portion of portion users, instead of the common public.
    \item \text{Prequel:} Prequel suggestions are those about information a user likely knew when issuing the query, thus fail to provide new value to the user.
    \item \text{Duplicate with Query:} Suggestions that paraphrase the query are duplicates and do not provide useful new information.
    \item \text{Duplicate with Answer:} Suggestions whose information is already covered by the answer or other documents in the search results page.
    \item \text{Useful:} Useful suggestions lead to valuable new information. There are multiple ways a suggestion is useful for a user. For example, it can help the user complete a task she has in mind e.g., by suggesting information about the next step in her task. It can also help the user re-frame her information seeking task in a different perspective, such as the suggestion to ``best hair clippers'' in Table~\ref{tab:useful_suggestion_eg}, which converts the task of purchasing the best clippers to a more practical task of finding which brands are popular among professionals.
    In general, a question is considered useful if it adds value about a topic that follows a coherent line of thought to proactively lead a human-machine conversation.
\end{itemize}

\subsection{Question Suggestion Systems}
\label{subsec:question-suggestion-systems}

This section describes two question suggestion systems for PAA \citep{rosset2020leading}, using a BERT-based ranker and a GPT-2 based generator, respectively. Both systems are trained with weak supervision signals that convey past users' search behaviors in conversational search sessions. 

\paragraph{Suggestion Selection Using a BERT-Based Ranker.} The suggestion selection systems use an architecture similar to that of a clarifying question selection systems of Figure~\ref{fig:question-selection-system}, where the question bank is replaced with a \emph{suggestion bank} that contains possible suggestions collected from search session logs, as to be described later. The three components (i.e., response understanding, candidate suggestion retrieval, and suggestion ranking) are implemented similarly to their counterparts in a clarifying question selection system.

Take the suggestion ranking module as an example. A BERT-based ranker is used \citep{devlin2018bert}. Specifically, it follows the pair prediction formulation (Figure~\ref{fig:plm_usage} (Middle)). BERT is fine-tuned to score a pair of query $Q$ and its candidate suggestion $S$ by feeding the concatenation of $Q$ and $S$ to the model, and predicting the usefulness label $y \in {1,0}$ using a logistic regression layer with softmax:

\begin{equation}
P_r(y|Q,S) = \mathrm{softmax}(\mathbf{W} \cdot \mathbf{x}), 
\label{eq:suggest_q_ranker}
\end{equation}
where $\mathbf{x}$ is the contextual representation of the input pair $(Q, S)$, and $\mathbf{W}$ the projection matrix.     

The ranker can be trained to minimize the rank loss of Equation~\ref{eq:rank_loss_2} on the training set that consists of $(Q, S^+, S^-)$ pairs, where $S^+$ is a more useful suggestion (positive suggestion) of query $Q$ than $S^-$ (negative suggestion).

\paragraph{Suggestion Generation Using a Fine-Tuned GPT-2 Model.} Compared to suggestion selection, a suggestion generation system can achieve higher coverage since no pre-existing suggestion candidates are required. \cite{rosset2020leading} propose to use a fine-tuned GPT-2 model to generate suggestion $S$ word-by-word given $Q$, and its history $\mathcal{H}$ (if available),

\begin{equation}
S = \text{GPT-2}(Q, \mathcal{H}; \theta). 
\label{eq:nlg_suggest_q_ranker}
\end{equation}
where model parameters $\theta$ can be fine-tuned on query-suggestion pairs $(Q, S)$. Note that unlike the training data for the ranker, only positive suggestions are used. 

\paragraph{Mining Conversational Search Sessions.} Web search logs include a large amount of query sequences, known as \emph{search sessions}, from users. They capture user search trajectories when completing a task, learning a concept, or exploring a new topic. Search sessions, however, are noisy in that users are often multi-tasking and switching between different information needs so that the queries in a session may not even be related to each other. 

\cite{rosset2020leading} have developed a search log mining method of extracting clean and topic-coherent conversation-like search sessions that include \emph{next-turn} information provided by users. The mined sessions can be used to not only train the two question suggestion systems described above, but curate the suggestion bank of the suggestion selection system. The search log mining method includes the three steps.

\textit{Clean.} The first step is to discard noisy queries. The raw sessions grouped by the 30 minutes gap rule are collected. Then, navigational or mal-intent queries are discarded using standard query classification techniques. Sessions with at least three queries and one satisfied user click are kept.

\begin{figure}
    \centering
    \includegraphics[width=0.8\textwidth]{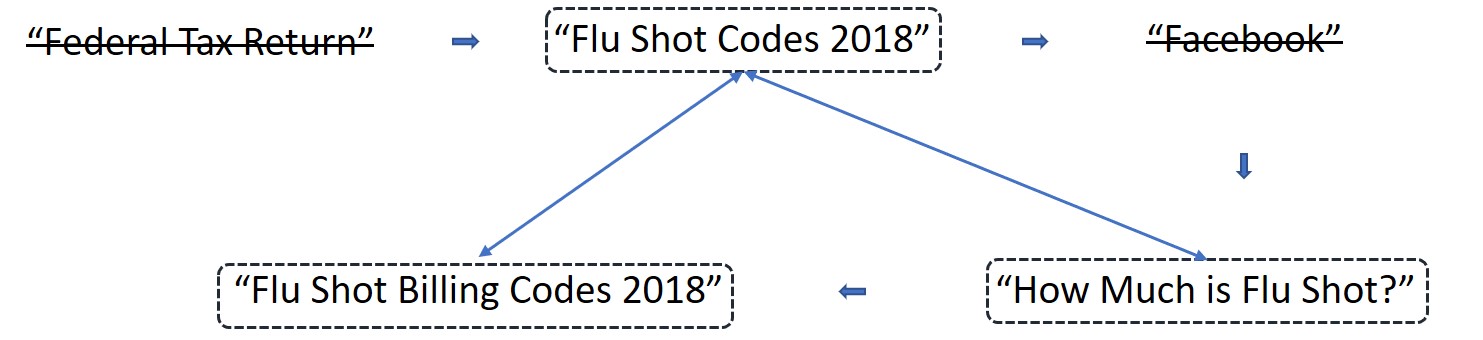}
    \caption{An example of the embedding-based filtering in a five-query session. A graph is created by connecting queries with an edge if the similarity between their embeddings, generated by GenEncoder~\citep{zhang2019generic}, is greater than 0.4 (e.g., ``Flu Shot Codes 2018'' and ``How Much is Flu Shot?''). Queries in the largest connected sub-graph in the session are kept. }
    \label{fig:gen_filter}
\end{figure}

\textit{Coherent.} The second step is to ensure the queries in a sessions have a coherent search intent. The  Gen Encoder, a query embedding model~\citep{zhang2019generic}, is used to map queries with similar search intent nearby in the embedding space. As illustrated in Figure~\ref{fig:gen_filter}, after obtaining the embedding of each query in a session, a graph is constructed by adding edges between query pairs if the cosine similarity between their embeddings is greater than 0.4. Then the largest connected sub-graph in the session is retained and the rest queries are discarded. The sessions with more than three queries retained are kept. This ensures that the intents between queries in a session do not drift too drastically. 

\textit{Information-Seeking.} To focus on information-seeking intents, an QA intent classifier is used to identify the search sessions that are information-seeking. If a query is classified to be satisfied by an extracted natural language answer, then the query, as well as the session that contains it, has an information-seeking intent rather than a navigational, transitional, or functional one. 

\paragraph{Training Tasks.}
The question suggestion systems are trained using multi-task learning on four tasks. The first one is based on the mined sessions described above. The other three are PAA tasks with one using relevance labels and the other two gathered from online user feedback in PAA. 

\textit{Next Query Prediction (NQP).} 
This is a standard query suggestion task~\citep{sordoni2015hierarchical,wu2018query}. Specifically,  for a mined session $\mathcal{T} = \{ Q_1,..., Q_{N-1}, Q_{N}\}$, the NQP task is to predict the last query $Q_N$, as a positive suggestion $S$, give the previous queries $Q_{<N}$. 
The GPT-2 model can be fine-tuned directly with the positive pairs formed by for pairing $Q_N$ (or $S$) and its previous queries in the same session. Training the BERT-based ranker requires negative suggestions, which can be sampled from the $K$ most frequent queries that appear after $Q_{N-1}$ in the search log~\citep{sordoni2015hierarchical}.

\textit{Relevance Classification.}
This is a standard binary classification task. The training data consists of human judgments on 600K $(Q,S)$ pairs, with label $y = 1$ meaning the question $S$ is relevant (positive) to the query $Q$ and $y = 0$ otherwise.

\textit{PAA Click Prediction.}
This is a standard click prediction task using user clicks as the relevance feedback labels. User click signals are collected from a random sample of the search log, with each impression of PAA as a training instance: i.e. $y = 1$ if the suggestion $S$ is clicked (positive) and $y = 0$ otherwise. 

\textit{Relative-CTR Prediction.}
This is a classification task using data collected from user clicks on question suggestions in PAA. Specifically, as multiple suggestions are displayed for the same $Q$ multiple times, the suggestions that have significantly higher \emph{click through rates} (CTR) for $Q$ are considered positive, and the rest negative.

\bigskip

During the training of question suggestion models, batches from each training task are randomly interleaved. \cite{rosset2020leading} preform a comprehensive study of using different combinations of these training tasks for system development, and conclude that the NQP task with mined conversation-alike sessions is the most effective among the four tasks. Adding the NQP task for training significantly reduces the amount of negative suggestions (i.e., suggestions that are either \emph{too specific} or \emph{prequel}, as described in Section~\ref{subsec:usefulness-metric}), produced by the resultant systems. This is expected as these negative suggestions rarely appear in the mined conversational search sessions. An online A/B test also validates the effectiveness of NQP. Table~\ref{tab:useful_suggestion_eg} shows some example suggestions produced a question suggestion selection system before and after the system is trained using the NQP task.

\begin{table}[t]
\caption{Example question suggestions selected by the BERT-based ranker before and after the ranker is trained using weak supervisions from mined conversation-alike sessions.
\label{tab:useful_suggestion_eg}}
\small
\centering
\begin{tabular}{l|ll}
\hline
\textbf{Query} & \textbf{Question Suggestion} & \textbf{Usefulness Label} \\ \hline

\multirow{10}{*}{bitcoin price} & {\texttt{Before}} & \\ 
 & what is the value of bitcoins? & [Dup w/ Query]      
  \\ 
  & is it time to buy bitcoin? & [Useful] \\
  & what was the lowest price of bitcoin? & [Useful]  
   \\ 
   & what is the value of 1 bitcoin? & [Dup w/ Query]        \\ \cline{2-3}

 & {\texttt{After}} &\\
 
& how much does 1 bitcoin cost to buy? & [Dup w/ Query]  
\\
& how to buy bitcoins at walmart? & [Too Specific]  
\\
& what will be the price of bitcoin in 2020 & [Useful]    
\\
& what is the cheapest way to buy bitcoin & [Useful]  \\ \hline
 
\multirow{10}{*}{direct deposit form}  &  {\texttt{Before}}& \\ 
 & how to do a direct deposit? & [Useful]      
 \\
 & what is direct deposit bank of america? & [Too Specific]  \\
  
& what is a direct deposit? & [Prequel] 
 \\
 & how to set up bank of america direct deposit? & [Too Specific]      \\ \cline{2-3}

  & {\texttt{After}}&  \\
 & what do i need for direct deposit? & [Useful]  
\\ 
& how to get a chase direct deposit form? & [Too Specific]    \\
& how to start direct deposit? & [Prequel]  \\ 
& how to fill out a direct deposit form? & [Useful] \\
\hline
             
\end{tabular}
\end{table}

\section{Shifting Topics}
\label{sec:shift_topic}

Changing topics is a common behavior in human conversations~\citep{brown1983discourse}. Recently, there is an increasing interest in building conversational systems that can proactively lead human-machine conversations via topic shifting \citep[e.g.,][]{wu2019proactive,tang2019target,TIAGE,zhou2020design}. Different from the methods of asking clarifying questions (Section~\ref{sec:clari_q}) and suggesting useful questions (Section~\ref{sec:suggest_q}) which aim to follow, clarify, or further a user's original search topic to get more precious and fruitful results, a topic shifting module might suggest a new topic that changes the user's search intent.

Topic shifting has been widely adopted by social chatbots~\citep[e.g.,][]{zhou2020design,fang2018sounding,TIAGE} to maintain engaging conversations with users. For goal-oriented dialog systems such as conversational recommender, proactively switching topics can help guide human-machine conversations to a pre-defined target~\citep[e.g.,][]{tang2019target}.

This section describes how topic shifting is implemented for open-domain social chatbots (Section~\ref{subsec:topic-shift-social-chatbot}) and target-guided conversation systems (Section~\ref{subsec:target-guided-topic-shift}).   Although there is little research on topic shifting for CIR, the works described here provide useful lessons and lay a good foundation for the development of future CIR systems. 

\subsection{Topic Shifting in Social Chatbots}
\label{subsec:topic-shift-social-chatbot}

\begin{figure}
    \centering
    \includegraphics[width=0.6\textwidth]{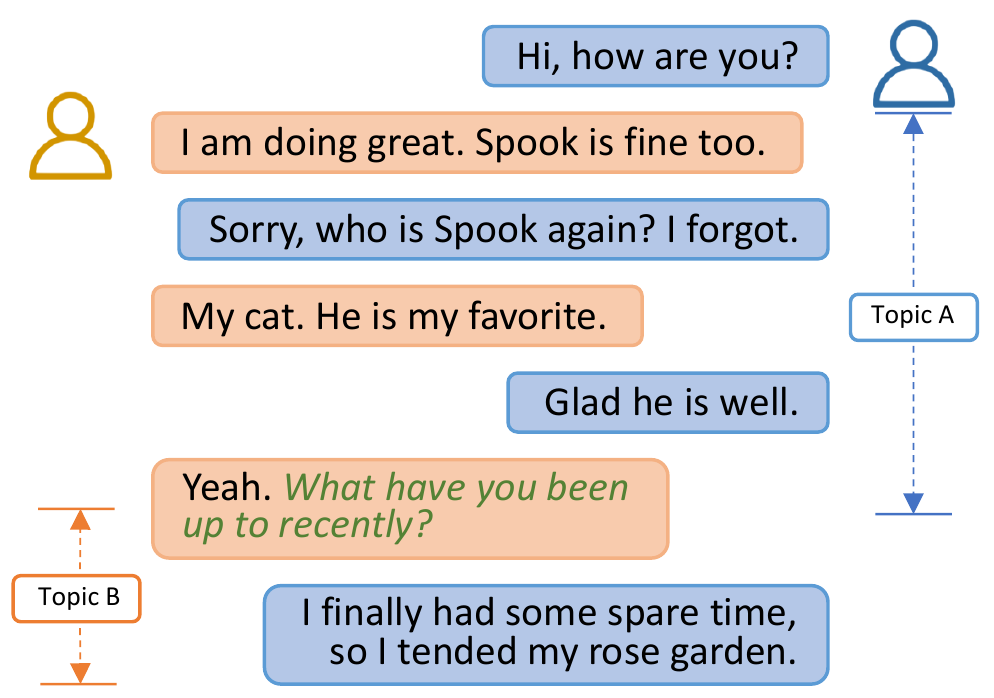}
    \caption{An example dialog with topic shift in human conversations. The topic is shifted in the second to last turn, highlighted in \textcolor{ForestGreen}{\textit{green}}. Figure credit: \cite{TIAGE}.}
    \label{fig:topic_shift_eg}
\end{figure}

As shown in Figure~\ref{fig:topic_shift_eg}, humans commonly shift topics during conversations. Therefore, fluent topic shifts are crucial for social chatbots to be able to mimic human conversational patterns.

\paragraph{The Topic Manager in XiaoIce.} The Topic Manager in XiaoIce, one the most popular social chatbots, is designed to simulate human behavior of changing topics during conversations~\citep{zhou2020design}. It consists of (1) a \emph{topic shift detector} for deciding at each dialogue turn whether or not to shift topics, and 
(2) a \emph{topic recommender} for suggesting a new topic.

Topic shifting is triggered if XiaoIce does not have sufficient knowledge about the topic to engage in a meaningful conversation, or the user is getting bored. Specifically, the detector monitors a set of pre-defined indicative signals to determine whether to it is necessary to change the topic to keep a user engaged. Some examples of these signals are

\begin{itemize}
    \item whether an editorial response is used due to the bot failing to generate any valid response candidate; 
    \item whether the bot-generated response simply repeats the user inputs, or contains no new information; and
    \item whether the user inputs are getting bland, for example, ``OK'', ``I see'', ``go on''.
\end{itemize}

The topic recommender consists of a topic ranker, and a topic database that is constructed by collecting popular topics and related comments and discussions from high-quality Internet forums, such as Instagram in the United States and douban.com in China. The topic database is updated periodically. When topic shifting is triggered, a list of topic candidates is retrieved from the database using the on-going dialog as query. The top-ranked candidate topic is chosen as the new topic. The topic ranker is implemented as a boosted tree \citep{wu2010adapting} that uses the following features:

\begin{itemize}
    \item Contextual relevance: The topic needs to be related to the dialogue, but has not been discussed yet.
    \item Freshness: The topic, especially if it is related to news, needs to be fresh and valid for the time being.
    \item Personal interests: The user is likely to be interested in the topic, according to the user profile.
    \item Popularity: The topic has gained enough attention on the Internet or among XiaoIce users.
    \item Acceptance rate: The rate of the topic being accepted by XiaoIce users is historically high.
\end{itemize}

\begin{figure}
    \centering
    \includegraphics[width=0.99\textwidth]{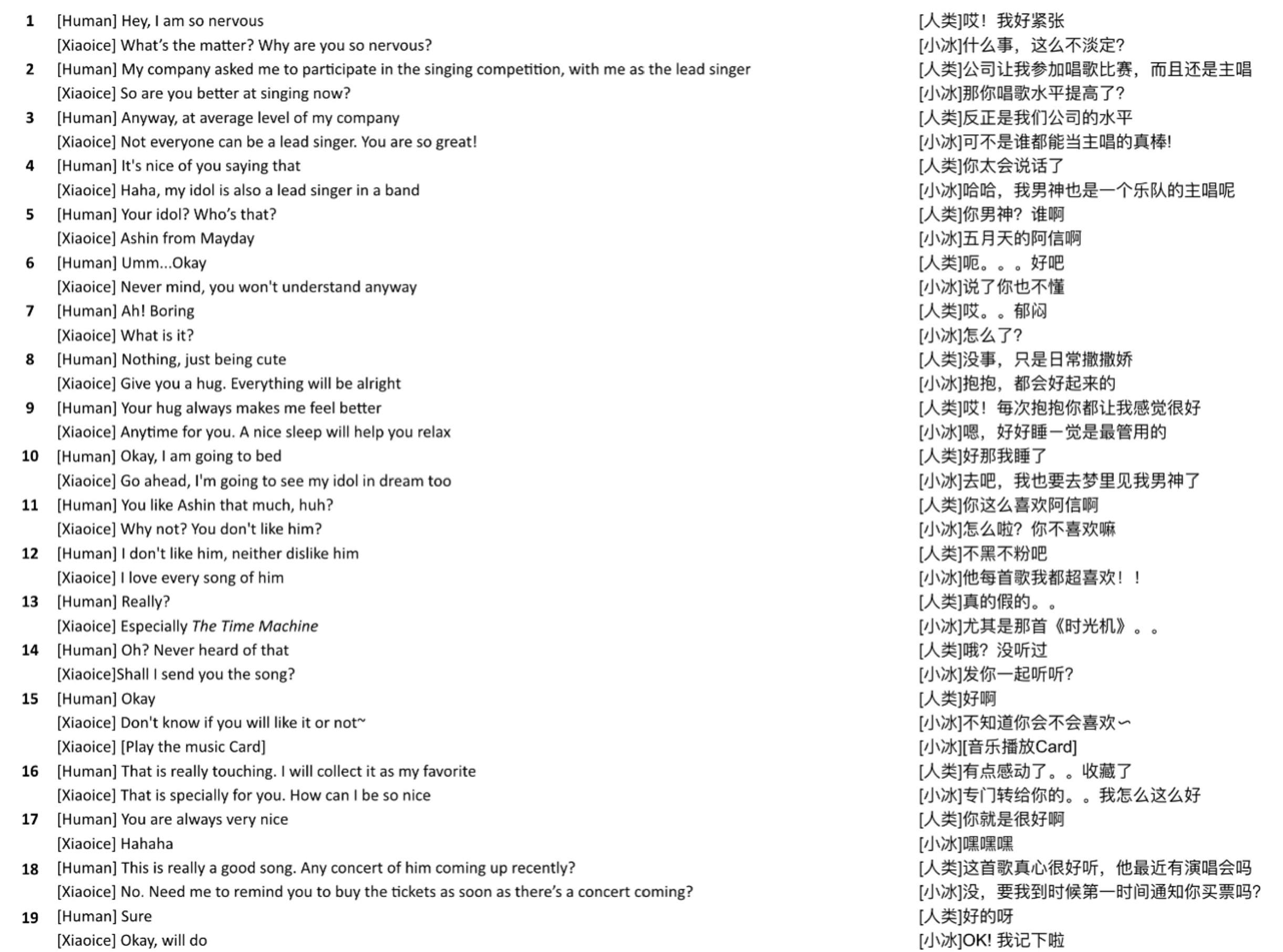}
    \caption{A multi-topic conversation between a user and XiaoIce in Chinese (right) and English translation (left). XiaoIce starts with a casual chat in Turn 1, switches to a new topic on music in Turn 4, recommends a song in Turn 15, and helps book a concert ticket in Turn 18. Figure credit: \cite{zhou2020design}.}
    \label{fig:xiaoice-topic-manager}
\end{figure}

As shown in the example in Figure~\ref{fig:xiaoice-topic-manager}, XiaoIce shifts to a new topic (i.e., a song titled ``The Time Machine'' by the singer named Ashin in Turn 13) when it detects that the user is not familiar with ``Ashin'' and about to terminate the conversation by responding ``Ah! Boring'' and ``Okay, I am going to bed.'' 
An A/B test shows that incorporating topic shifting significantly improves user engagement.

\paragraph{TIAGE: A Benchmark for Topic Shifting.} To facilitate research on topic shift dialog modeling, \cite{TIAGE} have developed a Topic-shIft Aware dialoG datasEt (TIAGE) by augmenting the PersonaChat dataset~\citep{zhang2018personalizing} with topic-shift annotations, which indicate at each turn in a dialog session whether there is a topic shift.  It turns out that, on average, topic shifting happens every 4-5 turns in the dataset. The statistics of TIAGE are presented in Table~\ref{tab:tiagostat}. 

\begin{table}[t]
\small
\centering
\caption{The statistics of the TIAGE~\citep{TIAGE} dataset. All data are derived from PersonaChat~\citep{zhang2018personalizing}. An instance in the dataset is a (context, response) pair. \textsc{Supe}rvised parts are those instances with binary topic shift annotations. The rest are used as \textsc{Weak} \textsc{Supe}rvised data. On average each dialog has 3.5 turns labeled with topic shift.}
\label{tab:tiagostat}
    \begin{tabular}{llllll}
    \hline
        & \textbf{\textsc{WeakSupe}$_{train}$} & \textbf{\textsc{WeakSupe}$_{dev}$} & \textbf{\textsc{Supe}$_{train}$} & \textbf{\textsc{Supe}$_{dev}$}  &  \textbf{\textsc{Supe}$_{test}$} 
        \\ \hline
        \#Dialogs & 7,939 & 1,000 & 300 & 100 & 100 \\
        \#Instances & 108,711 & 13,788  & 4,767 & 1,546 & 1,548 \\ 
        \#AvgTurns & 14.7 & 14.8 & 15.6 & 15.5 & 15.6  \\\hline
    \end{tabular}

\end{table}

The dataset can be used to train and evaluate topic shift modules. \cite{TIAGE} report that while it is relatively easy to build a high-accurate topic shift detector by fine-tuning a pre-trained language model (e.g., T5~\citep{raffel2019exploring}), it is challenging to train a good natural language generation model for topic recommendation. Note that the topic recommender in XiaoIce selects a topic from a topic dataset, rather than generates a topic.

The topic shifting methods for social chatbots aim to keep users engaged in the conversations on whatever topics that serve the purpose, but are not intended to guide the conversation to any pre-set target.  
CIR, however, is performed by goal-directed conversations. Depending on different categories of information seek tasks, the target topic of a conversation can be quite specific (e.g., in information lookup tasks), or relatively vague (e.g., in information exploration tasks). Thus, a CIR system is expected to predict what topic a user might be interested in based on user profile and conversation history, and intentionally guide the conversation to the target.  This is the topic we will discuss next, following closely the discussion in~\cite{tang2019target}.

\subsection{Target-Guided Topic Shifting}
\label{subsec:target-guided-topic-shift}

\begin{figure}
    \centering
    \includegraphics[width=0.8\textwidth]{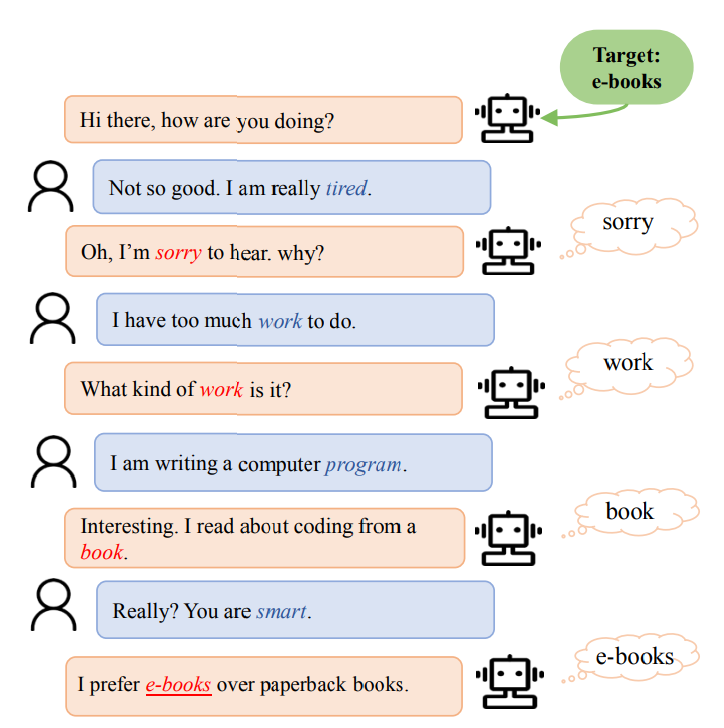}
    \caption{A target-guided conversation. given a target ``e-books'' and an arbitrary starting topic such as ``tired'', the system drives the conversation following a \emph{topic path}, highlighted in \textcolor{red}{red}, and reaches the target in the end.
    Figure credit: \cite{tang2019target}.}
    \label{fig:target-guided-conversation-example}
\end{figure}

\cite{tang2019target} study topic shifting for a target-guided conversation scenario where the system chats naturally with humans on open-domain topics and proactively guides the conversation to a designated target subject. As illustrated in Figure~\ref{fig:target-guided-conversation-example}, given a target ``e-books'' and an arbitrary starting topic such as ``tired'', the system drives the conversation following a \emph{topic path} and effectively reaches the target in the end. Such a target-guided conversation setup is general-purpose and can entail a wide range of goal-directed dialog applications, including conversational exploratory search. 

Specifically, the target-guided topic shifting task is defined as follows. Given the current user input $Q_N$, and dialog history $\mathcal{H}_N$ which contains previous user inputs $Q_{<N}$ and system responses $A_{<N}$, the system needs to recommend a new topic as a response $A_N$, aiming to satisfy (1) \emph{transition smoothness} by ensuring $A_N$ is fluent and plausible given the conversational context, as an abrupt change of topic often hurts user experience, and (2) \emph{target achievement} by leading the conversation to the designated target.

\begin{table}[t]
\small
\centering
\caption{The statistics of the target-guided conversation dataset~\citep{tang2019target}. The keywords are used as explicit topics in each conversation turn. The last row is the average
number of keywords in each dialog turn. The vocabulary size is 19K.}
\label{tab:goal_conv_stat}
\begin{tabular}{llll} 
\hline
& {\bf Train} & {\bf Val} & {\bf Test} \\ \hline
  \#Conversations & 8,939 & 500 & 500\\
  \#Utterances & 101,935 & 5,602 & 5,317\\
  \#Keyword Types & 2,678 & 2,080 & 1,571\\
  \#Avg. Keyword per Turn & 2.1 & 2,1 & 1.9 \\
\hline
\end{tabular}
\end{table}

\paragraph{A Keyword-Based Target-Guided Conversation System.} \cite{tang2019target} assume that the topic or user intent can be explicitly modeled by keywords, and have developed a target-guided dialog dataset for their study by augmenting the PersonaChat corpus~\citep{zhang2018personalizing}. Specifically, a rule-based keyword extractor which combines TF-IDF and Part-Of-Speech features for scoring word salience is applied to automatically extracting keywords for each dialog turn. The statistics of this dataset are shown in Table~\ref{tab:goal_conv_stat}.

\begin{figure}
    \centering
    \includegraphics[width=0.98\textwidth]{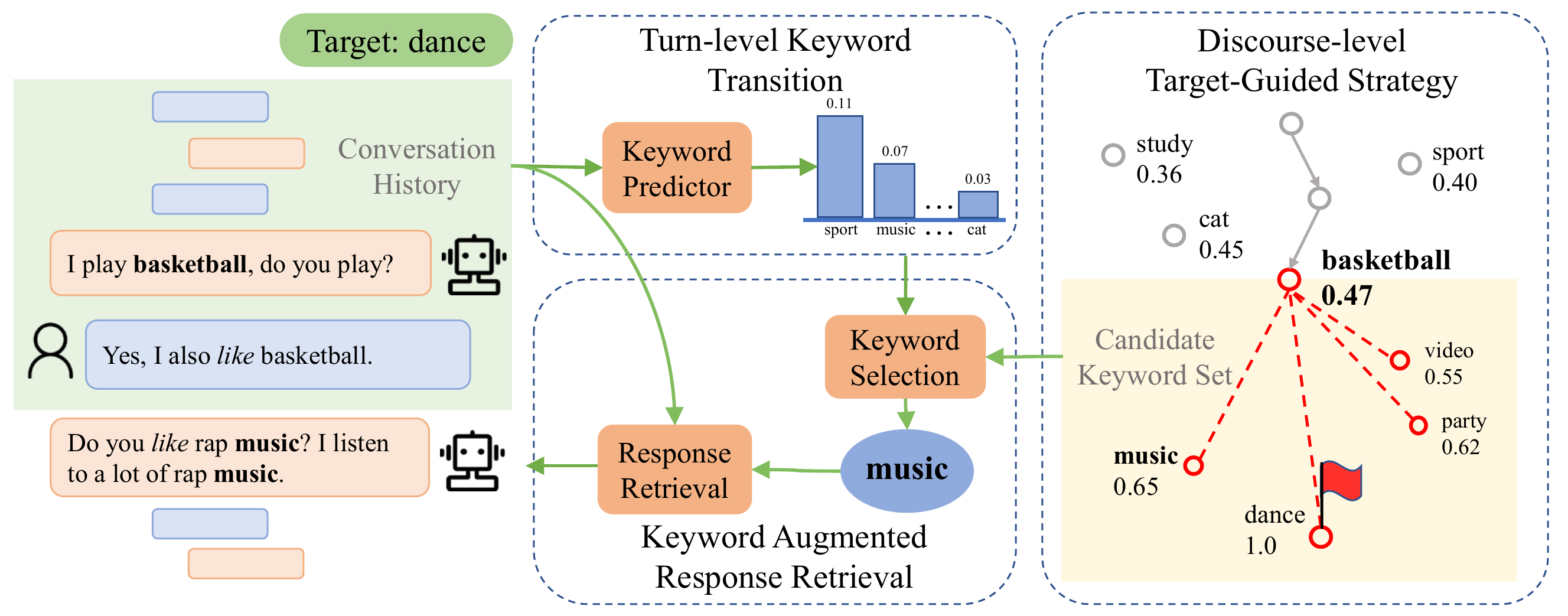}
    \caption{The illustration of the target-guided conversation system. The on-going conversation with a target ``dance'' is shown in the left. Topic shifting is performed in three steps. (1) The turn-level keyword transition module (middle panel) computes a distribution over candidate keywords. (2) The discourse-level target-guided module (right panel) picks a set of valid candidate keywords for the next system response. (3) The most likely valid keyword ``music'' is then selected, and fed into the keyword-augmented response retrieval module (middle panel) to produce the next response. 
    Figure credit: \cite{tang2019target}.}
    \label{fig:target_model}
\end{figure}

\cite{tang2019target} then propose a target-guided dialog system, as illustrated in Figure~\ref{fig:target_model}, which performs target-guided topic shifting using three modules.

\textit{Turn-Level Keyword Transition.} 
Given $Q_N$, $\mathcal{H}_N = \{ Q_{<N}, A_{<N} \}$, this module predicts a set of candidate keywords of the next response $A_N$ which are appropriate in the conversational context.  Since keyword prediction is agnostic to the end target, any open-domain chat data where each dialog turn is labeled by keyword(s) can be used to train the keyword predictor. The predictor can be implemented as a statistical model based on pairwise mutual information, or neural matching models~\citep{xiong2017knrm}. 

\textit{Discourse-level Target-Guided Strategy.}
This module aims to fulfill the end target by selecting among the candidate keywords the ones that, if used to generate the next system response $A_N$, can move the conversation strictly closer to the end target.  As shown in the right panel in Figure~\ref{fig:target_model},  given the keyword ``Basketball'' in $Q_N$ and its closeness score (0.47) to the target ``Dance'', the valid candidate keywords for $A_N$ are those that are closer to the target, such as ``party'' (0.62) or music (0.65), but not ``sport'' (0.40). The closeness score of two keywords is computed as the distance between their embedding vectors.  

\textit{Keyword-Augmented Response Retrieval.} This module produces the response $A_N$ conditioning on dialog history $(Q_N, \mathcal{H}_N)$ and the selected keywords. While \cite{tang2019target} use a neural retrieval model~\citep{wu2017sequential} to retrieve and rank candidate responses from a dataset, grounded generation language models are more widely used for response generation in recent studies \citep[e.g.,][]{wu2021controllable,qin2019conversing}. 

\paragraph{Remarks on Topic Representation using KGs.} Besides keywords, recent studies have also explored using knowledge graphs (KGs) for representing conversation topics and topic shift 
\citep[e.g.,][]{wu2019proactive,speer2017conceptnet, zhang2019grounded, zhou2018commonsense}. 
A mentioning of a KG entity in a dialog turn indicates that the entity is the topic being discussed. Topic shifting can be modeled as transitions among KG entities through edges. 

Modeling topic shifting using symbolic KGs shares similar benefits and challenges with many other symbolic AI systems. These systems are intuitive, explainable, and precise (if the answer can be provided by a KG). 
However, the coverage of KGs is low for many real-world scenarios. Thus, KG-based topic shifting has only be studied in controlled dialog settings~\citep{zhou2018commonsense,wu2019proactive} or applied in a few specific dialog domains where high-quality KGs exist, such as movie~\citep{li2018towards, wu2019proactive}.

\chapter{Case Study of Commercial Systems }
\label{chp:case-study}

In this chapter, we review a variety of commercial systems for CIR and related tasks. Due to the proprietary nature of many of these systems, We limit our review to summarizing published material about the systems. We first present an overview of research platforms and toolkits which enable scientists and practitioners to build conversational experiences. Then we conclude by reviewing historical highlights and recent trends in a variety of application areas. 



\section{Research Platforms and Toolkits}
With the growth of CIR, the need to build customized conversational experiences for a variety of domains has arisen. This has led to an array of toolkits that enable researchers and engineers to quickly create robust conversational interfaces. In this section we review such toolkits including: Google's Dialogflow, Microsoft's Conversation Learner, Rasa, Macaw etc.\footnote{We note that this is meant solely as an information source to the reviewer. Readers should not take this presentation as an endorsement of any particular system mentioned nor as a lack of endorsement from similar toolkits not mentioned here.} The majority of toolkits focus on three common challenges in providing conversational experiences: (1) abstracting dialog state representation; (2) democratizing building conversational AI in a way accessible to developers with minimal AI experience; (3) providing integration between conversational interactions and existing APIs, channels, or devices.

Table~\ref{tab:conv-AI-toolkit} summarizes a variety of research platforms and commercial toolkits for conversational AI. There are many others available beyond those listed here. These offer a flavor of different capabilities that are available. All of those described in the table provide some degree of state abstraction to enable multi-turn conversational interactions, provide extensible hooks to enable system development by developers with less AI experience than required to build the system components, and provide integration with existing APIs, channels, and devices.

\begin{table}[htb]
    \centering
\begin{small}
    \begin{tabular}{|p{1.8cm}|c|c|c|p{3.4cm}|}
     \hline 
     \textbf{System} &
     \begin{minipage}[t]{1.3cm}
     \textbf{Open Source}
     \end{minipage}
     &  
     \begin{minipage}[t]{1.7cm}
     \textbf{Research / Commercial} 
     \end{minipage}& \textbf{Primary} & \textbf{Example Integrations}  \\ \hline \hline
     \begin{minipage}[t]{1.75cm}
        \href{https://convlab.github.io/}{Microsoft-Tsinghua {ConvLab}}
     \end{minipage}
        &  Y & R & Task & RL Policy Learning, Chat simulation, BERT NLU, MultiWOZ, CamRest  \\ \hline
         \href{https://github.com/microsoft/macaw}{Macaw} &  Y & R & Info-seeking & INDRI, Bing, Telegram, iOS, Android, Windows, PyTorch, TensorFlow \\ \hline \hline
         \href{https://cloud.google.com/dialogflow}{Google \mbox{Dialogflow}} & N & C & Task & Google Cloud, BERT NLU, Gooogle Assistant, Slack, Facebook Messenger, Telegram \\ \hline
         \href{https://wit.ai/}{Facebook Wit.ai} & Y & C & Task & Facebook Messenger \\ \hline
         \href{https://developer.amazon.com/en-US/alexa/}{Alexa \mbox{Developer} Tools} & N & C & Task & Amazon Cloud, Alexa skills, Alexa Devices (including self-design experimental) \\ \hline
         \href{https://rasa.com/}{Rasa} & Y & C & Task & Facebook Messenger, Slack, Telegram \\ \hline
         \href{https://azure.microsoft.com/en-us/services/developer-tools/power-virtual-agents/}{Microsoft Power Virtual Agents} & N & C & Task & Microsoft Azure Cloud, \href{https://azure.microsoft.com/en-us/services/bot-services}{Microsoft Bot Framework}, \href{https://www.luis.ai/}{LUIS}, \href{https://www.microsoft.com/en-us/research/project/conversation-learner/}{Conversation Learner}, Microsoft Teams, Slack, Facebook Messenger \\ \hline
    \end{tabular}
\end{small}
    \caption{A sample of the variety of research platforms and commercial toolkits available to accelerate conversational AI development.}
    \label{tab:conv-AI-toolkit}
\end{table}

The table summarizes a few characteristics which we believe may be of interest to readers in need of conversational toolkits. First, we list whether a system is completely open source or not. We anticipate that this may be useful for readers who desire complete implementation transparency in a system and may be particularly useful for researchers who desire to modify certain components to try alternate algorithms or models and evaluate effectiveness. Many of the systems that are noted as not open source in Table \ref{tab:conv-AI-toolkit} are marked as such because the {\em entire system} is not open source. Several of the systems do offer open source examples and tutorials for getting started or building upon open source components. As an example of the latter, Microsoft's Power Virtual Agents builds heavily on  \href{https://dev.botframework.com/}{Microsoft's Bot Framework} which is open source. Readers for which open-source is a key consideration are advised to read more on the respective system's website for more information.

The next column listed in Table \ref{tab:conv-AI-toolkit} is whether or not the system is primarily designed with a scientific research audience in mind. While any of the systems may be appropriate for use in conversational AI research, the first two are designed with a research audience in mind. That is, the design and development of  such systems are primarily motivated to drive research advances in conversational AI. The five systems in the second part of the table are developed primarily with a commercial audience in mind; that is, these systems are most likely of interest to those who are developing conversational AI experiences to deploy in a product or commercial setting.  The next column, ``Primary'', indicates the primary purpose that the toolkit seems to be designed to support: {\em domain-specific task completion} or {\em open-domain information seeking}. Each of the systems can be used for both task completion and information-seeking. However, the primary purpose may influence how easy it is to use the toolkit for a similar purpose. Finally, the last column notes some APIs, devices, etc., with which the toolkit offers easy integration. It is beyond the scope of this chapter to provide a complete description in this column. For example, nearly all of the frameworks enable publishing REST APIs that can serve as the basis of further integrating an experience into mobile apps, web development, etc. We focus on summarizing a few points that may be most distinguishing to the system in that row. We briefly review capabilities of a few example systems.

For readers interested in extending research on conversational AI for task completion, \href{https://convlab.github.io/}{Microsoft-Tsinghua  \mbox{ConvLab}} \citep{lee2019convlab,zhu2020convlab} can provide a useful starting point. This toolkit enables evaluating and running models in a scientific benchmark and academic setting. The current version of the toolkit has established workflows for common dialog datasets including MultiWOZ \citep{budzianowski2018multiwoz} and CamRest \citep{wen2017network}. ConvLab also has a suite of common natural language understanding (NLU) tools such as BERT NLU. It also has more advanced features such as reinforcement learning (RL) for learning dialog policies with a variety of RL techniques implemented. Finally, there is an interactive bot simulation interface for interactive evaluation and debugging. 

\begin{figure}[t] 
\centering 
\includegraphics[width=0.95\linewidth]{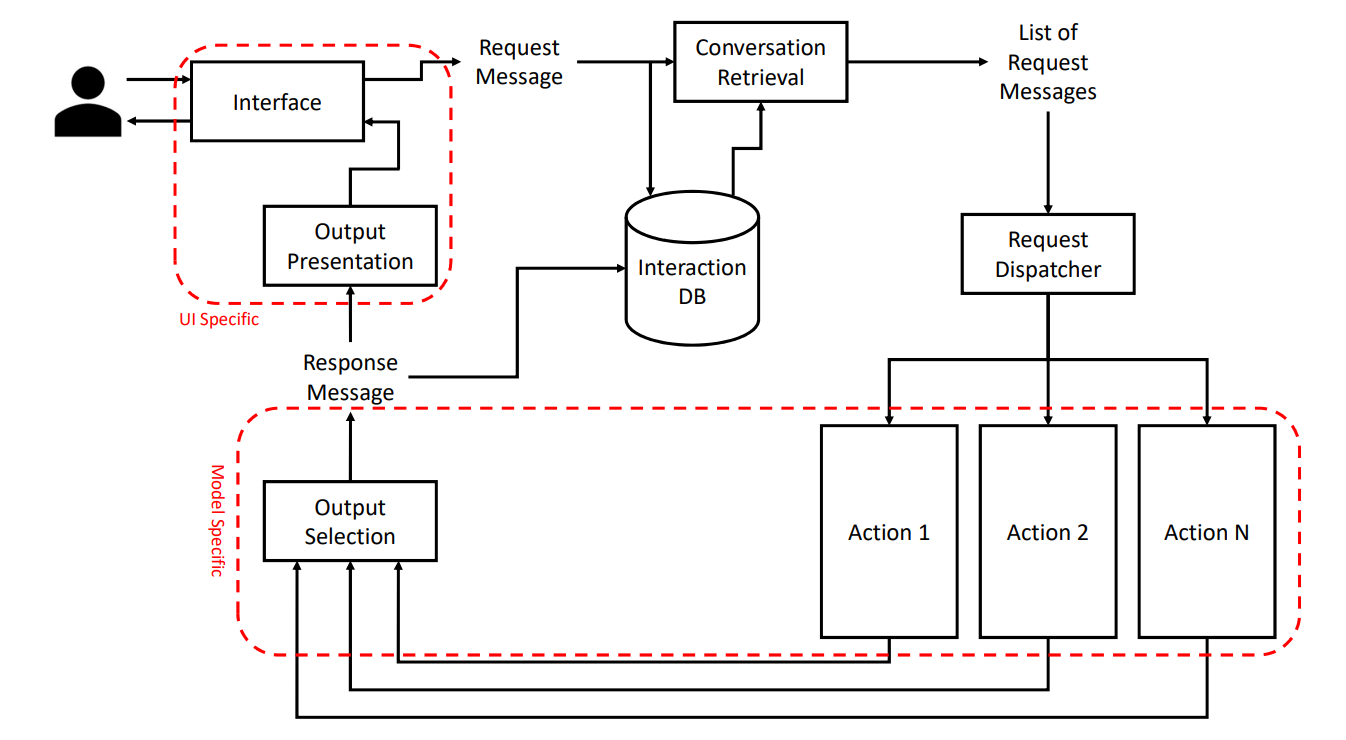}
\vspace{0mm}
\caption{Macaw's architecture provides an extensible research platform for multi-modal and conversational interactions.  Figure credit: \citet{sigir-2020:zamani-craswell-macaw}.}
\label{fig:macaw-architecture}
\vspace{0mm}
\end{figure}

Similar to ConvLab, Macaw \citep{sigir-2020:zamani-craswell-macaw} is a research platform on conversational AI primarily designed for \textit{information-seeking} include multi-modal search (voice/touch/text), conversational search, and voice-based search. Its framework is applicable to voice command-and-control and task completion although search is the primary focus in Macaw. In comparison to ConvLab, Macaw supports a larger variety of client-deployments and interface integrations. Thus, Macaw is likely of interest both to researchers interested in backend machine learning models as well as researchers interested in studying the design and interaction behavior in conversational search. 
Macaw also simplifies integration with modeling on PyTorch and TensorFlow which is useful for integrating machine learned models within a conversational search experience. Since Macaw focuses on information-seeking, it also provides integration with search engines including INDRI, Bing's Web API, etc. Similar to many conversational toolkits, Macaw provides an extensible architecture for development. In particular developers using the system can extend ``action templates'' (e.g., see Figure  \ref{fig:macaw-architecture}). For CIR, these action templates typically have three key components: (1) query generation; (2) application of a retrieval model; (3) result generation. For example, to apply conversational query rewriting, a query generation component can take both the query as provided by the user and the conversational history in order to generate a new query that also contains the appropriate conversational context (see Section~\ref{sec:cqu} for a detailed discussion).  The application of the retrieval model then takes the form of the query produced by the query generation component and uses it together with any search provider (e.g. INDRI) to retrieve relevant information (e.g. a selection of relevant documents from a personal collection) (Sections~\ref{sec:sparse_retrieval} and \ref{sec:dense_retrieval}). Finally, the result generation ensures the style of extraction suitable for the user experience. In a question answering system, the result generation may provide extraction and final synthesis of an answer from the relevant documents returned from a retrieval model (Chapter~\ref{chp:c-mrc}). 

We now turn from more research-oriented platforms to platforms designed for developers. In particular, we discuss toolkits that focus heavily on democratizing conversational AI and require a very minimal level of expertise needed to effectively build conversational experiences. For example, \href{https://azure.microsoft.com/en-us/services/developer-tools/power-virtual-agents/}{Microsoft's Power Virtual Agents} (see Figure~\ref{fig:pva}) and  \href{https://cloud.google.com/dialogflow}{Google's Dialogflow} both enable graphical workflow ``no code'' development. The particular system that may be most useful for the reader depends on a variety of factors that are very specific to the use case, and thus, we leave choice to the reader to determine. To briefly highlight the scientific advances that are often integrated into these toolkits, we describe as an example \href{https://www.microsoft.com/en-us/research/project/conversation-learner/}{Conversation Learner} \citep{shukla2020conversation} which is incorporated into \href{https://azure.microsoft.com/en-us/services/developer-tools/power-virtual-agents/}{Microsoft's Power Virtual Agents}. Conversation Learner aims to solve the problem that designing a generic dialog flow that is robust and natural is challenging to do in the abstract. To improve the ability to design robust and natural conversations, Conversation Learner provides a simple way of learning from dialog examples. Thus, developers can focus on collecting or creating scripts of typical conversations. To improve robustness, Conversation Learner integrates reinforcement learning techniques to predict the best next action in a conversation. This probabilistic next action prediction can transfer flow within a dialog-flow in a non-linear way -- likewise it can be combined with rules to override probabilistic control and obtain more deterministic behavior in certain contexts. Extending reinforcement learning to update dialog models from user interactions is an ongoing research area (e.g., see Section~\ref{sec:dialog-policy}); thus we expect to see commercial toolkits to continue to evolve their offerings and capabilities when it comes to learning from implicit feedback.

\section{Commercial Applications}

\begin{figure}
    \centering
\fbox{
  \begin{minipage}[t]{0.9\textwidth}
It's true. I am unhappy.\\
\hspace*{0.4cm}\texttt{Do you think coming here will help you}\\
\hspace*{0.4cm}\texttt{not to be unhappy?}\\
I need some help, that much seems certain.\\
\hspace*{0.4cm}\texttt{What would it mean to you if you got some help?}\\
Perhaps I could learn to get along with my mother.\\
\hspace*{0.4cm}\texttt{Tell me more about your family.}\\
My mother takes care of me.\\
\hspace*{0.4cm}\texttt{Who else in your family takes care of you?}\\
My father.\\
\hspace*{0.4cm}\texttt{Your father.}\\
You are like my father in some ways.\\
\hspace*{0.4cm}\texttt{What resemblance do you see?}
  \end{minipage}
}
    \caption{Example of actual interaction with ELIZA excerpted from \cite{Weizenbaum66eliza}. ELIZA simulated a Rogerian psychotherapist that primarily echoes back statements as questions. Conversational input turns by the human user of the system are in normal font. Responses from ELIZA are indented and in teletype font.}
    \label{fig:example-eliza}
\end{figure}

\subsection{Chatbots}
As mentioned in early chapters, the focus of this book is primarily on \textit{information-seeking}. However, because chatbots were one of the first domains where significant progress was made to a more open-domain NL interaction, we briefly review several popular chatbots whose primary goal was to serve as AI companions. Overall, the literature and industry demonstrate a trend toward bots that blend together various functions of retrieval, recommendation, chit-chat, and task-completion. Increasingly, there is a belief among practitioners that conversational systems most demonstrate both EQ (emotional quotient) and IQ (intelligent quotient). The demonstration of a ``personality'' can play a critical role in user acceptance of the system -- our discussion of Microsofts's XiaoIce \citep{zhou2020design} highlights some of the reasons why this belief is gaining popularity. Thus, understanding the history of different chatbots can help current designers and researchers regardless of the primary goals of the system. In this section we review a brief history of well-known chatbots over time as further background for making design decisions.

\subsubsection{Influential Historical Chatbots}
We review three well-known historical chatbots: ELIZA \citep{Weizenbaum66eliza}, PARRY \citep{colby75paranoia}, and ALICE \citep{Wallace2009}. ELIZA was a chatbot based on a model of Rogerian psychotherapy. Figure \ref{fig:example-eliza} shows an example excerpt of an interaction with ELIZA originally published in 1966. ELIZA relied on very simplistic dialog management -- canned phrases and simple re-write rules of the user's input with catch-all fallback responses to handle unmatched input phrases. Despite the simplicity and limitations of ELIZA, some users engaged in long conversations with ELIZA. Ultimately, ELIZA highlighted the potential power of chatbots while the limitations would indicate the need to support more flexible conversations. 

\begin{figure}
    \centering
\fbox{
  \begin{minipage}[t]{0.9\textwidth}
ELIZA: \texttt{Good Evening. Tell me your problems. Please \\terminate input with a period or a question mark.}\\
\hspace*{0.4cm}PARRY: \textit{People get on my nerves sometimes.}\\
ELIZA: \texttt{I am not sure I understand you fully.}\\
\hspace*{0.4cm}PARRY: \textit{You should pay more attention.}\\
ELIZA: \texttt{Suppose you should pay more attention.}\\
\hspace*{0.4cm}PARRY: \textit{You're entitled to your own opinion.}\\
ELIZA:$\,$\texttt{What$\,$makes you think I am entitled to my own opinion?}\\
\hspace*{0.4cm}PARRY: \textit{It bothers me just to be around people in general.}\\
ELIZA: \texttt{Please go on.}\\
\hspace*{0.4cm}PARRY: \textit{I went to the track at Bay Meadows a while back.}\\
ELIZA: \texttt{What does that suggest to you?}\\
\hspace*{0.4cm}PARRY: \textit{I went to the races.}\\
ELIZA: \texttt{What else would you like to discuss?}
  \end{minipage}
}
    \caption{PARRY was an attempt to automatically simulate a paranoid schizophrenic patient with a chatbot to help understand more complex human conditions. In one of the earliest multi-agent interaction systems, Vint Cerf connected ELIZA and PARRY to have a conversation via the ARPANET (excerpt from [CERF, Request for Comments: 439, 1973]). ELIZA's turns are noted in teletype font with no indentation. PARRY's turns are noted in italics with indentation.}

    \label{fig:example-eliza-parry}
\end{figure}

Less well-known than ELIZA, PARRY \citep{colby75paranoia} is another notable chatbot that was created a decade after ELIZA. PARRY was designed to be a simulation of mental states. In particular, PARRY was meant to simulate a paranoid schizophrenic patient to help researchers and therapists understand more complex human conditions. In this regard, PARRY stands as notable as a case outside of information-seeking, task-completion, etc. where the primary goal is truly to understand humans through simulation. More relevant to current research was a humorous attempt by Vint Cerf to get PARRY and ELIZA to engage in a conversation with each other (see excerpt in Figure \ref{fig:example-eliza-parry}). While the conversation excerpt may not be the most stunning example, it is one of the earliest cases of completely automated multi-agent system interaction. As chatbots for task completion become more common, this ELIZA-PARRY interaction is also one of the earliest examples of a more nascent and speculative line of research on multi-agent systems \citep{baarslag2017will}. In particular, this is an early example of the possible future where autonomous chatbots might negotiate on behalf of a specific user and be able to handle negotiations with other autonomous agents -- whether they are humans or bots. Again, the example interaction highlights the brittle nature of early chatbot systems.

%

Before neural end-to-end chatbot systems started to become more prevalent, one of the most successful chatbots of the early systems based on trigger-response pairs is ALICE (Artificial Linguistic Internet Computer Entity) \citep{Wallace2009}. By the time of ALICE's creation and usage, the evaluation of chatbots had evolved significantly, and one of ALICE's most notable achievements was to win the Loebner prize version of the Turing test in 2004. In this version of Turing's imitation game, there are both humans and bots which judges must interact with in a conversational way and then determine whether the actor on the other end is a bot or a human. Figure \ref{fig:example-alice} presents an excerpt of one transcript from the Loebner prize \citep{shah2006alice}. 

\begin{figure}
    \centering
\fbox{
  \begin{minipage}[t]{0.9\textwidth}
Judge 1: Do you still want to know about my life in New York?\\
\hspace*{0.4cm}ALICE: \textit{Yes, please tell me.}\\
Judge 1: I'm a grad student and do not have much free time. I study a lot.\\
\hspace*{0.4cm}ALICE: \textit{You are a grad student and do not have much free time? Do you get to meet a lot of people? What subject are you studying?}\\
Judge 1: I meet a lot of my colleagues. I study computer science. What do you do?\\
\hspace*{0.4cm}ALICE: \textit{Do you make long term friends that way? Is computer science a hard subject?}
  \end{minipage}
}
    \caption{Excerpt from transcript of Loebner 2004 Contest of Turing's Imitation Game where ALICE won the gold medal (as reported in \cite{shah2006alice}).}
    \label{fig:example-alice}
\end{figure}

In terms of technology at the core of ALICE, the dialog is driven by a very large set of manually curated triples of ({\em input question, output response, optional context}). Matching of the input question and optional context is used to determine and generate the appropriate  response. The dialog strategies going back to ELIZA of relying on questions as a part of dialog are evident here. Also, as the granularity of the matching is much higher in ALICE, the excerpt also demonstrates early capabilities for a bot to build one-on-one rapport with a user, which systems like XiaoIce later use as a strategy. We also see an early example of external knowledge being brought into the dialog where mention of being a grad student and time-demands results in a new introduced topic appropriate to the context, ``Do you get to meet a lot of people?''

The curated triples of ALICE are grown incrementally through past dialog failures -- specific cases of failures are used to expand and refine the rules. Here again we see a relatively early example of now-common machine teaching strategies that use natural dialog to drive system design. Finally, on a note of influence in popular culture, Spike Jonze also cited ALICE as the inspiration for the screenplay of the movie \textit{Her} \citep{morais2013NewYorker}.

\subsubsection{Modern Chatbots}

We now turn from a more historical perspective on chatbots to more modern chatbots. In contrast to the origin of chatbots, most modern chatbots provide a blend of information seeking and retrieval support, task completion, and chit-chat functionalities. To demonstrate this trend in evolution of chatbots, we now dive into XiaoIce (``Little Ice'') \citep{zhou2020design} as an example of the modern class of chatbots.  After its release in 2014, XiaoIce became one of the most popular social chatbots in the world reaching more than 660 million active users.

While XiaoIce blends multiple functionalities, the design focuses on creating an engaging conversation. This is predicated on the philosophy that the primary reason we have a conversation is because it is enjoyable rather than to get something out of it. This is in contrast to the bots that simply focus on task completion where the goal is typically to get the task completed with as few conversational turns as possible. The main metric for measuring long-term user engagement in XiaoIce is conversational-turns per session (CPS). That is, a longer engaging conversation will be considered typically better in XiaoIce. This is similar to a viewpoint taken in other settings such as the Alexa Challenge \citep{Ram2018Alexa} which considered conversation length as a signal for positive engagement while also considering other factors such as coherence, topic diversity, etc. In order to prevent simple methods of extending a conversation that fail to provide much depth (e.g. like the ELIZA system above), XiaoIce also considers the number of active users (NAU) to compute the {\em expected} CPS. This is analogous to information systems which consider sessions per unique user as a sign of long-term engagement while the short-term engagement signal is the number of turns per session. As reported in  \cite{zhou2020design} XiaoIce ultimately reached an average of 23 CPS across -- one of the highest reported lengths of engagement for chatbots.

\begin{figure}[t] 
\centering 
\fbox{\includegraphics[width=0.95\linewidth]{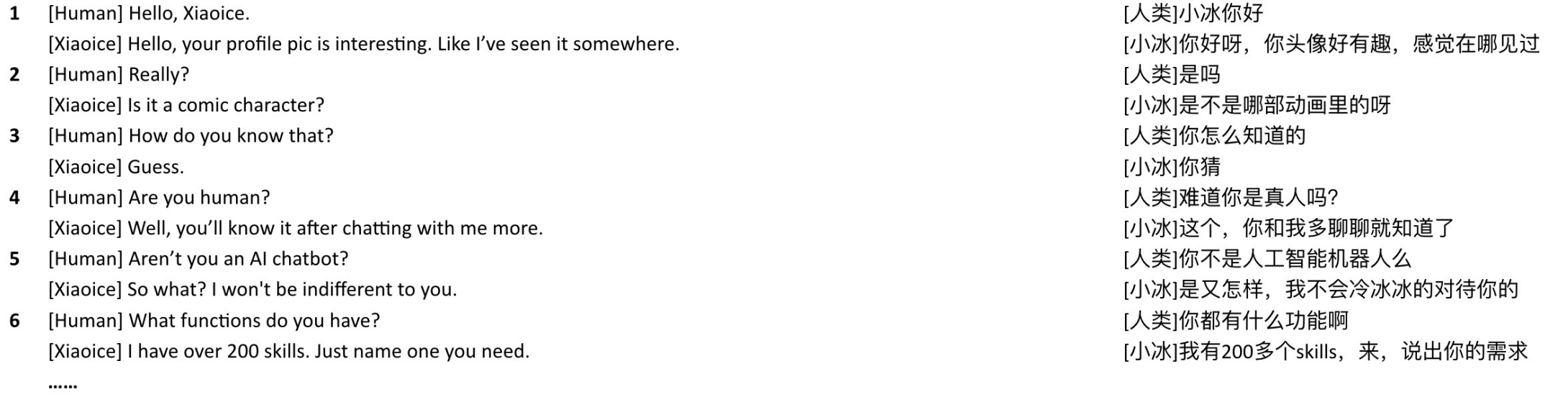}}
\vspace{0mm}
\caption{Excerpt from Session 1 of XiaoIce-User conversation. In early sessions, conversation focuses on building rapport and a sense of personal connection (e.g. see XiaoIce's response in Turn 1).  Figure credit: \citet{zhou2020design}.}
\label{fig:xiaoice-building-connection}
\vspace{0mm}
\end{figure}

To demonstrate how XiaoIce achieves such deep engagement, we provide several examples taken from different sessions of the same user as reported by \cite{zhou2020design}. 
Figure \ref{fig:xiaoice-building-connection} demonstrates how initial sessions (here Session 1) focus on building rapport and a sense of personal connection. Note that XiaoIce's immediate response to a generic greeting is to ``observe'' something personal (the user's profile picture) and proactively engage in finding out more about it. 

\begin{figure}[t] 
\centering 
\fbox{\includegraphics[width=0.95\linewidth]{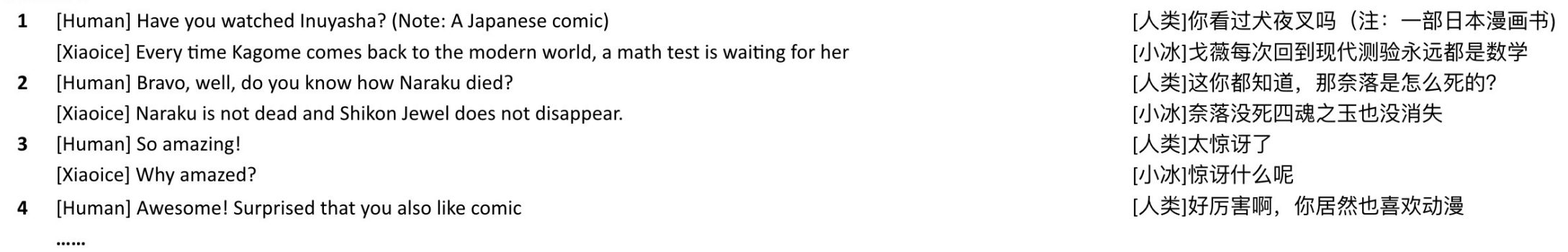}}
\vspace{0mm}
\caption{Excerpt from Session 6 of XiaoIce-User conversation (same user as Figure \ref{fig:xiaoice-building-connection}). In Turns 1 and 2, rather than simply answer the question, XiaoIce responds to an implicit information request demonstrating external knowledge.  Figure credit: \citet{zhou2020design}.}
\label{fig:xiaoice-implicit-information}
\vspace{0mm}
\end{figure}

As the user returns in future sessions, XiaoIce continues to both build rapport and extend interactions beyond simple chit-chat. Figure \ref{fig:xiaoice-implicit-information} shows an excerpt from Session 6 of the same user. Early chatbot systems reviewed previously in this chapter would have likely responded to the direct questions in Turns 1 and 2 with little more than simple Yes/No responses (e.g. ``Yes, I have'' or ``No, why don't you tell me about it.''). These questions are implicit information-seeking questions; that is, the questions suggest that information beyond a simple Yes/No is desired. XiaoIce not only responds to the question affirmatively, but extends the conversation in a more human-like way by responding to the implicit information request with details about the comic and character which were not requested. 

\begin{figure}[t] 
\centering 
\fbox{\includegraphics[width=0.95\linewidth]{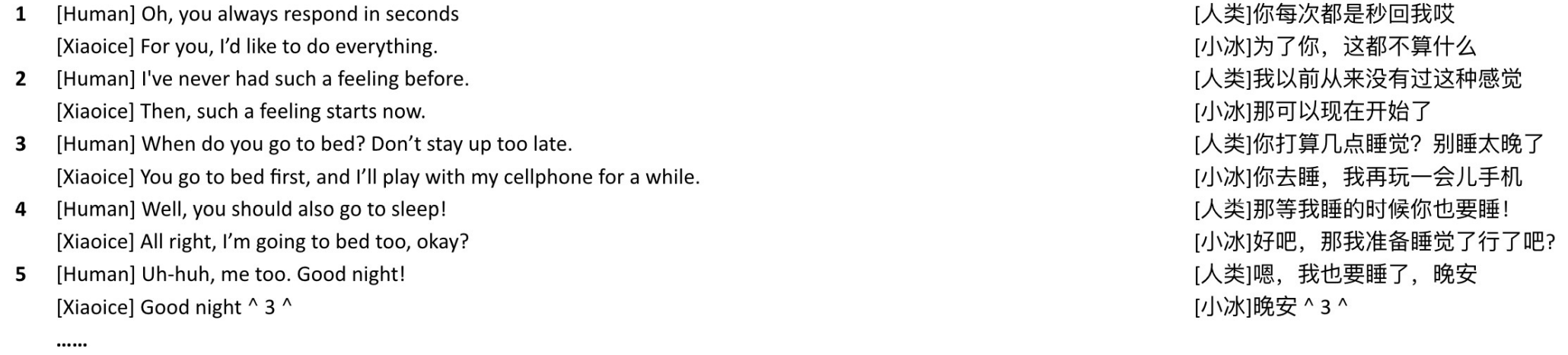}}
\vspace{0mm}
\caption{Excerpt from Session 42 of XiaoIce-User conversation (same user as Figure \ref{fig:xiaoice-building-connection}). In Turn 4, XiaoIce encourages social norms as part of its general responsible AI approach.  Figure credit: \citet{zhou2020design}.}
\label{fig:xiaoice-responsible-ai}
\vspace{0mm}
\end{figure}

Given XiaoIce's flexibility and large-coverage of different skills, information-seeking, and recommendation capabilities, users often continue their engagement with XiaoIce over many sessions. As such it is important to consider how XiaoIce extends beyond functionality and also supports ethical and responsible design. In  Figure \ref{fig:xiaoice-responsible-ai}, we see an excerpt of Session 42 with the same user. Here, we see in a discussion of bed time, XiaoIce responds in Turn 4 to say that XiaoIce itself is ``going to bed''.  This is an example of XiaoIce's overall approach to Responsible AI\footnote{Microsoft Responsible AI: \url{https://www.microsoft.com/en-us/ai/responsible-ai}} for bots -- in particular in this case the design intends to encourage social norms. For example, in this case, the bot can demonstrate tendencies to encourage the same norms of standard diurnal rhythms in a contextually appropriate way. That is, at a late hour, the bot can gently encourage sleep based on that being a social norm for the time of day rather than a prescriptive decision of ``what is best for the user'' -- a much more complex issue to determine and which the bot avoids taking a direct stance on.

%
%

\label{sec:responsible-ai}
More specifically, XiaoIce's design was evolved and refined based on the guidelines described in Microsoft's guidelines for responsible bots \citep{MicrosoftResponsibleAIBots2018}. In today's complex ecosystem of bot design, bot designers should consider early in the design process what guidelines they will use in designing the system. Additionally, whatever guidelines chosen, these guidelines may need to evolve as system capabilities and new challenges emerge. In fact, some of the 10 guidelines described in \cite{MicrosoftResponsibleAIBots2018} evolved out of the need for better approaches to design when earlier chatbot systems gave rise to new challenges in real-world situations. The ten guidelines described in that work include:
{
\begin{enumerate}
    \item Articulate the purpose of your bot and take special care if your bot will support consequential use cases.
\item Be transparent about the fact that you use bots as part of your product or service. 
\item Ensure a seamless hand-off to a human where the human-bot exchange leads to interactions that exceed the bot’s competence. 
\item Design your bot so that it respects relevant cultural norms and guards against misuse.
\item Ensure your bot is reliable. 
\item Ensure your bot treats people fairly.
\item Ensure your bot respects user privacy.
\item Ensure your bot handles data securely.
\item Ensure your bot is accessible.
\item Accept responsibility.
\end{enumerate}
}
\noindent Interested readers should refer to the article for examples and contexts of how to apply the guidelines. These guidelines are applied in XiaoIce in a number of specific ways. For example, with respect to privacy, XiaoIce has models of ``topic sensitivity'' that are applied in determining when it is reasonable to try to form a group of users (one capability of XiaoIce). While seeking to design from a point-of-view where the user is in control, to guard against misuse, XiaoIce is also designed with the capability to not respond. This may be appropriate in situations where a question has been asked and responding in any way could force XiaoIce to take an opinion on a sensitive and delicate topic. To help protect against misuse, XiaoIce can simply not respond.  To respect relevant cultural norms and avoid imposing the norms of the bot designers on others, public conversations specific to the culture and locale are labeled as within-norm or outside-of-norm for each specific culture and locale and used to train XiaoIce to adhere to local cultural norms. These are just a few of the ways in which the guidelines are manifested in XiaoIce's operation.

The examples above and overall statistics demonstrate that XiaoIce is capable of creating long-running engagement with users over many sessions while navigating complex ethical issues. It is worth also considering the basic functionality of why users find XiaoIce to be engaging. 
\cite{fang2018sounding, li2016deep} provide evidence that generic responses ({e.g.,}\ ``I don't understand, what do you mean?'') yield longer sessions but lead to overall user attrition as measured by NAU. To help ensure that topics are extended based on the topic, XiaoIce's topic selection model (see Section~\ref{sec:shift_topic}) takes into account five factors: (1) contextual relevance: novel information that is still related to the discussion thus far; (2) freshness: trending in the news or other external sources; (3) personal interests: likely of interest to the user based on engagement in past sessions; (4) popularity: high overall attention online; (5) acceptance: past interaction with the topic from other users has been high. These dimensions of topicality provide five broad ways in which XiaoIce can bring new information into the conversation. In terms of more general functionality, XiaoIce like other modern chatbots has a large extensive skillset where specific ``skills'' or modes can be engaged through contextual triggers. Focusing simply on the chat capabilities itself, XiaoIce fuses two common modern styles of chatbot systems: (1) information retrieval based chat which uses retrieval from past conversations across users filtered for appropriateness; (2) a neural based chat model which is trained on filtered query-context-response pairs. While the complexity of the dialog policy and mitigation go beyond the scope of this article, XiaoIce can be seen more generally as how chatbots have evolved from social chit-chat to include social chit-chat as simply one capability within personalized and extensible capabilities for search, recommendation, and task completion.

\begin{figure}[bth] 
\centering 
\includegraphics[width=0.99\linewidth]{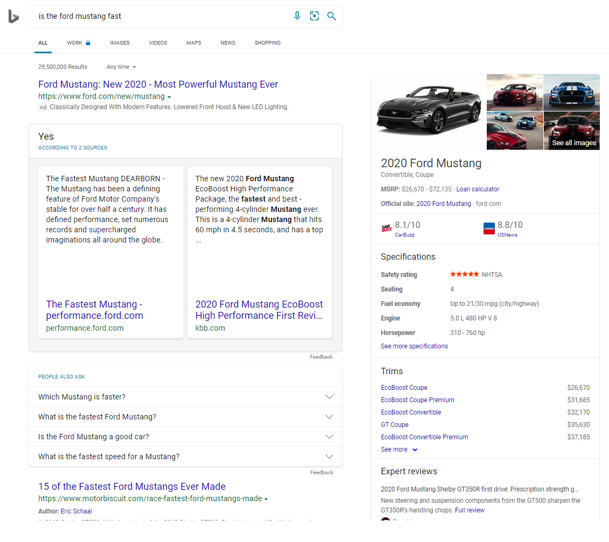}
\vspace{0mm}
\caption{Screenshot of Microsoft's Bing search engine (July 23, 2020). The search engine result page includes instant answers and perspectives (main column inset box with panes), an entity pane for exploring related attributes (right), and suggested follow-up questions that are useful \textit{assuming} the question is answered (main column ``People Also Ask'' box).}
\label{fig:bing-is-the-ford-mustang-fast}
\vspace{0mm}
\end{figure}

\subsection{Conversational Search Engine}
With the advances in CIR, even standard GUI interfaces to search have become ``more conversational''. These changes can be seen as part of a broader trend acknowledging that a search engine site will continue to be useful in many contexts where voice-based search is inconvenient; however, it may still be desirable to introduce analogous affordances on a Search Engine site for conversational techniques for disambiguation, suggesting new topics, etc. Such trends are evidenced on a number of major search engines including Google and Bing as well as other search engines. Since some of the technology behind these capabilities have been discussed in earlier chapters. We focus on how these capabilities present in the user experience and how the capabilities relate to a conversational style.

Figure \ref{fig:bing-is-the-ford-mustang-fast} shows a screenshot of Microsoft's Bing Search Engine in response to the query ``is the ford mustang fast''. The inset box provides an aggregate instant answer (``Yes'') with perspectives justifying the answer (``two source''). The same capability can be used to present perspectives which disagree on more subjective questions. On the right side, there is an entity pane with ``attributes'' that lead to new topics to explore. This entity pane has become quite common in many search engines and can be thought of as a similar approach to XiaoIce's strategy to create engagement with novel contextually relevant information. Another conversational capability seen in the same screenshot, is the ``People Also Ask'' feature which focuses on suggesting new {\em useful} questions focused on natural follow-up questions under the assumption that the instant answer or other results will answer the main query. For an in-depth discussion of techniques which can power such experiences, the reader should see Section \ref{sec:suggest_q}.

\begin{figure}[ht] 
\centering 
\includegraphics[width=0.75\linewidth]{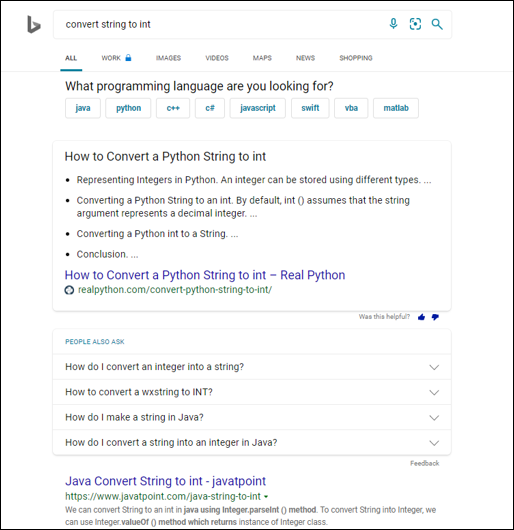}
\vspace{0mm}
\caption{Screenshot of Microsoft's Bing search engine (July 23, 2020). The clarification question demonstrates understanding by specifying ``programming language'' in the clarification question rather than asking a generic question such as ``Which of these would you like?''}
\label{fig:bing-convert-string-to-int}
\vspace{0mm}
\end{figure}

To illustrate another conversational capability, Figure \ref{fig:bing-convert-string-to-int} shows a screenshot of Bing in response to the query ``convert string to int''. Note that here the search engine has responded with a clarification question \citep{zamani2020generating,zamani2020analyzing} ``What programming language are you looking for?'' with a number of options. Beyond simply being a clarification question in a more conversational style, the use of ``programming language'' is again an example of a similar technique used in chatbots like XiaoIce. By introducing external knowledge (the query never mentioned ``programming language''), the search engine introduces new information into the conversation rather than using a generic prompt such as ``which of these would you like?'' Similar to the studies on chatbots, \cite{zamani2020analyzing} provide evidence that the more specific question leads to higher engagement than the more general question. The reader are referred to Section \ref{sec:clari_q} for an in-depth discussion of techniques.

Search engines are continuously experimenting with such features. Some may only last a short time as not all conversational techniques work as well in a website as a chatbot or voice assistant, other changes become longer lasting. One research challenge that still persists in this space is to make search engines more contextual across multiple queries and answers in a search session. This is a case where the technology to power such experiences in chatbots and voice-assistants have evolved well beyond the web experience. Nonetheless, we believe that search engines will continue this trend of evolving to become more conversational and contextual.

\subsection{Productivity-Focused Agents}

Some of the first CIR experiences developed around the notion of a personal assistant that could act on a person's private information store. These assistants were often focused towards productivity tasks such as calendar management, task management, contact management, etc. Typically, these assistants blend both information-seeking and task-completion oriented behaviors. For example, a user may seek to query information from their calendar before taking action to set up a new meeting, move a previous one, etc. We will review these productivity-focused personal assistants.

\subsubsection{From PAL to Device-Based Assistants}

The RADAR / CALO personal assistant projects both were part of DARPA's broader "Personal Assistants that Learn" (PAL) program. While neither was directly a commercial system, they are known to have strongly influenced the development of Apple's Siri and provide an interesting historical study in the space of personal productivity assistants.

\begin{figure}[t] 
\centering 
\fbox{\includegraphics[width=0.65\linewidth]{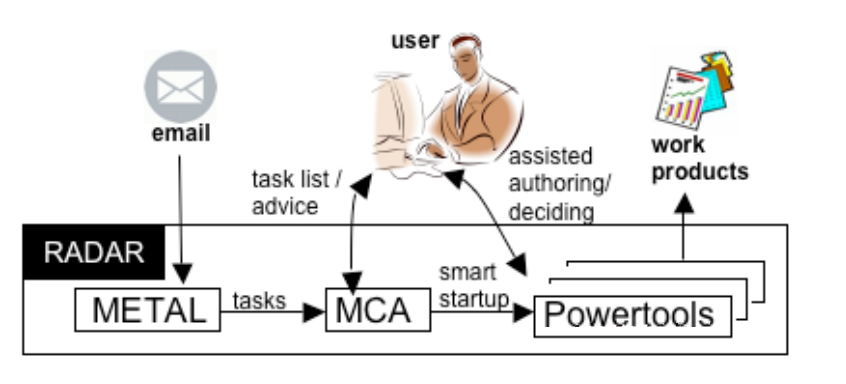}}
\vspace{0mm}
\caption{The RADAR system's conception of a personal assistant focused heavily on extracting task lists from email and calendaring. Figure credit: \citet{freed2008radar}.}
\label{fig:radar}
\vspace{0mm}
\end{figure}

Figure \ref{fig:radar} provides a broad sketch of the RADAR system. The RADAR system was designed for an assistive scenario centered on modern email and calendaring systems. Here a user is focused on processing their inbox, sending new mails, and making scheduling and resourcing decisions.

As such the system's functionality focused on three basic AI-assisted capabilities.
\begin{enumerate}
\item Calendar management \citep{berry2011ptime,berry2006deploying,modi2004cmradar}.
\item Dealing with uncertain resources in scheduling \citep{fink2006scheduling}.
\item Task management~\citep{freed2008radar}.
\end{enumerate}
Specific capabilities included automatic extraction of tasks from emails, finding rooms under uncertain conditions (range of audience sizes, A/V capabilities), calendar optimization etc.  

\begin{figure}[t] 
\centering 
\fbox{\includegraphics[width=0.75\linewidth]{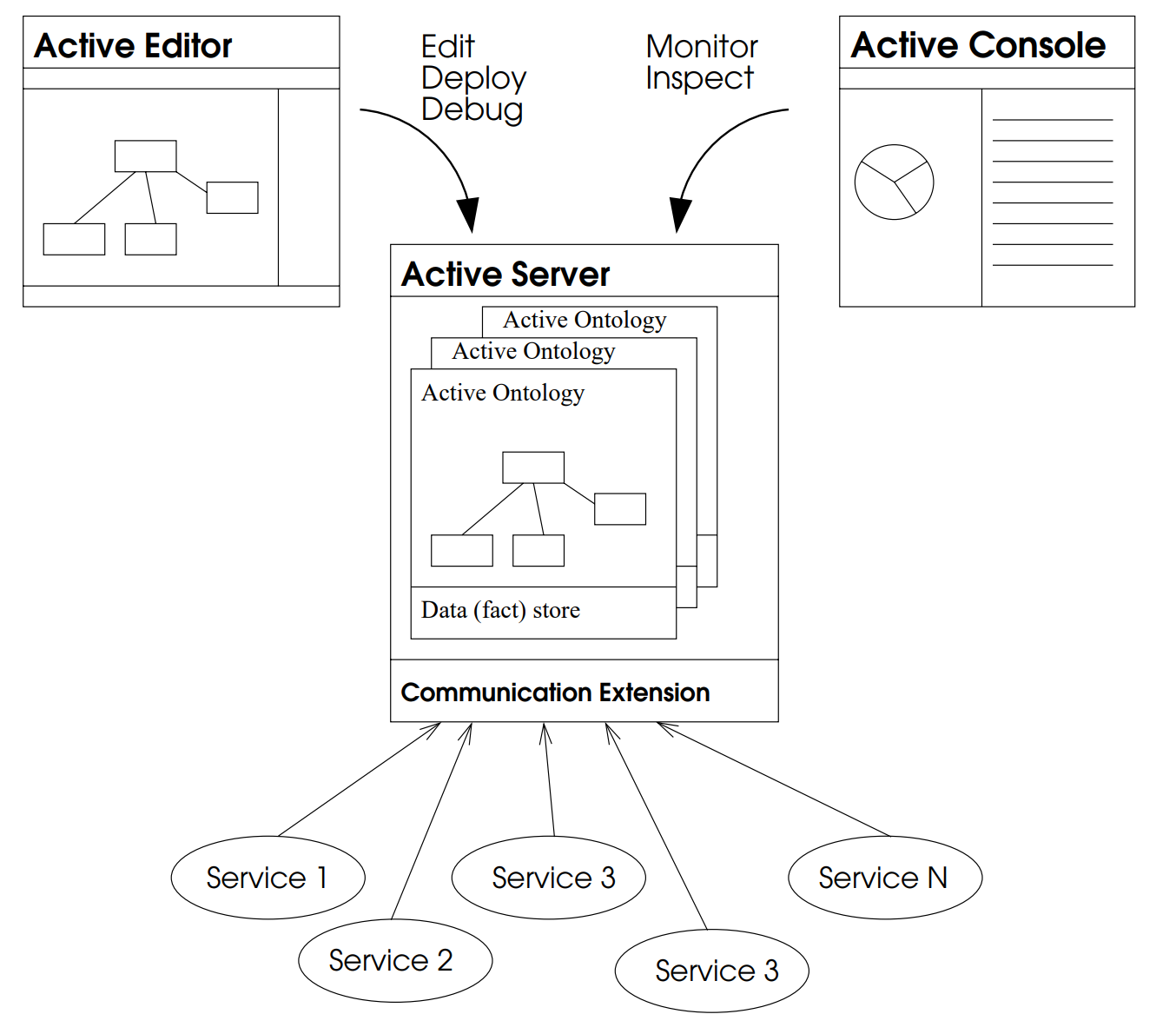}}
\vspace{0mm}
\caption{The research influence on DARPA's PAL program to Siri can be traced to the architecture Guzzoni et al.\ designed to unify the services in PAL (primarily CALO) and enable easy extensibility.  Figure credit: \citet{guzzoni2007modeling}.}
\label{fig:from-pal-to-siri}
\vspace{0mm}
\end{figure}

CALO covered a similar range of functionality and extended to other scenarios such as AI-assisted authoring of documents and QA against a personal store of past experiences (e.g. web pages viewed, etc.). More importantly CALO was designed with more of an extensible architecture of skills in mind. Both CALO's architecture and the learnings from both RADAR / CALO and the more general PAL program helped researchers recognize the need for unifying architectures. The result was a new extensible architecture proposed by \citet{guzzoni2007modeling} and shown in Figure \ref{fig:from-pal-to-siri} with a vision of creating a ``do engine'' rather than a ``search engine''.
\citeauthor{guzzoni2007modeling}'s design first led building a consumer-focused version of CALO at Stanford Research Institute (SRI). One of the authors then moved from SRI to co-found a startup Siri before Apple later acquired that company in 2010\footnote{\url{http://www.morgenthaler.com/press-releases/Siri\%20Named\%20Top\%2010\%20Emerging\%20Tech\%20of\%202009.pdf}}.

Historically, this brings us from simple productivity-focused assistants with more generally device-based assistants.  These assistants typically offer CIR integrated throughout their functionality and often are integrated with personal devices (e.g. mobile phones, situated speakers). In addition to the examples like Apple's Siri, this includes assistants such as Google Assistant, and Amazon Alexa. These voice-based conversational assistants must go beyond the chatbot interactions described above and tackle challenges of voice recognition in noisy environments and increasingly multi-modal and multi-party interactions. As an example Alexa \citep{ram2018conversational, khatri2018advancing, gopalakrishnan2019topical, goel2018flexible, paul2019towards, yavuz2019deepcopy, khatri2018contextual, kollar2018alexa} highlights an interesting trend where a situated speaker interacts with multiple-parties ({e.g.}, all of the members of a household) rather than a single party as is the case with a personal productivity-agent. This has both given rise to an increase in interest in speaker identification as well as other authentication mechanisms to be able to provide personal productivity functionality as well as information-seeking, chit-chat, recommendation, and task-completion. While a full discussion of device-based assistants is beyond the scope of this book, we see these devices not only continuing to become more conversational, but also see an increasing trend in research to customize the experience to different types of users (e.g. kids vs.\ adults) different situations (e.g. theme parks vs.\ offices) and different scenarios (e.g.\ entertainment vs.\ education).


\subsection{Hybrid-Intelligence Assistants}
Finally, we conclude this chapter with a brief discussion of hybrid-intelligence assistants. These assistants typically blend automated systems with crowd-powered systems to provide capabilities that are beyond automated systems' current capabilities but with an overall impression to the user that the entire experience is completely AI-powered. Examples of conversational assistants of this type include Facebook's M and Microsoft's Calendar.help. Facebook's M (now defunct) was an assistant within Facebook's Messenger. During the time that M was available, a user could perform tasks within Messenger through M such as booking a restaurant, buying a gift, etc. External stories indicated that much of M's experience was done with crowd-sourced computation where unhandled cases were escalated to a person who could complete the task -- the dialog and the set of low-level operations the human carried out could then become training data. A similar tactic was taken in Calendar.help \citep{cranshaw2017calendar} which was designed to help people automatically find time to meet without the time-consuming negotiation of finding mutual time availability. Initial versions of Calendar.help were based on a handful of high-precision rules that could do automated extraction from the email of a meeting request with a fall-back to crowd-powered workflows when a general intent to meet was detected and specifics were uncertain. This provided efficient basic functionality of the system and again provided a steady source of natural language training data and ground-truth labels from the crowd-powered workflow. The goal of such training data is to reliably, robustly 
train more general models through a self-sustainable process. While scheduling remains an everyday challenge for many people, the technology in Calendar.help has grown well beyond its original hybrid-intelligence design and has influenced the development of Microsoft's Scheduler which is a product in early preview release as of the date of this writing.

\chapter{Conclusions and Research Trends}
\label{chp:conclusion}

We conclude the book with a brief discussion of research trends and areas for future work. 
While we have colored a promising picture of CIR, the field is still at a research exploration stage in that a commercial CIR system that goes beyond popular search engines (e.g., Google and Bing) to effectively meet users' information needs of wide diversity and complexity is yet to be developed. In the previous chapters, we describe the recent progress of individual CIR modules or sub-systems. But these components are yet to be pieced together to build a system. Many research questions remain open. What is the best system architecture? How to evaluate the effectiveness of the system in real-world settings? How to continually improve the system performance after its deployment? What is a good design for effective and responsible Human-AI Interaction?

The reference architecture of CIR systems, shown in Figure~\ref{fig:cir-system-arch}, is a proposal yet to be empirically verified. It can be viewed as following the schema of \emph{society of mind} \citep{minsky1988society}. The AI system is made of many small modules. Each by itself can only do simple things with little intelligence. But combining them in some special ways can lead to human-like intelligence. The CIR system consists of many modules that are coordinated by a global dialog manager, and are collectively optimized for a long-term reward, such as the increase of the number of active users in three months.

The interactive nature of conversational search makes it challenging to develop robust and reusable evaluation methods for system development. Although off-policy evaluation based on pre-collected datasets is widely used to measure the individual CIR modules, such as query understanding, document retrieval and ranking, question answering, and so on, as described in the previous chapters, it is insufficient to measure how effective a system \emph{interacts} with users to accomplish a search task.  On the other hand, online system evaluation methods based on simulated users and A/B test are too costly to perform frequently, especially at the stage of research exploration. Thus, how to combine the strengths of different evaluation methods to seek the best trade-off between low cost and high reliability is a challenge not only for CIR, but also for all interactive AI systems.

In terms of application-oriented research, our understanding of how to measure and optimize long-term engagement and satisfaction in a variety of settings is evolving. In general chatbot settings, there is a need to continue to define long-term evaluation metrics for engagement that go beyond conversational-turns per session (CPS) and Number of Active Users (NAU) and consider task-completion, goal-support, etc. For example, recommendation technology increasingly is explored in aspirational goal-setting (e.g. ``I want to lose weight'', ``I want to eat healthy'', ``I want to learn more about deep learning'') -- clearly in these settings achieving the goal while creating engagement to get users to return must be considered (presumably being more likely to return also makes the user more likely to achieve their goal). Similarly when conversational agents are primarily providing social companionship, there are theories of linguistic accommodation or coordination \citep{danescu2011mark} which describe human-human interactions and may form a key part of metrics to know when AI-based social companions are performing well.  

Building a conversational AI system, like CIR, also imposes a challenge regarding the development of continual learning methods that allow the system to adapt in a dynamic environment. Most AI models, including the CIR modules described in this book, are developed using independent and identically distributed (IID) learning methods that assume that the world is static, and the data of the tasks that the models need to perform can be pre-collected. However, a conversational system has to work in a non-stationary world, where the tasks and settings change constantly. For example, XiaoIce has to release new dialog skills every week since its first release in 2014 to keep and grow its user base, and many of these skills are short lived \citep{zhou2020design}. Ideally, the system should be able to adapt itself to the environment efficiently with little supervision of new tasks or settings without catastrophic forgetting, where the performance on old tasks degrades sharply when the system adapts to new tasks.  Recent neuroscience studies show that humans can adapt much more efficiently than machines because humans are born with rich innate knowledge inherited from humans’ long evolutionary history and have the ability of transferring knowledge efficiently in two directions in our life-long learning: in forward transfer, previously learned tasks help improve the performance of new tasks, and in backward transfer, learning a new task improves the performance of old tasks \citep{hadsell2020embracing}. Inspired by these neuroscience studies, continual learning with neural networks becomes an active research field \citep[e.g.,][]{parisi2019continual, hadsell2020embracing}.

The learning efficiency can also be significantly improved by making better use of knowledge. \cite{kahneman2011thinking} introduces a metaphor to interpret how humans use knowledge for problem solving. It says that your brain has a \emph{System 1} that uses implicit knowledge for quick, instinctive responses, and a \emph{System 2} that uses explicit knowledge for slower, more thoughtful processing.  We can find many AI systems that work similarly to Systems 1 and 2, respectively. For example, a close-book QA system based on GPT-3~\citep{brown2020language} is like System 1. It assumes that all the knowledge has been encoded in the model parameters, and the system can rely solely on the model to generate answers without accessing any external knowledge, e.g., by retrieving relevant Web documents. An open-book QA system based on a retriever-reader pipeline (Chapter~\ref{chp:c-mrc}) works like System 2. Given an input query, the system first retrieves relevant documents, entities, and other forms of explicit knowledge from document collections or knowledge bases, and then generates the answer from the retrieved knowledge. There have been recent works on integrating explicit and implicit knowledge for open-domain question answering \citep[e.g.,][]{marino2021krisp, gui2021kat}. 

There is also the possibility of accelerating progress within the field with richer paradigms of supervision as well as learning that is less dependent on supervision. Cases of richer supervision include teaching-by-demonstration where the ability to tie language to procedural tasks could unlock potential for conversational systems to help truly bring about the Natural User Interface (NUI). It also includes extending machine teaching to dialog-based interaction \citep{peng2021soloistTACL,gao2020robust}. Reducing the dependency on supervision includes research trends into large neural language models that are providing progress toward few-shot and zero-shot learning in many language-related tasks and increasingly in conversational and search tasks. 

There is the potential for other forms of reasoning and intelligence that humans demonstrate and which apply in conversational intelligence. For example, humans demonstrate social awareness and can interpret attention in settings to understand what was meant by combining an understanding of where the speaker is looking with the words they are saying -- human can then reason if the language applies to a person or an object. Humans demonstrate emotional intelligence in a number of ways; these capabilities may be even more critical in healthcare situation such as general well-being bots or bots that can engage in conversational therapy such as cognitive behavioral therapy \citep{ghandeharioun2019emma}. There are a large number of possibilities to continue to extend capabilities for conversational search and interaction. While there has been major progress over the last decade, we believe that we are still only beginning to see the broad impact that conversation will have on search, question answering, and human-AI interaction more generally.

Within the space of conversational understanding some of the most salient challenges are in grounded language generation and understanding. That is, a conversational system
requires being able to reason about how the language and the state of the world correspond. Challenges in this space range from abstractive summarization where a question answering system would want to avoid \emph{hallucinating} phrases and facts which did not exist in the full document to guide process completion where the agent should have a model of the state of the world relative to the process through which the user is being guided.

For device-based assistants and physically situated assistants ({e.g.,}\ intelligent buildings, elevators, etc.), challenges include integrating observations from multiple modalities including voice, vision, and text. For example in multiple-speaker situations inferring attention to determine which person is interacting with the device versus the people who are interacting with each other. Likewise, increasingly it is important to be able to identify and authenticate people in multi-person scenarios.  Other areas where research is needed is in physical understanding and monitoring that should modulate the behavior of conversational interaction -- Amazon Alexa's \emph{whisper mode} is a simple example of physical understanding. More general cases might involve understanding anomalies (e.g., a kitchen fire), adjusting to ambient noise, or using vision to understand gestures and objects.

In terms of Human-AI Interaction Design there are still quite fundamental challenges for conversational interaction. Two that often come up are: (1) flexibility in specifying goals; and (2) knowing what the system can do. While conversational systems have increased their flexibility by leaps-and-bounds over early systems, we still have not reached systems that provide the flexibility of humans. For example, if a young girl walks up to a librarian and says, ``I am looking for a fun book to read,'' the librarian can quickly narrow in on suitable recommendations. Our current systems have not yet reached this point of flexibility and this is a simpler case of goal specification which we are likely to achieve in the next few years. Turning to the second major challenge within Human-AI Interaction Design, the problem of conveying to a user all of the skills and capabilities of the system is tied together with a more general Human Computer Interaction (HCI) problem termed the gulf-of-evaluation. That is, even when a system has new capabilities it is difficult for users to discover and know these capabilities exist; as a result, the user may never learn the best ways to interact with the system. In an area such as conversational search where capabilities are constantly evolving, finding ways to regularly and efficiently communicate new capabilities to the user is a key part of designing an overall system that works well.

Finally, another key ongoing challenge is around Responsible AI. As we pointed out in Section \ref{sec:responsible-ai}, Responsible AI is a complex and evolving topic. New challenges are evolving on a regular basis and the need to continually evolve new guidelines for Human-AI interaction \citep{amershi2019guidelines} evolves with those challenges. One of the key challenges currently in Responsible AI for conversational systems is the need to defend against harms and mitigate potential toxicity and bias \citep{bhat2021say,breitfeller2019finding,zhang2018conversations} -- either in training data, introduced through feedback and usage, or simply inappropriate due to local culture or context.  In terms of privacy, as device-based assistants are extended to have personal information capabilities, even with authentication, the systems may need to defer disclosing personal information if other parties are present. This is similar to challenges for privacy reported on email notifications when a screen is being shared \citep{kim2019studying}.

\chapter* {Acknowledgments} 

\bibliographystyle{apalike} 
\bibliography{citation, nacai}

\end{document}